\documentclass[aps,prb,twocolumn,notitlepage]{revtex4-2}
\usepackage{listings}
\usepackage{times}
\usepackage{amsmath}
\usepackage{amsfonts}
\usepackage{amssymb}
\usepackage{dsfont}
\usepackage{graphicx}
\usepackage{natbib}
\usepackage{epsfig}
\usepackage{verbatim}
\usepackage{bm}
\usepackage{mathrsfs}
 
\usepackage{yfonts}

\usepackage{cancel}
\usepackage{adjustbox}
\usepackage{pst-all}
\usepackage{leftidx}
\usepackage{mathtools}
\usepackage{tikz}
\usepackage{tikz-cd}
\usetikzlibrary{matrix}
\usetikzlibrary{decorations.markings}
\usepackage{braket}
\usepackage{multirow}
\usepackage[scr=boondox]{mathalpha}

\DeclarePairedDelimiter\floor{\lfloor}{\rfloor}

\def\H{\mathcal{H}}
\def\F{\mathcal{F}}
\def\Z{\mathbb{Z}}
\def\R{\mathbb{R}}
\def\E{\mathbb{E}}

\def\A{\mathcal{A}}

\newcommand{\coho}[1]{\textswab{#1}}

\DeclareMathOperator{\im}{im}

\newtheorem{theorem}{Theorem}[section]

\usepackage[normalem]{ulem}

\begin{document}
	\tikzset{->-/.style={decoration={
				markings,
				mark=at position #1 with {\arrow{>}}},postaction={decorate}}}

\title{Classification of fractional quantum Hall states with spatial symmetries}
\author{Naren Manjunath}
\author{Maissam Barkeshli}
\affiliation{Department of Physics, Condensed Matter Theory Center, and Joint Quantum Institute,
  University of Maryland, College Park, Maryland 20742, USA}

\begin{abstract}
  Fractional quantum Hall (FQH) states and their closely related cousins, quantum spin liquids (QSL), are paradigmatic examples of \it symmetry-enriched topological states \rm (SETs). In addition to the intrinsic topological order, which is robust to arbitrary symmetry-breaking perturbations, they possess \it symmetry-protected topological invariants\rm, such as fractional charge of anyons and fractionally quantized Hall conductivity, which require $U(1)$ charge conservation symmetry.
  In this paper, we develop a comprehensive theory of symmetry-protected topological invariants for FQH states with continuum or crystalline spatial symmetries, which applies to both Abelian and non-Abelian topological states, by using a recently developed framework of $G$-crossed braided tensor categories (BTCs) for SETs. Specifically, we consider clean FQH systems with $U(1)$ charge conservation, magnetic translational, and spatial rotational symmetries, both in the continuum and for all $5$ orientation-preserving crystalline space groups in two spatial dimensions, allowing arbitrary rational magnetic flux per unit cell, and considering the case where symmetries do not permute anyon types. In the continuum, we find that symmetry fractionalization is characterized by the fractional charge and fractional angular momentum of the anyons, while the quantized response theory contains three distinct known invariants: Hall conductivity, shift, and angular momentum of curvature sources. We provide a derivation of the relation between the filling fraction and the Hall conductivity contained entirely within the framework of $G$-crossed BTCs, without relying on Galilean invariance. In the crystalline setting, which also applies to fractional Chern insulators and QSLs, we find that symmetry fractionalization is fully characterized by a generalization to non-Abelian states of the charge, spin, discrete torsion, and area vectors, which specify fractional charge, angular momentum, linear momentum, and fractionalization of the translation algebra for each anyon. The latter two have no analog in the continuum, while the discrete torsion vector is only non-trivial for $2$,$3$, and $4$-fold rotational symmetry. The fractionally quantized response theory contains $9$ terms, which attach, in various topologically protected ways, charge, linear momentum, and angular momentum to magnetic flux, lattice dislocations, disclinations, corners, and units of area. These are characterized by the Hall conductivity, discrete version of the shift, angular momentum of disclinations, fractionally quantized charge and angular momentum polarizations, a quantized torsional response, and charge, angular momentum, and linear momentum filling fractions. We provide a derivation within the $G$-crossed BTC framework of a generalized Lieb-Schulz-Mattis formula relating the charge filling to the Hall conductivity and flux per unit cell. We provide systematic formulas for topological invariants that fully characterize SETs with the above symmetries in terms of the data of the $G$-crossed BTC; this gives, for example, a new definition of the Hall conductivity in terms of $G$-crossed BTC data. We also systematically provide solutions of the $G$-crossed BTC equations for the symmetry groups under consideration. As a byproduct of our analysis, we also derive the classification of (2+1)D symmetry-protected topological (SPT) states for orientation-preserving space groups with $U(1)$ charge conservation symmetry and in the presence of a magnetic field. 
\end{abstract}

\maketitle
\tableofcontents

\section{Introduction}
\label{Sec:Intro}

The fascinating quantized properties of fractional quantum Hall (FQH) systems \cite{Wenbook} are broadly a result of the interplay between their intrinsic topological order and a global symmetry. The intrinsic topological order is characterized by the braiding and fusion properties of topologically non-trivial quasiparticles and is robust to arbitrary perturbations of the system, irrespective of any symmetries \cite{wang2008,nayak2008}. The presence of a global symmetry such as charge conservation or spatial rotational symmetry endows the FQH system with additional \it symmetry-protected \rm topological invariants. The fractionally quantized Hall conductivity and the fractional electric charge of quasiparticles, for example, correspond to quantized topological invariants that are well-defined only in the presence of $U(1)$ charge conservation symmetry. With continuous translational and spatial rotational symmetry, FQH systems also possess a shift and spin vector \cite{Wen1992shift}, which are closely related to a quantized Hall viscosity \cite{Avron1995hvisc,Read2009hvisc}, and which yield an additional set of symmetry-protected invariants that can be used to distinguish FQH states. FQH systems are therefore a paradigmatic example of \it symmetry-enriched \rm topological phases of matter (SETs).

A natural question is to understand the full set of symmetry-protected topological invariants of FQH states, given the full group of global symmetries of the system. In the case of clean, isotropic FQH states in the continuum, the past few years have seen intense study of the coupling of FQH states to continuum geometry  \cite{Cho2014,Gromov2014,Gromov2015,Bradlyn2015,Klevtsov2015}. The quantized geometric response is characterized by a set of quantized symmetry-protected invariants associated with the continuum spatial symmetry of the unpertubed system (e.g. in flat space) combined with $U(1)$ charge conservation. A natural question is whether these studies have found the full set of symmetry-protected topological invariants of clean, continuum FQH states. At the very least, these studies do not provide a complete account of the fractionalization of spatial rotational symmetry in non-Abelian FQH states (i.e. a generalization of the spin vector used in Abelian FQH states to non-Abelian FQH states). 

Moreover, clean FQH states can also arise in lattice systems with crystalline space group symmetries, where they are often referred to as fractional Chern insulators (FCIs). While FCIs have been the subject of intense theoretical study \cite{parameswaran2013,Bergholtz2013FCI}, there has been little work in systematically identifying all possible crystalline symmetry-protected topological invariants. These questions are particularly relevant given the recent experimental realization of FCIs in graphene \cite{Spanton2018FCI}, where experimental observation of non-trivial quantized geometrical responses may be within reach.

FQH states on a lattice can not only possess symmetry-protected invariants that are discrete analogs of those in the continuum, but they can also possess symmetry-protected invariants that have no analog in the continuum \cite{Manjunath2020}. To fully understand this physics, we need to complete a program to systematically identify all possible topological invariants independently for each symmetry group of physical interest. 

Recently, a systematic theory to fully characterize SETs has been developed using $G$-crossed braided tensor categories (BTCs) \cite{Barkeshli2019}. Roughly speaking, $G$-crossed BTCs are defined by a set of data that determines the combined braiding and fusion properties of anyons and symmetry defects. Topological invariants are associated with gauge-invariant combinations of the G-crossed data, while different SETs correspond to gauge inequivalent solutions of the $G$-crossed BTC consistency equations. The main purpose of this paper is to apply the theory of $G$-crossed BTCs to fully characterize and classify FQH states with space group symmetries and to develop a comprehensive understanding of their symmetry-protected topological invariants. 

In this paper, we consider symmetry groups that consist of $U(1)$ charge conservation symmetry, magnetic translational symmetry, and spatial rotational symmetry. More specifically, we consider both the cases of (1) continuum spatial symmetries, where the global symmetry group is a central extension of the Euclidean group by $U(1)$, which includes the magnetic translation algebra corresponding to a non-zero magnetic field, and (2) crystalline space group symmetries, where the global symmetry group is a central extension of an orientation-preserving crystalline space group symmetry by $U(1)$, specified by a fixed magnetic flux per unit cell. 

Our results give a classification of SETs with the above symmetries, and thus are applicable to FQH states and quantum spin liquids \cite{savary2017} with the above symmetries. We mathematically define a FQH state to be any (2+1)D topologically ordered state of matter consistent with the above symmetries, and which has a non-zero fractionally quantized Hall conductivity. As such, our results give a classification of FQH states with the above spatial symmetries \footnote{We do not consider spontaneous symmetry breaking, which is an additional feature of some experimentally observed FQH states.}.

We restrict to the special case where discrete symmetries of the system do not permute topologically distinct quasiparticle types \cite{barkeshli2012,barkeshli2013,Barkeshli2019}, leaving the case with permutations to future work.

In the case of Abelian topological phases of matter with global symmetry $G = U(1) \times G_{\text{space}}$, where $G_{\text{space}}$ is an orientation-preserving crystalline space group symmetry, Ref. \cite{Manjunath2020} recently developed a systematic classification and quantized response theory using crystalline gauge fields and Abelian Chern-Simons theory, assuming symmetries do not permute anyons. Here we generalize these results in two directions. First, our results generalize those of Ref. \cite{Manjunath2020} to the case of non-zero magnetic fields, where $G$ becomes a non-trivial central extension of $G_{\text{space}}$ by $U(1)$. Secondly, our results generalize those of Ref. \cite{Manjunath2020} to the case of non-Abelian topological phases of matter, for which the $G$-crossed BTC is a more direct description of the topological properties of the system than CS gauge theory. 

Our main results are as follows. We provide a classification of SETs with the above symmetries, including the non-commuting nature of magnetic translations, by computing the relevant cohomology groups (see Table \ref{Table:OverallClassif}) and then explicitly classifying and presenting the distinct solutions of the G-crossed BTC consistency equations. We provide general formulas for symmetry-protected topological invariants in terms of gauge-invariant combinations of the $G$-crossed BTC data, for all of the above symmetries (see Tables \ref{Table:ContSymmClassif} and \ref{Table:SpaceGrpClassif}). We further describe the physical meaning of these topological invariants, which lead to both fractional quantum numbers of the quasiparticles and fractionally quantized responses (for a summary see Tables \ref{Table:ContSymmClassif} and \ref{Table:SpaceGrpClassif}).

The fractional quantum numbers of the quasiparticles are characterized in the continuum by two quantities: the fractional electric charge and fractional orbital angular momentum. In the discrete setting, the fractional quantum numbers are characterized by four quantities: the fractional electric charge, fractional orbital angular momentum, fractional linear momentum, and fractionalization of the translation algebra. In particular, our general formulas provide a way to generalize the spin vector, defined previously for Abelian FQH states, to non-Abelian FQH states and also to discrete rotational symmetries. They also allow us to generalize the discrete torsion vector and area vector, introduced in Ref. \cite{Manjunath2020} for crystalline SETs with Abelian topological order, to non-Abelian FQH states and also to the case of non-zero magnetic flux per unit cell.

In the continuum, the fractionally quantized responses are given by the Hall conductivity, shift, and fractional angular momentum of sources of curvature. In the discrete setting, the fractionally quantized responses are given by discrete analogs of the continuum responses, together with several additional responses summarized in Table \ref{Table:SpaceGrpClassif} \cite{Manjunath2020}. These include a discrete analog of the shift, the angular momentum of disclinations, quantized charge and angular momentum polarizations, a quantized torsional response, and quantized charge, angular momentum, and linear momentum per magnetic unit cell. Importantly, our results give category theoretic definitions of nearly all of the topologically invariant responses and fractional quantum numbers, such as the Hall conductivity, which might eventually lead to new ways of extracting these invariants from ground state wave functions \cite{dehghani2020,cian2020}. 

In addition, using the G-crossed BTC framework, we show how continuous magnetic translation symmetry alone can be used to relate the Hall conductivity and the filling fraction, \it without using Galilean invariance. \rm 
Moreover, in the presence of discrete magnetic translation symmetry, we use the G-crossed BTC framework to derive a generalized Lieb-Schultz-Mattis (LSM) formula relating the Hall conductivity to the filling per unit cell, which was previously derived using flux insertion arguments \cite{Lu2017fillingenforced}.

Our general classification results provide an independent, systematic framework to show that the gravitational response theories discussed previously \cite{Wen1992shift,Gromov2014,Gromov2015,Bradlyn2015,Klevtsov2015,Cho2014} are exhaustive, once the symmetry fractionalization class is specified.

In a number of specific cases such as the bosonic Laughlin, Moore-Read \cite{Moore1991}, and Read-Rezayi \cite{read1999} topological orders, we provide an explicit counting of the number of distinct SETs for the case where $G = \Z_M$ appropriate for $M$-fold rotational symmetries and $G = U(1)\leftthreetimes_{\phi} [\Z^2\rtimes \Z_M]$ appropriate to discrete magnetic translations and rotational symmetry (see Table \ref{Table:ZMrelabelings} and \ref{Table:relabelingFullExamples}). We assume that the integer part of the Hall conductivity and the charge and angular momentum filling fractions are fixed. With these assumptions, we find 1024 distinct SET phases with the topological order of the bosonic Moore-Read Pfaffian for the case of a square lattice, where the symmetry group is $G = U(1)\leftthreetimes_{\phi} [\Z^2\rtimes \Z_4]$. For the $SU(2)_3$ Read-Rezayi state, which contains Fibonacci anyons, the same count gives 576 distinct SETs (Table \ref{Table:relabelingFullExamples}). 

When our results are applied to the case of trivial intrinsic topological order, we obtain a classification of symmetry-protected topological (SPT) phases \cite{senthil2015} for each orientation-preserving space group symmetry combined with $U(1)$ charge conservation and arbitrary magnetic flux per unit cell. Such a classification has not been studied explicitly to date.

\subsection{Classification method}

Our method of characterizing and ultimately classifying topological phases of matter with symmetry is based on the mathematical framework of $G$-crossed BTCs \cite{Barkeshli2019}. The idea is to first fix a particular intrinsic topological order, which is described mathematically by a unitary modular tensor category (UMTC) $\mathcal{C}_{{\bf  0}}$. $\mathcal{C}_{{\bf  0}}$ consists of a finite set of topologically distinct anyons together with the algebraic data -- the $F$ and $R$ symbols -- which capture the braiding and fusion properties of the anyons. Given $\mathcal{C}_{{\bf  0}}$ and the symmetry group of the system, $G$, we then consider the properties of the symmetry defects (i.e. the symmetry fluxes) associated with $G$. The braiding and fusion properties of the symmetry defects, and their interplay with the anyons, are captured by a $G$-crossed BTC, denoted $\mathcal{C}_G^\times$. The data of $\mathcal{C}_G^\times$ should be interpreted as a set of algebraic data that characterize the essential algebraic properties of extended operators that create and transport anyons and symmetry defects (see e.g. Ref. \onlinecite{kawagoe2020} for a microscopic definition of $F$ and $R$ symbols of $\mathcal{C}_{\bf 0}$).

Given a particular $\mathcal{C}_{{\bf  0}}$ and $G$, there is a set of inequivalent possible $G$-crossed extensions, $\mathcal{C}_G^\times$, which can be obtained by systematically solving a set of consistency equations for the algebraic data that defines $\mathcal{C}_G^\times$, which we review in Sec. \ref{Sec:ReviewGxShort}.

There are an infinite number of different UMTCs, whose classification is an ongoing research direction related to the classification of rational conformal field theories (see e.g. Ref \cite{rowell2009}). However given a fixed UMTC $\mathcal{C}_{{\bf  0}}$, the classification of the distinct $G$-crossed extensions is a significantly simpler problem. In this paper, we will study the classification of distinct $G$-crossed extensions for a fixed $\mathcal{C}_{{\bf  0}}$ for the symmetry groups $G$ which are of relevance to the FQH problem in the continuum and on the lattice. In some cases, by fixing $\mathcal{C}_{{\bf  0}}$ to correspond to well-established topological orders, such as the Laughlin, Moore-Read, or Read-Rezayi states, we obtain an explicit counting of the number of distinct possible SETs (see Table \ref{Table:relabelingFullExamples} and \ref{Table:ZMrelabelings}).

We note that while the study of FQH states is often advanced through the study of model wave functions, we are interested in the classification of distinct gapped quantum phases of matter, which correspond to equivalence classes of many-body wave functions. Distinct (2+1)D gapped phases of matter with symmetry are distinguished by their topological properties, which are encapsulated in the mathematical framework of $G$-crossed BTCs. Consequently, we do not study model wave functions, but rather the $G$-crossed BTCs with the symmetry group $G$ that is of interest. Obtaining model wave functions for each possible choice of $\mathcal{C}_G^\times$ is an interesting and important problem.

\begin{table*}[t]
	\centering
	\begin{tabular} {l |l |l}
		\hline
		Symmetry, $G$ & Symmetry fractionalization, $\mathcal{H}^2(G, \mathcal{A})$, & Defect classes, $\mathcal{H}^3(G, U(1))$ \\
		
		\hline
		$U(1)$ & $\mathcal{A}$ & $\mathbb{Z}$ \\
		\hline
		$U(1)\leftthreetimes \mathbb{R}^{2}$ & $\A$ & $\Z$ \\     
		\hline
		$U(1) \leftthreetimes \mathbb{E}^2$ & $\mathcal{A} \times \A$ & $\mathbb{Z}^3$ \\
		\hline
		$ U(1)\leftthreetimes_{\phi}\Z^2$ & $\mathcal{A} \times \A$ & $\mathbb{Z}^2$ \\
		\hline
		$U(1) \times \mathbb{Z}_M$ & $\mathcal{A} \times (\mathcal{A}/M\A)$ & $\mathbb{Z} \times \mathbb{Z}_M^2$\\ 
		\hline
		$U(1)\leftthreetimes_{\phi} [\Z^2\rtimes \Z_2]$ & $\mathcal{A} \times \A \times (\A / 2\A) \times (\Z_2^2 \otimes \A)$ & $\mathbb{Z}^2 \times \mathbb{Z}_2^7$ \\
		$U(1)\leftthreetimes_{\phi} [\Z^2\rtimes \Z_3]$ & $\mathcal{A} \times \A \times (\A / 3\A) \times (\Z_3 \otimes \A)$  & $\mathbb{Z}^2 \times \mathbb{Z}_3^5$\\
		$U(1)\leftthreetimes_{\phi} [\Z^2\rtimes \Z_4]$ & $\mathcal{A} \times \A \times (\A / 4\A) \times (\Z_2\otimes \A)$  & $\mathbb{Z}^2 \times \mathbb{Z}_4^3 \times \mathbb{Z}_2^2$\\
		$U(1)\leftthreetimes_{\phi} [\Z^2\rtimes \Z_6]$ & $\mathcal{A} \times \A \times (\A / 6\A) $  &  $\mathbb{Z}^2 \times \mathbb{Z}_6^3$\\
		\hline 
	\end{tabular}
	\caption{Cohomology groups classifying symmetry fractionalization and defect classes for various symmetries relevant for the FQH effect, with $\A$ defined as the Abelian group characterizing fusion of Abelian anyons. $\mathbb{E}^2 = \R^2 \rtimes SO(2)$ is the special Euclidean group consisting of continuous translations and spatial rotations of the plane, while $U(1)\leftthreetimes_{\phi} [\Z^2\rtimes \Z_M]$, for $M = 2,3,4,6$, denotes the group of $U(1)$ charge conservation, magnetic translations (associated to the flux $\phi$ per unit cell) and $M$-fold point group rotations. Note that the classification is the same for every rational value of $\phi$. The notation $G \leftthreetimes H$ denotes a non-trivial central extension of $H$ by $G$. The groups with the $\otimes$ symbol are defined in Eq. \ref{Eq:K_M_Otimes_defn}. } 
	\label{Table:OverallClassif}
\end{table*}

\subsubsection{Applicability of $G$-crossed BTC to spatial symmetries}
\label{Sec:ApptoSpSymm}

An important question is the applicability of the $G$-crossed BTC theory when $G$ contains spatial symmetries. The $G$-crossed BTC can be thought of as prescribing the rules for coupling a (2+1)D topological quantum field theory (TQFT), specified by $\mathcal{C}_{\bf 0}$ and the chiral central charge $c$, to (flat) non-trivial background gauge fields of $G$. That is, by defining the TQFT on (flat) principal $G$ bundles. This corresponds to defining the TQFT with an internal symmetry $G$. This raises the question of whether such a theory is applicable when $G$ is a spatial symmetry of the microscopic system for which the TQFT is a long-wavelength description. It is a well-motivated assertion that spatial symmetries can be treated as internal symmetries when studying the classification of SETs. Below we provide a brief (non-rigorous) discussion of the justification for this (see also Ref. \onlinecite{Manjunath2020} for a similar discussion).

Let us consider a TQFT with an internal symmetry $G_{IR}$, i.e. a TQFT that can be defined on $G_{IR}$ bundles. The full symmetry of such a TQFT is of the form $\mathcal{G}_{IR} = G_{IR} \times \text{Diff}(M)$, where $\text{Diff}(M)$ is the diffeomorphism group of the space-time manifold $M$. Now let $G$ be the full global symmetry of the system of interest. The action of $G$ in the TQFT is via a group homomorphism $\alpha: G \rightarrow G_{IR} \times \text{Diff}(M)$. If ${\bf g} \in G$ is a spatial symmetry such as a spatial translation or rotation, then $\alpha({\bf g})$ restricts to a corresponding isometry group element of $\text{Diff}(M)$. Moreover, $\alpha({\bf g})$ will also generally be non-trivial when restricted to $G_{IR}$. The most general possibility is that $G_{IR} = G$, and $\alpha$ is the identity homomorphism when restricted to $G_{IR}$. Distinct SETs with symmetry group $G$ should therefore correspond to distinct ways of coupling the TQFT to an internal symmetry $G$. We note that this understanding is actually more general than TQFTs, and applies to any system which is described at long wavelengths by a QFT, in which case $\text{Diff}(M)$ is replaced by the space-time symmetry of the QFT. In all known examples, spatial symmetries in a lattice model map in the low energy effective QFT description to a combination of internal symmetries and isometries of the spatial symmetry group of the QFT. See, for example, the action of translation symmetry in spin-1/2 chains and the corresponding action in the Luttinger liquid description \cite{giamarchi2003}.

The expectation that spatial symmetries should be treated as internal symmetries in the classification of topological phases with symmetry was formalized as the ``crystalline equivalence principle'' in Ref. \cite{Thorngren2018}. In the case of symmetry-protected topological states (SPTs), which are symmetric topological phases with no intrinsic topological order, the crystalline equivalence principle has been well-tested by comparing the group cohomology classification with more direct classifications of crystalline SPTs developed in Refs. \cite{Song2017,Huang2017,Song2020}.

Therefore, in this work we assume that SETs can be classified by $G$-crossed BTCs, even when $G$ contains spatial symmetries. A mathematically rigorous formulation of the notion of a gapped phase of matter and a justification that UMTCs and G-crossed BTCs, together with the chiral central charge, can fully characterize and classify gapped phases of matter in (2+1)D is an important open mathematical problem.

\subsubsection{Continuity and finiteness of $G$}

A further complication of the application of $G$-crossed BTCs to our problem is the fact that in the cases we consider in this paper, $G$ is a group extension involving continuous groups, corresponding to $U(1)$ charge conservation, $SO(2)$ spatial rotational symmetry, and $\R^2$ continuous translational symmetry. When studying crystalline space group symmetries, $G$ contains an infinite discrete subgroup, $\Z^2$, corresponding to discrete translational symmetry.

The consistency equations developed in Ref. \cite{Barkeshli2019} apply equally well to such continuous and/or infinite groups. However when $G$ has continuous components, it is not clear what continuity properties to require of the algebraic data of the $G$-crossed BTC. Some natural choices include requiring the algebraic data to be either piecewise continuous or measurable functions of the group elements of $G$ (requiring the data to be continuous or smooth functions of $G$ is too restrictive, as it does not reproduce known results, as seen in the example $G = U(1)$). We will see that for the symmetry groups considered in this paper, both requirements of either piecewise continuous or measurable are equivalent, so that the issue is easily resolved (see Appendix \ref{Sec:Cont_meas}). Presumably this is the case for all physically relevant symmetry groups, although we do not have a completely general proof.  

A second complication is that when $G$ is continuous but not compact, it is \it a priori \rm possible that there could be a continuous parameter family of inequivalent solutions to the $G$-crossed BTC data. This is related to the well-known fact that topological effective actions can sometimes have non-quantized coefficients (such as the theta term in Maxwell theory in even space-time dimensions). While we do not directly run into this problem for the symmetry groups of interest in this work, we expect that any two solutions belonging to the same continuous family should be regarded as defining equivalent SETs. 

We note that while the classification of $G$-graded extensions of fusion categories and $G$-crossed braided extensions of braided fusion categories has been studied in the mathematics literature \cite{ENO2010,etingof2015,jones2020}, the analysis is typically restricted to the case where $G$ is finite \cite{ENO2010,etingof2015} or otherwise $G$ is always treated as a discrete group (i.e. with the discrete topology) \cite{jones2020}.
In the case where symmetries do not permute anyons, which is the case considered in this paper, Ref. \cite{Barkeshli2019} explicitly solved the $G$-crossed BTC consistency equations in full generality, without imposing any finiteness or discreteness requirements on $G$. The result is that distinct $G$-crossed BTCs can be related to each other by elements of the cohomology groups $\mathcal{H}^2(G, \mathcal{A})$ and $\mathcal{H}^3(G, U(1))$. Here $\mathcal{A}$ is the finite Abelian group associated with fusion of the Abelian anyons of $\mathcal{C}_{\bf 0}$. The continuity properties of the algebraic data of the $G$-crossed BTC is then identical to the continuity properties required of the cochains for the stated cohomology groups.

When $G$ is a finite or compact Lie group, one has $\mathcal{H}^3(G, U(1)) \cong \mathcal{H}^4(G, \mathbb{Z})$. We show that for the non-compact symmetry groups we consider in this paper, one can give a precise meaning to $\mathcal{H}^3(G, U(1))$ modulo the continuous part, and that this is isomorphic to $\mathcal{H}^4(G, \Z)$ (see Appendix \ref{H4H3cont}). Therefore $\mathcal{H}^4(G, \Z)$ is the natural group to consider for classifying SETs using $G$-crossed BTCs. (Note also that for measurable (Borel) group cohomology, we have $\mathcal{H}^4(G, \Z) \cong H^4(BG, \Z)$, where $H$ denotes the singular cohomology of $BG$, the classifying space of $G$ \cite{Stasheff1978} ).

Moreover, we show that $\mathcal{H}^3(G, U(1)) \cong \mathcal{H}^4(G, \Z)$ for the non-compact symmetry groups considered in this paper, although our derivation assumes the applicability of the Lyndon-Hochcshild-Serre spectral sequence to measurable group cohomology with continuous coefficients, for which we are unaware of a rigorous mathematical theorem. 

In the main text below we will refer to $\mathcal{H}^3(G, U(1))$ instead of $\mathcal{H}^4(G, \Z)$, as the former is more directly related to the algebraic data of the $G$-crossed BTC. 

Note that other proposals for characterizing and classifying SETs and SPTs have revolved around ``gauging the symmetry'' (i.e. promoting the background gauge field to a dynamical gauge field) and studying the resulting topological order (see eg. Ref. \cite{Levin2012}). Such procedures are in general inapplicable for continuous and/or infinite symmetries (they are also inapplicable for anti-unitary symmetries, which are not considered in this paper) \footnote{For compact, continuous symmetries such as $G = U(1)$, one major problem is that promoting the background gauge field to a dynamical gauge field leads to confinement. In the special case where the response theory for the background gauge field contains a Chern-Simons term, the confinement issue can be avoided, and it may be possible to  develop an algebraic theory of gauging compact continuous $G$. However such a mathematical framework has not to our knowledge been systematically developed.}. In contrast, the $G$-crossed BTC framework of Ref. \onlinecite{Barkeshli2019} is applicable to continuous and/or infinite symmetry groups as well as finite symmetry groups.

\subsubsection{Relation to other classification approaches}

There have been a number of approaches in the past to develop classifications of FQH states and more generally SETs, which we briefly comment on.

A general method to classify SETs is via the projective symmetry group (PSG) approach  \cite{Wen2002QOrders}, which is based on a projective construction of mean field theories (sometimes referred to as parton mean-field theories). The relation between the $G$-crossed BTC approach and the PSG approach was discussed in detail in Ref. \cite{Barkeshli2019}. The $G$-crossed BTC approach directly describes the physical processes that distinguish different SETs -- namely the properties of the anyons and the symmetry defects -- and gives a systematic framework to characterize distinct SETs and obtain a complete set of topological invariants. In contrast, the PSG approach is biased because it depends on a choice of parton decomposition and subsequent mean-field ansatz, of which there are infinitely many; extracting the intrinsic topological order and SET data given a particular mean field ansatz is in general difficult. Moreover, many distinct parton mean-field theories describe equivalent SETs due to hidden dualities, and it is a difficult problem to understand all of these redundancies without passing to a more direct analysis using $G$-crossed BTCs. 

The $G$-crossed BTC framework is also directly related to exactly solvable models, either in (2+1)D for non-chiral topological orders \cite{heinrich2016,cheng2016model} or at the (2+1)D surface of (3+1)D SPTs for general topological orders \cite{Bulmash2020}. However the $G$-crossed BTCs do not directly give model (2+1)D wave functions for chiral topological orders. Therefore the PSG framework will still be useful for constructing variational many-body wave functions for SETs.

A distinct approach to systematically classifying FQH wave functions is in terms of the pattern of zeros \cite{wen2008poz,wen2008poz2,barkeshli2009,barkeshli2010}, vertex algebra \cite{Moore1991,read1999,lu2010vertex}, or Jack polynomials \cite{bernevig2008}. These approaches classify model FQH wave functions that are exact ground states of a certain class of model Hamiltonians, and which can be written as a correlation function of vertex operators in a vertex algebra. As such, they do not directly and systematically classify SETs and their associated symmetry-protected topological invariants. It is also not expected that all SETs for a given intrinsic topological order can be obtained from such an approach. It is an interesting open question to understand to what extent these approaches can describe different SETs.

\subsection{Summary of main results}
\label{Sec:NTSummary}

      To obtain the classification of SETs, we assume that the intrinsic topological order, which is robust in the absence of any symmetries, is fixed. Mathematically, this corresponds to
      a choice of the pair $(\mathcal{C}_{\bf 0}, c)$, where $\mathcal{C}_{\bf 0}$ is the UMTC and $c$ is the chiral central charge. For bosonic systems, $\mathcal{C}_{\bf 0}$ determines $c \text{ mod
      } 8$. For fermionic systems, $\mathcal{C}_{\bf 0}$ is spin modular, and determines $c \text{ mod } 1/2$.

      Our mathematical results are formulated and most complete for bosonic
      systems. As such, in the remainder of this paper we will restrict to bosonic systems. Many of our results carry over to the fermionic case as well, however we leave
      a comprehensive analysis of the fermionic case for future work. 

      Once the intrinsic topological order described by $(\mathcal{C}_{\bf 0}, c)$ is fixed, we carry out the classification of SET phases through the following steps. Note that
      throughout this paper we consider the case where symmetries do not permute anyon types.

\begin{enumerate}
\item Compute symmetry fractionalization classification, $\mathcal{H}^2(G, \mathcal{A})$. This determines the possible fractional charge assignments of the anyons.
  
\item Compute the defect classification, $\mathcal{H}^3(G, U(1))$. This group structure determines how distinct possible braiding and fusion properties of the symmetry defects can be related to each other, once symmetry fractionalization has been fixed. $\mathcal{H}^3(G, U(1))$ is also equivalent to the classification of SPTs, so this freedom can be understood as ``stacking'' an SPT to a given SET to obtain a (possibly) different SET. Alternatively, it can be understood as adding a Dijkgraaf-Witten term \cite{dijkgraaf1990} to the topological action for the background $G$ gauge field. 

\item Determine all possible solutions to the consistency equations of the $G$-crossed BTC, using the results from the computation of $\mathcal{H}^2(G, \mathcal{A})$ and $\mathcal{H}^3(G, U(1))$.
  
\item Determine invariants of the G-crossed braided tensor category that distinguish inequivalent solutions of the $G$-crossed BTC consistency equations. Then compute reduction of $\H^2(G, \A)$ and $\mathcal{H}^3(G, U(1))$. These invariants can be associated with physically meaningful, fractionally quantized responses of the system to symmetry defects. 
\end{enumerate}
The results for steps (1) and (2) are listed in Table \ref{Table:OverallClassif} for the various symmetry groups considered in this paper. The computations summarized in Table \ref{Table:OverallClassif}, which treat the full magnetic translation group combined with spatial rotations, are an important result of this paper and are explained in Appendix \ref{Sec:Coho_calc} using the Lyndon-Hochshild-Serre spectral sequence. 

Importantly, the SET classification is upper bounded by the cardinality of the group $\H^2(G,\A)\times\H^3(G,U(1))$. In some cases, different symmetry fractionalization classes $\H^2(G,\A)$ may be physically equivalent and can be related to each other by relabeling anyons. Furthermore, upon fixing a given symmetry fractionalization class, changing the defect class by an element of $\mathcal{H}^3(G, U(1))$ (i.e. stacking an SPT state) may not change the SET and may instead be accounted for by a relabeling of the symmetry defects. Therefore the correct classification needs to account for these potential equivalences, which reduce the number of distinct states from $|\H^2(G,\A)\times\H^3(G,U(1))|$. This is where the G-crossed BTC is indispensable -- the G-crossed BTC must be used to fully characterize inequivalent SETs and obtain topological invariants that can detect these redundancies. 

Note that in this paper, for the last step of the classification we will focus on the reduction of $\H^3(G,U(1))$. The reduction of $\H^2(G, \mathcal{A})$ is already easy to understand.

Below we will summarize briefly our main results regarding the symmetry-protected topological invariants that appear for various symmetry groups. Our explicit solutions of the $G$-crossed consistency equations for the various symmetry groups of interest are presented in Appendix \ref{Sec:Gxsolns} and will not be reviewed in the following summary.

\begin{table*}[t]
	\centering
	\begin{tabular} {l|l |p{0.3\textwidth} |l|l|l}
          \hline
          \multicolumn{6}{c}{Summary for $G = U(1) \leftthreetimes \E^2$, continuous magnetic translations and rotations} \\
          \hline
           \multicolumn{6}{c}{}\\
          \multicolumn{6}{c}{Symmetry fractionalization, $\H^2(U(1) \leftthreetimes \E^2, \A) = \A \times \A$} \\ \hline
          	Fractional quantum number & Anyon & Description & Required symmetry & Classification & Eq. No. \\
\hline
		$\{e^{2\pi i Q_a} \} = \{M_{a,v}\}$ & $v$ & Anyon induced by $2\pi$ flux insertion, determines fractional charge & $U(1)$ & $\A$ & \ref{Invt_SF_v} \\
		$\{e^{2\pi i L_a} \} = \{M_{a,s}\}$ & $s$ & Anyon induced by $2\pi$ curvature flux insertion, determines fractional angular momentum & $SO(2)$ & $\A$ & \ref{Invt_SF_s_cont} \\
          \hline
        \end{tabular}
    \begin{tabular} {l |l|l}
           \multicolumn{3}{c}{} \\ \hline
          \multicolumn{3}{c}{Defect (SPT) classification, $\H^3( U(1) \leftthreetimes \E^2, U(1) ) = \Z^3$} \\ \hline
         Invariant & Required symmetry & Classification \\
          \hline
		$k_1 = \floor{\bar{\sigma}_H/2} $ & $U(1)$  & $\Z$ \\
		$k_2 = \floor{\mathscr{S}}$ & $U(1) \times SO(2)$ & $\Z$ \\
          $k_3 = \floor{\ell_s/2}$ & $SO(2)$ & $\Z$ \\
          \hline
        \end{tabular}
        \begin{tabular} {l |p{0.5\textwidth} |l|l}
                    \multicolumn{4}{c}{} \\ \hline
          \multicolumn{4}{c}{Fractionally quantized physical responses} \\ \hline
          Label & Physical description & Required symmetry & Eq. No. \\
          \hline
		$\bar{\sigma}_H = Q_v + 2k_1$ & Hall conductivity, determines charge induced by magnetic flux & $U(1)$  & \ref{Invt_D_sigmaH} \\
		$\mathscr{S} = v \star s + k_2$ & Shift, determines charge induced by curvature flux and angular momentum induced by magnetic flux & $U(1) \times SO(2)$ & \ref{Invt_D_Shift_cont} \\
          $\ell_s = L_s + 2k_3$ & Angular momentum induced by inserting curvature flux & $SO(2)$ & \ref{Eq:D-k3} \\ \hline
          $\nu=\bar{\sigma}_H$ & Filling (charge per magnetic unit cell) & $U(1)\leftthreetimes \R^2$ & \ref{Invt_Ex_nu_cont} \\
          \hline
        \end{tabular}

	\caption{Summary of the parameters that completely describe SET phases with continuous magnetic translation and plane rotation symmetry. Charge conservation symmetry is denoted as $U(1)$, while plane rotation symmetry is denoted as $SO(2)$. We have defined the $\star$ product: $\star: \mathcal{A} \times \mathcal{A} \rightarrow [0,1)$, such that $M_{a,b} = e^{2\pi i a \star b}$, where $a \star b \in [0,1)$. Recall also that $Q_v = 2 h_v \in [0,2)$ and $L_s = 2 h_s \in [0,2)$ are defined modulo $2$. The 'Eq. No.' column references the equation number for the $G$-crossed invariant corresponding to each parameter.}\label{Table:ContSummary}\label{Table:ContSymmClassif}
\end{table*}

\subsubsection{Symmetries under consideration}

We are interested in orientation-preserving space group symmetries in the presence of a magnetic field. Before discussing the topological invariants for different symmetries, we will make some remarks on the definition of these symmetry groups. A more extensive discussion is given in Appendix \ref{Sec:FQHSymm}. 

In the continuum, the magnetic translation group consists of the operators $\hat{T}_{{\bf r}_1}$, $\hat{T}_{{\bf r}_2}$, and $U_c(\theta)$ with the relations
\begin{align}
\hat{T}_{{\bf r}_1} \hat{T}_{{\bf r}_2} = U_c\left( \frac{{\bf r}_1 \times {\bf r}_2}{2l_B^2} \right) \hat{T}_{{\bf r}_1 + {\bf r}_2}
  \end{align}
  Here $U_c(\theta)$ is the symmetry operator corresponding to a $U(1)$ rotation by an angle $\theta$. For a many-body quantum system, this is represented as 
  $U_c(\theta) = e^{i \theta \hat{N}}$ where $\hat{N}$ is the total number operator. The symmetry group is therefore a central extension of the continuous translation group, $\R^2$, by $U(1)$. We denote this magnetic translation group by $U(1) \leftthreetimes_{l_B} \R^2$. Note that by rescaling space, we see that groups with different values of $l_B$ can be related to each other; therefore we drop the subscript $l_B$ and denote the magnetic translation group as $U(1) \leftthreetimes \R^2$. Central extensions labeled by different $l_B$ are all isomorphic to each other, so as far as the symmetry group is concerned, no information is lost by dropping the subscript. 

With continuous spatial rotations, we replace $\R^2$ with the Euclidean group $\E^2 = \R^2 \rtimes SO(2)$. In the presence of a magnetic field, we then consider the central extension $U(1) \leftthreetimes \E^2$.

  The discrete case is similar. The discrete magnetic translations with a flux $\phi$ per unit cell are given by
  \begin{align}
\hat{T}_{{\bf r}_1} \hat{T}_{{\bf r}_2} = U_c\left( \phi \frac{{\bf r}_1 \times {\bf r}_2}{2} \right) \hat{T}_{{\bf r}_1 + {\bf r}_2},
  \end{align}
  where now ${\bf r}_1$ and ${\bf r}_2$ are discrete lattice vectors. Note that in this case since the lattice vectors are discrete, we cannot rescale space, so we keep the subscript $\phi$ in $U(1) \leftthreetimes_\phi G_{\text{space}}$. 

We note that the full space group symmetry we have considered is a global symmetry of an infinite plane, but not that of closed surfaces. However we note that in the thermodynamic limit, we can define whether a system on a closed surface possesses these space group symmetries by comparing the reduced ground state density matrices $\rho_{\mathcal{R}}$ and $\rho_{\mathcal{R}'}$ for patches $\mathcal{R}$ and $\mathcal{R}'$ related to each other by an element of the space group (see Appendix \ref{Sec:FQHSymm}). 
  
\subsubsection{General results on $G$-crossed invariants}
\label{Sec:Relabel_disc}

Explicit forms of the $G$-crossed invariants studied in this work are given throughout Section \ref{Sec:SummaryOfResults} and these results are summarized in Tables \ref{Table:InvtSummary} and \ref{Table:LatticeInvts}. The general method used to obtain these invariants is summarized in Section III A. The derivations of the general $G$-crossed identities involved are given in Appendix \ref{Sec:SymFracInvt} and \ref{Sec:DefectInvts}. These results are applied to obtain specific invariants for different symmetry groups in Appendix \ref{Sec:Invts}. Here we will present the main takeaways from Section \ref{Sec:GenResInvts} without any explicit formulas.

The first general result is regarding invariants for symmetry fractionalization, which fixes the $\mathcal{H}^2(G,\mathcal{A})$ freedom in defining a $G$-crossed extension $\mathcal{C}_G^\times$ given a UMTC $\mathcal{C}_{\bf 0}$. In general, the symmetry fractionalization class is specified by a set of $d$ Abelian anyons $f_1,f_2,\dots f_d$, subject to a set of equivalence relations (the value of $d$ depends on $G$). The invariants for the symmetry fractionalization class give the quantities $M_{f_i,a}$ in terms of the $G$-crossed data (see Eq. \ref{FaDef} - \ref{fDef}). $M_{f_i,a}$ is the phase obtained by a full braid between an arbitrary (possibly non-Abelian) anyon $a$ and $f_i$. Physically, the computation involves the insertion of a quantum of symmetry flux, and the invariant computes the braiding phase between $a$ and the anyon $f_i$ associated to the symmetry flux quantum. We believe that this procedure can be used to find symmetry fractionalization invariants for arbitrary $G$. 

The second general result is regarding formulas in terms of the $G$-crossed data for topological invariants that physically describe fractionally quantized responses.
These invariants fix the remaining $\mathcal{H}^3(G, U(1))$ freedom in fixing $\mathcal{C}_G^\times$, thus fixing the symmetry defect fusion and braiding properties, once symmetry fractionalization is fixed. Physically, these invariants measure the fractional quantum numbers of the symmetry defects, which defines the response theory. One of our general results shows how to obtain such invariants associated to a $U(1)$ or $\Z_m$ subgroup of $G$. Such invariants can always be obtained in terms of a formula involving the $G$-crossed modular $T$-matrix, defined in Ref. \cite{Barkeshli2019} (see Eq. \ref{InDef},\ref{InCharge},\ref{PureSPTInvtFormula}). Simple variations on this formula also give the mixed defect invariants (i.e. mixed responses) for $\Z_m \times \Z_n$ symmetries (see Eq. \ref{Eq:MixedTInvt},\ref{InQgh}). This approach allows us to completely characterize the defect response for the examples in this paper, apart from the $U(1)$ charge, linear and angular momentum per magnetic unit cell.

A separate formula (see Eq. \ref{nuHDef}) allows us to determine the charge filling per magnetic unit cell $\nu$, in both continuum and discrete FQH systems. Although they are well-motivated by group cohomology and topological field theory, we have not been able to find $G$-crossed invariants for the linear and angular momentum per magnetic unit cell; this problem is left for future work.   

We note that the $G$-crossed BTC contains a number of ambiguities: (1) gauge transformations of the $G$-crossed data, referred to as vertex basis and symmetry action gauge transformations, and (2) relabelings of the topologically distinct anyons and symmetry defects. Let us refer to expressions that are invariant under (1) as gauge-invariant quantities of the $G$-crossed BTC data. Expressions that are invariant under both (1) and (2) will be referred to as absolute invariants. In general, it is possible to write down gauge-invariant quantities, and these are mainly what we present for all symmetry groups. However writing down absolute invariants does not appear to be possible for generic symmetry groups; rather, in general the best we can do is list a collection of gauge-invariant quantities subject to possible equivalences arising from relabeling anyons and symmetry defects. For specific types of symmetry groups, such as $U(1)$ or $\Z_M$, it is possible to write down absolute invariants, however in these cases this can only be done with precise knowledge of the group $\mathcal{A}$ and the braiding statistics of $\mathcal{C}_{\bf 0}$. In this paper, by ``invariant'' we generally refer to gauge-invariant quantities of the $G$-crossed BTC data. Furthermore, when we use the terminology ``physical responses,'' we generally refer to the quantum numbers of defects defined in terms of these gauge-invariant quantities. These responses are not generally absolute invariants when the relevant symmetry is discrete, because the quantum numbers of defects are ambiguous up to the quantum numbers of anyons that can be attached to those defects, which corresponds to the ambiguity under relabeling symmetry defects.

\subsubsection{$G = U(1)$, charge conservation}

For $G=U(1)$, the topological invariants are fully specified as follows.

The classification of symmetry fractionalization is given by
\begin{align}
\H^2(U(1), \A) = \A.
\end{align}
  This class is fully determined by a choice of anyon $v \in \A$, referred to as the vison (sometimes referred to as the fluxon), which is the Abelian anyon associated to adiabatic $2\pi$ flux insertion. $v$ determines the fractional $U(1)$ charge $Q_a$ of each anyon $a$ (including non-Abelian anyons),
\begin{align}
  e^{i 2\pi Q_a} = M_{v,a},
\end{align}
where $M_{v,a}$ is the phase obtained by a full braid between $v$ and $a$. We can obtain $v$ uniquely by measuring the full set of fractional charges $\{e^{2\pi i Q_a}\}$ and knowing the braiding statistics. A formula for $e^{2\pi i Q_a}$ in terms of the data of the $G$-crossed BTC is given in Eq. \ref{Invt_SF_v} and which we reproduce here:
\begin{align}
e^{2\pi i Q_a} =  \left(R^{0_{{\bf  g}},a}R^{a,0_{{\bf g}}}\right)^p\prod_{j=1}^{p-1} \eta_a( {\bf g},{\bf g}^j ) 
\end{align}
Here $R$ and $\eta$ are part of the data of the $G$-crossed BTC, reviewed in Sec. \ref{Sec:ReviewGxShort}, and $0_{\bf g}$ labels a choice of (Abelian) ${\bf g}$-symmetry defect. 
We have defined ${\bf g} \in U(1)$ such that ${\bf g}^p = 1$ and $p = \text{lcm}(n_1, \cdots, n_r)$, assuming the fusion of Abelian anyons forms the group $\mathbb{Z}_{n_1} \times \dots\times \mathbb{Z}_{n_r}$. 

The defect classification is given by
\begin{align}
\H^3(U(1), U(1)) = \Z.
\end{align}

The fractionally quantized $U(1)$ response is the Hall conductivity, $\sigma_H$. In units where $e = \hbar = 1$, we define
\begin{align}
\bar{\sigma}_H = 2\pi \sigma_H.
\end{align}
In what follows, we will refer to $\bar{\sigma}_H$ as simply the Hall conductivity. 

The Hall conductivity $\bar{\sigma}_H$ changes by an even integer when we change the defect class by the generator of $\H^3(U(1), U(1)) = \Z$.
Therefore the even integer part of $\bar{\sigma}_H$ fixes the defect class. Since no two values of $\bar{\sigma}_H$ are physically equivalent, there is no reduction in the defect classification in this case.

The fractional part of the Hall conductivity is given by the charge of the vison,
\begin{align}
\bar{\sigma}_H = 2(h_v + k_1) = Q_v + 2k_1,
\end{align}
where $k_1$ is the integer part that is determined by the $\H^3(U(1), U(1))$ freedom, and we have defined $e^{2\pi i h_v} = \theta_v$, with $h_v \in [0,1)$, where $\theta_v$ is the topological twist of the anyon $v$. In this notation, $e^{2\pi i Q_v} = M_{v,v} = \theta_v^2 = e^{4\pi i h_v}$. We then define
\begin{align}
Q_v = 2h_v \in [0,2),
\end{align}
so that $Q_v$ is well-defined modulo $2$. 
  
One of the results of this paper is an explicit formula for the Hall conductivity in terms of the data of the $G$-crossed BTC (see Eq. \ref{Invt_D_sigmaH} - \ref{I2n}), which we reproduce below:
\begin{align}
e^{2\pi i \bar{\sigma}_H/(2n)} = \theta_{0_{\bf g}}^n \prod_{j=0}^{n-1} \eta_{0_{\bf g}}({\bf g}, {\bf g}^j) .
\end{align}
$\theta$ and $\eta$ are part of the defining data of the $G$-crossed BTC that we review in Sec. \ref{Sec:ReviewGxShort}, $n$ is any integer, and ${\bf g} \in U(1)$ has order $n$: ${\bf g}^n = 1$. $0_{\bf g}$ here is a ${\bf g}$-defect that is continuously connected to the trivial excitation (by adiabatically turning on the $U(1)$ flux), which is well-defined for large enough $n$. This gives a new categorical definition of the Hall conductivity in any topological order with $U(1)$ symmetry. 

\subsubsection{$G = U(1) \leftthreetimes \R^2$, continuous magnetic translational symmetry}

Here we consider the case of a clean FQH system, where we have $U(1)$ charge conservation together with continuous translation symmetry in two dimensions. The presence of a magnetic field implies that we should consider magnetic translations, which do not commute. Thus the group is a central extension of $\R^2$ by $U(1)$, which we denote as $G = U(1) \leftthreetimes \R^2$. We do not assume spatial rotational symmetry for this example.

The $G$-crossed invariants for continuum translational and rotational symmetries of the FQH system are summarized in Table \ref{Table:InvtSummary}. In this case, the presence of the continuous translations does not change the SET classification. Our computations in Appendix \ref{Sec:Coho_calc_cont} show that
\begin{align}
  \H^2( U(1) \leftthreetimes \R^2 , \A) &= \A
\nonumber \\
  \H^3( U(1) \leftthreetimes \R^2 , U(1)) &= \Z.
  \end{align}

  The presence of translation symmetry implies that the system now has a uniform $U(1)$ charge density set by the filling fraction $\nu$, which can be interpreted as the charge per magnetic unit cell. We derive a formula for $\nu$ in terms of the data of the $G$-crossed BTC (see Eq. \ref{Invt_Ex_nu_cont}), which we reproduce below:
  \begin{align}
  e^{2\pi i \nu/n} = \frac{\eta_{0_{{\bf g}}}({\bf x},{\bf y})}{\eta_{0_{{\bf g}}}({\bf y},{\bf x})},
  \end{align}
  where $n$ is any integer, ${\bf g}$ is a pure $U(1)$ rotation such that ${\bf g}^n = 1$ and ${\bf x}, {\bf y} \in \R^2$ are pure translations that span a magnetic unit cell. As above, $0_{\bf g}$ is a ${\bf g}$-symmetry defect that is continuously connected to the trivial excitation by adiabatically turning on the $U(1)$ flux. 

Moreover, the $G$-crossed BTC framework can be used to prove the identity
\begin{align}
  \label{HallcondFilling}
e^{2\pi i \bar{\sigma}_H} = e^{2\pi i \nu} 
\end{align}
Using the $G$-crossed formalism, in all cases where we have obtained particular explicit solutions to the $G$-crossed BTC equations (see Appendix \ref{Sec:Gxsolns}),
we can verify the stronger result that
\begin{align}
   \label{HallcondFilling2}
\bar{\sigma}_H = \nu .
\end{align}
Note that this is a rather remarkable result, because it does not rely on Galilean invariance. The usual argument for Eq. \ref{HallcondFilling},\ref{HallcondFilling2} requires Galilean invariance \cite{girvin1999quantum}, which is not a conventional global symmetry of a quantum system because it is a space-time symmetry; to our knowledge such a relation has not been established without assuming Galilean invariance. 

\subsubsection{$G = U(1) \leftthreetimes \mathbb{E}^2$, continuous magnetic translational and rotational symmetry}

Next we consider the Euclidean group $\mathbb{E}^2 = \R^2 \rtimes SO(2)$. In the presence of a magnetic field, the symmetry group of the system is a central extension of the Euclidean group by $U(1)$, corresponding to the continuum magnetic translation algebra reviewed in Appendix \ref{Sec:FQHSymm}.  

The presence of the spatial rotational symmetry $SO(2)$ now adds additional symmetry fractionalization and defect classes.

In this case we have
\begin{align}
  \H^2( U(1) \leftthreetimes \mathbb{E}^2 , \A) = \A \times \A.
  \end{align}
  The first factor is from the $U(1)$ charge fractionalization, while the second factor is due to fractionalization of the $SO(2)$ spatial rotational symmetry, which corresponds to a fractional orbital angular momentum. These symmetry fractionalization classes are therefore specified by two Abelian anyons
  \begin{align}
    (v, s) \in \A \times \A,
    \end{align}
  where $v$ is the vison discussed in the $U(1)$ case. 

The anyon $s$ is more subtle. Within the $G$-crossed BTC formalism, $s$ is the anyon obtained under insertion of a unit flux of the $SO(2)$ symmetry. However from the point of view of the microscopic theory, $2\pi$ curvature flux is not trivial and in fact changes the topology of the manifold according to the Gauss-Bonnet theorem. This is related to the issue discussed in Sec. \ref{Sec:ApptoSpSymm}, where the $G$-crossed BTC is characterizing different ways of coupling the intrinsic topological order to an internal symmetry of the TQFT. The spatial symmetries of the microscopic system map to a combination of internal symmetries of the TQFT and spatial isometries of the spatial manifold on which the TQFT is defined.

The choice of $s$ determines the fractional orbital spin of the anyons:
\begin{align}
e^{i 2\pi L_a} = M_{s,a},
\end{align}
where $L_a$ is the fractional orbital spin. Similar to the case of the fractional charge $Q_a$, we provide an explicit expression for $e^{2\pi i L_a}$ in terms of the data of the $G$-crossed BTC in Eq. \ref{Invt_SF_s_cont}. 
The total spin of an anyon can be defined by the Aharonov-Bohm phase acquired by adiabatically transporting an anyon $a$ around a curvature angle $\Omega$:
\begin{align}
\gamma_{AB}(\Omega) = e^{i S_a \Omega} . 
\end{align}

The total spin $S_a$ is given by (see also Refs. \cite{gromov2016,EINARSSON1995})
\begin{align} \label{Eq:totalspin_anyon}
e^{2\pi i S_a} = \theta_a e^{2\pi i L_a} .
\end{align}
Here $\theta_a$ is the topological twist of the anyon $a$, which is a property of the UMTC defining the intrinsic topological order. The additional contribution $L_a$ thus arises from fractionalization of the spatial rotational symmetry. 

The defect classification is given by (see Appendix \ref{Sec:Coho_calc_cont})
\begin{align}
\H^3(  U(1) \leftthreetimes \mathbb{E}^2  ,U(1)) \cong \Z^3. 
\end{align}
The defect classification determines the integer part of various physical quantized responses. In particular, we have three distinct fractionally quantized physical responses: the Hall conductivity $\sigma_H$, the shift $\mathscr{S}$, and $\ell_s$. These correspond to independent gauge invariant quantities in the $G$-crossed BTC. Explicit formulas for $\mathscr{S}$ and $\ell_s$ in terms of the data of the $G$-crossed BTC are given in Eq. \ref{Invt_D_Shift_cont} and Eq. \ref{Invt_D_ell_s} (see also Table \ref{Table:ContSummary}). 

They can also be understood in terms of the well-known effective response theory of FQH states on curved space:
\begin{align}
\mathcal{L}_{\text{response}} = \frac{\sigma_H}{2} A d A + \frac{\mathscr{S}}{2\pi} \omega d A + \frac{\ell_s}{4\pi} \omega d \omega + \mathcal{L}_{\text{anom}} ,
\end{align}
where $A$ is the background gauge field for the $U(1)$ charge conservation symmetry, and $\omega$ is a background $SO(2) \cong U(1)$ gauge field corresponding to the spatial component of the spin connection. 

To express these quantized responses in terms of the fractional quantum numbers, we define the $\star$ product:
\begin{align}
  \star: \mathcal{A} \times \mathcal{A} \rightarrow [0,1),
  \nonumber \\
e^{2\pi i a \star b} = M_{a,b}.
\end{align}
We then have:
\begin{align}
\mathscr{S} = v \star s + k_2 ,
\end{align}
where $k_2 \in \Z$. The integer $k_2$ can be chosen arbitrarily, and contributes one of the $\Z$ factors in the defect classification.

The shift of a FQH state defined on a closed surface of Euler characteristic $\chi$ with $N_\Phi$ flux quanta uniformly piercing the surface and $N$ particles is defined by the relation
\begin{align}
N = \bar{\sigma}_H N_{\Phi} + \mathscr{S} \chi . 
\end{align}
Note that the quantity usually referred to as the shift in the FQH literature is defined by $\mathcal{S} = 2 \mathscr{S} / \bar{\sigma}_H$, so that $N_\Phi = \bar{\sigma}_H^{-1} N - \mathcal{S} \chi/2$. Here we instead refer to $\mathscr{S}$ as the shift, which is well-defined even when $\bar{\sigma}_H$ vanishes. 

$\mathcal{L}_{\text{anom}}$ is a gravitational CS term proportional to the chiral central charge $c$, which arises from the framing anomaly of CS theory \cite{Witten1989}. Setting all but the spatial component of the spin connection to zero, $\mathcal{L}_{\text{anom}} = -\frac{c}{48\pi} \omega d\omega$. $\ell_s$, together with the chiral central charge $c$, determines the angular momentum of a curvature flux \cite{Klevtsov2015}.

$\ell_s$ can be determined in terms of the angular momentum $L_s$ of $s$:
\begin{align}
\ell_s = L_s + 2 k_3,
\end{align}
where $L_s = 2h_s \in [0,2)$ is defined modulo $2$, since we can define $h_s$ modulo $1$ via $\theta_s = e^{2\pi i h_s}$. 

The three independent integer-valued invariants associated with $\H^3(  U(1) \leftthreetimes \mathbb{E}^2  ,U(1)) \cong \Z^3$ correspond to the integer parts of the fractionally quantized responses, $\bar{\sigma}_H/2$, $\mathscr{S}$, and $\ell_s/2$, as shown in Table \ref{Table:ContSummary}. 

It is important to note that there can be additional invariants which are not independent due to constraints on the various parameters specifying the FQH state. For example, in a system with continuous translation symmetry we can also define a quantized filling per magnetic unit cell; this is however constrained by symmetry to equal the Hall conductivity.

\subsubsection{$G = U(1) \leftthreetimes_\phi \Z^2$, discrete magnetic translational symmetry}

Systems with discrete magnetic translation symmetry are associated to a flux per unit cell given by $\phi = p/q$, and a filling, which we define as the $U(1)$ charge per magnetic unit cell, $\nu$. Note that the charge per unit cell, which is often also called the filling, is denoted here by $\nu/q$.

The symmetry fractionalization is given by
\begin{align}
\H^2(U(1) \leftthreetimes_\phi \Z^2, \A) = \A \times \A . 
\end{align}
Therefore, the symmetry fractionalization is completely characterized by two Abelian anyons,
\begin{align}
(v, m) \in \A \times \A. 
\end{align}
$v$ is the vison which determines the fractional $U(1)$ charge. $m$ is the anyon per unit cell \cite{Jalabert1991,sachdev1999translational,Essin2013SF,Essin2014spect,Cheng2016,Sachdev_2018}, which has no analog in the case with continuous symmetries.
Adiabatically transporting an anyon $a$ around a unit cell gives rise to a phase
\begin{align}
e^{2\pi i \tau_a} = M_{a,m} ,
\end{align}
which can be understood as the fractionalization of the translation algebra. 

The defect classification is
\begin{align}
\H^3( U(1) \leftthreetimes_\phi \Z^2 , U(1)) \cong \H^4(U(1) \leftthreetimes_\phi \Z^2 ,\Z) \cong  \Z^2. 
\end{align}
These correspond to the integer parts of the physical responses, as follows.

The fractionally quantized physical responses of the system are fully characterized by the Hall conductivity $\bar{\sigma}_H$ and the charge per magnetic unit cell, $\nu$. Note that the filling can be interpreted as a response, in the sense that the total charge changes when the number of magnetic unit cells is changed. As discussed previously, we can write the Hall conductivity as given in Eq. \ref{Invt_D_sigmaH}. From an effective response theory written using crystalline gauge fields, the filling $\nu$ can be read off in terms of $\bar{\sigma}_H$ as follows (see Table \ref{Table:SpaceGrpClassif} and Sec \ref{Sec:MagTransSumm})
\begin{align}
  \label{genLSM1}
  \nu &= p (Q_v + 2k_1) + q (Q_m + k_6)
  \nonumber \\
  &= q(\phi \bar{\sigma}_H + Q_m+ k_6) 
\end{align}
Note that, importantly, we define $Q_v = 2h_v \in [0,2)$, but $Q_m = v \star m \in [0,1)$ since we can only define $Q_m$ through $e^{2\pi i Q_m} = M_{v,m}$. 
An explicit formula for $\nu$ given the data of the $G$-crossed theory is given in Eq. \ref{Invt_D_nuC}, which we reproduce below:
\begin{align}
 e^{\frac{2\pi i }{n}\nu} = \frac{\eta_{0_{\bf g}} ({\bf r_1},{\bf r_2})}{\eta_{0_{\bf g}}({\bf r_2},{\bf r_1})}  ,
\end{align}
where ${\bf r_1}$ and ${\bf r_2}$ correspond to pure translation group elements that span a magnetic unit cell and ${\bf g}$ is a pure $U(1)$ rotation with ${\bf g}^n = {\bf 0}$. $0_{\bf g}$ is the symmetry defect continuously connected to the trivial particle, which is well-defined for large enough $n$.

The defect classes classified by $\H^3( U(1) \leftthreetimes_\phi \Z^2 , U(1)) = \Z^2$ are therefore characterized by two integer invariants $k_1$ and $k_6$, corresponding to the integer parts:
\begin{align}
  k_1 &= \floor{\bar{\sigma}_H/2},
  \nonumber \\
  2 p k_1 + q k_6 &= \nu - (p Q_v + q Q_m) .
\end{align}
The contribution $(p Q_v + q Q_m)$ can be understood as the contribution to the filling arising from symmetry fractionalization. 
Changing $k_1 \rightarrow k_1 + 1$ corresponds to stacking with a (bosonic) IQH state, which changes the Hall conductivity by an even integer.
Changing $k_6 \rightarrow k_6 + 1$ corresponds to changing the charge per unit cell by 1; this means that the charge per magnetic unit cell $\nu$ is shifted by $q$.
The total contribution to the charge per magnetic unit cell from the parameters $k_1$ and $k_6$ equals $2p k_1 + q k_6 = q(2\phi k_1 + k_6)$.

The fractional part of $\nu$ is therefore determined by a generalized LSM type relation, 
\begin{align}
  \label{discreteLSM}
e^{2\pi i \nu} = e^{2\pi i q (Q_m + \phi \bar{\sigma}_H)} .
\end{align}
Eq. \ref{discreteLSM} was derived in Ref. \cite{Lu2017fillingenforced} and also in Ref. \cite{Matsugatani2018} assuming an additional rotation symmetry. Here we rederive this relation in a different way, entirely within the $G$-crossed BTC framework. Whenever we can write down explicit solutions to the $G$-crossed BTC equations, we can verify the stronger result, Eq. \ref{genLSM1}. Although expected from crystalline gauge theory, to our knowledge, this stronger result has not been stated or rigorously proven in previous work. Obtaining a completely general proof of this result within the $G$-crossed BTC formalism is an interesting problem which we leave for future work.

\subsubsection{$G = \Z_M$, discrete rotational symmetry}

Next we temporarily drop the charge conservation symmetry and consider $G=\Z_M$. This can correspond to an internal symmetry with $M$ arbitrary, or to a rotation point group symmetry with $M = 2,3,4,6$. It is useful to study this case because it helps us isolate which topological invariants can be associated to purely discrete rotational symmetry alone, as opposed to mixed invariants that rely on rotational symmetry together with other symmetries.

When the $\Z_M$ symmetry corresponds in the microscopic system to a discrete rotational symmetry of a lattice, the elementary $\Z_M$ flux physically corresponds to an elementary disclination of angle $2\pi/M$, while unit $\Z_M$ charge corresponds to a unit angular momentum. 

The symmetry fractionalization classification is given by
\begin{align}
\H^2(\Z_M, \A) \cong \A/M\A . 
\end{align}
In other words, the symmetry fractionalization is specified by an equivalence class $[s] \in \A / M\A$, where $s \in \A$ is a representative anyon. $s$ can be thought of as the Abelian anyon obtained by inserting $M$ elementary $\Z_M$ fluxes. However if $s = s'^M$ for some $s' \in \A$, then the symmetry fractionalization class is trivial because it can be completely accounted for by attaching the anyon $s'$ to each elementary $\Z_M$ flux, which can be done by adjusting local energetics. Therefore, the anyons $s$ and $s s'^M$ specify the same fractionalization class. Consequently the non-trivial symmetry fractionalization classes are classified by $\A / M \A$, where $M\A := \{s'^M, s' \in \A\}$. For $\A = \prod_{i=1}^k \Z_{n_i}$, we have $\A/\A^M = \prod_{i=1}^k \Z_{(M,n_i)}$, where $(n,m) := \text{gcd}(n,m)$. 

The symmetry fractionalization implies that each anyon carries a fractional $\Z_M$ charge $L_a$, where 
\begin{align}
e^{2\pi i L_a} = M_{a,s} . 
\end{align}
Note that different fractional values of $L_a$ may actually be physically equivalent since $s \sim s s'^M$. Therefore, the invariants correspond to the set $\{e^{2\pi i L_a}\}$, with the equivalence
\begin{align}
\{e^{2\pi i L_a}\} \sim \{e^{2\pi i L_a} M_{a,s'^M} \}, \;\; \forall s' \in \A
\end{align}
An explicit formula for $e^{2\pi i L_a}$ in terms of the $G$-crossed data is given in Eq. \ref{Invt_SF_s_disc}.

In the case where $\Z_M$ is a discrete rotational symmetry, $L_a$ corresponds to a fractional angular momentum. The total Aharonov-Bohm phase $\gamma_{AB}$ obtained by an anyon $a$ adiabatically transported around $M$ elementary disclinations is then expected to be 
\begin{align}
\gamma_{AB} = \theta_a e^{2\pi i L_a },
\end{align}
which is a discrete analog of the continuum result, Eq. \ref{Eq:totalspin_anyon} \cite{gromov2016}. The contribution from the topological twist $\theta_a$ arises when an anyon encircles a $2\pi$ source of curvature, which arises because an elementary $2\pi/M$ disclination acts as a source of $2\pi/M$ curvature in the TQFT (see e.g. Ref. \onlinecite{Biswas2016} for numerical evidence of this for integer quantum Hall states). 

The defect classification is
\begin{align}
\H^3(\Z_M,U(1)) \cong \Z_M.
\end{align}
There are naively $M$ distinct defect classes. These $M$ classes correspond to changing the $\Z_M$ charge of an elementary $\Z_M$ flux by an even integer.

For this symmetry group, the third step of the classification in general gives a non-trivial reduction of $\H^3(\Z_M, U(1))$. That is, the classification of SETs is in general reduced from the naive estimate of $|\H^2(G,\A)\times\H^3(G,U(1))| = (|\A/M\A|\times M)$. Physically, this means that one can stack a nontrivial $\Z_M$ SPT onto a system, but then relabel the symmetry defects so that all the topological properties (fusion, braiding, etc) correspond exactly to those of the original system.

The symmetry fractionalization and the defect class combine together to define a physical response, which is a fractionally quantized $\Z_M$ charge of an elementary $\Z_M$ flux. When $\Z_M$ is a lattice rotation symmetry, this contributes a fractionally quantized angular momentum $\ell_s/M$ of an elementary disclination. This fractional angular momentum is given by
\begin{align}
  \label{lsEq}
  \ell_s = 2(h_s + k_3) = L_s + 2k_3
\end{align}
where we have introduced the topological spin $h_s \in [0,1)$, defined by $\theta_s = e^{2\pi i h_s}$, and $k_3$ is an integer specifying the defect class. The fractional orbital angular momentum of $s$ is defined by $L_s = 2 h_s \in [0,2)$, in analogy with the way the charge of the vison, $Q_v = 2 h_v$, was defined while discussing the Hall conductivity. However, importantly $\ell_s$ is not completely invariant. Under gauge transformations of the $G$-crossed BTC (referred to as vertex basis and symmetry action gauge transformations), $\ell_s$ only stays invariant modulo $2M$. A formula for $e^{\pi i \ell_s/M}$ is given in terms of the data of the $G$-crossed theory in Eq. \ref{Invt_D_ell_s}.

Note that $e^{\pi i \ell_s/M}$ is also not an absolute invariant in general, as we discuss in more detail in Appendix \ref{Sec:relabeling}. Physically, this is because the fractional angular momentum of a disclination can change by binding an anyon to it. Therefore the topologically invariant response is defined by $e^{\pi i \ell_s/M}$ up to a certain equivalence determined by the fractional angular momentum of the Abelian anyons.  An absolute invariant can be obtained by raising $e^{\pi i \ell_s/M}$ to a suitable power that depends on the group structure of $\mathcal{A}$ and the braiding statistics, as well as the symmetry fractionalization class. 

Mathematically, the invariant used to obtain $\ell_s$ depends on the choice of a particular $\Z_M$ defect $0_{\bf g}$. It is however possible to relabel the defects such that $0_{\bf g} \rightarrow a_{\bf g} = a \times 0_{\bf g}$ without changing the data associated to $0_{\bf g}$, for certain special values of the anyon $a$. The $G$-crossed invariant evaluated with $0_{\bf g}$ and $a_{\bf g}$ thus may give two different values of $e^{i \pi \ell_s/M}$, corresponding to different values of $k_3$, which equally well characterize the defect class. Values of $k_3$ related in this way must be treated as physically equivalent, and therefore we have a redundancy in the counting of SETs. These issues are discussed in detail in Appendix \ref{Sec:relabeling_ZM}. For concrete results of the final counting of SETs in some well-known topological orders, see Table \ref{Table:ZMrelabelings}.

Furthermore, note that the total angular momentum of a disclination is expected to receive an additional contribution proportional to the chiral central charge, arising from the gravitational CS term (see e.g. Ref. \onlinecite{Manjunath2020} for a recent discussion).

      \begin{table*}[t]
	\begin{tabular}{l|l|p{0.4\linewidth}|l|l|l}
          \hline          
          \multicolumn{6}{c}{Summary for $G = U(1)\leftthreetimes_{\phi} [\Z^2\rtimes \Z_M]$, discrete magnetic translations and rotations} \\
          \hline
           \multicolumn{6}{c}{}\\
                  \multicolumn{6}{c}{Symmetry fractionalization, classified by $H^2(U(1)\leftthreetimes_{\phi} [\Z^2\rtimes \Z_M],\A) = \A \times \A \times \A/M\A \times (K_M \otimes \A)$} \\ \hline
          Fractional quantum numbers & $\mathcal{H}^2$ class & Description & Required symmetry &Classification &Eq. No.  \\
          \hline
                 $\{ e^{2\pi i Q_a}\} = \{M_{a,v} \}$ & $v$ & Inserting $2\pi$ flux induces $v$, specifies fractional $U(1)$ charge & $U(1)$ & $\A$& \ref{Invt_SF_v}\\
                 $\{ e^{2\pi i L_a}\}= \{M_{a,s} \} $ & $[s]$ & $M$ elementary disclinations fuse to $s$, specifies fractional orbital angular momentum &$\Z_M$ & $\A/M\A$ &\ref{Invt_SF_s_disc} \\
                 $\{ e^{i 2\pi \vec{P}_a \cdot ({\bf b} - \,^h {\bf b})}\} = \{M_{a, \vec{t}\cdot {\bf b}}\}$ & $[t_x, t_y]$ & Elementary dislocation with total Burgers vector ${\bf b} - \,^h {\bf b}$ induces the anyon $\vec{t}\cdot {\bf b}$, specifies fractional linear momentum & $\Z^2\rtimes\Z_M$ & $K_M\otimes\A$& See Table \ref{Table:LatticeInvts}\\ 
          $\{e^{2\pi \tau_a} \} =  \{ M_{a,m} \}$ & $m$ & Anyon per unit cell, specifies fractionalization of translation algebra & $\Z^2$ & $\A$ & \ref{Invt_SF_m} \\ \hline  
		\end{tabular}
       
      	\begin{tabular} {l |l|l}
      		\multicolumn{3}{c}{} \\
      		\hline
      		\multicolumn{3}{c}{Defect (SPT) classification, $\H^3( [U(1) \leftthreetimes_{\phi} \Z^2]\rtimes \Z_M, U(1) ) = \Z^2 \times \Z_M^3 \times K_M^2$} \\ \hline
      		Invariant & Required symmetry & Classification \\
      		\hline
      		$k_1 = \floor{\bar{\sigma}_H/2}$ & $U(1)$  & $\Z$ \\
      		$k_2 = \floor{\mathscr{S}} $ & $U(1) \times \Z_M$ & $\Z_M$ \\
      		$k_3 = \floor{\ell_s/2}$ & $\Z_M$ & $\Z_M$ \\
      		$\vec{k}_4 = \floor{ {^{(1-h)}} \vec{\mathscr{P}}_c }$ & $ [U(1) \leftthreetimes_{\phi} \Z^2]\rtimes \Z_M$  & $K_M$ \\
      		$\vec{k}_5 = \floor{ {^{(1-h)}} \vec{\mathscr{P}}_s }$ & $\Z^2\rtimes\Z_M$  & $K_M$ \\
      		$k_6=\nu/q -\phi \sigma_H - v \star m$ & $U(1) \times \Z^2$  & $\Z$ \\
      		$k_7=\nu_s/q - \phi \mathscr{S} - s \star m$  & $\Z^2\rtimes\Z_M$  & $\Z_M$ \\ \hline
      	\end{tabular}
	\begin{tabular}{l|p{0.4\linewidth}|l|l}
          \multicolumn{4}{c}{} \\
          \hline
          \multicolumn{4}{c}{Fractionally quantized physical responses } \\ \hline
          Response coefficient & Physical description &  Required symmetry & Eq. No.  \\
          \hline
                  $\bar{\sigma}_H = Q_v + 2 k_1 $ &Hall conductivity & $U(1)$ & \ref{Invt_D_sigmaH}\\
                  $\mathscr{S} = v \star s + k_2 \mod M$  & Discrete shift: angular momentum of flux and charge of disclination. & $U(1)\times\Z_M$ & \ref{Invt_D_Shift_disc}\\
                  $\ell_s = L_s + 2 k_3 \mod 2M$  & Angular momentum of a disclination & $\Z_M$ &  \ref{Invt_D_ell_s}\\
                  $\vec{\mathscr{P}}_c = {^{(1-h)^{-1}}}(\vec{t} \star v + \vec{k}_4) \mod \Z^2$ & Quantized charge polarization: charge of dislocation and momentum of flux & $[U(1)\leftthreetimes_{\phi} \Z^2]\rtimes\Z_M$ & See Table \ref{Table:LatticeInvts}\\
          $\vec{\mathscr{P}}_{s} ={^{(1-h)^{-1}}}(\vec{t} \star s + \vec{k}_5)\mod \Z^2$  & Quantized angular momentum polarization: angular momentum of dislocation & $\Z^2\rtimes\Z_M$ & See Table \ref{Table:LatticeInvts} \\
          $\nu = q(\phi \bar{\sigma}_H + v \star m + k_6) $ & Charge per magnetic unit cell & $U(1)\leftthreetimes_{\phi}\Z^2$ & \ref{Invt_D_nuC}\\
          $\nu_s = q(\phi \mathscr{S} + s \star m + k_7) \mod M$  & Angular momentum per magnetic unit cell  & $\Z^2\rtimes\Z_M$ & Not determined \\ \hline
          $\Pi_{ij}$  & Quantized torsional response: momentum of dislocation &  $\Z^2\rtimes\Z_M$ & \ref{Eq:THV}\\
          $\vec{\nu}_p$ & Momentum per magnetic unit cell &  $\Z^2\rtimes\Z_M$ & Not determined\\ \hline
			\hline
		\end{tabular}
                \caption{Summary of the invariants that completely describe SET phases with $G = U(1)\leftthreetimes_{\phi} [\Z^2\rtimes \Z_M]$, corresponding to $U(1)$ charge conservation, discrete magnetic translation and $\Z_M$ point group rotation symmetry. The system is assumed to have a flux $\phi = p/q$ per unit cell. We have defined $a \star b$ such that $M_{a,b} = e^{2\pi i a \star b}$, where $a \star b \in [0,1)$. $h$ is a $2 \times 2$ matrix representing the generator of the $2\pi/M$ rotation, written in the lattice basis. The group $K_M$ is defined as follows: $K_2 \cong \Z_2\times\Z_2, K_3\cong\Z_3,K_4\cong\Z_2,K_6\cong\Z_1$. The entries associated with $\Pi_{ij}$ and $\vec{\nu}_p$ are separated from the rest as they are already fully specified by the other topological invariants; they do not have a corresponding SPT term and arise purely from the topological order and symmetry fractionalization. The ``Eq. No.''
                  column refers to the equation in the main text that defines the invariant in terms of the data of the $G$-crossed BTC. The quantities $\mathscr{S}, \ell_s, \vec{\mathscr{P}}_s, \nu_s,\vec{\nu}_p,\Pi_{ij}$ are in general not completely invariant, but can be subject to some further equivalences arising from relabeling defects; these equivalences depend sensitively on the group $\mathcal{A}$, the braiding statistics, and the symmetry fractionalization class.}\label{Table:SpaceGrpClassif}
      \end{table*}

\subsubsection{$G = U(1)\leftthreetimes_{\phi} [\Z^2\rtimes \Z_M]$, discrete magnetic translational and rotational symmetry}

Finally we consider FQH systems with a symmetry group consisting of $U(1)$ charge conservation, discrete magnetic translations which form an algebra determined by the flux $\phi$ per unit cell, and $\Z_M$ point group rotation symmetry, for $M=2,3,4,6$. This symmetry group describes a large class of fractional Chern insulators (FCIs) and FQH states in the presence of a periodic potential. 

The symmetry fractionalization and defect classifications are summarized in Table \ref{Table:SpaceGrpClassif}, along with the interpretation of the various parameters.

The symmetry fractionalization classification is given by 
\begin{equation}
\H^2(G,\A) \cong \A\times\A\times(\A/M\A)\times(K_M\otimes\A) .
\end{equation}
Here the group $K_M = \Z^2 / \,^{(1 - h)}\Z^2$, where $h$ denotes the generator of $2\pi/M$ rotations and acts on lattice vectors, which are elements of $\Z^2$, through a $2 \times 2$ matrix written in the lattice basis. The notation $\,^{1 -h}\Z^2$ refers to the set of vectors generated by ${\bf r} - \,^h {\bf r}$, where ${\bf r}$ generates $\Z^2$. $K_M$ is related to the conjugacy classes of defects with disclination angle $2\pi/M$, associated with the fact that the Burgers vector of an impure $2\pi/M$ disclination is only defined modulo a $2\pi/M$ rotation, as has been noted in previous works studying disclination defects on a lattice (see for example Refs. \cite{Gopalakrishnan2013,Benalcazar2014,Li2020disc,geier2020bulkboundarydefect}). $K_M$ can also be understood as a finite group grading on Burgers vectors in the presence of $M$-fold rotational symmetry, as discussed in Ref. \cite{Manjunath2020}.

We find that $K_M = \Z_2^2, \Z_2, \Z_2, \Z_1$ for $M = 2,3,4,6$ respectively. The group $K_M \otimes \A$ is mathematically defined in Appendix \ref{Sec:GrpCohIntro}.  For $\A=\prod_{i=1}^k \Z_{n_i}$, we have
\begin{align}
  K_2 \otimes \A &= \prod_{i=1}^k \Z_{(2,n_i)}^2 \nonumber \\
  K_3 \otimes \A &= \prod_{i=1}^k \Z_{(3,n_i)} \nonumber \\
  K_4 \otimes \A &= \prod_{i=1}^k \Z_{(2,n_i)} \nonumber \\
  K_6 \otimes \A &= \Z_1 \label{Eq:K_M_Otimes_defn}
\end{align}

A particular symmetry fractionalization class is specified by the equivalence classes of anyons
\begin{equation}
(v,m,[s],[t_x, t_y]) \in \A\times\A\times(\A/M\A)\times(K_M\otimes\A) .
\end{equation}
The anyons $v$, $m$, and $s$ have been discussed above: they define $U(1)$ charge fractionalization, the anyon per unit cell, and $\Z_M$ rotational symmetry fractionalization, respectively.

The anyons $t_x,t_y \in \A$ give a generalization of the discrete torsion vector recently introduced in Ref. \onlinecite{Manjunath2020}. $t_x$, $t_y$ are subject to an equivalence relation, which gives rise to the equivalence class $[t_x,t_y] \in (K_M\otimes\A) $ (see also Section \ref{Sec:FullSpaceGrpSummary}). The class $[t_x,t_y]$ is associated to a mixed symmetry fractionalization class involving translational and rotational symmetry; this particular type of symmetry fractionalization requires both discrete translational and rotational symmetries, although it is only non-trivial for 2-,3-, and 4-fold rotational symmetry. 

The symmetry fractionalization class $[t_x,t_y]$ associates a fractional linear momentum to each anyon, as follows. Consider a defect with Burgers vector ${\bf b} = (b_x, b_y)$ (this can correpond to a pure dislocation, i.e. a defect with zero disclination angle, or to an 'impure' disclination, which additionally has a nonzero disclination angle). Letting $h$ be the elementary $2\pi/M$ rotation operator, consider also the rotated defect with Burgers vector ${^{h}}{\bf b}$. Let $\gamma_{a,{\bf b}}$ be the Berry phase accumulated by braiding the anyon $a$ around the defect with Burgers vector ${\bf b}$. Then, we have
\begin{equation}\label{RefEq:Momentum}
\frac{\gamma_{a,{\bf b}}}{\gamma_{a,{^{h}}{\bf b}}} = M_{a,\vec{t} \cdot {\bf b}} ,
\end{equation}
where we have defined the Abelian anyon $\vec{t} \cdot {\bf b} = t_x^{b_x} t_y^{b_y}$. In other words, the symmetry fractionalization is defined by associating the anyon $\vec{t}\cdot {\bf b}$ to
the dislocation with Burgers vector ${\bf b} - \,^{h} {\bf b}$. The braiding phase between an anyon and a dislocation can be viewed as defining the charge under translations, which is the linear momentum. Therefore, we can view
\begin{align}
  \label{fracMom}
 e^{i 2\pi \vec{P}_a \cdot ({\bf b} - \,^{h} {\bf b})} = M_{a,\vec{t} \cdot {\bf b}} 
\end{align}
  as defining a fractional linear momentum $\vec{P}_a$, modulo certain equivalences. One equivalence is due to the fact that $\vec{P}_a$ can only be defined via Eq. \ref{fracMom}, dotted into the vector ${\bf b} - \,^h {\bf b}$, for any integer vector ${\bf b}$. 

  A second equivalence for $\vec{P}_a$ arises because different choices of $\vec{t}$ can describe the same symmetry fractionalization class.  
  Specifically, the symmetry fractionalization is trivial if it can be completely accounted for by binding an anyon to the elementary dislocations, as this can be done trivially by adjusting local energetics. Therefore, we have the equivalences
  \begin{align}
 \{ e^{i 2\pi \vec{P}_a \cdot ({\bf b} - \,^{h} {\bf b})} \} \sim \{ e^{i 2\pi \vec{P}_a \cdot ({\bf b} - \,^{h} {\bf b})} M_{a,\vec{\chi} \cdot ({\bf b} -^{h} {\bf b})}\} ,
  \end{align}
  for any $\chi_x,\chi_y \in \mathcal{A}$.
  This leads to a classification by the group $K_M \otimes \A$, as we discuss in Appendix \ref{Sec:Coho_calc_gspace}.

  We present explicit formulas for $e^{i 2\pi \vec{P}_a \cdot ({\bf b} - \,^{h} {\bf b})} $ in terms of the data of the $G$-crossed BTC, as summarized in Table \ref{Table:LatticeInvts}, and in Eqs. \eqref{Eq:P_a2},\eqref{Eq:P_a3},\eqref{Eq:P_a4}, for $M=2,3,4$ respectively. These equations are written for specific choices of ${\bf b}$, and give the part of $\vec{P}_a$ which is invariant under the equivalences described above.

The defect (SPT) classification is (see Appendix \ref{Sec:Coho_calc_gspace})
\begin{equation}
  \label{defectClass}
\H^3(G,U(1)) \cong \H^4(G,\Z) \cong \Z^2\times\Z_M^3\times K_M^2
\end{equation}
Changing the defect class corresponds to changing certain integer parts of the fractionally quantized responses to lattice dislocations and disclinations. These responses are essentially the same as those discussed recently in Ref. \cite{Manjunath2020}, which are summarized in Table \ref{Table:SpaceGrpClassif}, and which we briefly review below. The $G$-crossed invariants used to extract these responses are summarized in Table \ref{Table:LatticeInvts}. Importantly, not all of these different $\H^3$ classes give physically inquivalent SETs. The gauge-invariant quantities associated with the fractionally quantized responses can be used to determine when changing the defect class may in fact yield a physically equivalent SET. 

The fractionally quantized responses can be separated according to the relevant symmetries involved. For the group $U(1) \times \Z_M$ alone, the quantized responses include the Hall conductivity $\sigma_H$, the discrete shift $\mathscr{S}$, and the disclination angular momentum, $\ell_s$. Including the discrete magnetic translation symmetry then also adds the $U(1)$ charge filling (charge per magnetic unit cell) $\nu$, the angular momentum filling $\nu_s$, quantized charge polarization $\vec{\mathscr{P}}_c$, quantized angular momentum polarization $\vec{\mathscr{P}}_s$, quantized torsional response (the momentum of a dislocation) $\Pi_{ij}$, and the linear momentum filling, $\vec{\nu}_p$.

As in the continuum case, the above physical responses can be conveniently represented in terms of an effective response theory involving background gauge fields for the symmetry group. This was explained in the Abelian case for the case of zero flux per unit cell in Ref. \cite{Manjunath2020}. In this paper we extend the results to the non-Abelian case and with non-zero flux per unit cell. Below we briefly summarize the response theory, and leave a detailed derivation for Appendix \ref{Sec:Eff_Ac}. We refer the reader to Ref. \cite{Manjunath2020} for additional discussion on the physical meaning of these quantized responses. 

\begin{table*}[t]
	\centering
	\begin{tabular} {l |l |l|l|l|l|l}
		\hline 
		\multicolumn{7}{c}{Classification of $SU(2)_k$ topological orders with $G=U(1)\leftthreetimes_{\phi}[\Z^2\rtimes\Z_M]$} \\ \hline
		Anyon model & $\A$ &$M$ & $\H^2(G,\A)$ & $\H^3(G,U(1))$ &Naive SET count ($k_1,k_6,k_7$ fixed)& Reduced SET count ($k_1,k_6,k_7$ fixed)\\
		\hline
		
		\multirow{4}{*}{$SU(2)_k$, $k$ odd} &$\Z_2$& 2 &$\Z_2^5$ &$\Z^2\times\Z_2^7$ &2048 &400 \\
		&$\Z_2$& 3 &$\Z_2^2$ &$\Z^2\times\Z_3^5$ &729 &729 \\
		 &$\Z_2$& 4 &$\Z_2^4$ &$\Z^2\times\Z_2^2\times\Z_4^3$ &1024 &576 \\
		 &$\Z_2$& 6 &$\Z_2^3$ &$\Z^2\times\Z_6^3$ &288 &162 \\  
		\hline
		\multirow{4}{*}{$SU(2)_k$, $k$ even} &$\Z_2$& 2 &$\Z_2^5$ &$\Z^2\times\Z_2^7$ &2048 &2048 \\
		&$\Z_2$& 3 &$\Z_2^2$ &$\Z^2\times\Z_3^5$ &729 &729 \\
		&$\Z_2$& 4 &$\Z_2^4$ &$\Z^2\times\Z_2^2\times\Z_4^3$ &1024 &1024 \\
		&$\Z_2$& 6 &$\Z_2^3$ &$\Z^2\times\Z_6^3$ &288 &288 \\  
		\hline
	\end{tabular}
	\caption{Results of the relabeling analysis for $SU(2)_k$ topological orders with $G=U(1)\leftthreetimes_{\phi}[\Z^2\rtimes \Z_M]$. Note that $SU(2)_1$ describes the topological order of the bosonic $1/2$ Laughlin state, $SU(2)_2$ describes the bosonic Moore-Read Pfaffian state, and $SU(2)_k$ for $k > 2$ describes the bosonic Read-Rezayi states.  See Appendix \ref{Sec:relabeling_full} for further details regarding the derivation. 
          To obtain the SET counts, we have fixed $k_1$, $k_6$, and $k_7$, which are the integer parts of the Hall conductivity, and the charge and angular momentum fillings. If we do not fix $k_7$, we do not have a rigorous analysis within the $G$-crossed BTC formalism because we do not have a formula for the angular momentum filling $\nu_s$ derived using the $G$-crossed BTC data. However if we use the formula for $\nu_s$ predicted by the effective response action and allow $k_7$ to vary, the SET counts all get multiplied by an additional factor of $M$, implying that there is no additional reduction in the count due to $k_7$. 
        }\label{Table:relabelingFullExamples}
\end{table*}  

The response theory is defined in terms of background gauge fields $A$, $\vec{R}$, and $C$ associated with the $U(1)$, $\Z^2$, and $\Z_M$ symmetries, respectively. However, because of the non-commuting nature of the group elements, they should be viewed together as a gauge field $(A, \vec{R}, C)$ for the full symmetry group $G = U(1)\leftthreetimes_{\phi} [\Z^2\rtimes \Z_M]$. 
The response theory is then written as follows:
\begin{widetext}
  \begin{align}
    \label{resTheoryFull}
	\mathcal{L}_{\text{eff}} &= \frac{\sigma_H}{2} A \cup dA + \frac{\mathscr{S}}{2\pi} A \cup dC + \frac{\ell_s}{4\pi} C \cup dC  
	+ \frac{\vec{\mathscr{P}}_c }{2\pi} \cdot (A \cup d\vec{R}) +\frac{\vec{\mathscr{P}}_s}{2\pi} \cdot (C \cup d\vec{R}) 
	+ \frac{1}{2\pi}\left(\nu A + \nu_s C\right) \cup A_{XY} \nonumber \\
	& + \frac{\Pi_{ij}}{4\pi} R_i \cup d R_j + \frac{\vec{\nu}_p}{2\pi} \cdot \vec{R} \cup A_{XY} + \frac{\alpha}{4\pi} A_{XY} \cup d^{-1} A_{XY} + \mathcal{L}_{\text{anom}},
	\end{align}
\end{widetext}
where the term $\mathcal{L}_{\text{anom}}= -\frac{c}{48\pi} C \cup dC$ with chiral central charge $c$ is due to the framing anomaly \cite{Witten1989,Manjunath2020} (note we assume that the spin connection of the space-time manifold has only a spatial component, which is set by the rotation gauge field $C$). The gauge fields are defined so $A$ and $C$ are the lifts of $U(1)$ and $\Z_M$ to $\R$ and $\frac{2\pi}{M} \Z$, respectively, and the action is independent of the precise choice of lift. Here we have defined the Lagrangian on an arbitrary triangulation of the space-time, and $\cup$ refers to the cup product of cohomology. The term $\frac{1}{2\pi} dC$ represents the flux of $C$ and physically describes the disclination density. The term $\frac{1}{2\pi}{^{(1-h)^{-1}}} d\vec{R}$ represents the gauge-invariant part (mod 1) of the Burgers vector of a lattice defect \cite{Manjunath2020}. The term $A_{XY}$ is defined in terms of the $\vec{R}$ and $C$ gauge fields, and its integral over space counts the number of unit cells in the lattice. The main difference between the above action and the one written down in Ref. \cite{Manjunath2020} is that the parameters associated to $A_{XY}$ depend on $\phi$, because each unit cell is associated to the flux $\phi$ (see Appendix \ref{Sec:Eff_Ac-gspace} for a derivation). The other parameters are $\phi$-independent. 

The discrete shift $\mathscr{S}$ \cite{Biswas2016,Han2019,Liu2019ShiftIns,Li2020disc}, which is the analog of the shift in continuum FQH states, is protected by the discrete rotational symmetry. It associates a charge of $\mathscr{S}/M = (v \star s +k_2)/M$ to an elementary $2 \pi/M$ disclination and a corner angle of $2\pi/M$. 
Presumably the fractionally quantized charge localized to a corner angle on the boundary of the system remains well-defined only if the edge theory is gapped. We can also define the angular momentum of a $2\pi/M$ flux as $\mathscr{S}/M$: since we have the relation $e^{2\pi i \mathscr{S}} = M_{v,s}$, this implies that the angular momentum of a $2\pi$ flux is indeed that of the anyon $v$, as expected. This response defines a notion of fractional ``higher order'' topological phases, for both Abelian and non-Abelian FQH states \cite{You2018HOTI,Benalcazar2019HOTI,Rasmussen2020HOSPT}. 

The term $\ell_s$ which contributes a fractional angular momentum of $\ell_s/M$ to an elementary $2\pi/M$ disclination, is reviewed above (see Eq. \ref{lsEq}).

The response $\vec{\mathscr{P}}_c$ defines a quantized charge polarization compatible with rotational invariance \cite{Manjunath2020}. The component $\mathscr{P}_{c,i}$ of the charge polarization can be given various interpretations: (i) the charge per unit length on a system with boundary along the $i$ direction, (ii) the fractional charge associated to an elementary dislocation in the $i$ direction, or (iii) the $i$th component of the linear momentum of a $2\pi$ instanton. Note that a nontrivial Burgers vector can also be associated to a disclination: such disclinations are sometimes referred to as 'impure disclinations'. This response is also computed in free fermion systems in Ref. \cite{Li2020disc}; the linear and angular momenta of $2\pi$ instantons have also been studied in Refs. \cite{Song_2019monopole,Song_2020monopole} in the context of Dirac spin liquids, which raises the interesting question of how these results, which apply to gapped topological phases, can be extended to gapless phases.

Similarly, $\vec{\mathscr{P}}_s$ defines a quantized angular momentum polarization. The component $\mathscr{P}_{s,i}$ of the angular momentum polarization can be interpreted as (i) the angular momentum per unit length on a system with boundary along the $i$ direction, or (ii) the angular momentum associated to an elementary dislocation in the $i$ direction. 

The quantized torsional response $\Pi_{ij}$, which is only non-trivial for $M = 2,3,4$, associates a fractionally quantized momentum $p_i = \sum_j \Pi_{ij} b_j$ to a defect with dislocation Burgers vector ${\bf b}$. After making some simplifying assumptions, we derive a formula for $\Pi_{ij}$ modulo $1/M$  in terms of the data of the $G$-crossed theory using Eq. \ref{Eq:THV}. This term is related to the ``torsional Hall viscosity'' studied Ref. \onlinecite{Hughes2011thv,Hughes2013thv}, however in those contexts the response term was found to be non-quantized, since the theories studied were coupled to continuum geometry, rather than a discrete background gauge field as appropriate for crystalline space group symmetries. In contrast, in our theory the torsional response $\Pi_{ij}$ is indeed quantized. 

The response coefficient $\nu_s$ associates an angular momentum $\nu_s$ per magnetic unit cell, while $\vec{\nu}_p$ associates a linear momentum $\vec{\nu}_p$ per magnetic unit cell to the system. These definitions are motivated by crystalline gauge theory and group cohomology; we have however not been able to find invariants for $\nu_s,\vec{\nu}_p$ in terms of $G$-crossed BTC data, and leave this problem for future work. Note that $\vec{\nu}_{p}$ is only specified by the topological field theory modulo the same equivalences that exist for the linear momentum of $m$, $\vec{P}_{m}$. Also, note that $\Pi$ and $\vec{\nu}_p$ do not receive any pure SPT contribution; they appear only due to the non-trivial topological order and symmetry fractionalization.

As discussed in Ref. \cite{Manjunath2020}, certain terms, such as $\Pi_{ij}$ and $\vec{\nu}_p$ may receive an additional quantized contribution that is topologically trivial (corresponding to a coboundary term in the group cohomology classification of topological terms), but nonetheless may have physical consequences. Such contributions, if they exist, cannot be determined from the $G$-crossed BTC. It is not clear whether they do in fact physically arise and will not be studied further here.

We note that the last term in Eq. \ref{resTheoryFull} arises from the existence of an anyon per unit cell; it is not clear how to physically interpret it as a response, and as such we will ignore it. Its effect is completely accounted for by specifying the anyon per unit cell. 

The $G$-crossed identities studied in this work only allow us to determine directly the properties of $U(1)$ fluxes and defects with a nonzero disclination angle, where several identical copies of these defects fuse to give an anyon. In particular, we cannot directly obtain the charge, angular momentum or linear momentum of a pure dislocation defect, which does not have this property. However, we can always treat a dislocation as a dipole of two disclinations with opposite disclination angles. The symmetry charges associated to a dislocation are then obtained by summing the symmetry charges associated to the two disclinations which form the dipole. This procedure is consistent in the following sense: the individual disclinations can be arbitrarily chosen, but as long as their fusion product is fixed, the charge of a dislocation is well-defined up to the charge of arbitrary anyons, which can be attached to the dislocation by adjusting local energetics. (See Section \ref{Sec:PropArbDefects} for further discussion of these issues, and Appendix \ref{Sec:Invts-gspace} for the mathematical derivations.) 

We make two general remarks about the SET classification. First, note that the classification is independent of the flux per unit cell; therefore each of these responses can be observed at any value of $\phi$. Secondly, as in the case of pure $\Z_M$ symmetry, we find that the total number of SET phases is in fact different from the naive estimate of $|\H^2(G,\A)\H^3(G,U(1))|$, and the exact reduction depends sensitively on the topological order as well as the value of $M$. Using $G$-crossed BTC data, the reduction in the SET count has been computed explicitly for the bosonic Laughlin, Moore-Read, and Read-Rezayi FQH topological orders with $G = U(1)\leftthreetimes_{\phi} [\Z^2\rtimes \Z_M]$ as summarized in Table  \ref{Table:relabelingFullExamples}. We have also computed the exact SET counting with simply $G = \Z_M$ symmetry for a variety of topological orders, as summarized in Table \ref{Table:ZMrelabelings}.

\subsection{Possible experimental applications}

Although the main objective of this work is to obtain a mathematically complete description of different symmetry-enriched FQH states, our analysis has nonetheless pointed out several new phenomena, particularly in systems with space group symmetry, that can potentially be realized with available experimental technology. In this section we will briefly point out some concrete measurements that could be performed, and mention some promising experimental platforms. We will primarily focus on the measurement of dislocation and disclination charges; note that Ref. \cite{Manjunath2020} has a detailed discussion of the mathematically allowed response properties of the 1/2 Laughlin topological order and of gapped $\Z_2$ quantum spin liquids, with topological order described by a $\Z_2$ gauge theory ($\Z_2$ toric code), with the symmetry $G=U(1)\times G_{\text{space}}$.

Our work indicates that certain novel phenomena can be studied by measuring the fractional charge in lattice FQH systems, as we now summarize. There are two distinct ways in which fractional charges appear in lattice FQH systems: they can appear in conjunction with anyons, or they can be bound to lattice defects such as dislocations, disclinations and corners. For example, in the 1/2 Laughlin state on a lattice with magnetic translations and either $\Z_2$ or $\Z_4$ point group rotation symmetry, the quasiholes can carry a half-charge independent of any spatial symmetries, while dislocations can also carry integer or half-integer charges independent of the topological order. Therefore we might naively think that the minimal charge that can be measured in a system with both anyons and dislocation defects must be half-integer. However, our work predicts that due to the interplay of the space group symmetry with the topological order, in the above mentioned systems a dislocation can in fact carry a $1/4$ charge (see the discussion of this response in \cite{Manjunath2020}). This happens when the form of symmetry fractionalization which we refer to as the 'torsion vector' $\vec{t}$ is nontrivial.

A similar phenomenon can be observed in the 1/3 Laughlin state on a lattice with $\Z_3$ point group rotation symmetry. Here the minimal anyon charge as well as the minimal dislocation charge from symmetry alone is $1/3$, however if our formalism is applied to the 1/3 Laughlin state, the minimal charge bound to a dislocation should be 1/9. In general, such effects occur when there is some commensuration between the group of Abelian anyons and the $\Z_M$ point group symmetry.

The charge associated to disclination or corner angles is an independent invariant which is a discrete analog of the shift (see the response denoted by $\mathscr{S}$), and is also necessary to characterize the FQH state. Note that if we can measure the charge at an arbitrary disclination, we can also measure the charge at an arbitrary dislocation, since any dislocation can be treated as a dipole of two disclinations with equal and opposite Frank angles, but unequal Burgers vectors.

Below we mention some experimental platforms in which lattice integer and fractional quantum Hall effects have been successfully realized, and in which it may be possible to measure dislocation and disclination charges in the near future. 

\paragraph*{FCIs in twisted bilayer graphene:} Fractional Chern insulator (FCI) states have been extensively studied in previous numerical work (\cite{parameswaran2013,Bergholtz2013FCI}) and were recently observed experimentally in twisted bilayer graphene (TBG) aligned with a hexagonal boron nitride substrate \cite{Spanton2018FCI}. This has motivated several recent works studying the properties of FCIs in TBG with and without a magnetic field \cite{Abouelkomsan2020FCI,Ledwith2020FCI,Repellin2020FCI}. It may be possible to prepare a 1/3 Laughlin state with $\Z_3$ point group rotation symmetry in such a platform; this would be a simple state in which the fractional charges bound to dislocations and disclinations can be studied, which would give an experimental probe of the torsion vector, spin vector, shift, and quantized charge polarization predicted by theory. It would be interesting to study the interplay with symmetry of more complicated topological orders, which can be accessed using TBG according to the numerical evidence from the works above. Related 2D systems such as twisted transition metal dichalcogenides (e.g. $\text{WSe}_2$) also appear to have Chern bands as well as strong interactions in their moire lattices \cite{Pan2020}, and could also serve as model systems to study some of our predictions.   

\paragraph*{FCIs in optical lattices:} In recent years there has been substantial effort aimed towards numerically simulating bosonic fractional Chern insulators, in particular the 1/2 Laughlin state, in optical lattices \cite{Lukin2005OL,Palmer2006OL,hafezi2007,Dalibard2012OFL}. Previous work includes studies on adiabatic state preparation \cite{barkeshli2015,He2017OL,Motruk2017FCI} and characterization of quasiholes \cite{Raciunas2018FCIOp,Dong2018FCI}, in various lattice geometries \cite{Liu2015FCI}; on the experimental front, the Chern number of a Chern band of ultracold bosons was first measured in Ref. \cite{Aidelsburger2015}. Proposals exist for measuring the anyon charge using quantum gas microscopy \cite{Raciunas2018FCIOp}; similar techniques could potentially allow us to measure the fractional charge at lattice defects. 

\paragraph*{Chern insulators in photonic crystals} Photonic crystals (reviewed in Ref. \cite{Ozawa2019TopPhot}), in which the spatial periodicity of the material properties can result in a photonic band struture with nonzero Chern number, are a well-established platform for realizing analog IQH states, and therefore offer promise in experimentally realizing the bosonic SPT states studied in this work. In recent work \cite{Li2018topphot}, a photon zero mode bound to a dislocation was predicted and also experimentally measured in a square lattice geometry. We expect that such measurements should be possible in other lattice geometries and also for disclination defects. Photonic materials with continuum symmetries are also of much interest. In another work \cite{Schine2019} a Landau level of photons was prepared in a continuum system with a conical curvature defect, and the fractional number excess at the conical tip (i.e. the shift response in the photonic system) was experimentally measured. 

In the above discussion, we have restricted our attention to the fractional charge at lattice defects, for which the available experimental imaging technology is most developed. It is an open and interesting question to develop probes that could image angular momentum or linear momentum in a manner that can give access to the other invariants studied in this work. We expect that optical lattices will offer the most feasible way to make such measurements in bosonic systems.

\subsection{Organization of paper}

Below we explain the organization of the paper. 

Section \ref{Sec:Review} gives a brief review of the $G$-crossed BTC formalism of Ref. \cite{Barkeshli2019} and the crystalline gauge theory developed in Ref. \cite{Manjunath2020} for Abelian topological phases, which we use to provide a topological effective action for the response theories.  

Section \ref{Sec:SummaryOfResults} presents the major results of the paper but without proofs. First, in Section \ref{Sec:GenResInvts}, we discuss general identities in terms of the $G$-crossed data that are repeatedly used in this paper to derive the formulas for topological invariants. The rest of Section \ref{Sec:SummaryOfResults} summarizes the topological invariants for different symmetry groups. The tables in this Section list the different invariants characterising the symmetry fractionalization and the topological response. They also summarize the mathematical classification, formulas for the $G$-crossed invariants, and the physical interpretation of these invariants. In this section we also write down topological effective actions that describe Abelian SET phases with these symmetries, in terms of crystalline gauge fields. We conclude and discuss future directions in Section \ref{Sec:Discussion}. 

The technical calculations of this paper are organized in the appendices.

In Appendix \ref{Sec:FQHSymm} we review the derivation of the group multiplication laws for FQH systems with magnetic translation and spatial rotation symmetries, both in the continuum and on the lattice. These symmetries being usually defined on the infinite plane, we also briefly discuss how to define them on compact manifolds whose global topology may be inconsistent with the symmetry.

In Appendix \ref{Sec:GCReview}, we review in more detail the mathematical formulas and consistency relations among the $G$-crossed data used in this paper, and also review gauge transformations of the data. We review a general solution to the $G$-crossed data for arbitrary $G$, which can be written for non-permuting symmetries. We then describe, with proofs, the general procedures to construct invariants for symmetry fractionalization and for the defect response. These constructions are new to this paper. 

Appendix \ref{Sec:Cont_meas} describes how, for our examples, the SET classification is the same whether the $G$-crossed data are assumed to be measurable or piecewise continuous. Appendix \ref{Sec:Gxsolns} lists solutions to the $G$-crossed BTC equations for each of the symmetry groups in this paper. In Appendix \ref{Sec:Invts}, we use $G$-crossed identities to construct specific topological invariants for the symmetry fractionalization and defect classes in each case.

In Appendix \ref{Sec:LSM}, we discuss the derivation of LSM-type relations between the filling and Hall conductivity using $G$-crossed invariants, both in the continuum and on the lattice. In Appendix \ref{Sec:relabeling}, we discuss how to account for redundancies in the SET classification due to possible relabelings of the symmetry defects. 

The cohomology calculations underlying the $G$-crossed BTC results are given in Appendix \ref{Sec:Coho_calc}. The derivation of effective actions for the symmetry groups studied in this paper, in terms of crystalline gauge fields, is given in Appendix \ref{Sec:Eff_Ac}. Finally, in Appendix \ref{Sec:GrpCohIntro} we review the group cohomology results used in this paper, and in Appendix \ref{app_sseq} we review the method of spectral sequences that is required in order to study systems with magnetic translation symmetry.

\section{Review}
\label{Sec:Review}
\subsection{Brief review of G-crossed BTC framework}
\label{Sec:ReviewGxShort}
\subsubsection{Overview}

The $G$-crossed braided tensor category (BTC) theory is a mathematical framework to characterize and classify the different topological phases that can be realized when a topologically ordered system is endowed with a symmetry described by the group $G$, such that the ground state preserves the symmetry \cite{Barkeshli2019}. The starting point is a unitary modular tensor category $\mathcal{C}_{\bf 0}$, which characterizes the braiding and fusion properties of the anyons via a self-consistent set of algebraic data, the $F$ and $R$ symbols. $\mathcal{C}_{\bf 0}$, together with the chiral central charge $c$, are believed to fully characterize intrinsic topological orders in (2+1)D, in the absence of symmetry. 

The $G$-crossed BTC $\mathcal{C}_G^\times$ keeps track of the anyons and the topologically inequivalent symmetry defects, and their combined fusion and braiding properties. This is specified by a consistent set of symbols $\{F,R,U,\eta\}$. The $G$-crossed BTC thus contains two additional symbols: the $U$ symbol specifies how symmetry group elements act on the fusion and splitting spaces in the space of topological states, while the $\eta$ symbol defines symmetry fractionalization. 

Below we give a brief description of the essentials of the $G$-crossed BTC framework. Additional details are presented in Appendix \ref{Sec:GCReview}. A more complete account, using the same notation, can be found in Ref. \onlinecite{Barkeshli2019}.  

\subsubsection{The basic $G$-crossed data}

\textit{Anyon model:} The anyon model, described by a UMTC $C_{{\bf 0}}$, consists of a full description of the topological phase without $G$ symmetry \cite{wang2008,nayak2008}. Below we briefly review some of the key defining properties of $\mathcal{C}_{\bf 0}$ and set notation. Ref. \onlinecite{Barkeshli2019} contains a more in-depth review aimed at physicists, using the same notation. Recently Ref. \onlinecite{kawagoe2020} has shown how the defining data of a UMTC can be obtained from the microscopic properties of a quantum state of matter. 

The UMTC $\mathcal{C}_{\bf 0}$ contains a finite set of topological charge labels, $a$, referred to as anyons. For anyons $a,b,c$, there exists a set of vector spaces, referred to as fusion spaces, $V_{ab}^c$, and their dual ``splitting'' spaces $V_c^{ab}$. The dimensions of these vector spaces give the fusion coefficients:
\begin{align}
N_{ab}^c = \text{dim } V_{ab}^c = \text{dim } V_c^{ab} . 
\end{align}
States in these spaces are depicted graphically as follows
\begin{equation}
\left( d_{c} / d_{a}d_{b} \right) ^{1/4}
\pspicture[shift=-0.6](-0.1,-0.2)(1.5,-1.2)
  \small
  \psset{linewidth=0.9pt,linecolor=black,arrowscale=1.5,arrowinset=0.15}
  \psline{-<}(0.7,0)(0.7,-0.35)
  \psline(0.7,0)(0.7,-0.55)
  \psline(0.7,-0.55) (0.25,-1)
  \psline{-<}(0.7,-0.55)(0.35,-0.9)
  \psline(0.7,-0.55) (1.15,-1)	
  \psline{-<}(0.7,-0.55)(1.05,-0.9)
  \rput[tl]{0}(0.4,0){$c$}
  \rput[br]{0}(1.4,-0.95){$b$}
  \rput[bl]{0}(0,-0.95){$a$}
 \scriptsize
  \rput[bl]{0}(0.85,-0.5){$\mu$}
  \endpspicture
=\left\langle a,b;c,\mu \right| \in
V_{ab}^{c} ,
\label{eq:bra}
\end{equation}
\begin{equation}
\left( d_{c} / d_{a}d_{b}\right) ^{1/4}
\pspicture[shift=-0.65](-0.1,-0.2)(1.5,1.2)
  \small
  \psset{linewidth=0.9pt,linecolor=black,arrowscale=1.5,arrowinset=0.15}
  \psline{->}(0.7,0)(0.7,0.45)
  \psline(0.7,0)(0.7,0.55)
  \psline(0.7,0.55) (0.25,1)
  \psline{->}(0.7,0.55)(0.3,0.95)
  \psline(0.7,0.55) (1.15,1)	
  \psline{->}(0.7,0.55)(1.1,0.95)
  \rput[bl]{0}(0.4,0){$c$}
  \rput[br]{0}(1.4,0.8){$b$}
  \rput[bl]{0}(0,0.8){$a$}
 \scriptsize
  \rput[bl]{0}(0.85,0.35){$\mu$}
  \endpspicture
=\left| a,b;c,\mu \right\rangle \in
V_{c}^{ab},
\label{eq:ket}
\end{equation}
where $\mu=1,\ldots ,N_{ab}^{c}$. $d_a$ is the quantum dimension of $a$, given by the largest eigenvalue of the fusion matrix $[N_a]_b^c := N_{ab}^c$,
and the factors $\left(\frac{d_c}{d_a d_b}\right)^{1/4}$ are a normalization convention for the diagrams. Physically, the splitting diagram above can be interpreted as an anyon $c$ which is split into two anyons $a$ and $b$ by the action of a ``splitting operator'' \cite{kawagoe2020}. The index $\mu$ corresponds to distinct possible ways this can happen to yield orthogonal many-body states. 

The fusion coefficients are encapsulated in the fusion rules written as:
\begin{equation}
a \times b = \sum\limits_{c \in \mathcal{C}_{{\bf 0}}} N^c_{ab} c .
\end{equation}
The theory contains a unique ``trivial'' particle $0$, which physically corresponds to topologically trivial excitations. For each anyon $a$, we have a unique conjugate $\bar{a}$, such that
\begin{align}
a \times \bar{a} = 0 + \cdots .
\end{align}
We can define fusion and splitting spaces for multiple anyons by decomposing into fusion channels. For example,
\begin{align}
V^{abc}_d \simeq \bigoplus_e V_e^{ab} \otimes V_d^{ec} \simeq \bigoplus_f V_f^{bc} \otimes V_{d}^{af} ,
\end{align}
which follows from associativity of the fusion rules. The $F$-symbols are unitary maps encoding the basis transformations between the above decompositions; they are written in components as $[F_d^{abc}]_{(e,\alpha,\beta)(f,\mu,\nu)}$, where the Greek indices refer to the fusion channels, which run over the basis states of the splitting spaces. In a diagrammatic calculus that corresponds to moving, splitting, and braiding anyons in time (with the time direction oriented vertically) (see e.g. Ref. \onlinecite{kawagoe2020} for a recent discussion), the $F$-symbol can be depicted graphically as follows:
\begin{equation}
\pspicture[shift=-1.0](0,-0.45)(1.8,1.8)
\small
\psset{linewidth=0.9pt,linecolor=black,arrowscale=1.5,arrowinset=0.15}
\psline(0.2,1.5)(1,0.5)
\psline(1,0.5)(1,0)
\psline(1.8,1.5) (1,0.5)
\psline(0.6,1) (1,1.5)
\psline{->}(0.6,1)(0.3,1.375)
\psline{->}(0.6,1)(0.9,1.375)
\psline{->}(1,0.5)(1.7,1.375)
\psline{->}(1,0.5)(0.7,0.875)
\psline{->}(1,0)(1,0.375)
\rput[bl]{0}(0.05,1.6){$a$}
\rput[bl]{0}(0.95,1.6){$b$}
\rput[bl]{0}(1.75,1.6){${c}$}
\rput[bl]{0}(0.5,0.5){$e$}
\rput[bl]{0}(0.9,-0.3){$d$}
\scriptsize
\rput[bl]{0}(0.3,0.8){$\alpha$}
\rput[bl]{0}(0.7,0.25){$\beta$}
\endpspicture
= \sum_{f,\mu,\nu} \left[F_d^{abc}\right]_{(e,\alpha,\beta)(f,\mu,\nu)}
\pspicture[shift=-1.0](0,-0.45)(1.8,1.8)
\small
\psset{linewidth=0.9pt,linecolor=black,arrowscale=1.5,arrowinset=0.15}
\psline(0.2,1.5)(1,0.5)
\psline(1,0.5)(1,0)
\psline(1.8,1.5) (1,0.5)
\psline(1.4,1) (1,1.5)
\psline{->}(0.6,1)(0.3,1.375)
\psline{->}(1.4,1)(1.1,1.375)
\psline{->}(1,0.5)(1.7,1.375)
\psline{->}(1,0.5)(1.3,0.875)
\psline{->}(1,0)(1,0.375)
\rput[bl]{0}(0.05,1.6){$a$}
\rput[bl]{0}(0.95,1.6){$b$}
\rput[bl]{0}(1.75,1.6){${c}$}
\rput[bl]{0}(1.25,0.45){$f$}
\rput[bl]{0}(0.9,-0.3){$d$}
\scriptsize
\rput[bl]{0}(1.5,0.8){$\mu$}
\rput[bl]{0}(0.7,0.25){$\nu$}
\endpspicture
.
\end{equation}

The $R$-symbols, written in components as $[R_c^{ab}]_{\mu\nu}$, are unitary matrices that describe half-braids between $a$ and $b$, which fuse to $c$. They are defined via the the following
diagram:
\begin{equation}
\pspicture[shift=-0.65](-0.1,-0.2)(1.5,1.2)
\small
\psset{linewidth=0.9pt,linecolor=black,arrowscale=1.5,arrowinset=0.15}
\psline{->}(0.7,0)(0.7,0.43)
\psline(0.7,0)(0.7,0.5)
\psarc(0.8,0.6732051){0.2}{120}{240}
\psarc(0.6,0.6732051){0.2}{-60}{35}
\psline (0.6134,0.896410)(0.267,1.09641)
\psline{->}(0.6134,0.896410)(0.35359,1.04641)
\psline(0.7,0.846410) (1.1330,1.096410)	
\psline{->}(0.7,0.846410)(1.04641,1.04641)
\rput[bl]{0}(0.4,0){$c$}
\rput[br]{0}(1.35,0.85){$b$}
\rput[bl]{0}(0.05,0.85){$a$}
\scriptsize
\rput[bl]{0}(0.82,0.35){$\mu$}
\endpspicture
=\sum\limits_{\nu }\left[ R_{c}^{ab}\right] _{\mu \nu}
\pspicture[shift=-0.65](-0.1,-0.2)(1.5,1.2)
\small
\psset{linewidth=0.9pt,linecolor=black,arrowscale=1.5,arrowinset=0.15}
\psline{->}(0.7,0)(0.7,0.45)
\psline(0.7,0)(0.7,0.55)
\psline(0.7,0.55) (0.25,1)
\psline{->}(0.7,0.55)(0.3,0.95)
\psline(0.7,0.55) (1.15,1)	
\psline{->}(0.7,0.55)(1.1,0.95)
\rput[bl]{0}(0.4,0){$c$}
\rput[br]{0}(1.4,0.8){$b$}
\rput[bl]{0}(0,0.8){$a$}
\scriptsize
\rput[bl]{0}(0.82,0.37){$\nu$}
\endpspicture
.
\end{equation}

The consistency of fusion and braiding operations is enforced via the pentagon equation, which can be schematically written as $FF = \sum FFF$, and the hexagon equations, which can be represented as $RFR = \sum FRF$ and $R^{-1}FR^{-1} = FR^{-1}F$.

The $F$ and $R$ symbols can be used to define the gauge-invariant data of the UMTC. This includes the topological $S$ matrix, which is a unitary non-degenerate symmetric matrix $S_{ab}$, and the topological twists $\theta_a$.

An anyon $a$ is Abelian if the fusion outcome with any other anyon $b$ is unique. That is, $N_{ab}^c \neq 0$ for exactly one choice of $c$, given any other anyon $b$. Such anyons $a$ form an Abelian group under fusion, denoted as $\mathcal{A}$, which plays a central role in the classification of symmetry fractionalization. 

\textit{Notation}: For later convenience we will define two quantities below. First, for an anyon $a$, we define the topological spin
\begin{align}
h: \A \rightarrow [0,1),
\end{align}
as defined by the relation
\begin{equation}\label{Eq:Defn_h}
e^{2\pi i h_a} := \theta_a.
\end{equation}
For Abelian anyons, we have $\theta_a = R^{aa}$. 

Second, we define the product
\begin{align}
  \star: \A\times \A \rightarrow [0,1)
\end{align}
between two Abelian anyons $a,b \in \mathcal{A}$ by the relation
\begin{equation}\label{Eq:Defn_star}
e^{2\pi i a \star b} := M_{a,b} = R^{ab} R^{ba} .
\end{equation}
Note that for Abelian anyons $a, b \in \mathcal{A}$, the fusion outcome of $a \times b$ is fixed, so $R^{ab}$ is short-hand for $R^{ab}_{a\times b}$. Furthermore, note that we have $2h_a \mod 1= a \star a \mod 1$, but the two sides need not be equal as real numbers since
$2 h_a \in [0,2)$.

In the above, $e^{2\pi i a \star b}$ is a symmetric bilinear form from $\mathcal{A} \times \mathcal{A} \rightarrow U(1)$, and $e^{2\pi i h_a}$ for Abelian $a$ is its quadratic refinement. 

Note that the double braid
\begin{align}
M_{ab} := R^{ab}_{a\times b} R^{b a}_{a \times b},
\end{align}
is a $U(1)$ phase whenever $a \in \mathcal{A}$, for any $b$. 
  
\textit{Anyon permutations and action on topological state space:}
The first important piece of data in specifying the way the intrinsic topological order interacts with the global symmetry group $G$ is a group homomorphism $[\rho]: G \rightarrow \text{Aut } \mathcal{C}_{\bf 0}$. Here $\text{Aut } \mathcal{C}_{\bf 0}$ is the group of intrinsic symmetries of $\mathcal{C}_{\bf 0}$, which corresponds to the group of permutations on the anyons which keep their topological properties invariant (up to gauge transformations). These are sometimes referred to as topological or anyonic symmetries. Mathematically, $\text{Aut } \mathcal{C}_{\bf 0}$ contains the group of braided auto-equivalences of $\mathcal{C}_{\bf 0}$, although it can also contain anti-unitary and parity-reversing symmetries. In this paper we will completely ignore the anti-unitary or parity-preserving part and assume always that the maps in question are unitary, parity-preserving.

A particular element $\rho_{\bf g}$  of the equivalence class $[\rho_{\bf g}]$ is an invertible map from the UMTC $\mathcal{C}_{\bf 0}$ back to itself. In particular, this has an action on the anyons:
\begin{align}
\rho_{\bf g} : a \rightarrow \,^{\bf g} a,
\end{align}
which is the same for every member of the equivalence class in $[\rho_{\bf g}]$. $\rho_{\bf g}$ further has a unitary action on the fusion and splitting spaces:
\begin{align}
\rho_{\bf g}: V_{ab}^c \rightarrow V_{\,^{\bf g}a, \,^{\bf g}b}^{\,^{\bf g} c} ,
\end{align}
which in a particular basis is written as
\begin{align}
  \rho_{\bf g} | a,b;c,\mu \rangle =\sum_\nu  [U_{\bf g}(\,^{\bf g} a,  \,^{\bf g}b; \,^{\bf g} c)] _{\mu \nu} |\,^{\bf g}a \,^{\bf g}b; \,^{\bf g} c, \nu \rangle,
\end{align}
where $|a,b;c, \mu\rangle \in V_{c}^{ab}$, $\mu = 0,\cdots N_{ab}^c - 1$. The maps $\rho_{\bf g}$ only form a group up to a special class of ``do-nothing'' maps referred to as natural isomorphisms. Changing $\rho_{\bf g}$ by a natural isomorphism is referred to as a symmetry action gauge transformation, which is further reviewed in Appendix \ref{Sec:GCReview}. 

 In this paper, we will always consider the case where the group homomorphism $[\rho]$ is trivial, in which case there is a gauge in which we can set $U = 1$ whenever $a,b,c$ are all anyons.

\begin{figure}
	\includegraphics[width=\linewidth]{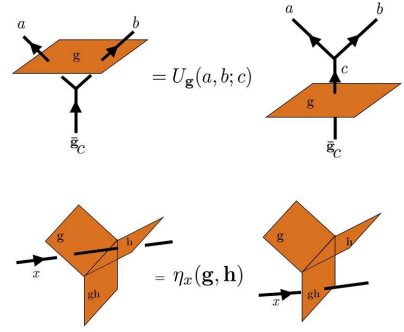}
	\caption{Diagrammatic representations of the actions of the $U$ and $\eta$ symbols. Anyon lines are black and symmetry defect sheets are orange. }\label{Fig:U_eta}
      \end{figure}

 The action $U_{\bf g}$ on the fusion and splitting spaces can be represented diagrammatically by consider a codimension-1 sheet, labeled ${\bf g}$, passing through the anyon vertex, as shown in Fig. \ref{Fig:U_eta}.

 \textit{Symmetry fractionalization}: The second important piece of data which determines the SET phase is the symmetry fractionalization class. Below we provide a description in the specific case under consideration in this paper, where symmetries do not permute anyons, so that the map $\rho$ is trivial.

Consider a many-body state of the microscopic system whose long wavelength description is in terms of the UMTC under consideration, and which contains $n$ anyons, written as $\ket{\Psi_{a_1,\dots , a_n}}$. It is an important property of the topological phase that such a state is only partially described by the Hilbert space of the TQFT, i.e. by the vector spaces defined by the UMTC $\mathcal{C}_{\bf 0}$. 

The action of the global symmetry on $\ket{\Psi_{a_1,\dots , a_n}}$ can be decomposed into local unitary operations in the neighbourhood of the $a_j$ as follows:
\begin{equation}
R_{{\bf g}}\ket{\Psi_{a_1,\dots , a_n}} = \displaystyle\prod_{j=1}^n U_{{\bf g}}^{(j)} \ket{\Psi_{a_1,\dots , a_n}}.
\end{equation}
Here $R_{\bf g}$ is the representation of ${\bf g}$ on the many-body Hilbert space of the microscopic system. We note that in the case with anyon permutations, where the map $\rho$ is non-trivial, the RHS has an additional action of $\rho_{\bf g}$, which acts only in the topological state space associated with the above many-body quantum state. In this case, the above decomposition may be in conflict with associativity for certain choices of $[\rho]$, as quantified by the symmetry-localization obstruction $[\coho{O}] \in \mathcal{H}^3_{\rho}(G, \mathcal{A})$ \cite{Barkeshli2019,barkeshli2018}. Such an obstruction indicates that the TQFT has a 2-group symmetry where $G$ and $\mathcal{A}$ are intrinsically intertwined in a way which is incompatible with $G$ being an ordinary global symmetry of the microscopic system. 

Without loss of generality \cite{Barkeshli2019}, we can consider the operators $R_{\bf g}$, which have support over the whole system, to form a linear representation of $G$. Nevertheless, the operators $U_{\bf g}^{(j)}$ need not in general form a linear representation of $G$; the manner in which they fail to do so is referred to as symmetry fractionalization. 

Let us define $^{{\bf g}}U_{{\bf h}}^{(j)} = R_{{\bf g}}U_{{\bf h}}^{(j)}R_{{\bf g}}^{-1}$. One can show that in general (assuming $[\rho]$ is trivial), we have 
\begin{equation}
U_{{\bf g}}^{(j)} (^{{\bf g}}U_{{\bf h}}^{(j)})\ket{\Psi_{a_1,\dots , a_n}} = \eta_{a_j}({\bf g},{\bf h}) U_{{\bf gh}}^{(j)}\ket{\Psi_{a_1,\dots , a_n}} ,
\end{equation}
where $\eta_a({\bf g}, {\bf h})$ is a $U(1)$ phase. We have that $\eta_{0}({\bf g},{\bf h}) = 1$ and $\eta_a({\bf g},{\bf 0}) = \eta_a({\bf 0},{\bf h}) = 1$ ({\bf 0} being the identity element in $G$), for any topological charge $a$.

$\eta_a$ must be compatible with the fusion rules:
$\eta_a({\bf g}, {\bf h}) \eta_b({\bf g}, {\bf h}) = \eta_c({\bf g}, {\bf h})$ if $N_{ab}^c > 0$, due to which we can always set \cite{Barkeshli2019}
\begin{align}
\eta_a({\bf g}, {\bf h}) = M_{a,\mathfrak{w}({\bf g}, {\bf h})},
\end{align}
for an $\mathcal{A}$-valued $2$-cocyle $\mathfrak{w}({\bf g}, {\bf h})$. Under a local redefinition of the local operators $U_{\bf g}^{(j)}$, $\mathfrak{w}({\bf g}, {\bf h})$ changes by an $\mathcal{A}$-valued 2-coboundary. Therefore distinct symmetry fractionalization patterns are characterized and classified by $\mathcal{H}^2(G, \mathcal{A})$.

We note that in the case where permutations $[\rho]$ are non-trivial, $\eta_a({\bf g}, {\bf h})$ can still be defined, however the relation with the local operators $U_{\bf g}^{(j)}$ is more non-trivial; moreover, $\mathfrak{w}({\bf g}, {\bf h})$ is no longer a $2$-cocyle. However distinct choices of $\eta_a({\bf g}, {\bf h})$ and therefore $\mathfrak{w}({\bf g}, {\bf h})$ are related to each other by an $\mathcal{A}$-valued $2$-cocycle $[\mathfrak{t}] \in \mathcal{H}^2_\rho(G, \mathcal{A})$. In this case, $\mathcal{H}^2_\rho(G, \mathcal{A})$ classifies, while $\{\eta_a({\bf g}, {\bf h})\}$ characterizes, symmetry fractionalization.

The $\eta$ symbols for the anyons can also be defined diagrammatically in terms of passing the anyon world-lines through tri-junctions of the codimension-1 symmetry defect sheets, as shown in Fig. \ref{Fig:U_eta}.

     Note also that the description above is for the case where $G$ acts as an on-site symmetry in the microscopic system; the generalization of the discussion to time-reversal and translation symmetries is discussed in Ref. \cite{Barkeshli2019}. We are not aware of a similar discussion starting from the microscopic many-body state that applies directly to spatial rotational symmetries; nevertheless, as discussed in Sec. \ref{Sec:ApptoSpSymm}, we ultimately treat spatial and internal symmetries on equal footing in the algebraic $G$-crossed BTC formalism. 

\textit{Topological charges and extrinsic defects}: Just as the fundamental objects in $\mathcal{C}_{\bf 0}$ are anyons with associative fusion and braiding, the objects in $\mathcal{C}_G^{\times}$ are comprised of anyons and symmetry defects. For each group element ${\bf g}$, there is a set $\mathcal{C}_{{\bf g}}$ containing topological charges written as $a_{{\bf g}}$. The set $\mathcal{C}_{\bf 0}$ is the set of anyons; thus $a_{\bf 0} = a$. The topological charge $a_{{\bf g}}$ is called a ${\bf g}$-defect, understood as a localized symmetry flux. Such a flux is non-dynamical and can be introduced by an appropriate deformation of the system Hamiltonian, while remaining in the same SET phase. However, there can be multiple defects associated to a ${\bf g}$-flux that are still topologically distinct from each other, in the sense that no local operation can convert one such flux to the other. This happens for example when an anyon is bound to a given ${\bf g}$-flux: no local operator can convert the original flux into a composite of that flux and an anyon.

The $G$ defects form a $G$-graded fusion category, and thus can be fused with each other as follows:
\begin{equation}
a_{{\bf g}}\times b_{{\bf h}} = \sum\limits_{c \in \mathcal{C}_{{\bf gh}}} N_{a_{\bf g} b_{\bf h}}^{c_{\bf gh}} c_{\bf gh}
\end{equation}

It is often useful to designate a special ${\bf g}$-defect to be labeled as $0_{\bf g}$. In general there is no unique way to specify $0_{{\bf g}}$. This means that one may be able to relabel $0_{{\bf g}}$ by another defect $a_{{\bf g}}$ of the same quantum dimension. Therefore the $G$-crossed data corresponding to two seemingly different SET phases may actually be equivalent upon such a relabeling of topological charges. This must be kept in mind while computing the final SET classification.

In particular, in the case where symmetries do not permute the anyons so $[\rho]$ is trivial, we can always pick a choice of $0_{\bf g}$ such that
\begin{align}
a_{\bf g} = a_{\bf 0} \times 0_{\bf g} = 0_{\bf g} \times a_{\bf 0} ,
\end{align}
and
\begin{align}
0_{\bf g} \times 0_{\bf h} = \mathfrak{w}({\bf g}, {\bf h}) \times 0_{\bf gh} . 
\end{align}
Redefining $0_{\bf g} \rightarrow \chi({\bf g}) \times  0_{\bf g}$ for an Abelian anyon $\chi({\bf g}) \in \mathcal{A}$ corresponds to changing $\mathfrak{w}({\bf g}, {\bf h})$ by an $\mathcal{A}$-valued $2$-coboundary. 

We can now define the generalized $F$-symbols $[F^{a_{{\bf g}}b_{{\bf h}}c_{{\bf k}}}_{e_{{\bf ghk}}}]$ and the generalized $R$-symbols $[R^{a_{{\bf g}}b_{{\bf h}}}_{c_{{\bf gh}}}]$ which encode several of the fusion and braiding properties of the symmetry defects. The braiding involving $G$-defects is referred to as a $G$-crossed braiding because such braidings do not generally preserve the topological charges: braiding $b_{\bf h}$ through a $a_{\bf g}$ branch sheet results in the defects $\,^{\bf g}b_{\bf h} = b'_{\bf g h g^{-1}}$ and $a_{\bf g}$.

The action $\rho_{\bf g}$ defined above on the UMTC $\mathcal{C}_{\bf 0}$ is extended to the full $G$-crossed BTC,
which means that $U_{\bf g}$ is then extended to be defined for the full $G$-crossed BTC. We consider the diagram
\begin{eqnarray}
\label{eq:GcrossedU}
\psscalebox{.6}{
	\leavevmode\pspicture[shift=-1.7](-0.8,-0.8)(1.8,2.4)
	\small
	\psset{linewidth=0.9pt,linecolor=black,arrowscale=1.5,arrowinset=0.15}
	\psline{->}(0.7,0)(0.7,0.45)
	\psline(0.7,0)(0.7,0.55)
	\psline(0.7,0.55)(0.25,1)
	\psline(0.7,0.55)(1.15,1)	
	\psline(0.25,1)(0.25,2)
	\psline{->}(0.25,1)(0.25,1.9)
	\psline(1.15,1)(1.15,2)
	\psline{->}(1.15,1)(1.15,1.9)
	\psline[border=2pt](-0.65,0)(2.05,2)
	\psline{->}(-0.65,0)(0.025,0.5)
	\rput[bl]{0}(-0.4,0.6){$x_{\bf k}$}
	\rput[br]{0}(1.5,0.65){$^{\bf \bar k}b$}
	\rput[bl]{0}(0.4,-0.4){$^{\bf \bar k}c$}
	\rput[br]{0}(1.3,2.1){$b$}
	\rput[bl]{0}(0.0,2.05){$a$}
	\scriptsize
	\rput[bl]{0}(0.85,0.35){$\mu$}
	\endpspicture
}
&=&
\sum_{\nu} \left[ U_{\bf k}\left(a, b; c\right)\right]_{\mu \nu}
\psscalebox{.6}{
	\leavevmode\pspicture[shift=-1.7](-0.8,-0.8)(1.8,2.4)
	\small
	\psset{linewidth=0.9pt,linecolor=black,arrowscale=1.5,arrowinset=0.15}
	\psline{->}(0.7,0)(0.7,0.45)
	\psline(0.7,0)(0.7,1.55)
	\psline(0.7,1.55)(0.25,2)
	\psline{->}(0.7,1.55)(0.3,1.95)
	\psline(0.7,1.55) (1.15,2)	
	\psline{->}(0.7,1.55)(1.1,1.95)
	\psline[border=2pt](-0.65,0)(2.05,2)
	\psline{->}(-0.65,0)(0.025,0.5)
	\rput[bl]{0}(-0.4,0.6){$x_{\bf k}$}
	\rput[bl]{0}(0.4,-0.4){$^{\bf \bar k}c$}
	\rput[bl]{0}(0.15,1.2){$c$}
	\rput[br]{0}(1.3,2.1){$b$}
	\rput[bl]{0}(0.0,2.05){$a$}
	\scriptsize
	\rput[bl]{0}(0.85,1.35){$\nu$}
	\endpspicture
}
\end{eqnarray}
where one can also choose $a$, $b$, and $c$ to be defects with non-trivial $G$-grading. Here, the worldlines of the defects $a_{\bf g}$ are understood to be codimension-2 boundaries of codimension-1 symmetry defect sheets labeled by ${\bf g}$, which go into the page. 

As discussed above, since the symmetry action on anyons is trivial, $U_{\bf g}$ is trivial when restricted to the anyons. However, the symmetry may still permute defects, for example if $G$ is a non-Abelian symmetry. Therefore $U_{{\bf  g}}$ can still be non-trivial when acting on the symmetry defect fusion and splitting states.

Similarly, the $\eta$ symbol also gets extended to be defined for the full $G$-crossed BTC, corresponding to the following diagram:

\begin{align}
\psscalebox{.6}{
	\pspicture[shift=-1.7](-0.8,-0.8)(1.8,2.4)
	\small
	\psset{linewidth=0.9pt,linecolor=black,arrowscale=1.5,arrowinset=0.15}
	\psline(-0.65,0)(2.05,2)
	\psline[border=2pt](0.7,0.55)(0.25,1)
	\psline[border=2pt](1.15,1)(1.15,2)
	\psline(0.7,0.55)(1.15,1)	
	\psline{->}(0.7,0)(0.7,0.45)
	\psline(0.7,0)(0.7,0.55)
	\psline(0.25,1)(0.25,2)
	\psline{->}(0.25,1)(0.25,1.9)
	\psline{->}(1.15,1)(1.15,1.9)
	\psline{->}(-0.65,0)(0.025,0.5)
	\rput[bl]{0}(-0.5,0.6){$x$}
	\rput[bl]{0}(0.5,1.3){$^{\bf \bar g}x$}
	\rput[bl]{0}(1.6,2.1){$^{\bf \bar h \bar g}x$}
	\rput[bl]{0}(0.4,-0.35){$c_{\bf gh}$}
	\rput[br]{0}(1.3,2.1){$b_{\bf h}$}
	\rput[bl]{0}(0.0,2.05){$a_{\bf g}$}
	\scriptsize
	\rput[bl]{0}(0.85,0.35){$\mu$}
	\endpspicture
}
=
\eta_{x}\left({\bf g},{\bf h}\right)
\psscalebox{.6}{
	\pspicture[shift=-1.7](-0.8,-0.8)(1.8,2.4)
	\small
	\psset{linewidth=0.9pt,linecolor=black,arrowscale=1.5,arrowinset=0.15}
	\psline(-0.65,0)(2.05,2)
	\psline[border=2pt](0.7,0)(0.7,1.55)
	\psline{->}(0.7,0)(0.7,0.45)
	\psline(0.7,1.55)(0.25,2)
	\psline{->}(0.7,1.55)(0.3,1.95)
	\psline(0.7,1.55) (1.15,2)	
	\psline{->}(0.7,1.55)(1.1,1.95)
	\psline{->}(-0.65,0)(0.025,0.5)
	\rput[bl]{0}(-0.5,0.6){$x$}
	\rput[bl]{0}(1.6,2.1){$^{\bf \bar h \bar g}x$}
	\rput[bl]{0}(0.4,-0.35){$c_{\bf gh}$}
	\rput[br]{0}(1.3,2.1){$b_{\bf h}$}
	\rput[bl]{0}(0.0,2.05){$a_{\bf g}$}
	\scriptsize
	\rput[bl]{0}(0.85,1.35){$\mu$}
	\endpspicture
}
\qquad
.
\end{align}

The set of consistency equations that must be satisfied by the data $\{F, R, U, \eta\}$ of the $G$-crossed BTC is summarized in Appendix \ref{Sec:GCReview}. Furthermore, the data $\{F, R, U, \eta\}$ is subject to two distinct types of gauge transformations, referred to as symmetry action and vertex basis gauge transformations, which are also summarized in Appendix \ref{Sec:GCReview}. 

\subsubsection{Classification}

As mentioned above, symmetry fractionalization is classified by $\H^2(G,\mathcal{A})$, which in the case where symmetries do not permute anyons can be understood as distinct possible $\mathcal{A}$-valued 2-cocycles $\mathfrak{w}({\bf g}, {\bf h})$, modulo $\mathcal{A}$-valued $2$-coboundaries that correspond to redefining $0_{\bf g}$.

We can define $\mathfrak{w}$ purely in terms of the symmetry action on the anyons. However, in order to define the full theory $\mathcal{C}_G^{\times}$, we need to show that $[\mathfrak{w}]$ is compatible with the fusion and braiding of ${\bf g}$-defects. In general, each symmetry fractionalization class defines an obstruction class $[\mathcal{O}] \in \H^4(G,U(1))$, which is referred to as a symmetry fractionalization anomaly, or a 't Hooft anomaly. If $[\mathcal{O}]$ is nontrivial, there is no solution to the full $G$-crossed consistency equations, and the fractionalization class $[\mathfrak{w}]$ cannot be associated to an SET phase. Physically, a solution with a non-trivial $\mathcal{H}^4$ obstruction is understood as an anomalous SET \cite{Barkeshli2020Anomaly,Bulmash2020,Chen2014}. Viewing $\H^4(G,U(1))$ as the classification of SPT phases of $G$ in (3+1) dimensions, we say that the state with the symmetry fractionalization class $[\mathfrak{w}]$ can only be realized on the surface of the SPT corresponding to the class $[\mathcal{O}]$, but not in a standalone (2+1)D system. This has been demonstrated in full generality in Ref. \onlinecite{Bulmash2020}. In the present work all our choices of $G$ have trivial $\H^4(G,U(1))$.

When the obstruction class $[\mathcal{O}]$ is trivial, we can finally solve the $G$-crossed consistency equations and obtain the full description of the SET phase. Given one solution $\mathcal{C}_G^{\times}$, it has been proved (within our usual assumptions) that a complete set of solutions to the $G$-crossed consistency equations can be obtained by stacking an arbitrary (2+1)D $G$-SPT state onto the SET given by $\mathcal{C}_G^{\times}$. Therefore the set of gauge-inequivalent solutions for a particular symmetry fractionalization class is given by a torsor over $\H^3(G,U(1))$, the group of $G$-SPT phases in (2+1)D; each solution of this type is said to belong to a unique defect class. 

Although mathematically complete, this procedure in general has redundancies: different defect classes may turn out to describe the same phase upon relabelling the defects in the respective theories, so the correct defect classification is in fact a torsor over a subgroup of $\H^3(G,U(1))$; this subtlety must be taken into account when determining the final SET classification. 

\subsection{Brief review of crystalline gauge theory}
\label{Sec:ReviewCGT}

In this section we will briefly describe the crystalline gauge fields that were defined in Ref \cite{Manjunath2020} (and to which the reader is referred for a more detailed treatment). In this work they are used to write down effective actions for Abelian topological orders enriched by an orientation-preserving spatial symmetry group $G$. The approach based on topological actions is particularly useful in describing the various types of topological response allowed in these systems. Therefore we will use it in parallel with the $G$-crossed treatment to gain additional insight into the behaviour of clean quantum Hall systems in the topological limit. 

Consider a $(2+1)$D space-time manifold $M = \Sigma^2 \times \mathbb{R}$, where $\Sigma^2$ is the space on which the clean lattice system is defined. We fix an arbitrary triangulation of $M$ and we define on the links a gauge field valued in the symmetry group $G$. For example, we define a $U(1)$ gauge field $A_{ij}$ on the link $ij$ of the triangulation, with the link directed towards $j$ (with $A_{ij} = -A_{ji}$ and $A_{ij} \sim A_{ij} + 2\pi$). More generally, the gauge field $B$ places a group element lifted to the real numbers on each link $ij$; this group element is written as $B_{ij}$. When $G$ is a crystalline symmetry group with discrete $\Z^2$ translations and $\Z_M$ point group rotations, we write $B_{ij} = (\vec{R}_{ij},C_{ij})$, where $\frac{1}{2\pi}\vec{R} \in \Z^2$ and $C \in \frac{2\pi}{M} \Z$. When $G$ is a central extension of such a space group by $U(1)$, we define $B_{ij} = (A_{ij},\vec{R}_{ij},C_{ij})$ where $A$ is now the $U(1)$ component of the full $G$ gauge field. Since these gauge fields are non-Abelian, the product $B_{ij}B_{kl}$ is defined by the group multiplication law of $G$, and due to this, the components $A,\vec{R}$ and $C$ are not independent.

The topological effective Lagrangian is
\begin{align}
 \mathcal{L} =- \frac{1}{4\pi} a^I \cup K_{IJ} da^J +\mathcal{L}_{frac}+\mathcal{L}_{SPT}.
 \end{align}
The non-degenerate $D \times D$ symmetric integer matrix $K$, which couples the dynamical $U(1)$ gauge field $a^I$, characterizes the intrinsic topological
order \cite{Wenbook}. As in the case of the symmetry gauge fields discussed above, the gauge field $a^I$ is a lift of a gauge field configuration of $U(1)$ group elements to $\mathbb{R}$. Topologically distinct quasiparticles correspond to integer vectors $\vec{l} \sim \vec{l} + K \vec{\Lambda}$, where $\vec{l}, \vec{\Lambda} \in \Z^D$.
The quasiparticles form an Abelian group $\A = \Z_{n_1}\times\dots\times\Z_{n_D}$ under fusion, where the $n_i$ are the diagonal entries in the Smith normal form of $K$.

The terms in $\mathcal{L}_{frac}$ can be given a general interpretation as follows: they associate flux of the symmetry group $G$ to anyons, and therefore collectively describe the symmetry fractionalization class of the SET. In our examples these fluxes are of four types, and are defined on 2-simplices of the triangulation as follows: (i) the $U(1)$ flux $dA$, where $d$ is the simplicial coboundary operator, i.e. $dA[012] = A_{01} + A_{12} - A_{20}$; (ii) the $\Z_M$ flux $dC$, which corresponds to the disclination density; (iii) the flux $d\cancel{\vec{R}} = (1-h)^{-1} d\vec{R}$, whose fractional part represents the local, gauge-invariant part of the dislocation density. Here $h$ is the $2 \times 2$ matrix representation of the generator of the point group rotations, written in the lattice basis. This matrix implements the physical fact that the allowed choices for the gauge-invariant dislocation flux are quantized by rotation symmetry (see Ref \cite{Manjunath2020} for further explanation); and (iv) the area flux $A_{XY}$, which measures the number of unit cells in a spatial slice of the manifold $M$. The terms in $\mathcal{L}_{frac}$ couple the above symmetry fluxes to the internal gauge fields.

The terms in $\mathcal{L}_{SPT}$ involve only the symmetry gauge fields. They can also be given a general interpretation: in each SPT term, a symmetry charge is associated to one of the above symmetry fluxes.

As is familiar from the continuum CS theory for $U(1)$ symmetry, the internal gauge fields can be integrated out to give an effective response theory. Such a theory encodes the topological response in terms of the SPT action as well as contributions due to symmetry fractionalization (for example, in a theory with only $U(1)$ charge conservation symmetry, the response coefficient is the Hall conductivity, which has an integer part contributed by an IQH state, and a fractional part coming from the symmetry fractionalization, as specified by the charge vector of the theory). The various examples studied in later sections will highlight this feature.   

\section{Topological invariants from G-Crossed BTCs and response theories}
\label{Sec:SummaryOfResults}

In this section we summarize in detail the results of the $G$-crossed BTC analysis. We derive formulas for the invariants that characterize distinct $G$-crossed BTCs, and thus provide symmetry-protected topological invariants that characterize distinct SETs. These invariants were briefly summarized in Section \ref{Sec:NTSummary}. 

The SET can be physically characterized by its topologically quantized response properties, which are most easily studied by writing down topological field theories in terms of background gauge fields associated with the global symmetry. We have done this for Abelian topological orders described by a $K$ matrix. As discussed in the preceding section, in general, the effective action is given by $\mathcal{L} = -\frac{1}{4\pi} K_{IJ} a^I \cup d a^J + \mathcal{L}_{frac} + \mathcal{L}_{SPT}$. Here $a^I$ represent internal dynamical gauge fields associated to the topological order. The terms in $\mathcal{L}_{frac}$ describe symmetry fractionalization by coupling the fields $a^I$ to the background symmetry gauge fields, while those in $\mathcal{L}_{SPT}$ involve only the symmetry gauge fields. The approaches based on the $G$-crossed theory and the effective action together provide a unified understanding of the SET phase and the invariants that characterize it.  
Both the $G$-crossed BTC and effective action approaches require a knowledge of the cohomology groups of $G$ as well as a representative set of cocycles for symmetry fractionalization and for $G$-SPTs. These details, which are more technical, will be discussed in the appendices.

We note that in all our examples, $\mathcal{H}^4(G, U(1))$ vanishes. Therefore, there are no symmetry fractionalization anomalies to be concerned about. Furthermore, since the symmetries do not permute anyon types, the symmetry localization obstruction \cite{Barkeshli2019,barkeshli2018}, valued in $\H^3(G, \A)$, also vanishes. 

\subsection{General results on invariants in $G$-crossed BTCs}
\label{Sec:GenResInvts}

In this paper, we repeatedly use certain general identities to construct $G$-crossed invariants distinguishing the symmetry fractionalization and defect classes. These invariants are summarized in Tables \ref{Table:InvtSummary} and \ref{Table:LatticeInvts} for the continuum and lattice cases respectively. A description of these formulas and their physical meaning is provided below. The mathematical derivation of the identities can be found in Appendix \ref{Sec:GCReview}.

\begin{table*}[t]
	\centering
	\begin{tabular} {l |p{0.4\textwidth} |p{0.4\textwidth}}
		\hline
	  Invariant & Formula & Details \\ \hline
	  \multicolumn{3}{c}{Symmetry fractionalization invariants, $\A = \Z_{n_1}\times\dots\times\Z_{n_r};  p = \text{lcm}(n_1,\dots , n_r)$} \\ \hline
	  $\{Q_a\}$& $e^{2\pi i Q_a} = M_{v,a} = \F_a( {\bf  g}_1,\dots , {\bf  g}_p)$ & ${\bf  g}_1 = \dots = {\bf  g}_p = (e^{i 2\pi/p},{\bf  0},1)$ \\ 
	  $\{L_a\}$& $e^{2\pi i L_a} = M_{s,a} = \F_a( {\bf  g}_1,\dots , {\bf  g}_p)$ & ${\bf  g}_1 = \dots = {\bf  g}_p = (1,{\bf  0},e^{i 2\pi/p})$ \\ \hline
	  \multicolumn{3}{c}{Defect invariants} \\ \hline
	  $\bar{\sigma}_H$ & $e^{2\pi i \frac{\bar{\sigma}_H}{2p}} = \mathcal{I}_p(0_{{\bf  g}})$  & ${\bf  g} =(e^{i 2\pi/p},{\bf  0},1)$  \\
	  $\mathscr{S}$ & $e^{2\pi i \frac{\mathscr{S}}{p}} = \frac{\mathcal{I}_p(0_{{\bf gh}})}{\mathcal{I}_p(0_{{\bf g}})\mathcal{I}_p(0_{{\bf h}})}$  & ${\bf  g} =(e^{i 2\pi/p},{\bf  0},1);{\bf h} =(1,{\bf  0},e^{i 2\pi/p}) $  \\
	  $\ell_s$ & $e^{2\pi i \frac{\ell_s}{2p}} = \mathcal{I}_p(0_{{\bf h}})$  & ${\bf h} =(1,{\bf  0},e^{i 2\pi/p})$  \\
		\hline
		\multicolumn{3}{c}{Additional quantized invariants} \\ \hline
		$\nu$ & $e^{2\pi i \frac{\nu}{n}} = \frac{\eta_{0_{{\bf  g}}}({\bf r}_1,{\bf r}_2)}{\eta_{0_{{\bf  g}}}({\bf r}_2,{\bf r}_1)}$  & ${\bf  g} =(e^{i 2\pi/n},{\bf  0},1)$; ${\bf r}_1,{\bf r}_2$ span a magnetic u.c. \\\hline
	\end{tabular}
	\caption{List of $G$-crossed invariants studied in this paper for $G= U(1) \leftthreetimes \mathbb{E}^2$. A general group element is written as ${\bf  g} = (e^{2\pi i z},{\bf r},e^{2\pi i h})$, where the three components refer to $U(1)$, translation and rotation symmetries respectively. }\label{Table:InvtSummary}
\end{table*}  

\subsubsection{Symmetry fractionalization invariants}

In general, the classification of symmetry fractionalization is of the form $\H^2(G,\A) \cong \A^d / \sim$, meaning that the symmetry fractionalization classes are specified by a set of $d$ anyons
\begin{align}
  (f_1,\dots , f_d) \in \A^d,
  \end{align}
modulo some equivalence relation $\sim$. Here the value of $d$ depends on $G$. Therefore, identifying the $f_i$ will allow us to fix the symmetry fractionalization class. The $f_i$ can in turn be found by measuring the quantities $M_{f_i,a}$ for each anyon $a$. Below we discuss a general prescription to write down formulas for the quantities $M_{f_i,a}$ in terms of the $G$-crossed data.

Let us fix a set of defects $0_{\bf g}$ for every ${\bf g} \in G$, which are Abelian, in the sense that they have unit quantum dimension, $d_{0_{\bf g}} = 1$. 
Any two such Abelian defects $0_{{\bf  g}}$ and $0_{{\bf  h}}$ fuse to give the defect $0_{{\bf gh}}$ and an anyon $\mathfrak{w}({\bf  g},{\bf h})$, according to the relation
\begin{equation}
0_{{\bf  g}}\times 0_{{\bf h}} = 0_{{\bf gh}} \times \mathfrak{w}({\bf  g},{\bf h}).
\end{equation}

We can generalize this fusion rule to an arbitrary number of defects $0_{{\bf  g}_i}, i = 1, \dots , n$: 
\begin{widetext}
\begin{equation}
0_{{\bf  g}_n}\times \dots \times 0_{{\bf  g}_1}= \mathfrak{w}({\bf  g}_2,{\bf  g}_1)\times \mathfrak{w}({\bf  g}_3,{\bf  g}_2{\bf  g}_1)\times\dots\times \mathfrak{w}({\bf  g}_n,{\bf  g}_{n-1}\dots {\bf  g}_1) \times 0_{{\bf  g}_n\dots {\bf  g}_1}.
\end{equation}
\end{widetext}
The anyon on the rhs can be measured in terms of $G$-crossed data. Suppose we perform the sequence of symmetry operations ${\bf  g}_1, \dots , {\bf  g}_n$ in the neighbourhood of $a$. In terms of the $G$-crossed data, we have the following identity (see Appendix \ref{Sec:SymFracInvt} for a proof):

\begin{align}
 \label{FaDef}
& \mathcal{F}_a({\bf g}_1, \dots , {\bf g}_n) \nonumber \\
&:= \frac{\left(\prod\limits_{i=1}^{n} R^{a, 0_{{\bf g}_i}} R^{ 0_{{\bf g}_i},a}\right)}{ R^{a, 0_{{\bf  g}_n \dots {\bf  g}_1}} R^{ 0_{{\bf  g}_n \dots {\bf  g}_1,a}}} \prod\limits_{i=1}^{n-1}\eta_{a}({\bf g}_{i+1},{\bf g}_i\dots {\bf g}_1) \nonumber \\ &=  M_{a,\prod_{i=1}^{n-1}\mathfrak{w}({\bf g}_{i+1},{\bf g}_i\dots {\bf g}_1)}
\end{align} 

Now we claim that for each $f_j$, we can find a sequence of operations $\{{\bf  g}_i\}$ such that
\begin{align}
  \label{fDef}
  f_j = \prod_{i=1}^{n-1}\mathfrak{w}({\bf g}_{i+1},{\bf g}_i\dots {\bf g}_1).
  \end{align}
  For example, when the ${\bf  g}_i$ correspond to $U(1)$ rotations by the angle $2\pi/n$ for $i = 1,2,\dots , n$, the anyon $f$ coresponds to the vison $v$ which is induced by a $2\pi$ flux insertion. When the ${\bf  g}_i$ correspond to a sequence of translations which transport $a$ around a lattice unit cell, $f$ corresponds to the anyon $m$ associated to each unit cell of the lattice. We do not present here a general method to obtain the correct sequence $\{{\bf  g}_i\}$ associated to each $f_j$, for arbitrary groups $G$; in our examples this has been done on an individual basis, by inspection.

  The above identity depends on a particular choice of the defect $0_{{\bf  g}}$. However, this choice is not necessarily canonical. Consider changing $\coho{w}$ by a coboundary, $\mathfrak{w}({\bf  g},{\bf h})\rightarrow \mathfrak{w}({\bf  g},{\bf h}) \times \chi({\bf  g}) \chi({\bf h})\overline{\chi({\bf gh})}$, for $\chi({\bf  g}) \in \A$.
Then,
\begin{align}
  \label{fGauge}
  f_i \rightarrow f_i \times \overline{\chi({\bf g}_n\dots {\bf  g}_1)}\times \prod_{i=1}^{n} \chi({\bf  g}_i).
  \end{align}

Such a change is effectively the same as relabeling the defects $0_{{\bf  g}} \rightarrow 0_{{\bf  g}}\times \chi({\bf  g})$. Therefore the anyon associated to the fusion of defects has an ambiguity under a relabeling of the defects. In general, this means that different choices of $\{f_i\}$ may correspond to the same symmetry fractionalization class. This redundancy is in fact the origin of the equivalence relation in the definition of $\H^2(G,\A)$. 

The fractional quantum numbers which determine the symmetry fractionalization class are therefore specified by the set
\begin{align}
\{M_{a, f_i}\}, \;\; i = 1, \cdots, d, \;\; \forall a \in \mathcal{C}_0
\end{align}
subject to the equivalence relation of Eq. \ref{fGauge} for $f_i$.

\subsubsection{Physical responses and defect (SPT) invariants}

Here we discuss methods to obtain invariant formulas for the gauge invariant quantities that capture the fractionally quantized physical responses. Once the symmetry fractionalization is fixed, these quantized physical responses then fix the freedom associated with $\mathcal{H}^3(G, U(1))$, which then completely characterizes the symmetry defect properties. Unlike in the case of symmetry fractionalization, we do not know a completely general procedure to determine these defect invariants. However, all the defect invariants presented in this paper have been obtained using two general formulas. 

First, consider a defect $0_{{\bf  g}}$ such that
\begin{align}
  \label{order0g}
  0_{{\bf  g}}^n = f
\end{align}
for some integer $n$. To obtain nontrivial invariants, $f$ should correspond to some combination of the anyons $f_i$ which characterize the symmetry fractionalization. The defect invariant is then written as 
\begin{align}
  \label{InDef}
  \mathcal{I}_n (0_{{\bf g}}) := [\mathcal{T}^n]^{ ({\bf g}, {\bf 0})}_{0_{\bf g} 0_{\bf g}}
=  \theta_{0_{\bf g}}^n \prod_{j = 0}^{n-1} \eta_{0_{\bf g}}({\bf g}, {\bf g}^j).
\end{align}
Here $\mathcal{T}$ refers to the $G$-crossed modular $\mathcal{T}$ matrix, defined in Ref. \cite{Barkeshli2019}. As we explain below, the invariant $\mathcal{I}_n(0_{\bf g})$ measures one half of the ${\bf g}$-symmetry charge that is associated to the ${\bf  g}$-defect (${\bf g}$-flux) $0_{\bf g}$. That is, if we let $Q^{({\bf g})}_{0_{\bf g}}$ be the charge of $0_{\bf g}$ under symmetry transformations generated by ${\bf g}$, we can write
\begin{align}
  \label{InCharge}
\mathcal{I}_n(0_{\bf g}) = e^{2\pi i Q^{({\bf g})}_{0_{\bf g}} / 2}
\end{align}
For example, if ${\bf  g}$ denotes $\frac{1}{n}$th of a complete $U(1)$ rotation, the above invariant measures $\frac{1}{2} \times \frac{\bar{\sigma}_H}{n}$, i.e. $\mathcal{I}_n(0_{\bf g}) = e^{i 2\pi \bar{\sigma}_H/2n}$. As such, the argument of $\mathcal{I}_n(0_{\bf g})$ can be associated with the fractionally quantized physical responses of the system, such as the Hall conductivity.

In Appendix \ref{Sec:GCReview} it is shown that in general, when $0_{{\bf  g}}^n = f$, we have
\begin{align}\label{PureSPTInvtFormula}
(\mathcal{I}_n)^{2n}(0_{{\bf  g}}) &= M_{f,f} = e^{2\pi i f \star f},
\end{align}
which implies
\begin{align}
\mathcal{I}_n(0_{\bf g}) = e^{i \pi \frac{f \star f + k}{n}} .
\end{align}
The integer part $k \in \Z_{2n}$ characterizes the defect class. 

Using Eq. \ref{PureSPTInvtFormula}, we can understand Eq. \ref{InCharge} as follows. Suppose $n$ copies of the flux $0_{\bf g}$ fuse to a single flux quantum (i.e. $n$ is the minimal order of ${\bf g}$). We know that $M_{f,f}$ is the phase obtained by adiabatically transporting $n$ copies of the flux $0_{{\bf g}}$ around another set of $n$ copies of $0_{\bf g}$ and completing a full braid. Now the fractional ${\bf g}$-charge $Q_{0_{\bf g}}^{({\bf g})}$ of $0_{\bf g}$ can be measured by taking $0_{\bf g}$ around a unit flux quantum, i.e. taking $0_{\bf g}$ around a set of $n$ copies of $0_{\bf g}$. Therefore taking $n$ copies of $0_{\bf g}$ around another $n$ copies of $0_{\bf g}$ should give
\begin{equation}
e^{2\pi i n Q^{({\bf g})}_{0_{\bf g}}} = M_{f,f} = \mathcal{I}_n^{2n}(0_{\bf g}). 
\end{equation}
This justifies Eq. \ref{InCharge}.

In the example where ${\bf  g}$ is a pure $U(1)$ symmetry element, this means that $(\mathcal{I}_n)^{2n}(0_{{\bf  g}}) = M_{v,v} = e^{2\pi i v \star v}$, where $v$ is the vison. Therefore, we must have $\mathcal{I}_n (0_{{\bf g}}) = e^{\pi i \frac{v \star v + k}{n}} $ for some $k \in \Z_{2n}$. The integer $k \text{ mod } 2n$ (which is even for bosonic systems) represents the contribution of bosonic integer quantum Hall states to $\bar{\sigma}_H$, and characterizes the defect class. In the $U(1)$ case, we can make $n$ arbitrarily large, so that the defect class is actually characterized by an integer $k \in \Z$. In the absence of topological order, $k$ parametrizes the $U(1)$ SPT state. 

$\mathcal{I}_n$ is invariant under the gauge transformations of the $G$-crossed theory discussed in Appendix \ref{Sec:GCReview}. Therefore, if we fix the defect $0_{{\bf  g}}$, this formula is fully invariant. However, it is in general not invariant under relabelings of defects. From Eq. \eqref{order0g}, we see that a relabeling of the defect $0_{{\bf  g}} \rightarrow 0_{\bf g} \chi({\bf g})$, for $\chi({\bf g}) \in \A$, will generally change the obtained value of $k$. Conversely, changing the integer $k$ by certain integers may be compensated for by a relabeling of $0_{\bf g}$. This corresponds to the known ambiguity that changing the $G$-crossed data by different elements of $\mathcal{H}^3(G, U(1))$ may lead to the same SET upon a relabeling of defects. In other words, the classification of distinct SETs requires a reduction of $\mathcal{H}^3(G, U(1))$. In Appendix \ref{Sec:relabeling} we use this redundancy in the defect invariant to determine the correct counting of SET phases for different lattice FQH systems.

Eq. \eqref{PureSPTInvtFormula} can be further utilized to determine mixed defect invariants. We will demonstrate the basic idea for the example of $G=\Z_m\times\Z_n$. Similar ideas can be used to find mixed defect classes associated to a $\Z_m\times\Z_n$ subgroup of $G$, if such defect classes exist within the group cohomology classification. These ideas will be used to calculate several mixed defect invariants in this paper.  
 
Suppose $G=\Z_m \times \Z_n$. Since $\H^3(\Z_m\times\Z_n,U(1)) \cong \Z_m\times\Z_n\times\Z_l$, we see that there are $l = \text{gcd}(m,n)$ mixed defect classes. Suppose the $\Z_n$ and $\Z_m$ subgroups are generated by ${\bf g}$ and ${\bf h}$, respectively, and we have the defect fusion rules
\begin{align}
  0_{\bf g}^n &= p \in \A
  \nonumber \\
  0_{\bf h}^m &= q \in \A.
  \end{align}
Let $L = \text{ lcm}(m,n)$ and $l = \gcd(m,n)$. Also suppose that the system does not have mixed symmetry fractionalization associated to the $\Z_m\times\Z_n$ symmetry (this will be true for all examples considered in this paper), so that we can pick a gauge in which 
\begin{equation}
0_{\bf g}\times 0_{\bf h} = 0_{\bf gh}.
\end{equation}
Then it can be shown (see Appendix \ref{Sec:DefectInvts}) that the defect $0_{\bf gh}$ is of order $L$, and consequently
\begin{equation}\label{MixedSPTInvtFormula}
\left( \frac{\mathcal{I}_L(0_{{\bf  g}{\bf h}})}{\mathcal{I}^{n/l}_m(0_{{\bf  h}})\mathcal{I}^{m/l}_n(0_{{\bf g}})} \right)^{L} = \left(\frac{\theta_{p\times q}}{\theta_p \theta_q}\right)^{L/l} = M_{p,q}^{L/l} . 
\end{equation}

Therefore this invariant has the form
\begin{align}\label{Eq:MixedTInvt}
\frac{\mathcal{I}_L(0_{{\bf g}{\bf h}})}{\mathcal{I}^{n/l}_m(0_{{\bf  g}})\mathcal{I}^{m/l}_n(0_{{\bf h}})} = e^{2\pi i \frac{p \star q + k}{l}}.
\end{align}
Now the integer $k \in \Z_l$ parametrizes the defect class; in the absence of topological order, it gives the mixed SPT invariant. In the simplest case, where $L=l=m=n$, this becomes

\begin{align}
\frac{\mathcal{I}_n(0_{{\bf g}{\bf h}})}{\mathcal{I}_n(0_{{\bf  g}})\mathcal{I}_n(0_{{\bf h}})} = e^{2\pi i \frac{p \star q + k}{n}}.
\end{align}

Eq. \eqref{Eq:MixedTInvt} also has a natural physical interpretation in terms of flux braiding. Suppose we raise Eq. \eqref{Eq:MixedTInvt} to the power $2L$. The terms on the lhs can then be interpreted as follows:
\begin{enumerate}
	\item The term $\mathcal{I}_L^{2L}(0_{\bf gh})$ in the numerator measures the braiding phase between $L$ copies of $0_{\bf g}$ and $0_{\bf h}$ with another set of $L$ copies of $0_{\bf g}$ and $0_{\bf h}$. \\
	\item The term $\mathcal{I}^{2Ln/l}_m(0_{{\bf  g}}) = \mathcal{I}^{2m (n/l)^2}_m(0_{{\bf  g}})$ in the denominator measures the braiding phase between $m (n/l)$ copies of $0_{\bf g}$ and another set of $m(n/l)$ copies of $0_{\bf g}$. Since $mn/l=L$, this braiding phase is cancelled by an equal contribution from the braiding of $0_{\bf g}$ defects in the numerator. Simlarly, the contribution from the second term in the denominator is cancelled by the braiding of $0_{\bf h}$ defects in the numerator.
\end{enumerate}
The remaining contribution can be written as $e^{i 2 \theta}$, where $e^{i \theta}$ is the braiding phase between $L$ copies of $0_{\bf g}$ and $L$ copies of $0_{\bf h}$. Note that the braiding phase between $0_{\bf g}$ and $m$ copies of $0_{\bf h}$ defines $e^{2\pi i Q^{({\bf h})}_{0_{\bf g}}}$, the ${\bf h}$-charge of $0_{\bf g}$. Now $e^{i 2 \theta}$ measures the phase due to $2L^2/m$ such braids. Thus we can write
\begin{equation}
  \label{InQgh}
\left( \frac{\mathcal{I}_L(0_{{\bf  g}{\bf h}})}{\mathcal{I}^{n/l}_m(0_{{\bf  h}})\mathcal{I}^{m/l}_n(0_{{\bf g}})} \right)^{2L} = e^{2\pi i \frac{2L^2}{m}Q^{({\bf h})}_{0_{\bf g}}}
\end{equation}
We conclude that the mixed defect invariant measures the fractional part of $\frac{L}{m} Q^{({\bf h})}_{0_{\bf g}}$. In fact, we can use the rhs of Eq. \eqref{Eq:MixedTInvt}, along with the identity $L/m=n/l$, to write
\begin{equation}
\frac{n}{l} Q^{({\bf h})}_{0_{\bf g}} \mod 1= \frac{p \star q + k}{l} \mod 1.
\end{equation}

By interchanging ${\bf g}$ and ${\bf h}$ in the argument, we can also see that the invariant measures the fractional part of $\frac{L}{n} Q^{({\bf g})}_{0_{\bf h}}$. When $m=n=L$, the invariant directly gives $Q^{({\bf h})}_{0_{\bf g}}=Q^{({\bf g})}_{0_{\bf h}}$. 

The second general formula that we use to obtain fractionally quantized responses applies to systems with translation symmetry and a magnetic field, either discrete or continuous. In general it measures a generalized filling, $\nu_{H}$, defined as the $H$ symmetry charge per magnetic unit cell. Consider any defect $0_{{\bf  h}}$, where ${\bf h}^n = {\bf 0}$, such that ${\bf h}$ generates a $\Z_n$ subgroup $H \cong \Z_n \subseteq G$. 
Consider also the translations ${\bf r}_1$ and ${\bf r}_2$, which are assumed to span a single magnetic unit cell. In particular we require that ${\bf r_1},{\bf r_2}$ commute with each other, as well as with $H$. Then the formula for this invariant is
\begin{equation}
  \label{nuHDef}
e^{2\pi i \nu_{H} / n} = \frac{\eta_{0_{{\bf  h}}}({\bf r}_1,{\bf r}_2)}{\eta_{0_{{\bf  h}}}({\bf r}_2,{\bf r}_1)}.
\end{equation}

Let
\begin{align}
\mathfrak{b} := \frac{\coho{w}({\bf r}_1, {\bf r}_2)}{\coho{w}({\bf r}_1, {\bf r}_2)} \in \A
\end{align}
be the anyon per magnetic unit cell, which is a gauge-invariant quantity (i.e. invariant under coboundary operations). 
In general, this will consist of the anyons in each unit cell, and also the anyon associated to the flux quantum contained in the magnetic unit cell. We can show that if
\begin{align}
0_{{\bf  h}}^n = f,
\end{align}
then
\begin{equation}
  e^{2\pi i \nu_H} = \left(\frac{\eta_{0_{{\bf  h}}}({\bf r}_1,{\bf r}_2)}{\eta_{0_{{\bf  h}}}({\bf r}_2,{\bf r}_1)}\right)^n = M_{f,\mathfrak{b}}.
\end{equation}
Therefore this invariant takes the form 

\begin{equation}
e^{2\pi i \nu_H/n} = e^{2\pi i \frac{f \star \mathfrak{b} + k}{n}}.
\end{equation}
The integer $k$ in this case parametrizes the defect (SPT) class modulo $n$, and represents the integer part of the total $H$-charge associated to each magnetic unit cell. As in the case of the formulas discussed above, this quantity is invariant under the gauge transformations of the theory but may have an ambiguity upon relabeling $0_{{\bf  h}}$.

Next we will study the different symmetry groups of interest one by one. Proofs and further derivations of the results stated below can be found in Appendix \ref{Sec:Invts}; Tables \ref{Table:InvtSummary} and \ref{Table:LatticeInvts} summarize the $G$-crossed invariants for the continuum and discrete cases respectively.

\subsection{$G = U(1)$} 
\label{Sec:U1Summary}

The classification of SETs with $U(1)$ symmetry and a fixed intrinsic topological order is $\A \times \Z$, before accounting for relabelings of anyons and defects. The factor $\A$ comes from the symmetry fractionalization, $\H^2(U(1), \A) = \A$. The factor of $\Z$ comes from the defect class $\H^3(U(1), U(1)) = \Z$, and specifies the even integer part of the Hall conductivity. 

For Abelian FQH states, this data can be conveniently encoded via an Abelian Chern-Simons theory where the coupling to the background $U(1)$ field $A$ is given by
\begin{align}
\mathcal{L}_{frac} = \frac{1}{2\pi}  q_I a^I \cup dA ; \qquad \mathcal{L}_{SPT} = \frac{2k_1}{4\pi} A \cup d A ,
\end{align}
where $q_I$ is the charge vector. The Hall conductivity is then given by
  \begin{align}
    \bar{\sigma}_H = \vec{q}^T K^{-1} \vec{q} + 2k_1,
    \end{align}
  where
  \begin{align}
    \mathcal{L}_{eff} = \frac{\bar{\sigma}_H}{4\pi} A \cup dA
    \end{align}
  is the effecive response theory obtained by integrating out the internal gauge fields. The different possible choices of charge vector $\vec{q}$, after accounting for relabelings of the $K$ matrix theory, correspond to different choices of the symmetry fractionalization class. Thus there are at most $|\mathcal{A}| = \det K$ distinct choices. For general $G$ the number of distinct fractionalization classes is less than $|\mathcal{A}|$ because of the possibility of relabeling the anyons that can identify the different elements of $\mathcal{H}^2(U(1), \mathcal{A})$.

With ${\bf g_i} = e^{2\pi i z_i}$, we can pick a gauge in which the symmetry fractionalization cocycle is
 
\begin{equation}
w({\bf g_1},{\bf g_2}) =v^{z_1+z_2-[z_1+z_2]} 
\end{equation} 
where $[z_1+z_2] := z_1+z_2 \mod 1$.
Note that in this notation, the anyon $v^0$ is the trivial topological charge, which here we label as $1$; previously we also referred to the trivial topological charge using the label $0$. 

The vison $v$ (which is conventionally represented as $\vec{q}$ in the $K$ matrix theory) is the anyon associated to the insertion of one quantum of $U(1)$ flux. This is reflected in the structure of the above cocycle, which takes the values $v$ or $1$ depending on whether the total flux inserted through the symmetry operations ${\bf g}$ and ${\bf h}$ is greater than or less than one flux quantum. Therefore the choice $v \in \A$ determines the symmetry fractionalization class. 

For non-Abelian FQH states, describing the distinct possible fractional charge assignments of quasiparticles using CS theory is much more cumbersome. However, using the G-crossed BTC formalism, we can provide a topological invariant:
\begin{align}\label{Invt_SF_v}
e^{2\pi i Q_a} &= \mathcal{F}_a({\bf g}_1 = {\bf g} ,\dots , {\bf g}_p = {\bf g}) \nonumber \\
&:= \left(R^{0_{{\bf  g}},a}R^{a,0_{{\bf g}}}\right)^p\prod_{j=1}^{p-1} \eta_a( {\bf g},{\bf g}^j )  = M_{v,a}.
\end{align}
This formula defines $Q_a$, the fractional $U(1)$ charge of the quasiparticle $a$. Here $a$ is any anyon, not necessarily Abelian, and ${\bf g} = e^{2\pi i/p}$.

In the above equation, we can take $p$ to be a common multiple of $(n_1, ..., n_r)$, where $\A = \prod_{i=1}^r \Z_{n_i}$, in which case one can show that Eq. \eqref{Invt_SF_v} is invariant under any relabeling of the defects, $0_{\bf g} \rightarrow 0_{\bf g} \chi({\bf g})$, for $\chi({\bf g}) \in \A$. On the other hand, if we use a generic integer $p > 1$, then we pick a canonical $0_{\bf g}$ that is continuously connected to the identity anyon, which can be made precise by demanding that $R^{b,0_{\bf g}}$ and $R^{0_{\bf g},b}$ approach 1, for any anyon $b$, as ${\bf g}\rightarrow {\bf 0}$. As we discuss in Appendix \ref{Sec:Cont_meas}, since the data of the $G$-crossed BTC can equivalently be taken to be piecewise continuous as a function of group elements in $G$, it follows that for ${\bf g}$ close enough to the identity, there is a choice of $0_{\bf g}$ that is continuously connected to the trivial anyon $0$. Therefore for large enough $p$, there is a canonical choice of $0_{\bf g}$ with ${\bf g} = e^{2\pi i/ p}$.

The different defect classes, whose ambiguity is specified by $\mathcal{H}^3(U(1), U(1)) = \mathbb{Z}$, can be distinguished from the following invariant quantity in the $G$-crossed BTC:
\begin{align}\label{Invt_D_sigmaH}
\mathcal{I}_n(0_{{\bf  g}}) := [\mathcal{T}^n]^{ ({\bf g}, {\bf 0})}_{0_{\bf g} 0_{\bf g}} = \theta_{0_{\bf g}}^n \prod_{j = 0}^{n-1} \eta_{0_{\bf g}}({\bf g}, {\bf g}^j) ,
\end{align}
where ${\bf g} =e^{2\pi i/n}$ and $n$ is any integer. 

To obtain an absolute invariant of the $G$-crossed BTC, we take $0_{\bf g}$ as the unique ${\bf g}$-defect that is continuously connected to the identity particle, which as discussed above is well-defined as long as $n$ is taken to be sufficiently large. 
Eq. \eqref{Invt_D_sigmaH} in general gives us an infinite set of invariants, as we can pick ${\bf g}$ to be any root of unity, and thus $n$ can be any integer. Since we know that different defect classes are related by bosonic integer quantum Hall states, we expect that $\mathcal{I}_n$ should be related to the Hall conductivity, which is the physical quantity that distinguishes these states. This is indeed the case; we can show (see Appendix \ref{Sec:Invts-U1}) that
\begin{equation}
\label{I2n}
  (\mathcal{I}_n)^{2n}(0_{\bf g}) = e^{2\pi i \bar{\sigma}_H}
\end{equation}
Note that setting ${\bf g} = e^{2\pi i /n}$, then changing the defect class by the generator of $\mathcal{H}^3(U(1), U(1))$, changes the above invariant as
\begin{align}
\mathcal{I}_n \rightarrow \mathcal{I}_n e^{2\pi i /n} ,
\end{align}
while physically changing the Hall conductivity $\bar{\sigma}_H \rightarrow \bar{\sigma}_H + 2$. The fractional part of the Hall conductivity is determined by the symmetry fractionalization class through the relation 
\begin{align}
e^{2\pi i Q_v } = e^{2\pi i \bar{\sigma}_H}
\end{align}

\subsection{$G = U(1) \leftthreetimes \mathbb{E}^2$}

\label{Sec:EucGrpSummary}

The FQH system in the continuum possesses, in addition to $U(1)$ charge conservation symmetry, also $SO(2) \cong U(1)$ continuous rotational symmetry and magnetic translation symmetry. The continuum magnetic translation group, which we denote as $U(1)\leftthreetimes \mathbb{R}^{2}$, is a central extension of the continuous translation group $\mathbb{R}^{2}$, by $U(1)$ as indicated by the short exact sequence
\[
1\rightarrow U(1)\rightarrow U(1)\leftthreetimes_{l_B}\mathbb{R}^{2}\rightarrow \mathbb{R}^{2}\rightarrow 1,
\]
with the group multiplication law
\begin{equation}
(e^{2 \pi i z_1},{\bf r_1}) (e^{2 \pi i z_2},{\bf r_2})=(e^{2 \pi i(z_1+z_2+\frac{{\bf  r_1}\times {\bf  r_2}}{2l_B^2})},{\bf r_1}+{\bf r_2})
\end{equation}
(see Appendix \ref{Sec:FQHContSymm} for a derivation). Since groups defined with different values of the magnetic length $l_B$ are isomorphic under a rescaling of space, we will drop the subscript $l_B$ in the following discussion. The full symmetry of FQH states $G = [U(1) \leftthreetimes \mathbb{R}^{2}] \rtimes SO(2) $ is then captured by the following split extension:
\begin{equation}
1\rightarrow U(1)\leftthreetimes \mathbb{R}^{2}\rightarrow G\rightarrow SO(2)\rightarrow1,
\end{equation}
where $e^{2\pi i h} \in SO(2)$ acts on $U(1)\leftthreetimes \mathbb{R}^{2}$ by rotating the translation component ${\bf r}$
of $(e^{2\pi i z},{\bf r},e^{2\pi i h})$ by an angle $2\pi h$. This action is summarized by the following group multiplication law: 
\begin{widetext}
	\begin{equation}
	(e^{2\pi i z_1},{\bf r_1}, e^{2\pi i h_1})  (e^{2\pi i z_2},{\bf r_2}, e^{2\pi i h_2}) = (e^{2\pi i (z_1+z_2+ \frac{{\bf r_1}\times {^{h_1}}{\bf r_2}}{2 l_B^2})}, {\bf r_1} + {^{h_1}} {\bf r_2} , e^{2\pi i (h_1+h_2)})    
	\end{equation}
\end{widetext}
Note that we can represent the symmetry group in two equivalent ways depending on the order in which the group extensions are carried out: $G =  [U(1) \leftthreetimes \mathbb{R}^{2}] \rtimes SO(2) \simeq  U(1) \leftthreetimes [\mathbb{R}^{2} \rtimes SO(2) ]$.

In this case we find (see Appendix \ref{Sec:Coho_calc_cont})
\begin{align}
\mathcal{H}^2(U(1) \leftthreetimes \mathbb{E}^2, \A) =\A\times\A
\nonumber \\
\mathcal{H}^3(U(1) \leftthreetimes \mathbb{E}^2, U(1)) = \mathbb{Z}^3 . 
\end{align}
Interestingly, these cohomology groups are independent of the value of the magnetic field. Moreover, the cohomology groups are also independent of the translations; one would obtain the same results for just $U(1) \times SO(2)$ symmetries (i.e. for $U(1)$ charge conservation and $SO(2) \cong U(1)$ spatial rotation symmetries). However, in the absence of spatial rotational symmetry, the cohomology groups reduce to those with just the $U(1)$ charge conservation symmetry, discussed in the previous section.

The continuum field theory for Abelian FQH systems is generally written \cite{Cho2014,Gromov2014,Gromov2015} in terms of a 1-form $U(1)$ gauge field $A$ (the usual vector potential) and the 1-form $SO(2)$ spin connection $\omega$. The full action, including the symmetry fractionalization, is $\mathcal{L} = \mathcal{L}_0 + \mathcal{L}_{frac} + \mathcal{L}_{\text{SPT}}$, where 
\begin{align}
  \label{effectiveLcont}
	\mathcal{L}_0 &= -\frac{1}{4\pi} K_{IJ}  a^I da^J \\
	\mathcal{L}_{frac} &= \frac{q_I}{2\pi} a^I dA + \frac{s^I}{2\pi} a^I d\omega \\
	\mathcal{L}_{\text{SPT}} &= \frac{k_1}{4\pi} A dA  + \frac{k_2}{2\pi}  A d \omega + \frac{k_3}{4\pi} \omega d\omega
	\end{align}
where for bosonic systems, $k_1,k_3$ are assumed to be even integers and $k_2$ is any integer. 

The factor of $\mathcal{A}$ coming from $U(1)$ charge fractionalization and the factor of $\Z$ coming from the Hall conductivity were discussed in the preceding section, and can be obtained from the same formulas, Eq. \ref{Invt_SF_v} and \ref{Invt_D_sigmaH}, after setting both translations and rotations to zero. However in this case we have an additional constraint. The translational symmetry allows us to define a filling fraction $\nu$, which is the charge per magnetic unit cell. This is measured as follows: let ${\bf x},{\bf y}$ be pure translations in the $x$ and $y$ directions which span a magnetic unit cell, and let ${\bf  g} = (e^{i 2\pi/n},{\bf  0},1)$ (note that we use $1$ for the identity element of the rotation group and ${\bf 0}$ as the identity element of the translation component). Define 
\begin{equation}\label{Invt_Ex_nu_cont}
e^{2\pi i \nu} := \left(\frac{\eta_{0_{{\bf g}}}({\bf x},{\bf y})}{\eta_{0_{{\bf g}}}({\bf y},{\bf x})}\right)^n.
\end{equation}
This allows us to define the filling $\nu$ in terms of the projective phase associated to adiabatically transporting a $2\pi/n$ $U(1)$ flux around a magnetic unit cell.

Now one has the condition (see Appendix \ref{Sec:LSM_cont}) that
\begin{align}
e^{2\pi i \nu} = M_{v,v} = e^{2\pi i \bar{\sigma}_H}.
\end{align}
Although we do not have a formal proof using $G$-crossed identities, under certain assumptions on the anyon $F$ and $R$ symbols (see Appendix \ref{Sec:Gxsolns}) we can construct particular solutions to the $G$-crossed equations, and in all these cases verify that 
\begin{align}
e^{2\pi i \nu/n} = e^{2\pi i \sigma_{H}/n}, 
\end{align}
for every $n > 0$. This allows us to verify the well-known equation
\begin{align}
\nu = \bar{\sigma}_H.
\end{align}
The fact that $\nu = \bar{\sigma}_{H}$ is well-known from Galilean invariance, however it has not been proven more generally from translational symmetry alone.

The second factor of $\mathcal{A}$ is associated with fractionalization of the $SO(2) \simeq U(1)$ spatial rotational symmetry. The associated invariant is also given by the general $U(1)$ symmetry fractionalization formula, this time after setting the charge conservation and translation components to zero:
\begin{align}\label{Invt_SF_s_cont}
  e^{2\pi i L_a} &= \left(R^{0_{{\bf  g}},a}R^{a,0_{{\bf g}}}\right)^p\prod_{j=1}^{p-1} \eta_a( {\bf g},{\bf g}^j ) , \quad {\bf  g} = (1,{\bf  0},e^{i 2\pi/p})                   
\end{align}
As in the case of the fractional charge $Q_a$, in the above equation, we can take $p$ to be a common multiple of $(n_1, ..., n_r)$, where $\A = \prod_{i=1}^r \Z_{n_i}$, in which case one can show that Eq. \eqref{Invt_SF_s_cont} is invariant under any relabeling of the defects, $0_{\bf g} \rightarrow 0_{\bf g} \chi({\bf g})$, for $\chi({\bf g}) \in \A$.  On the other hand, if we use a generic integer $p > 1$, then we pick a canonical $0_{\bf g}$ that is continuously connected to the identity anyon as ${\bf g} \rightarrow {\bf 0}$. 

$L_a$ can be understood as a fractional orbital angular momentum of the anyon $a$. Under adiabatic transport of the anyon $a$ around a region of curvature $\Omega$ the wave function picks up a phase 
\begin{align}
\gamma_{AB} = e^{i S_a \Omega} ,
\end{align}
where
\begin{align}
e^{2\pi i S_a} = \theta_a e^{2\pi i L_a} . 
\end{align}
Such a calculation was performed explicitly for Abelian topological phases in \cite{gromov2016,EINARSSON1995}; we are not aware of such a calculation for non-Abelian topological orders. 

The $\mathcal{H}^3(G, U(1)) = \mathbb{Z}^3$ classification can be understood through the effective SPT action, $\mathcal{L}_{SPT}$ defined in Eq. \ref{effectiveLcont}.

      The total response theory, which can be obtained in the Abelian case by integrating out the dynamical $U(1)$ gauge fields, is of the form
	\begin{align}
	\mathcal{L}_{\text{Response}} &=  \frac{\bar{\sigma}_H}{4\pi} A dA + \frac{\mathscr{S}}{2\pi}  \omega dA + \left(\frac{\ell_s}{4\pi} - \frac{c}{48\pi}\right) \omega d\omega. 
	\end{align}
 The last contribution is proportional to the chiral central charge $c$, and arises from framing anomaly \cite{Witten1989,Gromov2015}. To obtain it we have kept only the spatial component of the space-time spin connection. 

 The $G$-crossed formulas for $\bar{\sigma}_H$ and $\ell_s$, which is the angular momentum associated to a conical defect of angle $2\pi$, can be obtained from the formula for the $U(1)$ defect class, Eq. \eqref{Invt_D_sigmaH}. Specifically, we define $ {\bf g} = (e^{2\pi i /n},{\bf 0},1), {\bf h} = (1,{\bf 0},e^{2\pi i /n})$, and obtain
 \begin{align}
 e^{i \pi \bar{\sigma}_H/n} := \mathcal{I}_n(0_{{\bf g}}) &= e^{2\pi i\left(h_v+k_1 \right)\frac{1}{n}}	\\
 e^{i \pi \ell_s/n} := \mathcal{I}_n(0_{{\bf h}}) &= e^{2\pi i\left(h_s+k_3 \right)\frac{1}{n}}\label{Eq:D-k3}
 \end{align}
Here we have used the topological spin $h_a \in [0,1)$, defined in terms of the topological twist $\theta_a = e^{2\pi i h_a}$.  
      
The mixed defect class is measured by the shift $\mathscr{S}$. To find the corresponding $G$-crossed invariant, we define
\begin{align}\label{Invt_D_Shift_cont}
e^{i 2\pi \mathscr{S}/n} := \frac{\mathcal{I}_n(0_{{\bf gh}})}{\mathcal{I}_n(0_{{\bf g}})\mathcal{I}_n(0_{{\bf h}})} &= e^{2\pi i\left(v\star s+k_2 \right)\frac{1}{n}}	
\end{align}
Note that the usual definition of the shift $\mathcal{S}$ in the FQH literature is related to $\mathscr{S}$ as follows:
\begin{align}
\mathcal{S} = \frac{2\mathscr{S}}{\bar{\sigma}_H}. 
\end{align}

\begin{table*}
	\centering
	\begin{tabular} {l |l |l|l|p{0.3\textwidth}|p{0.3\textwidth}}
		\hline
		Topological order & $\A$ & $R^{ab}_{a+b}$ & $\H^2(G,\A)$ & $P$ & No. of distinct defect classes\\
		\hline
		$1/N$ Laughlin ($N$ even)& $\Z_N$& $e^{\frac{\pi i a b}{N}}$ & $\Z_{d}$ & $\frac{2d}{(2d,2s+cd)}$ & $\frac{M\times (2d,2s+cd)}{2d}$\\
		$SU(2)_k$ Read-Rezayi&$\Z_2$ & $e^{\frac{\pi i k a b}{2}}$ & $\Z_{(M,2)}$ & $\frac{2}{(2,k (s+M/2))}$ if $M \in 2\Z$; 1 otherwise & $\frac{M\times (2,k(s+M/2))}{2}$ if $M \in 2\Z$; $M$ otherwise \\
	$\Z_2$	toric code&$ \Z_2\times\Z_2$ & $e^{\frac{\pi i  }{2} a^T \sigma_x b}$ & $\Z^2_{(M,2)}$ & $\frac{2}{(s_1,s_2)}$ if $M \in 2\Z$; $1$ otherwise& $\frac{M}{2}$ if $M \in 2\Z$ and $s_1=s_2=1$; $M$ otherwise\\
		$\Z_N^{(p)}$ anyons, $N$ even&$\Z_N$ & $e^{\frac{2 \pi i p a b}{N}}$ &  $\Z_{d}$ & $\frac{2d}{(2d,p(2s+cd))}; (2p,N) = 1$ & $\frac{M}{(M,P)}$\\
		$\Z_N^{(p)}$ anyons, $N$ odd&$\Z_N$ & $e^{\frac{2 \pi i p a b}{N}}$ &  $\Z_{d}$ & $\frac{2d}{(2d,p(2s+cd))}; (p,N) = 1$ & $\frac{M}{(M,P)}$\\
		\hline
	\end{tabular}
	\caption{The effect of relabelings on the defect classification for various topological orders with $G = \Z_M$. The defect classification prior to considering relabelings is a torsor over $\H^3(\Z_M,U(1)) \cong \Z_M$; but the true classification is a torsor over a subgroup of $\Z_M$, denoted as $\Z_{M/(M,P)}$. The quantity $\mathcal{I}^P_M(0_{\bf h})$ is an absolute SET invariant. For a given symmetry fractionalization class, the integer $P$ depends sensitively on the group structure and braiding data of $\A$. We have also defined $d = \gcd(M,N)$ and $c = \frac{MN}{d^2}$, while the Abelian anyon $s$ is a representative of $\H^2(G,\A)$. For details of the derivation, see Appendix \ref{Sec:relabeling_ZM}.}\label{Table:ZMrelabelings}
\end{table*}

\subsection{$G = \Z_M$}
\label{Sec:ZMSumm}
Although systems with $G = \Z_M$ do not describe FQH states, they provide an instructive example of how the SET classification based on $\H^2(G,\A)$ and $\H^3(G,U(1))$ can be reduced after taking into account defect relabelings, as we discuss in detail in Appendix \ref{Sec:relabeling}. This example will also help us build towards our subsequent analysis of FQH states with space group symmetries. 

Each symmetry fractionalization class corresponds to a choice of $s = 0_{\bf h}^M$: throughout this section, we take ${\bf h} = e^{2\pi i/M}$ to be the generator of the group $\Z_M$. The defect $0_{{\bf h}}$ is understood as an elementary $\Z_M$ flux; inserting $M$ such fluxes induces the anyon $s$. We use the symbol $s$ since in the context of point group symmetry this anyon is a discrete analog of the spin vector. However, there is a redundancy in the description of symmetry fractionalization classes, as there is no way to canonically define the elementary defect $0_{{\bf h}}$. We can equally define the elementary flux as $s'_{{\bf h}} = 0_{{\bf h}}\times s'$, in which case the anyon induced by inserting $M$ such defects is $s \times s'^M$. 

The inequivalent choices of $s$ are hence classified by $\mathcal{A}/M\mathcal{A}$, where $M\mathcal{A} = \{ a^M, a \in \mathcal{A}\}$. When $\mathcal{A} = \Z_{n_1}\times \dots \times \Z_{n_r}$, the symmetry fractionalization classes form a group $\mathcal{H}^2(\Z_M,\mathcal{A}) = \Z_{(M,n_1)}\times\dots\times\Z_{(M,n_r)}$.

In terms of a discrete background $\Z_M$ gauge field $C$, which is defined on a triangulation of the manifold $\mathcal{M}$ and is valued in $\frac{2\pi}{M}\Z$, we can write the following Abelian CS action capturing these phases:
\begin{align}
\mathcal{L}_{frac} = \frac{s_I}{2\pi} a^I \cup dC; \qquad \mathcal{L}_{SPT} =  \frac{k_3}{2\pi} C \cup dC
\end{align}

In the context of point group symmetries described by the above field theory, where a $\Z_M$ flux is simply a disclination, the $\A/M\A$ classification of spin vectors arises from the fact that $s$ is the anyon induced by inserting $M$ elementary disclinations; but we can always trivially associate an anyon $s'$ to each elementary disclination by adjusting local energetics. Therefore the anyons $s$ and $s\times s'^M$ describe the same symmetry fractionalization class. 

In the $G$-crossed theory, a representative $s$ of the symmetry fractionalization class is determined by the following formula, which also gives the fractional $\Z_M$ charge (angular momentum) $L_a$ of an anyon $a$:  
\begin{align}\label{Invt_SF_s_disc}
e^{2\pi i L_a} &= \mathcal{F}_a({\bf g}_1 = {\bf h} , \dots , {\bf g}_n = {\bf h}) \nonumber \\ &:=\left(R^{0_{\bf h},a}R^{a,0_{\bf h}}\right)^M\prod_{j=1}^{M-1} \eta_a( {\bf h},{\bf h}^j )  = M_{s,a}.
\end{align} 
If $M_{s,a}$ is known as a function of $a$, the value of $s$ can be determined. Note that the fractional $\Z_M$ charge of an anyon is not invariant for a given symmetry fractionalization class due to the ambiguity in $s$: for a given fractionalization class, the quantity $e^{2\pi i L_a}$ is determined only up to terms of the form $M_{s'^M,a}$ upon relabeling the defects. 

The defect classes form a torsor over $\mathcal{H}^3(\Z_M,U(1)) \cong \Z_M$; they can be measured using the formula 
\begin{align}\label{Invt_D_ell_s}
e^{\pi i \ell_s/M} :=\mathcal{I}_M(0_{{\bf h}})  = \theta_{0_{\bf h}}^n \prod_{j = 0}^{M-1} \eta_{0_{\bf h}}({\bf h}, {\bf h}^j) = e^{i 2\pi \frac{h_s + k_3}{M}},
\end{align}
where $k_3 \in \Z_M$ is the defect (SPT) invariant. We note that for $\Z_M$ SPTs characterized by a 3-cocycle $\alpha$, with $[\alpha] \in \H^3(\Z_M,U(1))$, this invariant takes the form $\prod\limits_{j=0}^{M-1} \alpha({\bf h}, {\bf h}^j, {\bf h})$ that has been discussed previously in Refs \cite{Tantiv2017,Tiwari2018}.

Next we discuss a redundancy in the defect classification. By making certain gauge choices, the remaining $G$-crossed data for $G=\Z_M$ are completely determined by the data of the UMTC and the symmetry fractionalization cocycle $\mathfrak{w}({\bf g_1},{\bf g_2})$ (see Appendix \ref{Sec:GeneralBTCSoln}). Under a relabeling which takes $0_{\bf h} \rightarrow s'_{{\bf h}} = 0_{\bf h}\times s'$, we have, for ${\bf g}_i = e^{2\pi i h_i/M}$,
\begin{align}
\mathfrak{w}({\bf g_1},{\bf g_2}) &= 0_{\bf g_1} 0_{\bf g_2} \overline{0_{\bf g_1g_2}} \nonumber \\ &\rightarrow 0_{\bf g_1} 0_{\bf g_2} \overline{0_{\bf g_1g_2}} \times s'^{h_1+h_2-[h_1+h_2]_M}.
\end{align}

Note that $s'^{h_1+h_2-[h_1+h_2]_M}$ equals either the trivial anyon or the anyon $s'^M$. Thus, whenever $s'^M$ is itself trivial, a relabeling of this type leaves the cocycle $\mathfrak{w}({\bf g_1},{\bf g_2})$, and hence the entire $G$-crossed data, invariant. However, the quantities $\mathcal{I}_M(0_{\bf h})$ and $\mathcal{I}_M(s'_{\bf h})$ may be unequal: they can correspond to different values of $k_3$, and hence different values of $\ell_s$. These values of $k_3$ parametrize the same defect class, because we can always relabel $0_{{\bf h}}$ as $s'_{{\bf h}}$ without changing the symmetry fractionalization anyon $s$, or the other $G$-crossed data. 

Although the above argument involves only a subset of all possible relabelings, it accounts for all the nontrivial equivalences among defect classes. This can be independently verified for Abelian SET phases using relabelings in the field theory \cite{Lu2016,Manjunath2020}. The procedure here thus gives the complete set of equivalences even in SETs with non-Abelian anyon models. An example of how the classification is reduced is given in Table \ref{Table:ZMrelabelings}, where we consider some common anyon models with $\Z_M$ symmetry.

\subsection{$G = U(1) \leftthreetimes_\phi \Z^2$}
\label{Sec:MagTransSumm}

The symmetry $G = U(1)\leftthreetimes_{\phi}\Z^2$ is given by discrete magnetic translations, whose associated many-body operators in a system of $N$ particles form the algebra
\begin{align}
T_{\bf{x}}T_{\bf{y}} = T_{\bf{y}} T_{\bf{x}} e^{i \phi N} ,
\end{align}
where $\phi$ is the flux per unit cell, and $\bf{x}$ and $\bf{y}$ are the basis vectors of the lattice. The symmetry group in this case can be considered a central extension of $\mathbb{Z}^2$ by the $U(1)$ charge conservation symmetry, as encapsulated by the short
exact sequence
\begin{align}
1 \rightarrow U(1) \rightarrow  U(1)\leftthreetimes_{\phi}\Z^2 \rightarrow \mathbb{Z}^2 \rightarrow 1.
\end{align}
Different values of $N$ correspond to different representations of the same group. With group elements given by ${\bf  g} = (e^{2\pi i z},\bf r)$, the group multiplication law is given in symmetric gauge by
\begin{equation}
(e^{2\pi i z_1},{\bf r_1}) (e^{2\pi i z_2},{\bf r_2}) = (e^{2\pi i (z_1 + z_2 + \frac{\phi}{2} {\bf r_1}\times{\bf r_2})},{\bf r_1}+{\bf r_2})
\end{equation}

We can construct a background gauge field for $G$ symmetry in terms of the pair $(A,\vec{R})$, where $A \in \mathbb{R}/2\pi\Z$ and $\frac{1}{2\pi}\vec{R} = \frac{1}{2\pi}(X,Y) \in \Z^2$. A topological field theory for Abelian FQH states with $G$ symmetry can now be written in terms of this gauge field as
\begin{align}
	\mathcal{L}_{frac} =& \frac{q_I}{2\pi} a^I \cup dA + \frac{m_I + \phi q_I}{2\pi} a^I \cup A_{XY}\label{Eq:MT-frac}\\
  \mathcal{L}_{SPT} =& \frac{k_1}{2\pi} A \cup dA + \frac{k_6 + 2\phi k_1}{2\pi} A \cup A_{XY}
              \nonumber \\
           &   + \frac{\phi^2 k_1}{2\pi} d^{-1}(A_{XY} \cup A_{XY})\label{Eq:MT-SPT},
	\end{align}
where we have defined the area element $A_{XY} = \frac{1}{4\pi}(X \cup Y - Y \cup X)$. The appearance of terms explicitly depending on $\phi$ is a direct consequence of the group multiplication law, which mixes translations into the $U(1)$ components (see Appendix \ref{Sec:Eff_Ac-magtrans} for a derivation).

Now we find 
\begin{align}
\mathcal{H}^2( U(1) \leftthreetimes_\phi \Z^2 , \mathcal{A}) &= \mathcal{A} \times \mathcal{A}
\nonumber \\
\mathcal{H}^3( U(1) \leftthreetimes_\phi \Z^2 , U(1)) &= \mathbb{Z}^2 . 
\end{align}
In this case it is possible to show explicitly that $\H^3(U(1) \leftthreetimes_\phi \Z^2 , U(1)) \cong \H^4(U(1) \leftthreetimes_\phi \Z^2 , \Z)$. Importantly, the SET classification does not depend on $\phi$, even though the $G$-crossed data, the effective actions, and the allowed values of the invariants all depend on $\phi$.

Each symmetry fractionalization class can be understood in terms of two anyons: (i) the vison $v$, which is the anyon associated to the insertion of $2\pi$ flux, as discussed previously, and (ii) the anyon per unit cell $m$. If the magnetic translation algebra is trivial (i.e. if $\phi=0$), we can define $m$ in a gauge-invariant manner through the formula $m = \mathfrak{b}({\bf x},{\bf y}) :=\mathfrak{w}({\bf x},{\bf y})\overline{\mathfrak{w}({\bf y},{\bf x})}$, where ${\bf x},{\bf y}$ are the elementary translations.

When $\phi = p/q$ is fractional, $m$ cannot be defined in terms of pure translations alone. Assuming that ${\bf r}_1,{\bf r_2}$ are magnetic translations that span a $q \times 1$ magnetic unit cell, we can write (see Appendix \ref{Sec:Coho_calc_magtrans}):
\begin{align}
  \label{bEq}
\mathfrak{b}({\bf r}_1,{\bf r}_2) = v^p \times m^q . 
\end{align}
$m$ is an anyon that can be associated to a single unit cell, as we discuss below. Eq. \ref{bEq} can be understood as the anyon associated to a magnetic unit cell, which takes the above form because 
each magnetic unit cell contains $q \phi = p$ flux quanta (associated to $v^p$) and $q$ unit cells (associated to $m^q$).

The anyon $m$ can be measured by applying the following sequence of elementary translation operations to an anyon $a$: ${\bf g}_1 = (1,{\bf x}); {\bf g}_2 = (1,{\bf y}); {\bf g}_3 = (1,{\bf -x}); {\bf g}_4 = (1,{\bf -y})$, so that $\Phi={\bf g}_4{\bf g}_3{\bf g}_2{\bf g}_1 = (e^{i \phi},{\bf  0})$. In terms of the associated projective phases, we can write the invariant as
\begin{widetext}
\begin{align}\label{Invt_SF_m}
\F_a({\bf g}_1,{\bf g}_2,{\bf g}_3,{\bf g}_4) & := \frac{\prod\limits_{i=1}^{4}R^{a, 0_{{\bf g}_i}} R^{ 0_{{\bf g}_i},a}}{ \left( R^{a, 0_{\Phi}} R^{ 0_{\Phi},a}\right)}  \eta_{a}({\bf g}_4,{\bf g}_1{\bf g}_2{\bf g}_3)\eta_{a}({\bf g}_3,{\bf g}_1{\bf g}_2)\eta_{a}({\bf g}_2,{\bf g}_1) = M_{a, m}
\end{align}
\end{widetext}

The symmetry operations take $a$ around a unit cell containing the anyon $m$, and the above invariant measures the braiding between them. From the group multiplication law, the process of going around the unit cell by a sequence of translations is equivalent to inserting a flux $\phi$, and transporting the anyon around a region containing this flux introduces an additional phase. This extra phase is cancelled by the denominator in Eq. \ref{Invt_SF_m}.

Similar to the case with continuous translation symmetry, it is possible to define a filling corresponding to the charge per magnetic unit cell $\nu$, which in the $G$-crossed BTC is given by the formula
\begin{align}\label{Invt_D_nuC}
e^{\frac{2\pi i }{n}\nu} = \frac{\eta_{0_{\bf g}} ({\bf r_1},{\bf r_2})}{\eta_{0_{\bf g}}({\bf r_2},{\bf r_1})} ,
\end{align}
where ${\bf g} = (e^{2\pi i/n},{\bf 0})$ for an integer $n$, and $0_{{\bf g}}$ is a $U(1)$ flux. The vectors $\bf r_1$ and $\bf r_2$ must be chosen to span a \textit{magnetic} unit cell, in order for this formula to be gauge-invariant. The charge per unit cell, defined by the $\Z^2$ translations, is then $\frac{\nu}{q}$.

The filling is related to the Hall conductivity and the anyon per unit cell as follows:
\begin{align}\label{DiscreteMagTransFillingRelation}
e^{2\pi i \nu} &= M_{v,m^q} \times e^{2\pi i q \phi \bar{\sigma}_H} 
\end{align}
Since $\nu$ is typically fixed in terms of the microscopic definition of the system, Eq. \ref{DiscreteMagTransFillingRelation} can be viewed as a constraint on the fractional statistics and symmetry fractionalization class in terms of the charge and flux per unit cell. As such, this can be viewed as a strengthened version of a Lieb-Schulz-Mattis constraint \cite{Cheng2016,Lu2017fillingenforced}.
Eq. \ref{DiscreteMagTransFillingRelation} was first shown in \cite{Lu2017fillingenforced} using flux insertion arguments. In Appendix \ref{Sec:LSM_lattice} we provide a proof of Eq. \ref{DiscreteMagTransFillingRelation} within the framework of the G-crossed BTC.

Note that whenever we can write down explicit solutions to the $G$-crossed BTC equations, we can verify the stronger result
\begin{equation}
\nu = q (\phi \bar{\sigma}_H + Q_m + k_6).
\end{equation}
Eq. \eqref{DiscreteMagTransFillingRelation} is a weaker result obtained by equating the two sides of this equation only modulo 1. To our knowledge, this stronger result has not been stated or rigorously proven in previous work. Obtaining a completely general proof of this result entirely within the framework of $G$-crossed BTCs is an interesting problem which we leave for future work.

The stronger LSM constraint can also be read off from the topological actions in Eqs. \eqref{Eq:MT-frac} and \eqref{Eq:MT-SPT} after integrating out the internal gauge fields. First we obtain $\bar{\sigma}_H = Q_v + 2k_1 = 2 (h_v + k_1)$ (the response theory coefficient of $A \cup dA$) and then we obtain
\begin{align}
  \frac{\nu}{q} &= v \star m + 2\phi  (h_v + k_1) + k_6 , 
\end{align}
which is the response theory coefficient of $A \cup A_{XY}$. However, this is not a fully rigorous derivation because the response theory has fractional coefficients and is not gauge-invariant. 

The defect classification $\mathcal{H}^3(U(1) \leftthreetimes_\phi \Z^2, U(1)) = \Z^2$ has two contributions associated with the integers $k_1$ and $k_6$, which correspond to changing the Hall conductivity by an even integer and the charge per unit cell by an integer. 

\subsection{$G = U(1) \leftthreetimes_\phi [\Z^2 \rtimes \Z_M]$}
\label{Sec:FullSpaceGrpSummary}

\begin{table*}[t]
	\centering
	\begin{tabular} {l |p{0.4\textwidth} |p{0.4\textwidth}}
		\hline
		Invariant & Formula & Details \\ \hline
		\multicolumn{3}{c}{Symmetry fractionalization invariants, $\A = \Z_{n_1}\times\dots\times\Z_{n_r};  p = \text{lcm}(n_1,\dots , n_r)$} \\ \hline
		$\{Q_a\}$& $e^{2\pi i Q_a} = M_{v,a} = \F_a( {\bf  g}_1,\dots , {\bf  g}_p)$ & ${\bf  g}_1 = \dots = {\bf  g}_p = (e^{i 2\pi/p},{\bf  0},1)$ \\ 
		$\{L_a\}$& $e^{2\pi i L_a} = M_{s,a} = \F_a( {\bf  g}_1,\dots , {\bf  g}_M)$ & ${\bf  g}_1 = \dots = {\bf  g}_M = (1,{\bf  0},e^{i 2\pi/M})$ \\
		$\{\tau_a \} $ & $e^{2\pi i \tau_a} = M_{m,a} = \F_a({\bf  g}_1,\dots , {\bf  g}_4)$ & ${\bf  g}_1 = (1,{\bf  x},1);{\bf  g}_2 = (1,{\bf  y},1);{\bf  g}_3 = (1,{\bf  -x},1);{\bf  g}_4 = (1,{\bf  -y},1)$ \\
		$\{\vec{P}_a\}$& We define $e^{2\pi i \vec{P}_a \cdot {^{1-h}}{\bf r}} := M_{\vec{t}\cdot {\bf r},a}$ & \\
		$\hspace{0.75cm}M=2$& $ M_{\vec{t}\cdot {\bf r},a} = \frac{\F_a( {\bf  g},{\bf  g})}{\F_a( {\bf h}, {\bf h})}$ & ${\bf  g} = (1,{\bf  r},e^{2\pi i/2}), {\bf h} = (1,{\bf  0},e^{2\pi i/2})$  \\
		$\hspace{0.75cm}M=3$& $ M_{a, \vec{t}\cdot ({\bf x}-{\bf y})} = \frac{\mathcal{F}_a({\bf g},{\bf g},{\bf g})}{\mathcal{F}_a({\bf k},{\bf k},{\bf k})} = M_{a, t_x\overline{t^2_y}}$ & ${\bf  g} = (e^{2\pi i z},{\bf  x},e^{2\pi i/3}), {\bf  k} = (e^{2\pi i z},{\bf  y},e^{2\pi i/3})$ \\
		$\hspace{0.75cm}M=4$& $ M_{\vec{t}\cdot {^{1+h}}{\bf r},a} = \frac{\F_a( {\bf  g},{\bf  g})}{\F_a( {\bf  k}, {\bf  k})}= M_{a, (t_xt_y)^{r_x}\times (t_x\overline{t_y})^{r_y}}$ & ${\bf  g} = (1,{\bf  r},e^{2\pi i/2}), {\bf  k} = (1,{\bf  0},e^{2\pi i/2})$  \\
		
		\hline
		\multicolumn{3}{c}{Fractionally quantized responses} \\ \hline
		$\bar{\sigma}_H$ & $e^{2\pi i \frac{\sigma_H}{2p}} = \mathcal{I}_p(0_{{\bf  g}})$  & ${\bf  g} =(e^{i 2\pi/p},{\bf  0},1)$  \\
		$\mathscr{S} $ & $e^{2\pi i \frac{\mathscr{S}}{M}} = \frac{\mathcal{I}_M(0_{{\bf gh}})}{\mathcal{I}_M(0_{{\bf g}})\mathcal{I}_M(0_{{\bf h}})}$  & ${\bf  g} =(e^{i 2\pi/M},{\bf  0},1);{\bf h} =(1,{\bf  0},e^{i 2\pi/M}) $  \\
		$\ell_s$ & $e^{2\pi i \frac{\ell_s}{2M}} = \mathcal{I}_M(0_{{\bf h}})$  & ${\bf h} =(1,{\bf  0},e^{i 2\pi/M})$  \\
		$\nu$ & $e^{2\pi i \frac{\nu}{n}} = \frac{\eta_{0_{{\bf  g}}}({\bf r}_1,{\bf r}_2)}{\eta_{0_{{\bf  g}}}({\bf r}_2,{\bf r}_1)}$  & ${\bf  g} =(e^{i 2\pi/n},{\bf  0},1)$; ${\bf r}_1,{\bf r}_2$ span a magnetic u.c. \\
		$\nu_s$ & Not determined 
		& \\
		$\vec{\mathscr{P}}_c$ & & \\
		$\hspace{0.75cm}M=2$ & $ e^{i \pi\frac{\sigma_H}{2}}\times e^{2\pi i \frac{\mathscr{S} }{2}}\times e^{2\pi i \vec{\mathscr{P}}_c\cdot {\bf r}} = \frac{\mathcal{I}_2(0_{{\bf g}})}{\mathcal{I}_2(0_{{\bf k}})}$ & ${\bf g} = (e^{i \pi},{\bf r},e^{2\pi i/2}); {\bf k} = (1,{\bf r},e^{2\pi i/2})$; ${\bf g}^2 = {\bf k}^2 = {\bf 0}$\\
		$\hspace{0.75cm}M=3$& $e^{i \pi\sigma_H }\times e^{2\pi i \frac{\mathscr{S} }{3}}\times e^{2\pi i \frac{\nu}{3}}\times e^{2\pi i \frac{1}{3}\vec{\mathscr{P}}_c\cdot {^{2+h}}{\bf r}} = \frac{\mathcal{I}_3(0_{{\bf g}})}{\mathcal{I}_3(0_{{\bf k}})} $ & ${\bf g} = (e^{2\pi i \frac{2-\phi}{3}},{\bf r},e^{2\pi i/3}),{\bf k} = (e^{2\pi i \frac{1-\phi}{3}},{\bf r},e^{2\pi i/3})$; ${\bf g}^3 = {\bf k}^3 = {\bf 0}$\\
          $\hspace{0.75cm}M=4$& $e^{i\pi \frac{\sigma_H}{2}}\times e^{2\pi i \frac{\mathscr{S} }{2}}\times e^{2\pi i \frac{1}{2}\vec{\mathscr{P}}_c\cdot {^{1+h}} {\bf r}} = \frac{\mathcal{I}_2(0_{{\bf g}})}{\mathcal{I}_2(0_{{\bf k}})}$ &${\bf g} = (e^{i \pi},{\bf r},e^{2\pi i/2}),{\bf k} = (1,{\bf r},e^{2\pi i/2})$; ${\bf g}^2 = {\bf k}^2 = {\bf 0}$ \\
		$\vec{\mathscr{P}}_s$ & & \\
		$\hspace{0.75cm}M=2$&$ e^{2\pi i \frac{\vec{\mathscr{P}}_s\cdot {\bf r}}{2}} \times e^{\frac{i \pi}{2} (\vec{t}\cdot {\bf r})\star ((\vec{t}\cdot {\bf r}))} = \frac{\mathcal{I}_2(0_{\bf g})}{\mathcal{I}_2(0_{\bf h})}$ & ${\bf g} = (1,{\bf r},e^{i\pi}), {\bf h} = (1,{\bf 0},e^{i\pi})$; ${\bf g}^2 = {\bf h}^2 = {\bf 0}$\\
		$\hspace{0.75cm}M=3$ &$e^{2\pi i \frac{1}{3}\vec{\mathscr{P}}_s\cdot {^{2+h}}({\bf x}-{\bf y})} \times e^{2\pi i \frac{\mathscr{S}}{3}} \times e^{\frac{\pi i}{3} (t_x\star t_y + t_y \star t_x + t_y \star t_y)} \times e^{2\pi i z \vec{\mathscr{P}}_c\cdot {^{2+h}}({\bf x}-{\bf y})} = \frac{\mathcal{I}_3(0_{\bf g})}{\mathcal{I}_3(0_{\bf k})}$ & ${\bf g} = (e^{2\pi i \frac{1-\phi}{3}},{\bf x},e^{i2\pi/3}), {\bf k} = (e^{2\pi i \frac{1-\phi}{3}},{\bf y},e^{i2\pi/3})$; ${\bf g}^3 = {\bf k}^3 = {\bf 0}$ \\
		$\hspace{0.75cm}M=4$& $ e^{2\pi i \frac{\vec{\mathscr{P}}_s\cdot {^{1+h}}{\bf r}}{2}} \times e^{\frac{i \pi}{2} (\vec{t}\cdot {\bf r})\star ((\vec{t}\cdot {^{1+h}}{\bf r}))}= \frac{\mathcal{I}_2(0_{\bf g})}{\mathcal{I}_2(0_{\bf k})}$ &${\bf g} = (1,{\bf r},e^{i\pi}), {\bf k} = (1,{\bf 0},e^{i\pi})$ ; ${\bf g}^2 = {\bf k}^2 = {\bf 0}$\\
		\hline 
		$\Pi_{ij}$ &$e^{2\pi i M {\bf r}^T \Pi {\bf r'}} := \frac{\mathcal{I}_M(d_4)\mathcal{I}_M(d_1)}{\mathcal{I}_M(d_2)\mathcal{I}_M(d_3)}$ & $d_1 = 0_{(1,{\bf 0},e^{2\pi i/M})};d_2 = 0_{(1,{\bf r'},e^{2\pi i/M})}; d_3 = 0_{(1,{\bf r},e^{2\pi i/M})}; d_4 = 0_{(1,{\bf r+r'},e^{2\pi i/M})} $; we have assumed that $v,m$ are trivial in this formula \\
		$\vec{\nu}_p$ & Not determined   
		&  
		\\
		\hline
	\end{tabular}
	\caption{List of $G$-crossed invariants studied in this paper for $G= [U(1)\leftthreetimes \Z^2]\rtimes \Z_M$. A general group element is written as ${\bf  g} = (e^{2\pi i z},{\bf r},e^{2\pi i j/M})$, where the three components refer to $U(1)$, translation and rotation symmetries respectively. The group element ${\bf h}$ is the generator of $\Z_M$ rotations, and $h$ is its $2\times 2$ matrix representation. See Appendix \ref{Sec:Invts} for derivations.}\label{Table:LatticeInvts}
\end{table*} 

\subsubsection{Group structure and properties of defects}

Here we consider the orientation-preserving space group $G_{\text{space}} = \mathbb{Z}^2 \rtimes \mathbb{Z}_M$ with $M$-fold rotations. The full symmetry group is then a central extension of $G_{\text{space}}$ by $U(1)$: 
\begin{align}
1 \rightarrow U(1) \rightarrow G = U(1)\leftthreetimes_{\phi} [\Z^2 \rtimes \Z_M] \rightarrow G_{\text{space}} \rightarrow 1.
\end{align}
The group multiplication law (which is derived in Appendix \ref{Sec:FQHSymm}) can be written as follows:
\begin{widetext}
\begin{equation}
(e^{2\pi i z_1},{\bf r_1},e^{2\pi i h_1/M}) (e^{2\pi i z_2},{\bf r_2},e^{2\pi i h_2/M}) = (e^{2\pi i (z_1+z_2+ \phi w({\bf  r_1},^{h_1}{\bf  r_2}))}, {\bf r_1}+^{h_1}{\bf r_2},e^{2\pi i [h_1+h_2]_M/M})    
\end{equation}
\end{widetext}
Here the function $w({\bf r_1}, {\bf r_2})$ may be real valued but satisfies $w({\bf r_1}, {\bf r_2}) - w({\bf r_2}, {\bf r_1}) = {\bf r_1} \times {\bf r_2}$. We will choose $w({\bf r_1}, {\bf r_2}) = \frac{1}{2}{\bf r_1} \times {\bf r_2}$ so as to be invariant under overall rotations, i.e. to be in symmetric gauge.

The symbol ${^{h_1}}{\bf r_2}$ denotes a rotation of the vector ${\bf r_2}$ by the matrix corresponding to the point group element $h_1$. Now, a general symmetry defect can be written as $a_{\bf g}$, where ${\bf g} = (e^{2\pi i z},{\bf r},e^{2\pi i j/M})$. This defect can be written as $a \times 0_{\bf g}$, where $a \in \mathcal{C}_{\bf 0}$ is an anyon, and is understood as follows: it is associated to a disclination angle of $2\pi j/M$, a dislocation Burgers vector ${\bf r}$, and a localized $U(1)$ flux equal to $z \mod 1$ in units of the flux quantum. Furthermore, the defect $a_{\bf g}$ is obtained from the defect $0_{\bf g}$ by attaching the anyon $a$ to the defect core. 

Although the symmetry action does not permute the anyons, it does permute the defects. A defect $a_{\bf g}$ associated to the group element ${\bf g}$ can be conjugated by the group element ${\bf k}$ to obtain a defect $\rho_{{\bf k}}(a_{\bf g})$, which is associated to the group element ${\bf k g \bar{k}}$, where $\bar{{\bf  k}}$ denotes the inverse of ${\bf k}$. The symmetry action on defects comes from the fact that if ${\bf g}$ and ${\bf k}$ do not commute, the group element associated to the fusion of two defects $a_{{\bf g}}$ and $c_{{\bf k}}$ depends on the order in which they fuse. In our examples, we can always choose the explicit form of $\rho_{{\bf k}}(a_{\bf g})$ so that the following fusion relation is satisfied:
\begin{equation}
c_{\bf k} \times a_{\bf g} = \rho_{{\bf k}}(a_{\bf g}) \times c_{\bf k} 
\end{equation}
This leads to the following definition of $\rho$:
\begin{align}
  \rho_{{\bf k}}(a_{{\bf g}}) &= [a \mathfrak{w}({{\bf k}},{{\bf g}})\overline{\mathfrak{w}({\bf  k g \bar{k}},{\bf  k})}]_{{\bf  k g \bar{k}}}
  \nonumber \\
&= [a \mathfrak{w}({{\bf g}},{{\bf \bar{k}}})\overline{\mathfrak{w}({\bf \bar{k}},{\bf  k g \bar{k}})}]_{{\bf  k g \bar{k}}}
\end{align}  

The mathematical classification of symmetry fractionalization and defect classes in the discrete case is given in Table \ref{Table:SpaceGrpClassif}, and a summary of formulas for the different $G$-crossed invariants is given in Table \ref{Table:LatticeInvts}. Importantly, the fractional quantum numbers associated to two distinct defects $a_{\bf g}$ and $\rho_{{\bf k}}(a_{{\bf g}})$ should be the same up to the corresponding quantum numbers of arbitrary anyons, for any ${\bf k}$. This can be verified in each of the invariant formulas. The derivation of these invariants is given in Appendix \ref{Sec:Invts}.

\subsubsection{Symmetry fractionalization invariants}

For $G=U(1) \leftthreetimes_{\phi} [\Z^2 \rtimes \Z_M]$ there are four symmetry fractionalization parameters, given by the anyons $\{v,s,m,\vec{t}\}$ and a set of equivalence relations among them. The classification for general $M$ is as follows (see Appendix \ref{Sec:Coho_calc_gspace} for a derivation):
\begin{equation}
\H^2( U(1) \leftthreetimes_{\phi} [\Z^2 \rtimes \Z_M] ,\A) \cong \A\times\A\times(\A/M\A)\times(K_M\otimes\A) .
\end{equation}
The parameters $v,s ,m$ were discused above: the distinct choices of $v,s,m$ are classified by the groups $\A,\A/M\A,\A$ respectively, and these anyons can be measured using the same $G$-crossed identities as given previously, after setting rotation group elements to zero. 

For a $2\pi/M$ disclination with $M \in \{2,3,4,6\}$, we define the group $K_M$ as follows:
\begin{equation}
K_M := \frac{\Z^2}{{^{1-h}}\Z^2}
\end{equation}
where $h$, the generator of $2\pi/M$ rotations, has the usual rotation action on vectors in $\Z^2$ (i.e. the action of $h$ is represented by a $2\times 2$ matrix written in the lattice basis). We find that
\begin{align}
K_2 &= \Z_2^2 \nonumber \\
K_3 &= \Z_3 \nonumber \\
K_4 &= \Z_2 \nonumber \\
K_6 &= \Z_1. 
\end{align}
The group $K_M$, which is discussed in additional detail in Ref. \cite{Manjunath2020}, represents a finite group grading on Burgers vectors, which can physically be understood in terms of conjugacy classes of the translation component of $G$. Physically this arises because the Burgers vector of an ``impure'' disclination is only well-defined up to a $2\pi/M$ rotation. For example, when $M=2$, a $\pi$ disclination defect Burgers vector ${\bf b} = (b_x,b_y)^T$ falls into one of four conjugacy classes determined by the values of $b_x \mod 2, b_y \mod 2$. 

Consider the case with $M=2$. In this case, the $\pi$ disclination defect corresponding to ${\bf g} = (1,{\bf x},e^{i \pi})$ has the property that ${\bf g^2} = {\bf 0}$; hence two $0_{\bf g}$ defects must fuse to give an Abelian anyon, $0_{\bf g}^2 = a$. In certain cases this anyon can be completely accounted for by rotation symmetry fractionalization: we could also obtain $0_{\bf h}^2 = a$ for ${\bf h} = (1,{\bf 0},e^{i \pi})$. In the general case we have $0_{\bf g}^2 = 0_{\bf h}^2 \times t_x$ for some Abelian anyon $t_x$. Now if $t_x = a^2$ for some $a \in \A$, this anyon can be trivialized by relabeling $0_{\bf h} \rightarrow a_{\bf h}$. However, if $t_x \ne a^2$ for any $a \in \A$, then there is a nontrivial symmetry fractionalization anyon associated to the fusion of two disclination defects with ${\bf g} = (1,{\bf x},e^{i \pi})$. The anyon $t_x$ cannot be understood in terms of rotational symmetry fractionalization alone. The number of distinct choices for $t_x$ is therefore given by the group $\frac{\A}{\{a^2, a \in \A \}} = \A/2\A$. For $M=2$ we can follow the same logic and independently associate an anyon to the fusion product of two dislocation defects ${\bf g'} = (1,{\bf y},e^{i \pi})$, and the number of distinct assignments is again given by $\A/2\A$. The total number of distinct assignments is given by 
\begin{equation}
K_2 \otimes \A = (\A/2\A)\times (\A/2\A).
\end{equation}
The definition of $\otimes$ is reviewed in Appendix \ref{Sec:GrpCohIntro}. Although we cannot directly discuss pure dislocation defects using these ideas, we can nonetheless treat a dislocation as a dipole of two disclination defects with opposite disclination angles. Since two copies of a disclination defect with Burgers vector ${\bf b}$ fuse to give an anyon $t_x^{b_x} t_y^{b_y} := \vec{t}\cdot {\bf b}$, a dislocation with Burgers vector $(2b_x,2b_y)$ can also be associated to the anyon $\vec{t} \cdot {\bf b}$. 

The above discussion is extended to $M=3,4,6$ as follows. Given a $2\times 2$ integer matrix $W = \begin{pmatrix}
a & b \\ c & d
\end{pmatrix}$ and a pair of anyons $\vec{t} = (t_x,t_y)$, let ${^{W}}\vec{t}$ denote the pair of anyons $(t_x^a \times t_y^b, t_x^c\times t_y^d)$ (this defines the action of $W$ on the group $\A\times\A$). Then, we find in general that a dislocation with Burgers vector ${\bf b} - {^{h}}{\bf b}$ is associated to the anyon $\vec{t} \cdot {\bf b}$. (As in the case $M=2$ discussed above, this symmetry fractionalization can also be understood by considering the fusion product of disclinations with nonzero ${\bf b}$, and subtracting away the possible contributions from rotational symmetry fractionalization alone.) Two pairs $(t_x, t_y)$ and $(t_x', t_y')$ describe the same symmetry fractionalization if
\begin{align}
\vec{t} = \vec{t'} \times \vec{\chi} \times [{^{h^T}}\vec{\chi}]^{-1}.
\end{align}
Here $h^T$ refers to the transpose of the $2\times 2$ matrix $h$. Now the difference between the pairs $\vec{t}$ and $\vec{t}'$ corresponds to attaching the anyons $\vec{\chi}\cdot {\bf b}$ and $ \vec{\chi}\cdot {^{h}}{\bf b} = {^{h^T}}\vec{\chi}\cdot {\bf b}$ to the dislocations with Burgers vectors ${\bf b}$ and ${^{h}}{\bf b}$ respectively. This attachment can always be achieved by adjusting local energetics in the underlying system Hamiltonian; therefore the symmetry fractionalization class remains the same upon replacing $\vec{t}$ with $\vec{t'}$ according to the equation above.

The equivalence relation on $\vec{t}$ implies that the classification of $(t_x,t_y)$ is not simply $\A\times\A$: in fact, any pair of the form $(t_x,t_y)^T = {^{1-h^T}} (t'_x,t'_y)^T$ can be seen to belong to the trivial symmetry fractionalization class. The physically distinct pairs $(t_x,t_y) \in \A\times\A$ give a set of symmetry fractionalization classes forming the group $K_M \otimes \A$, where
\begin{align}
K_2 \otimes \A &= (\A/2\A)\times (\A/2\A) \nonumber \\
K_3 \otimes \A &= \A/3\A \nonumber \\
K_4 \otimes \A &= \A/2\A \nonumber \\
K_6 \otimes \A &= \Z_1 .
\end{align}

Just as the anyons $v$ and $s$ yield the fractional $U(1)$ charge and angular momentum of an anyon $a$ via mutual braiding, the anyon $\vec{t}$ allows us to define a fractional linear momentum $\vec{P}_a$ in terms of braiding anyons around dislocations, as follows. The phase obtained by braiding an anyon $a$ around a dislocation with Burgers vector ${^{1- h}} {\bf r}$, which is in the trivial $K_M$ grading, is given by
\begin{align}\label{Eq:P_a}
e^{2\pi i \vec{P}_a \cdot {^{1-h}}{\bf r}} := M_{a,\vec{t} \cdot {\bf r}} . 
\end{align}

We can obtain a formula for $M_{a,\vec{t} \cdot {\bf r}}$, which allows us to effectively measure $\vec{t}$, using the $G$-crossed data by relying on the ideas discussed above for $M=2$. We consider an `impure' disclination defect, i.e. a defect $0_{\bf g}$ with nonzero disclination angle and dislocation Burgers vector ${\bf b}$ (for $M=2$, we chose ${\bf g} = (1,{\bf x},e^{i \pi})$). We can measure the anyon induced by the fusion of $n$ such impure disclinations, with $n$ being some integer that divides $M$ (above we chose $n=2$). This is done using the invariant $\F_a({\bf g}_1={\bf g},\dots , {\bf g}_n = {\bf g})$. Next we subtract away the contributions coming from $U(1)$ and rotational symmetry fractionalization. These contributions are measured by the invariant $\F_a({\bf g}_1={\bf h},\dots , {\bf g}_n = {\bf h})$ for some suitably defined ${\bf h}$. (Above, this was done by considering ${\bf h} = (1,{\bf 0},e^{i \pi})$.) The symmetry fractionalization invariant which measures $\vec{t}$ thus takes the form

\begin{equation}
\frac{\F_a({\bf g}_1={\bf g},\dots , {\bf g}_n = {\bf g})}{\F_a({\bf g}_1={\bf h},\dots , {\bf g}_n = {\bf h})} ,
\end{equation} 
where ${\bf g}$ contains a nontrivial translation component (corresponding to a non-trivial dislocation Burgers vector), and the translation component of ${\bf h}$ is the identity (corresponding to a trivial dislocation Burgers vector); the other components of ${\bf g}$ and ${\bf h}$ are equal.

The explicit formulas for each $M$ are summarized in Table \ref{Table:LatticeInvts}, and a detailed derivation is discussed in Appendix \ref{Sec:Invts-gspace}. There are a few subtleties regarding the choice of ${\bf g}$ that are also discussed there in detail. 

For each anyon $a$, the invariant $\frac{\F_a({\bf g}_1={\bf g},\dots , {\bf g}_n = {\bf g})}{\F_a({\bf g}_1={\bf h},\dots , {\bf g}_n = {\bf h})}$ which measures $\vec{t}$ can also be related to the fractional linear momentum $\vec{P}_a$ of the anyon $a$. Note that there is some ambiguity in $\vec{P}_a$ as determined by these formulas, as we discuss below.

\begin{enumerate}
	\item For $M=2$, we expect from group cohomology that the symmetry fractionalization class is specified by the anyons $t_x,t_y$ modulo anyons $a^2$, where $a \in \A$. Thus the proposed invariant must measure $M_{a,\vec{t} \cdot {\bf r}}$ where ${\bf r}$ is arbitrary. Indeed, if we choose ${\bf g} = (1,{\bf r},e^{i \pi})$ and ${\bf h} = (1,{\bf 0},e^{i \pi})$, we obtain
	 \begin{equation}\label{Eq:P_a2}
	 \frac{\F_a({\bf g},{\bf g})}{\F_a({\bf h},{\bf h})} = M_{a,\vec{t} \cdot {\bf r}} = e^{2\pi i (2 \vec{P}_a \cdot {\bf r})}, 
	 \end{equation}
	 as desired. Note that only the components of $2\vec{P}_a$ are determined modulo 1: fixing the symmetry fractionalization class does not completely fix $\vec{P}_a$. \\
	\item For $M=3$, we expect from group cohomology that the symmetry fractionalization class is specified by the combination $t_xt_y$ (or equivalently $t_x \overline{t_y}^2$) modulo anyons of the form $a^3$, where $a \in \A$. Thus the proposed invariant must measure $M_{a,t_x \overline{t_y}^2}$. In this case, if we choose ${\bf g} = (e^{2\pi i z},{\bf x},e^{2\pi i/3})$ and ${\bf h} = (e^{2\pi i z},{\bf y},e^{2\pi i/3})$, we have 
	\begin{equation}\label{Eq:P_a3}
	\frac{\F_a({\bf g},{\bf g},{\bf g})}{\F_a({\bf h},{\bf h},{\bf h})} = M_{a,t_x \overline{t_y^2}} = e^{2\pi i (3\vec{P}_a \cdot ({\bf x} - {\bf y}))}. 
	\end{equation}
	The relation to $\vec{P}_a$ was obtained using the definition in Eq. \eqref{Eq:P_a}. (The value of $z$ is chosen so that ${\bf g}^3 = {\bf h}^3 = {\bf 0}$, but the final answer is independent of $z$.) Now, only the quantity $3\vec{P}_a \cdot ({\bf x} - {\bf y})= 3(P_{a,x} -P_{a,y})$, is determined modulo 1. Here we have used the matrix representation $h = \begin{pmatrix}
	0 & 1 \\ -1 & -1 \end{pmatrix}$. \\
	\item For $M=4$, we expect from group cohomology that the symmetry fractionalization class is specified by the combination $t_xt_y$ (or equivalently $t_x \overline{t_y}$) modulo anyons of the form $a^2$, where $a \in \A$. Thus the proposed invariant must measure $M_{a,t_x \overline{t_y}}$. If we choose ${\bf g} = (1,{\bf r},e^{i \pi})$ and ${\bf h} = (1,{\bf 0},e^{i \pi})$, we obtain
	\begin{equation}\label{Eq:P_a4}
	\frac{\F_a({\bf g},{\bf g})}{\F_a({\bf h},{\bf h})} = M_{a,(t_xt_y)^{r_x}\times (t_x\overline{t_y})^{r_y}} = e^{2\pi i (2 \vec{P}_a \cdot {\bf r})}.
	\end{equation}
	Again, we used Eq. \eqref{Eq:P_a} to relate the invariant to $\vec{P}_a$. Therefore the invariant does measure the symmetry fractionalization class, but only the components of $2\vec{P}_a$ are determined modulo 1. We have used the matrix representation $h = \begin{pmatrix}
	0 & 1 \\ -1 & 0 \end{pmatrix}$ in the above calculation.
\end{enumerate}

The more detailed aspects of the derivations above are given in Appendix \ref{Sec:Invts-gspace}. By choosing different values of ${\bf r}$, we can determine $t_x$ and $t_y$ up to equivalences. The values of $\vec{t}$ measured with these formulas will change when we relabel the defects or consider a different defect in the same conjugacy class. However, as we show in Appendix \ref{Sec:Invts-gspace}, this change is trivial and is accounted for precisely by the equivalence relation on $\vec{t}$. 

The full symmetry fractionalization data can be understood within an effective Lagrangian approach, which we summarize for Abelian topological orders below. 

\subsubsection{Effective action and response theory}

Different terms in the SET classification can also be associated to concrete physical responses via an effective action in terms of a background symmetry gauge field $(A,\vec{R},C)$ which obeys the above group multiplication law. For Abelian topological phases described by a $K$ matrix, the effective action (derived in Appendix \ref{Sec:Eff_Ac-gspace}) can be written as follows: 
\begin{widetext}
\begin{align} \label{FullEffAction_NonzeroFlux}
\mathcal{L}_{frac} &= \frac{1}{2\pi} a^I \cup (v_I dA + s_I dC + \vec{t}_I \cdot d \vec{\cancel{R}} + (m_I + \phi q_I) A_{XY} ) \nonumber \\
\mathcal{L}_{SPT} &= \frac{k_1}{2\pi} A \cup dA + \frac{k_2}{2\pi} A \cup dC + \frac{k_3}{2\pi} C \cup dC 
+ \frac{1}{2\pi} A \cup (\vec{k}_4 \cdot  d \vec{\cancel{R}})+\frac{1}{2\pi} C \cup (\vec{k}_5 \cdot d \vec{\cancel{R}}) 
+ \left(\frac{k_6 + 2\phi k_1}{2\pi} A + \frac{k_7 + \phi k_2}{2\pi} C\right) \cup A_{XY} \nonumber \\
&+ \frac{\phi^2 k_1}{2\pi} d^{-1}(A_{XY} \cup A_{XY}) + \frac{\phi}{2\pi} A_{XY} \cup (\vec{k}_4 \cdot  \vec{\cancel{R}})
\end{align}
\end{widetext}
This is the generalization to $\phi \neq 0$ of the result of Ref. \cite{Manjunath2020}. Here we have used $\vec{v}$ to denote the charge vector, to avoid confusion with the denominator of $\phi = p/q$ Also, we have defined $\vec{\cancel{R}} = {^{(1-h)^{-1}}}\vec{R}$. 

In our simplicial formulation, the above action is defined on a 3-manifold $M$ by demanding that (i) the gauge field $(A,\vec{R},C)$ is flat, and (ii) the partition function of the system is invariant under a retriangulation of $M$. Note that in the case with $\phi \ne 0$, some of the terms acquire an explicit dependence on $\phi$.

The terms in the SPT effective Lagrangian $\mathcal{L}_{SPT}$ are in direct correspondence with the group cohomology classification,
\begin{align}
\mathcal{H}^3( U(1) \leftthreetimes_\phi [\Z^2 \rtimes \Z_M ], U(1)) = \Z^2 \times \Z_M^3 \times K_M^2 .
  \end{align}
The coefficients $k_1, k_6 $ are $\Z$ valued, the coefficients $k_2, k_3, k_7$ correspond to the  $\Z_M^3$ part of the classification, and $\vec{k}_4, \vec{k}_5$ correspond to the $K_M^2$ part of the classification. We emphasize that different choices of the $k_i$ do not give distinct SETs in general, and one needs to compute a further reduction to fully classify distinct SETs. In the effective Lagrangian for Abelian topological orders presented above, this reduction can be computed by considering redefinitions of the internal dynamical gauge fields. 
  
Now we can integrate out the dynamical gauge fields $a^I$ and obtain the following response theory:

\begin{widetext}
	\begin{align}
	\label{responseAction}
	\mathcal{L}_{\text{eff}} &= \frac{\sigma_H}{2} A \cup dA + \frac{\mathscr{S}}{2\pi} A \cup dC + \frac{\ell_s}{4\pi} C \cup dC  
	+ \frac{\vec{\mathscr{P}}_c }{2\pi} \cdot (A \cup d\vec{R}) +\frac{\vec{\mathscr{P}}_s}{2\pi} \cdot (C \cup d\vec{R}) 
	+ \frac{1}{2\pi q}\left(\nu A + \nu_s C\right) \cup A_{XY} \nonumber \\
	& + \frac{\vec{\nu}_p}{2\pi q} \cdot \vec{R} \cup A_{XY} + \frac{\Pi_{ij}}{4\pi} R_i \cup d R_j + \frac{\alpha}{4\pi} A_{XY} \cup d^{-1} A_{XY} + \mathcal{L}_{\text{anom}} ,
	\end{align}
\end{widetext}
where $\mathcal{L}_{\text{anom}}$ is the gravitational CS term that arises from the framing anomaly. Since the gauge field $C$ is the gauge field associated with $SO(2)$ rotations, we set it equal to the spatial component of the spin connection that arises in the gravitational CS term, and set the other components of the space-time spin connection to zero (see also Ref. \onlinecite{Manjunath2020} for a discussion of this). We thus get
\begin{align}
\mathcal{L}_{\text{anom}} = - \frac{c}{48\pi} C \cup dC .
\end{align}

We have normalized the fillings $\nu, \nu_s, \vec{\nu}_p$ by $q$ so that they correspond to the $U(1)$ charge, angular momentum, and linear momentum per magnetic unit cell.

The individual responses associated to each of these terms are discussed in Table \ref{Table:SpaceGrpClassif}. For Abelian topological phases, the effective action and the associated response theory provide a complete characterization of the allowed SET phases. Although we do not write down the effective action for general non-Abelian topological orders, the response theory will take the same general form as in the Abelian case. In the $G$-crossed BTC, the response coefficients can be determined entirely from the fusion and braiding data associated to the Abelian anyons and defects.

An important result of our calculations is that the SET classification with $G$ symmetry for any value of flux $\phi$ is isomorphic to that of $G_{\text{space}}\times U(1)$, even though the $G$-crossed data and the effective action explicitly depend on $\phi$.

The $G$-crossed invariants are summarized in Table \ref{Table:LatticeInvts}. Below we discuss them in more detail. 
 
\subsubsection{Defect invariants}

Here we discuss the invariants which measure the defect class. From Appendix \ref{Sec:Coho_calc_gspace} we find that for $G=U(1)\leftthreetimes_{\phi}[\Z^2\rtimes \Z_M]$,
\begin{equation}
\H^3(G,U(1)) \cong \H^4(G,\Z) \cong \Z^2\times \Z_M^3 \times K_M^2.
\end{equation}
We have already discussed how to measure the Hall conductivity $\bar{\sigma}_H$, the filling per magnetic unit cell $\nu$ and the quantity $\ell_s$, which defines the angular momentum of an elementary $2\pi/M$ disclination. We further have a discrete analog of the shift $\mathscr{S}$, which is a mixed topological invariant of $U(1)$ and $\Z_M$ symmetry, which is defined modulo $M$. We define the shift as follows:
\begin{align}
\mathscr{S} &= k_2 + v \star s.
\end{align}
To obtain the corresponding $G$-crossed invariant, we set ${\bf g} = (e^{2\pi i/M},{\bf 0},1),{\bf h} = (1,{\bf 0},e^{2\pi i/M})$, and evaluate
\begin{align}\label{Invt_D_Shift_disc}
 e^{\frac{2\pi i}{M} \mathscr{S}} = \frac{\mathcal{I}_M(0_{\bf gh})}{\mathcal{I}_M(0_{\bf g})\mathcal{I}_M(0_{\bf h})} 
\end{align}
The fractional $U(1)$ charge associated to an elementary $2\pi/M$ disclination or to a corner of angle $2\pi$ is given by $\mathscr{S}/M$. The angular momentum of a $2\pi/M
$ $U(1)$ flux is given by $\mathscr{S}/M$. 

As with $\ell_s$ (which is defined modulo $2M$), $e^{2\pi i \mathscr{S}/M}$ as defined above may change under a defect relabeling, which physically means that the fractional $U(1)$ charge and angular momentum of a disclination are only well-defined modulo the charge and angular momentum of certain anyons that can be used to relabel the disclination defects without changing the other $G$-crossed BTC data (see Appendix \ref{Sec:relabeling}). 

Next we discuss fractionally quantized responses in which defects with nontrivial dislocation Burgers vectors possess a fractional charge and angular momentum. The fractional charge of a dislocation with Burgers vector ${\bf b}$ is given by $\vec{\mathscr{P}}_c \cdot {\bf b}$, where we define
\begin{align}
  \vec{\mathscr{P}}_{c} = {^{(1-h)^{-1}}}(v \star \vec{t} + \vec{k}_4).
\end{align}
$\vec{\mathscr{P}}_c$ can be interpreted as a fractionally quantized charge polarization, which also associates a momentum $\vec{\mathscr{P}}_c$ to a $2\pi$ instanton uniformly spread out throughout the system, and a boundary charge $\vec{\mathscr{P}}_c \cdot \hat{n}$ per unit length for a boundary along the $\hat{n}$ direction. Note that this response is nonquantized in the absence of rotational symmetry \cite{song2019electric}, because in that case one can add a term $dA \cup \vec{R}$ to the topological response theory with arbitrary coefficient and which is topologically invariant, but corresponds to a coboundary term and is therefore topologically trivial (and thus not fixed by the $G$-crossed theory).

An analogous fractional angular momentum polarization $\vec{\mathscr{P}}_s$, which gives the fractional angular momentum $\vec{\mathscr{P}}_s \cdot {\bf b}$ to a dislocation with Burgers vector ${\bf b}$ is given by
\begin{align}
  \vec{\mathscr{P}}_{s} = {^{(1-h)^{-1}}}(s \star\vec{t} + \vec{k}_5). 
\end{align}

 The method used to extract $\vec{\mathscr{P}}_c$ and $\vec{\mathscr{P}}_s$ is similar to the method for measuring the symmetry fractionalization parameter $\vec{t}$. We choose ${\bf g} = (e^{2\pi i z},{\bf r},e^{2\pi i/n})$ as an order $n$ element of $G$, where for $M=2,3,4$, we have $n = 2, 3, 2$ respectively. 
We measure the total symmetry charge associated to a ${\bf g}$-flux by computing the quantity $\mathcal{I}_n(0_{{\bf g}})$, and then subtract various contributions from the responses already discussed above in order to extract the desired contribution from $\vec{\mathscr{P}}_c $ and $\vec{\mathscr{P}}_s $. The explicit formulas which accomplish this are given in Table \ref{Table:LatticeInvts}, and a derivation is discussed in Appendix \ref{Sec:Invts-gspace}.

A formula for the filling $\nu$ was given in Eq. \eqref{Invt_D_nuC}. Analogously, we have the angular momentum filling $\nu_s$, defined as the total angular momentum per magnetic unit cell. 
From group cohomology, we expect that there is additionally an SPT state associated to an integer angular momentum per unit cell, given by $k_7 \mod M$. The total angular momentum per magnetic unit cell is then intuitively given by the sum of $q k_7$ and the fractional angular momentum of the anyon $v^p \times m^q$ in each magnetic unit cell. That is, we expect
\begin{align}
 \nu_s = q k_7 + q L_m + p L_v,
\end{align}
where recall $L_a$ is the angular momentum of $a$. Although the rhs does appear in the response theory as the coefficient of the term $C \cup A_{XY}$, we have not been able to find a corresponding $G$-crossed invariant.

In addition to the defect (SPT) invariants presented above, there are additional quantized responses that are not associated to SPTs but are nonetheless constrained by the symmetry fractionalization parameters discused above. Two responses of this type are the quantized torsional response $\Pi_{ij}$ and a momentum per unit area $\vec{\nu}_p$ in the ground state. 

The quantized torsional response $\Pi_{ij}$ is defined as the $i$ component of linear momentum associated to a dislocation whose Burgers vector is a unit vector in the $j$ direction. Thus, the momentum $p_i$ of a dislocation with Burgers vector ${\bf b}$ is given by $p_i = \sum_j \Pi_{ij} b_j$. 

From crystalline gauge theory, written in a $K$-matrix formalism, we obtained \cite{Manjunath2020}  
\begin{equation}
\Pi_{ij} = \sum_{k,l} {(1-h)^{-1}_{ik}} (\vec{t}_k^T K^{-1} \vec{t}_l) {(1-h)_{lj}^{-1}}.
\end{equation}
We will derive an invariant within the $G$-crossed theory which is consistent with this definition of $\Pi_{ij}$; however, this invariant will not unambiguously determine the components $\Pi_{ij}$. Indeed, just as the momentum of an anyon $\vec{P}_a$ was not fully determined by $G$-crossed invariants, we will see that only the components $M \Pi_{ij} \mod 1$ are fixed by this invariant.

We saw that we can determine the anyon momentum components $\vec{P}_a \cdot ({\bf b}-{^{h}}{\bf b}) \mod 1$. Therefore we naively expect that we should also be able to determine the dislocation momentum components $\Pi_{ij} ({b_j - {^{h}}{\bf b}}_j) \mod 1$. Our proposed invariant gives a coarser characterization than expected: for example, with $M=4$ we expect to be able to fix $2\Pi_{ij} \mod 1$ by taking ${\bf b} = (1,\pm 1)$, but this invariant only fixes $4\Pi_{ij} \mod 1$. It is possible that a finer characterization exists, but we have not derived it in this paper; this problem will be left for future work.  

Below we will describe the invariant. Define the defects 

\begin{align}
d_1 &= 0_{(1,{\bf 0},e^{2\pi i/M})}, \;\;\; d_1^M = s \nonumber\\ 
d_2 &= 0_{(1,{\bf r'},e^{2\pi i/M})}, \;\;\; d_2^M = s \times \vec{t} \cdot {^{M(1-h)^{-1}}} {\bf r'} \nonumber\\ 
d_3 &= 0_{(1,{\bf r},e^{2\pi i/M})}, \;\;\; d_3^M = s  \times \vec{t} \cdot {^{M(1-h)^{-1}}} {\bf r} \nonumber\\ 
d_4 &= 0_{(1,{\bf r+r'},e^{2\pi i/M})}, \;\;\; d_4^M = s  \times \vec{t} \cdot {^{M(1-h)^{-1}}}({\bf r} + {\bf r'})
\end{align}
For simplicity, we have assumed here that $v=m=0$. In the general case the formula below can still be used, but additional contributions from $v$ and $m$ need to be subtracted carefully in order to isolate $\Pi_{ij}$. 

We then define
\begin{equation}\label{Eq:THV}
e^{2\pi i M {\bf r}^T \Pi {\bf r'}} := \frac{\mathcal{I}_M(d_4)\mathcal{I}_M(d_1)}{\mathcal{I}_M(d_2)\mathcal{I}_M(d_3)}.
\end{equation}

Note that the exponent in the lhs has a factor of $M$. In Appendix \ref{Sec:Invts-magtrans} we use the ribbon property of anyons to show that

\begin{align}
\left(\frac{\mathcal{I}_M(d_4)\mathcal{I}_M(d_1)}{\mathcal{I}_M(d_2)\mathcal{I}_M(d_3)} \right)^M &=  M_{\vec{t} \cdot {^{M(1-h)^{-1}}} {\bf r}, \vec{t} \cdot {^{M(1-h)^{-1}}} {\bf r'}}.
\end{align}

In order to be consistent with the field theory result, we see that the rhs of the above equation should be equal to $e^{2\pi i M^2 \sum_{i,j} {\bf r}_i \Pi_{ij} {\bf r'}_j}$; this justifies the factor of $M$ in the lhs of Eq. \eqref{Eq:THV}. The proposed invariant therefore gives the value of 
\begin{equation}
M \vec{p} \cdot {\bf r} \mod 1 = M \sum_j {\bf r}_i \Pi_{ij} {\bf r'}_j \mod 1,
\end{equation}
where $p_i = \Pi_{ij} r_j'$ is the momentum of a dislocation with burgers vector ${\bf b} = {\bf r'}$. By choosing different values of ${\bf r}$, the above formula allows us to obtain the momentum $M p_i \mod 1$. 
Thus there is an ambiguity in the value of the fractionally quantized torsional response, similar to the ambiguity that exists for the momentum of an anyon.

Finally, the response theory also predicts a fractionally quantized momentum $\vec{\nu}_p$, corresponding to a fractional linear momentum per magnetic unit cell. This is understood as the momentum of the anyon $v^p \times m^q$ in each magnetic unit cell. The crystalline gauge theory gives the expected result
\begin{equation}
\vec{\nu}_p = p \vec{P}_v + q \vec{P}_m.
\end{equation}
From group cohomology we find that there are no additional SPT coefficients associated to this response: therefore the momentum filling is simply given by the total momentum of the anyons in each magnetic unit cell. However, similar to the angular momentum filling, we do not have a formula for this quantity in terms of $G$-crossed defect data.

\subsubsection{Fractional quantum numbers of arbitrary symmetry defects}
\label{Sec:PropArbDefects}

In the previous section, we saw that the invariants measuring the charge, angular momentum and linear momentum of certain specially chosen defects can be used to characterize SET phases in the $G$-crossed BTC formalism. An important related question is whether we can compute these symmetry properties for arbitrary defects $a_{\bf g}$, where ${\bf g} = (e^{2\pi i z},{\bf r},e^{2\pi i j /M})$, and $j = 0,\cdots, M-1$. This involves generalizing the formulas for the defect quantum numbers, to arbitrary $a$ as well as arbitrary ${\bf g}$. The generalization to arbitrary $a$ is straightforward: we simply add the quantum numbers of $0_{\bf g}$, which were discussed above, to those of $a$. However, the generalization to arbitrary ${\bf g}$ is less straigntforward. This problem is discussed below.

\paragraph*{Fractional $U(1)$ charge of defects}

First, consider the $U(1)$ charge of a defect $a_{\bf g}$. Note that $Q_{a_{\bf g}} = Q_a + Q_{0_{\bf g}}$ (with all quantities defined modulo 1), so the problem lies in determining $Q_{0_{\bf g}}$ for arbitrary ${\bf g}$. Now there is a natural way in which the fractional charge of such a defect can be measured if the defect satisfies $0_{\bf g}^p = f$, where $f$ is an Abelian anyon. (This condition requires that ${\bf g}^p = {\bf 0}$.) The idea is that if $p$ copies of $0_{\bf g}$ give the anyon $f$, then the fractional charge $Q_{0_{\bf g}}$ of $0_{\bf g}$ is $1/p$th that of $f$, which is in turn given by $Q_f \mod 1 = v \star f \mod 1$. The formulas for deriving $Q_{0_{\bf g}}$ follow from our earlier discussion in Section \ref{Sec:GenResInvts}; they apply to any combination of $U(1)$ fluxes and disclination defects with nonzero disclination angle.

However, not all defects are contained within the above discussion. For example, pure dislocation defects, corresponding to ${\bf g} = (1,{\bf r},1)$, cannot fuse with each other to give an anyon, because ${\bf g}^p$ is always nonzero in this case. Now we cannot directly apply the above formula. However, we can always express a pure dislocation defect as a fusion product of two disclination defects $a_{\bf g_1},b_{\bf g_2}$ with opposite disclination angles. We can measure the fractional charge of $a_{\bf g_1}$ and $b_{\bf g_2}$ separately; the charge of a pure dislocation is then obtained as the sum of the charges of $a_{\bf g_1}$ and $b_{\bf g_2}$.

If this procedure were completely well-defined, the $U(1)$ charge of a dislocation dipole obtained in this way would have the same value irrespective of the choice of $a_{\bf g_1}$ and $b_{\bf g_2}$. In general, however, we do not expect this. In fact, from crystalline gauge theory predictions \cite{Manjunath2020}, we only expect that the charge of a dislocation is well-defined up to the charges of the Abelian anyons in the system. This is because we can always attach anyons to a localized defect by adjusting the local energetics of the system.

\paragraph*{Fractional angular momentum of defects}

Next, we consider the fractional angular momentum of defects. First note that $L_{a_{\bf g}} = L_a + L_{0_{\bf g}}$ (with all quantities defined modulo 1), so the problem lies in determining $L_{0_{\bf g}}$ for arbitrary ${\bf g}$. It is straightforward to define the angular momentum for defects $0_{\bf g}$ where ${\bf g}$ belongs to the $U(1)\times\Z_M$ subgroup of $G$. In this case, the angular momentum $L_{0_{\bf g}}$ is obtained by a direct application of the formula for the mixed defect invariant, Eq. \eqref{Eq:MixedTInvt}.

By performing the explicit computations, we can verify the expected result that the angular momentum of the defect $0_{\bf g}$ for ${\bf g} = (e^{2\pi i k/n},{\bf 0},e^{2\pi i j/M})$ equals
\begin{align}
L_{0_{\bf g}} = j \frac{\ell_s}{M} + \frac{k}{n} \mathscr{S},
\end{align}
i.e., it is the sum of two angular momenta: that of the $U(1)$ flux component of $0_{\bf g}$, and the pure disclination component of $0_{\bf g}$. Note that $\ell_s$ and $\mathscr{S}$ are not absolute SET invariants: they may transform under defect relabelings, as we discuss in Appendix \ref{Sec:relabeling}.

When we try to generalize this procedure to defects with nonzero dislocation Burgers vector, we encounter the difficulty that even if ${\bf g}^n = {\bf 0}$ for some $n$, the anyon associated to $0_{\bf g}^n$ has in general a contribution from $m$ due to translation symmetry fractionalization.  
Accounting for this added subtlety requires a more elaborate analysis, which we do not pursue here. However, if $m=0$, the procedure is similar.

As in the case of the fractional charge, we cannot directly determine the angular momentum of a pure dislocation by using the usual defect invariants. In this case we define the angular momentum as the sum of the angular momenta of two disclinations whose dipole gives the dislocation with the correct Burgers vector. However, this definition has an ambiguity. When we consider different decompositions of this kind, we can only expect to obtain answers that differ by the angular momentum of arbitrary anyons, as was found in the crystalline gauge theory approach \cite{Manjunath2020}.

\paragraph*{Fractional linear momentum of defects}

Finally, we discuss how to measure the linear momentum of an arbitrary defect. Note that there is an extra ambiguity in the definition of linear momentum compared to those of $U(1)$ charge and angular momentum: the momentum $\vec{P}_a$ of an arbitrary anyon $a$ is only determined through the components $\vec{P}_a \cdot {^{1-h}}{\bf b} \mod 1$. Since the quantum numbers of defects are constrained by the quantum numbers of anyons due to symmetry fractionalization, we will only be able to determine the momentum of symmetry defects up to this constraint. 

There is no direct formula to measure the fractional momentum of $0_{\bf g}$ when $0_{\bf g}$ is a pure $U(1)$ flux. The only exact statement we can make is that the momentum of a $2\pi$ flux equals $\vec{\mathscr{P}}_c$, which can be determined unambiguously and characterizes the defect class associated to the charge polarization. Note that $\vec{\mathscr{P}}_c$ itself transforms under a relabeling of defects, and is not an absolute SET invariant (see Appendix \ref{Sec:relabeling_full} for a discussion). The momentum of a fractional flux can be defined indirectly by demanding that the momentum is additive as well as a continuous function of the flux, however the momentum of a fractional flux cannot be measured directly by a $G$-crossed invariant.   

Similarly, there is no direct formula to measure the momentum of an elementary disclination. The only exact statement we can make is that the momentum of a composite of $M$ elementary disclinations equals $\vec{\mathscr{P}}_s$, which characterizes the defect class associated to the angular momentum polarization. $\vec{\mathscr{P}}_s$ is also not an absolute SET invariant.

The fractional momentum of a dislocation is defined within a field theory formalism by the quantity $\sum_j\Pi_{ij} b_j \mod 1$, where ${\bf b}$ is the Burgers vector of the dislocation. However, the $G$-crossed invariant we have proposed only measures $M \Pi_{ij} \mod 1$. As discussed in the previous section, it is not clear whether this formula gives the sharpest possible characterization within the $G$-crossed theory.

In addition to these constraints, we note that when we try to compute the momentum of composite defects, the following subtleties encountered in the calculation of angular momentum are also present while calculating the linear momentum: 1) when the system has $m \ne 0$, the computations require a more elaborate analysis; and 2) The momentum of the fusion product $a_{\bf g}\times b_{\bf h}$ can be determined from those of $a_{\bf g}$ and $b_{\bf h}$ up to the momentum of arbitrary anyons.

\section{Discussion}
\label{Sec:Discussion}

In this paper we have used $G$-crossed BTCs to give a systematic classification and characterization of (2+1)D topological orders with $U(1)$ charge conservation symmetry, magnetic translational symmetry, and spatial rotational symmetry, both in the continuum and on the lattice, and in the presence of an arbitrary magnetic field (with rational flux per unit cell). Our results can thus be viewed as providing a classification of FQH states on a lattice, fractional Chern insulators, or quantum spin liquids, for all $5$ orientation-preserving space group symmetries on a lattice. Our results are comprehensive for bosonic topological phases. For fermionic states, the analog of $G$-crossed BTCs is not fully developed, although we expect only minor modifications in the presence of $U(1)$ charge conservation symmetry; for example the quantization rules of various quantized response properties may be slightly modified. 

Our work has uncovered a new class of symmetry fractionalization quantum numbers, which associates fractional linear momentum to the anyons, in addition to providing a general framework to characterize fractional charge, angular momentum, and fractionalization of the translation algebra for both Abelian and non-Abelian topological phases of matter. Furthermore, our work shows that for the symmetry groups considered here, these fractional quantum numbers fully classify all possible patterns of symmetry fractionalization.

Our study of the fractionally quantized responses has uncovered a large class of responses to lattice defects, and explicit formulas for these responses in terms of data of the $G$-crossed BTC have been obtained. The understanding of these responses in terms of $G$-crossed BTC data may provide a new way of extracting these properties from the properties of ground state many-body wave functions. It would be interesting to revisit the large class of fractional Chern insulators discovered in numerical simulations and in experiments \cite{parameswaran2013,Spanton2018FCI} and to fully characterize the associated SET invariants keeping the full space group symmetry in mind.

An important next step is to develop a more comprehensive understanding of model wave functions that can give rise to phases that possess the distinct symmetry fractionalization and response properties studied here, and to complement the mathematical analysis here with microscopic numerical calculations on model wave functions. We note that the methods of Ref. \onlinecite{cheng2016model,heinrich2016,Bulmash2020} can be used in principle to derive exactly solvable models in general for either non-chiral SET phases in purely (2+1)D \cite{cheng2016model,heinrich2016} or at the (2+1)D boundary of (3+1)D SPTs \cite{Bulmash2020}.

We have not obtained expressions for the linear and angular momentum per unit cell in terms of the $G$-crossed data, which we leave as a problem for future work. Furthermore, the $G$-crossed formulas for the quantized torsional response only determined $M\Pi_{ij} \mod 1$ as opposed to $\Pi_{ij} \mod 1$, where $M$ is the order of the rotation point group; we leave it as a problem for future work to provide a more complete formula. 

The explicit solutions to the $G$-crossed BTC equations that we presented were complete for the case of $G = U(1)$ and $G = \Z_M$. However for the rest of the symmetry groups considered in this paper, we made an additional technical assumption on the form of the solutions discussed in Appendix \ref{Sec:Gxsolns}. Consequently, our proof of $\nu = \bar{\sigma}_H$ holds for these classes of solutions (note that the relation $e^{2\pi i \nu} = e^{2\pi i \bar{\sigma}_H}$ was proven in general without making any assumptions).  To complete the analysis, it may be useful to find solutions to the $G$-crossed equations in complete generality without making any assumptions, which may then allow a completely general proof of the relation $\nu = \bar{\sigma}_H$ within the framework of $G$-crossed BTCs. Analogously, for lattice FQH systems, whenever we are able to write explicit solutions, we can prove that $\nu = q(v \star m + k_6) + q\phi \bar{\sigma}_H$ (Appendix \ref{Sec:LSM}). Without using explicit solutions, we can only prove this result modulo 1 within the framework of $G$-crossed BTCs.

It would be interesting to further generalize our results to larger continuous non-Abelian on-site symmetries, such as $SO(3)$ as appropriate for quantum spin liquids, and derive expressions for the associated quantized responses entirely in terms of the $G$-crossed BTC data. 

In our work we have assumed the case where symmetries do not permute anyon types. One can also consider the case where lattice symmetries permute distinct anyons types, in which case lattice dislocations or disclinations can become non-Abelian defects \cite{barkeshli2012a,barkeshli2013}. Moreover, including additional symmetries such as a layer permutation symmetry can also give rise to such non-Abelian twist defects. We leave a comprehensive classification of such possibilities for future work. 

We note that while FQH states break time-reversal $\mathcal{T}$ and reflection $\mathcal{R}$ symmetries, they preserve the combination. Therefore one can also study the role of $\mathcal{R} \mathcal{T}$ symmetry, and thus extend the classification in principle to all 17 space groups in two spatial dimensions with a flux $\phi$ per plaquette. However such an analysis will require a general systematic mathematical framework to treat anti-unitary and spatial reflection symmetries, which is currently only partially developed \cite{Barkeshli2019,barkeshli2019tr,Barkeshli2020Anomaly,Bulmash2020}.

Finally, we note that a non-interacting system projected to a single Landau level has a $W_\infty$ symmetry. It may be interesting to consider other unconventional subgroups of the $W_\infty$ symmetry  that may be preserved in the presence of specially chosen interactions, beyond the case $G = U(1) \leftthreetimes \mathbb{E}^2$.

\section{Acknowledgements}

We are grateful to Su-Kuan Chu for collaboration during the initial stages of this project, and specifically for helping with the computations of $\mathcal{H}^n(U(1) \leftthreetimes E^2, \Z)$ and with the mathematical reviews of Appendix \ref{Sec:GrpCohIntro}, \ref{app_sseq}. We thank Zhenghan Wang, Dan Freed, Meng Cheng, David Penneys, Corey Jones, Ryan Thorngren, and Dominic Else for discussions. We thank Sankar Das Sarma for comments on the draft. This work is supported by NSF CAREER (DMR- 1753240), an Alfred P. Sloan Research Fellowship, UMD startup funds, and the NSF Physics Frontier Center at the Joint Quantum Institute at UMD. 

\clearpage
\appendix

\section{Spatial symmetries of FQH states}
\label{Sec:FQHSymm}
\subsection{Group multiplication laws for $G= U(1)\leftthreetimes \mathbb{E}^2$ and $G = U(1)\leftthreetimes_{\phi} [\Z^2 \rtimes \Z_M]$}
\label{Sec:FQHContSymm}

In this Appendix, we review our notation for the symmetry of continuum and lattice FQH systems in the two-dimensional plane, in the presence of a uniform background magnetic field. We will focus on certain subtleties that may not fall in standard treatments of magnetic translation symmetry.  

First we study continuum FQH systems with $U(1)$ charge conservation and $\R^2$ translation symmetry. A general group element of the magnetic translation symmetry of an $N$-particle system is denoted ${\bf g}=(e^{2\pi i N z},{\bf r})$. 
In terms of symmetry operators it can be represented as
\begin{equation}
U({\bf g})=U_{c}(z)\tilde{T}_{{\bf r}},
\end{equation}
where the $U(1)$ symmetry operators $U_c(z)$ and the magnetic translation operators $\tilde{T}_{{\bf r}}$ are defined as follows:
\begin{widetext}
	\begin{align}
		U_{c}(z) &=\exp\left(i2\pi z\int d^{2}\mathbf{R}\,\rho(\mathbf{R})\right) = e^{i 2\pi z N}, \\
		\tilde{T}_{\mathbf{r}}&=\exp\left(\int d^{2}\mathbf{R}\,\psi^{\dagger}(\mathbf{R})\left[r_{x}\left(-i\nabla_{x}+A_{x}({\bf R})\right)
		+r_{y}\left(-i\nabla_{y}+A_{y}({\bf R})\right)\right]\psi(\mathbf{R})\right),
	\end{align}
      \end{widetext}
      where $\psi(\mathbf{R})$ is the quantum field describing the particles in the microscopic system, and $\rho(\mathbf{R})$ is the charge density.
Here $\vec{A}$ is the usual vector potential. Using the Baker-Campbell-Haussdorf formula, with $\vec{A}$ in symmetric gauge the group multiplication law for
${\bf g}_{1}=(e^{2\pi i N z_{1}},{\bf r_{1}})$ and ${\bf g}_{2}=(e^{2\pi i N z_{2}},{\bf r_{2}})$
is seen to be
\begin{align}
U({\bf g}_{1}) U({\bf g}_{2}) & =U_{c}(z_{1})\tilde{T}_{{\bf r_{1}}}U_{c}(z_{2})\tilde{T}_{{\bf r_{\text{2}}}}\nonumber \\
 & =U_{c}\left(z_{1}+z_{2}+\frac{({\bf r}\times{\bf r}')_{z}}{2l_{B}^{2}}\right)\tilde{T}_{{\bf r}+{\bf r'}}
\end{align}
or abstractly,
\begin{align}
&(e^{2\pi i N z_{1}},{\bf r}_1) (e^{2\pi i N z_{2}},{\bf r}_2)\nonumber \\
&= \left(e^{2\pi i N (z_{1}+z_{2}+\frac{({\bf r}\times{\bf r}')_{z}}{2l_{B}^{2}})},{\bf r}_1+{\bf r}_2\right).
\end{align}

Note that the magnetic translation groups corresponding to different values of $l_B$ are isomorphic, and that the isomorphism between the groups $U(1)\leftthreetimes_{l_B} \R^2$ and $U(1)\leftthreetimes_{l_B=1} \R^2$ is given by scaling the ${\bf r}$ variables by $l_B$. Therefore we will denote the magnetic translation group as $U(1)\leftthreetimes \R^2$, without loss of generality.

We also note that the $U(1)$ component of the group elements contains an overall factor of $N$. The $U_c$ symmetry operators form an $N$-particle representation of the group $U(1)$, and therefore the symmetry operators defined above form an $N$-particle representation of the magnetic translation group. The group cohomology of $U(1)\leftthreetimes \R^2$, and the resulting classification of SET phases, is however independent of $N$. As a matter of convention we will hereafter write the group multiplication law in the $N=1$ representation:
\begin{align}
&(e^{2\pi i z_{1}},{\bf r}_1)(e^{2\pi i z_{2}},{\bf r}_2)=(e^{2\pi i (z_{1}+z_{2}+\frac{{\bf r}\times{\bf r}'}{2l_{B}^2})},{\bf r}_1+{\bf r}_2).
\end{align}

We will be interested in the limit where the system size tends to infinity, i.e. where $N \rightarrow \infty$.

To summarize, we can view the magnetic translation group $U(1)\leftthreetimes \R^2$
as a central extension of the ordinary
translational symmetry group $\mathbb{R}^2$ by the charge conservation symmetry $U(1)$.
Formally, we have a short exact sequence
\begin{equation}
1\rightarrow U(1)\rightarrow U(1)\leftthreetimes \R^2\rightarrow\mathbb{R}^{2}\rightarrow 1,
\end{equation}
where the cocycle of the extension is
\begin{equation}
c(\mathbf{r}_{1},\mathbf{r}_{2})=\frac{({\bf r_{1}}\times{\bf r_{2}})}{2l_{B}^{2}}=\frac{r_{1x}r_{2y}-r_{1y}r_{2x}}{2l_{B}^{2}}.\label{eq:centralextensioncocycle2}
\end{equation}

The above derivation was done with $\vec{A}$ chosen in symmetric gauge. It has a seeming ambiguity: if we repeat this calculation in Landau gauge, we will obtain a different form of the group multiplication law. Specifically, the cocycle of the group extension will now equal

\begin{equation}
c'(\mathbf{r}_{1},\mathbf{r}_{2})=\frac{r_{2x}r_{1y}}{l_{B}^{2}}.\label{eq:centralextensioncocycle}
\end{equation}
Even though we have two seemingly different group multiplication laws defined above, they are physically equivalent. This is because the cocycles of the central extension are related by a coboundary: $c'(\mathbf{r}_{1},\mathbf{r}_{2})=c(\mathbf{r}_{1},\mathbf{r}_{2})+f(\mathbf{r}_{1})+f(\mathbf{r}_{2})-f(\mathbf{r}_{1}+\mathbf{r}_{2})$
with $f(\mathbf{r})=-r_{x}r_{y}/2l_{B}^{2}$. The above group multiplication law can be written for a general vector potential gauge choice in the following manner:
\begin{equation}
(e^{2\pi i z_{1}},{\bf r}_1) (e^{2\pi i z_{2}},{\bf r}_2) = (e^{2\pi i (z_1+z_2 + w({\bf r}_1,{\bf r}_2))},{\bf r}_1+{\bf r}_2)
\end{equation}
where, in order to satisfy the magnetic translation algebra, the quantity $w({\bf r}_1,{\bf r}_2)$ must be a cocycle representative of $\H^2(\mathbb{R}^2,U(1)) \cong \R$.

Now suppose the Hamiltonian is also rotationally symmetric, with the associated symmetry operators given by 
\begin{widetext}
\begin{equation}
U_{{\rm rot}}(h)=\exp\left(i2\pi h\int d^{2}\mathbf{R}\,\psi^{\dagger}(\mathbf{R})\left[R_{x}\left(-i\nabla_{y}\right)-R_{y}\left(-i\nabla_{x}\right)\right]\psi(\mathbf{R})\right).
\end{equation}
\end{widetext} 

To incorporate the rotational symmetry in the full operator algebra,
we first notice that the rotational symmetry has a nontrivial action on the
$\mathbf{r}$ component of ${\bf g} = (e^{2\pi i z},{\bf r},e^{2\pi i h})$ by conjugation:
\begin{equation}
U_{{\rm rot}}(h)\left(U_{c}(z)\tilde{T}_{{\bf r}}\right)U_{{\rm rot}}^{\dagger}(h)=U_{c}(z)\tilde{T}_{{^{h}}{\bf r}}
\end{equation}
where ${^{h}}{\bf r}$ is the result of rotating the
vector $\mathbf{r}$ by an angle $2\pi h$. Therefore $G$ is defined as the semi-direct product group $G=U(1)\leftthreetimes [\R^2\rtimes SO(2)]$ corresponding to the group extension
\begin{equation}
1\rightarrow U(1)\leftthreetimes \R^2\rightarrow G\rightarrow SO(2)\rightarrow 1
\end{equation}

We can now represent the symmetry operator associated to a general group element ${\bf g}=(e^{2\pi i z},{\bf r},e^{2\pi i h})$ as
\begin{equation}
U({\bf g})=U_{c}(z)\tilde{T}_{{\bf r}}U_{{\rm rot}}(h).
\end{equation}
The operator multiplication law for ${\bf g}_{1}=(e^{2\pi i z_1},{\bf r_{1}},e^{2\pi i h_1})$
and ${\bf g}_{2}=(e^{2\pi i z_2},{\bf r_{2}},e^{2\pi i h_2})$ is now obtained from another application of the Baker-Campbell-Haussdorf formula:
\begin{widetext}
\begin{align}
U({\bf g}_{1}) U({\bf g}_{2}) 
& =U_{c}\left(z_{1}+z_{2}+\frac{({\bf r}_{1}\times {^{h_1}}{\bf r_2})_{z}}{2l_{B}^{2}}\right)\tilde{T}_{{\bf r_{1}}+{^{h_1}}{\bf r_2}}U_{{\rm rot}}(h_{1}+h_{2})=U({\bf g}_1{\bf g}_2).
\end{align}
\end{widetext}

Abstractly, we have the following group multiplication law:
\begin{align}
&(e^{2\pi i z_1},{\bf r_{1}},e^{2\pi i h_1})(e^{2\pi i z_2},{\bf r_{2}},e^{2\pi i h_2})\nonumber \\ &=\left(e^{2\pi i(z_{1}+z_{2}+\frac{({\bf r}_{1}\times {^{h_1}}{\bf r_2})_{z}}{2l_{B}^{2}})},{\bf r_{1}}+{^{h_1}}{\bf r_2},e^{2\pi i (h_{1}+h_{2})}\right).
\end{align}

We have used symmetric gauge so that the cocycle associated to magnetic translations is rotationally symmetric: this will be convenient for future calculations.

The derivation of the group multiplication law for lattice FQH systems, with magnetic translation and point group rotation symmetry defined on the infinite 2D plane, is almost identical to the continuum derivation discussed above. The main difference is that the uniform magnetic field $B$ is expressed as a flux per unit cell, given by $\phi = B A /\phi_0$, where $A$ is the area of a unit cell and $\phi_0$ is the flux quantum. We choose units where $A = \phi_0 = 1$, and can thereore simply use $B=1/l_B^2 = \phi$. 

Since the above arguments for adding rotation symmetry in the continuum are valid for arbitrary $U(1)$ rotations and magnetic translations, they also hold when the rotation symmetry is broken down to a discrete $\Z_M$ subgroup, and the magnetic translation symmetry is broken down to a discrete $\Z^2$ subgroup. Defining a general group element as ${\bf g}_i = (e^{2\pi i z_i},{\bf r_{i}},e^{2\pi i h_i/M})$, where $h_i \in \Z/M\Z$, we have 

\begin{align}
&(e^{2\pi i z_1},{\bf r_{1}},e^{2\pi i h_1/M})(e^{2\pi i z_2},{\bf r_{2}},e^{2\pi i h_2/M})\nonumber \\ &=\left(e^{2\pi i(z_{1}+z_{2}+\phi \frac{({\bf r}_{1}\times {^{h_1}}{\bf r_2})}{2})},{\bf r_{1}}+{^{h_1}}{\bf r_2},e^{2\pi i (h_{1}+h_{2})/M}\right).
\end{align}
If there is no rotational symmetry, this can be written as
\begin{equation}
(e^{2\pi i z_1},{\bf r_1}) (e^{2\pi i z_2},{\bf r_2}) = (e^{2\pi i (z_1+z_2+ \phi w({\bf r_1},{\bf r_2}))}, \bf{r_1}+\bf{r_2}).    
\end{equation}
The function $w({\bf r_1},{\bf r_2})$ can be chosen arbitrarily, as long as $w({\bf r_1},{\bf r_2}) - w({\bf r_2},{\bf r_1}) = {\bf r_1}\times {\bf r_2}$. In a lattice model, the precise form of $w({\bf r_1},{\bf r_2})$ depends on a choice of gauge for the magnetic vector potential. For example, in the Landau gauge, we have $w({\bf r_1},{\bf r_2}) = x_1 y_2$; in symmetric gauge, we have $w({\bf r_1},{\bf r_2}) = \frac{1}{2}{\bf r_1}\times {\bf r_2}$. In specific calculations we will generally work in symmetric gauge, which allows us to easily generalize our results if we later impose rotation symmetry. However, when we restrict to magnetic translation symmetry alone, the final results will not depend on the choice of gauge, as long as we compute observables depending on the gauge-invariant combination $w({\bf r_1},{\bf r_2}) - w({\bf r_2},{\bf r_1})$.

\subsection{Defining the symmetry of FQH states on compact manifolds}

We briefly note that the continuum symmetry group discussed above is a global symmetry of a system only in flat Euclidean space, while the discrete symmetry group discussed above requires an infinite lattice embedded in Euclidean space. When the spatial manifold on which the system is defined is a compact space, such as a torus, then the above global symmetries defined in the standard way are inapplicable; for example, the torus topology is incompatible with continuous $SO(2)$ rotational symmetry.

We can redefine the spatial symmetry in such a case by considering a local action of the symmetry restricted to disk-like patches. We consider the limit where the patch size and the area of the total system are taken to infinity in such a way that the ratio of the two goes to zero. We consider the reduced density matrix $\rho_D$ on a disk-like patch $D$ with vanishing curvature. The spatial translation and rotational symmetries map $D$ to a different patch, $D'$. Whether a given state has the above symmetries then corresponds to whether the symmetry operators map $\rho_D$ to $\rho_{D'}$. 

\section{Further details of $G$-crossed BTC theory}
\label{Sec:GCReview}
	In this section we provide a more detailed introduction to the $G$-crossed BTC theory developed in Ref. \cite{Barkeshli2019}. In parts B1-B3 we review the mathematical relations and formulas used in this paper. The formulas are all written assuming anyons are not permuted by the symmetry. Parts \ref{Sec:SymFracInvt} and \ref{Sec:DefectInvts} are new to this paper, where we prove identities that will be used to generate topological invariants for our examples. 
	\subsection{Consistency relations and identities}
	We begin this section by discussing the manner in which the topological charges transform under symmetry action. Although we will assume that the global symmetry $G$ does not permute anyons, there is always a symmetry action on defects, arising from the group structure of $G$. Specifically, a symmetry operation ${\bf  h}$ applied to a ${\bf  g}$-defect $a_{{\bf  g}} \in\mathcal{C}_{{\bf  g}}$ gives a defect ${^{{\bf  h}}}a_{{\bf  g}}\in \mathcal{C}_{{\bf h g h^{-1}}}$, i.e. the group element which labels the defect is conjugated by ${\bf  h}$. Here we have introduced the notation
	\begin{align}
	{^{{\bf  h}}}(a_{{\bf  g}}) &= \rho_{{\bf  h}}a_{{\bf  g}} \\
	{^{{\bf  h}}}{\bf  g} &= {\bf h g h^{-1}} \\
	\bar{{\bf  g}} &= {\bf  g}^{-1}
	\end{align}
	This additional subtlety becomes important whenever ${\bf  g}$ and ${\bf  h}$ do not commute. The action $\rho_{{\bf g}}$ is a self-consistent extension of the action $\rho$ on anyons, to the full set of topological charges. We will state the explicit form of $\rho_{{\bf g}}$ that we use for computations in Section \ref{Sec:GeneralBTCSoln}. 
	
	The consistency relations between $G$-crossed braiding and fusion are a generalization of the hexagon equations, known as the heptagon equations. Throughout this section, let $a_{{\bf g}},b_{{\bf h}},c_{{\bf k}},d_{{\bf ghk}},e_{{\bf gk}},f_{{\bf gh}},g_{{\bf hk}}$ ($a,b,\dots , g$ in short) be topological charges and $\alpha,\beta,\dots$ represent different fusion pathways. The heptagon equations for clockwise and anticlockwise braiding exchanges are 
	\begin{widetext}
		\begin{align} \label{FullHeptEqns}
		\sum\limits_{\lambda,\gamma} [R_e^{ac}]_{\alpha\lambda} [F_d^{ac^{\bar{{\bf  k}}}b}]_{(e,\alpha,\beta)(f,\delta,\sigma)}[R_g^{bc}]_{\gamma\mu} &= \sum\limits_{f,\sigma,\delta,\eta,\psi} [F_d^{c^{\bar{{\bf  k}}}a^{\bar{{\bf  k}}}b}]_{(e,\alpha,\beta)(^{\bar{{\bf  k}}}f,\delta,\sigma)} 
		[U_{{\bf k}}(a,b;f)]_{\delta\eta}
		[R_d^{fc}]_{\sigma\psi} [F_d^{abc}]_{(f,\eta,\psi)(g,\mu,\nu)} \\
		\sum\limits_{\lambda,\gamma} [(R_e^{ca})^{-1}]_{\alpha\lambda} [F_d^{a^{\bar{{\bf  g}}}cb}]_{(e,\alpha,\beta)(f,\delta,\sigma)}[(R_g^{^{\bar{{\bf  g}}}cb})^{-1}]_{\gamma\mu} &= \sum\limits_{f,\sigma,\delta,\eta,\psi} [F_d^{cab}]_{(e,\alpha,\beta)(f,\delta,\sigma)} 
		\eta_c({\bf g},{\bf h})
		[(R_d^{cf})^{-1}]_{\sigma\psi} [F_d^{ab{^{\bar{{\bf  h}}}}{^{\bar{{\bf  g}}}}c}]_{(f,\eta,\psi)(g,\mu,\nu)}
		\end{align}
	\end{widetext}
	
	When $a_{{\bf g}},b_{{\bf h}}$ and $c_{{\bf k}}$ are also Abelian defects, the fusion products and fusion pathways become unique, and we can ignore subscripts without loss of clarity. When additionally, the group action on the defects is trivial, we can also drop the superscripts indicating this action. This occurs for example when ${\bf g},{\bf  h},{\bf  k}$ are all powers of the same group element ${\bf g_0}$. It is then possible to combine the above equations in the following manner:
	\begin{equation}\label{AbelianHeptEqn}
	\frac{U_{{\bf k}}(a,b;ab)}{\eta_{c}({\bf g},{\bf h})} = \frac{R^{ac}R^{ca}R^{bc}R^{cb}}{R^{ab,c}R^{c,ab}},
	\end{equation} 
	which will be useful in applications. Some other important consistency relations are the 2-cocycle condition for $\eta$, arising from the requirement that symmetry fractionalization be consistent in the $G$-crossed theory: given any topological charge $x$ we have
	\begin{equation}
	\eta_x({\bf g},{\bf h})\eta_x({\bf gh},{\bf k}) = \eta_{^{\bar{{\bf  g}}}x}({\bf h},{\bf k})\eta_x({\bf g},{\bf hk}).
	\end{equation}
	We also have the following useful relation between $\eta$ and $U$ factors:
	\begin{widetext}
	\begin{equation}\label{EtaUEqn}
	\eta_b({\bf k},{\bf l})\eta_a({\bf k},{\bf l}) \sum\limits_{\lambda}[U_{{\bf l}}(^{\bar{{\bf k}}}a,^{\bar{{\bf k}}}b;^{\bar{{\bf k}}}c)]_{\mu\lambda}[U_{{\bf k}}(a,b;c)]_{\lambda\nu} = [U_{{\bf kl}}(a,b;c)]_{\mu\nu}\eta_c({\bf k},{\bf l}).
	\end{equation}
	\end{widetext}
	
	Finally we have a composite identity known as the $G$-crossed ribbon property: 
	\begin{equation}
	\sum\limits_{\lambda} [R_{c_{{\bf gh}}}^{b_{{\bf h}}{^{\bar{{\bf  h}}}}a_{{\bf g}}}]_{\mu\lambda}
	[R_{c_{{\bf gh}}}^{a_{{\bf g}}b_{{\bf h}}}]_{\lambda\mu} = \frac{\theta_c}{\theta_a\theta_b} \frac{[U_{{\bf gh}}(a,b;c)]_{\mu\nu}}{\eta_a({\bf g},{\bf h})\eta_b({\bf h},{^{\bar{{\bf  h}}}}{\bf g})}
	\end{equation}
	where $\theta_a = \sum\limits_{c,\mu} \frac{d_c}{d_a} [R_c^{aa}]_{\mu\mu}$ is the topological twist of $a$ and $d_a$ is the quantum dimension of $a$.
	\subsection{Gauge transformations and defect relabelings}  
	The formulas in this section are all written for trivial $[\rho]$. There are two main types of gauge transformations in $G$-crossed BTCs: \\
	1. Fusion/splitting vertex basis gauge transformations: these allow us to redefine a basis state in the topological state space as
	\begin{equation}
	\widetilde{\ket{a,b;c;\mu}} = \sum\limits_{\mu'}[\Gamma^{ab}_c]_{\mu\mu'} \ket{a,b;c;\mu'}
	\end{equation}
	where $\Gamma^{ab}_c$ is unitary, and $\mu,\mu'$ label the different fusion paths. One must fix $\Gamma^{a0}_a = \Gamma^{0b}_b = \Gamma^{00}_0$. This operation results in the following transformation of the $G$-crossed data:
	\begin{widetext}
	\begin{align}
	[\tilde{F}_d^{abc}]_{(e,\alpha,\beta)(f,\mu,\nu)} &= \sum\limits_{\alpha',\beta',\mu',\nu'} [\Gamma_e^{ab}]_{\alpha\alpha'}
	[\Gamma_d^{ec}]_{\beta\beta'}
	[F_d^{abc}]_{(e,\alpha',\beta')(f,\mu',\nu')}
	[(\Gamma_f^{bc})^{-1}]_{\mu'\mu}
	[(\Gamma_d^{af})^{-1}]_{\nu'\nu} \\
	[\tilde{R}_{c_{{\bf gh}}}^{a_{{\bf g}}b_{{\bf h}}}]_{\mu\nu} &= \sum\limits_{\mu',\nu'} [\Gamma_c^{b{^{\bar{{\bf  h}}}}a}]_{\mu\mu'}[R_{c_{{\bf gh}}}^{a_{{\bf g}}b_{{\bf h}}}]_{\mu'\nu'}[(\Gamma_c^{ab})^{-1}]_{\nu'\nu} \\
	[\tilde{U}_{{\bf k}}(a,b;c)]_{\mu\nu} &= \sum\limits_{\mu',\nu'} [\Gamma_{{^{\bar{{\bf  k}}}}c}^{{^{\bar{{\bf  k}}}}a{^{\bar{{\bf k}}}}b}]_{\mu\mu'}
	[U_{{\bf k}}(a,b;c)]_{\mu'\nu'}
	[(\Gamma_c^{ab})^{-1}]_{\nu'\nu} \\
	\tilde{\eta}_x({\bf g},{\bf h}) &= \eta_x({\bf g},{\bf h})
	\end{align}
	\end{widetext}
	
	2. Symmetry action gauge transformations: These correspond to changing the symmetry action $\rho_{{\bf g}}$ to $\check{\rho}_{{\bf g}} = \Upsilon_{{\bf g}}\rho_{{\bf g}}$, where $\Upsilon$ is a natural isomorphism between symmetry actions, which are therefore physically equivalent. We fix $\gamma_0({\bf g}) = \gamma_a({\bf 0}) = 1$. The transformation enacted on the $G$-crossed data is as follows:
	\begin{align}
	[\check{F}_d^{abc}]_{(e,\alpha,\beta)(f,\mu,\nu)} &= [F_d^{abc}]_{(e,\alpha,\beta)(f,\mu,\nu)} \\
	[\check{R}_{c_{{\bf gh}}}^{a_{{\bf g}}b_{{\bf h}}}]_{\mu\nu} &= \gamma_a({\bf h}) [R_{c_{{\bf gh}}}^{a_{{\bf g}}b_{{\bf h}}}]_{\mu\nu} \\
	[\check{U}_{{\bf k}}(a,b;c)]_{\mu\nu} &= \frac{\gamma_a({\bf k})\gamma_b({\bf k})}{\gamma_c({\bf k})}  [U_{{\bf k}}(a,b;c)]_{\mu\nu} \\
	\check{\eta}_x({\bf g},{\bf h}) &= \frac{\gamma_x({\bf gh})}{\gamma_x({\bf g})\gamma_{{^{\bar{{\bf  g}}}}x}({\bf h})} \eta_x({\bf g},{\bf h})
	\end{align}
	In addition to these gauge transformations, it is possible to change the symmetry fractionalization class $[\mathfrak{w}] \in \H^2(G,\mathcal{A})$ by a 2-coboundary, and leave the topological data invariant. If we change $\mathfrak{w}({\bf g},{\bf h}) \rightarrow \hat{\mathfrak{w}}({\bf g},{\bf h}) = \mathfrak{w}({\bf g},{\bf h}) 
	\times f({\bf g})f({\bf h})\overline{f({\bf gh})}$ 
	where $f:G \rightarrow \mathcal{A}$ is a function with $f({\bf 0}) = 0$ (the identity particle), then in order to preserve the defect fusion rules we must also modify $a_{{\bf g}} \rightarrow \hat{a}_{{\bf g}} = a_{{\bf g}} \times f({\bf g})$. Hence, a change in the symmetry fractionalization cocycle is accompanied by a relabeling of defects. It is possible to achieve a relabeling $0_{{\bf g}} \rightarrow a_{{\bf g}}$ that preserves the form of $\mathfrak{w}({\bf g},{\bf h})$. In this case, the two defects are physically indistinguishable. The existence of relabelings of this type can reduce the classification of SET phases from $\H^3(G,U(1))$ to a subgroup, as discussed in Appendix \ref{Sec:relabeling}.
\subsection{General solution of $G$-crossed BTC equations for non-permuting symmetries}
\label{Sec:GeneralBTCSoln}
This section summarizes the results of Sec. X of Ref. \cite{Barkeshli2019}. To preface the discussion below, we note that our objective is to determine the complete data for $\{F, R, U, \eta\}$ in terms of three pieces of information: a set of defect $F$-symbols, denoted as $F^{0_{\bf g} 0_{\bf h} 0_{\bf k}}$, the symmetry fractionalization cocycle $\mathfrak{w}({\bf g},{\bf h})$, and the data of the UMTC $\mathcal{C}_{\bf 0}$. The defect $F$-symbols are determined by solving Eq. \eqref{F0g0h0kCondition} below, which relates them to the defect obstruction $\mathcal{O}$, as defined for a specific gauge choice in Eq. \eqref{DefObsGeneralForm}. The symmetry fractionalization cocycle can be obtained from group cohomology calculations, and we assume that the data of $\mathcal{C}_{\bf 0}$ are known. With this in hand, the complete data for $\{F, R, U, \eta\}$ are obtained using Eqs. \eqref{FsymbolEqn}, \eqref{RSymEq},\eqref{UkEqn},\eqref{etaEqn}.

One ingredient in the $G$-crossed BTC which we have not introduced previously is the symmetry localization obstruction $\mathfrak{O}({\bf g},{\bf h},{\bf k})$, which is a 3-cocycle representative of the group $\H^3_{[\rho]}(G,\A)$; however, for symmetries than do not permute anyons, this obstruction is always in the trivial class of $\H^3(G,\A)$, so we will not discuss it further. 

We start by choosing a gauge in which 
	\begin{align}\label{GaugeChoicesAnyonsNoPerm}
		U_{{\bf k}}(a_{\bf 0},b_{\bf 0}; c_{\bf 0}) &= 1 \\
		\mathfrak{O}({\bf g},{\bf h},{\bf k}) &= 1 \\
		\eta_{c_{\bf 0}}({\bf g},{\bf h}) &= M_{c_{{\bf 0}} \mathfrak{w}({\bf g},{\bf h})}
	\end{align}
	with $[\mathfrak{w}] \in \H^2(G,\mathcal{A})$ being the symmetry fractionalization class. $M_{ab}$ is equal to the mutual braiding statistics  $R^{ab}R^{ba}$ of $a$ and $b$ (for Abelian anyons). It can be shown that there is always an Abelian defect in each $\mathcal{C}_{{\bf g}}$: we choose one such defect and label it as $0_{{\bf g}}$. Every other defect in $\mathcal{C}_{{\bf g}}$ can be written as $a_{{\bf g}} = a\times 0_{{\bf g}}$ where $a \in C_{\bf 0}$. This accounts for all the defect types in $\mathcal{C}_{{\bf g}}$. Therefore in this case, $\mathcal{C}_{{\bf g}}$ is in bijection with $\mathcal{C}_{\bf 0}$. The defects $0_{{\bf g}}$ satisfy the fusion rule
	\begin{equation}
	0_{{\bf g}}\times 0_{{\bf h}} = 0_{{\bf gh}} \times \mathfrak{w}({\bf g},{\bf h})
	\end{equation} 
	where $[\mathfrak{w}]$ is the symmetry fractionalization class. When we wish to specify that $\mathfrak{w}({\bf g},{\bf h}) \in \mathcal{C}_{\bf 0}$, we will write it as $[\mathfrak{w}({\bf g},{\bf h})]_{\bf 0}$; however, if there is no risk of confusion we will drop the ${\bf 0}$ subscript. From the above equation, we can determine the general defect fusion rule
	\begin{align}
	a_{{\bf g}}\times b_{{\bf h}} &= \sum\limits_{c \in \mathcal{C}_{\bf 0}} N_{a b }^{c} [c\mathfrak{w}({\bf gh})]_{{\bf gh}}.
	\end{align}
	By demanding that the fusion rules are consistent with $G$-crossed braiding, the symmetry action on the topological charges can now be determined explicitly as follows:
	
	\begin{align}
	\rho_{{\bf k}}(a_{{\bf g}}) &= [a \mathfrak{w}({{\bf k}},{{\bf g}})\overline{\mathfrak{w}({\bf  k g \bar{k}},{\bf  k})}]_{{\bf  k g \bar{k}}} \\
	&= [a \mathfrak{w}({{\bf g}},{{\bf \bar{k}}})\overline{\mathfrak{w}({\bf \bar{k}},{\bf  k g \bar{k}})}]_{{\bf  k g \bar{k}}}.
	\end{align}
	
	We present the solution below assuming that $N_{ab}^c =1$ if $a,b$ fuse to give $c$, and $N_{ab}^c =0$ otherwise. We use the vertex basis and symmetry action gauge freedom to fix
	\begin{align}
	&[F^{a_{\bf 0} b_{\bf 0} 0_{{\bf k}}}]_{c_{\bf 0} b_{{\bf k}}} = [F^{a_{\bf 0} 0_{{\bf h}}b_{\bf 0} }]_{a_{{\bf h}}b_{{\bf h}}} = [F^{0_{{\bf g}}b_{\bf 0} 0_{{\bf k}}}]_{b_{{\bf g}}b_{{\bf k}}} \nonumber \\ &=[F_{[c\mathfrak{w}({\bf h},{\bf k})]_{{\bf hk}}}^{a_{\bf 0} 0_{{\bf h}}b_{{\bf k}}}]_{a_{{\bf h}}[b\mathfrak{w}({\bf h},{\bf k})]_{{\bf hk}}} = 
	R_{b_{{\bf g}}}^{0_{{\bf g}}b_{\bf 0}} = R^{a_{{\bf g}}0_{{\bf h}}} = 1.
	\end{align} 
	The final additional piece we need is a function $F^{0_{{\bf g}}0_{{\bf h}}0_{{\bf k}}}$. Mathematically, the functions $F^{0_{{\bf g}}0_{{\bf h}}0_{{\bf k}}}$ are solutions to the consistency condition
	\begin{equation}\label{F0g0h0kCondition}
	\frac{F^{0_{{\bf gh}}0_{{\bf k}}0_{{\bf l}}}F^{0_{{\bf g}}0_{{\bf h}}0_{{\bf kl}}}}{F^{0_{{\bf g}}0_{{\bf h}}0_{{\bf k}}}F^{0_{{\bf g}}0_{{\bf hk}}0_{{\bf l}}}F^{0_{{\bf h}}0_{{\bf k}}0_{{\bf l}}}} = \mathcal{O}({\bf g},{\bf h},{\bf k},{\bf l})
	\end{equation}
	where we define the defect obstruction $\mathcal{O}$ as follows:
	\begin{widetext}
	\begin{equation}\label{DefObsGeneralForm}
	\mathcal{O}({\bf g},{\bf h},{\bf k},{\bf l}) = \frac{F^{\mathfrak{w}({\bf g},{\bf h})\mathfrak{w}({\bf k},{\bf l})\mathfrak{w}({\bf gh},{\bf kl})}}
	{F^{\mathfrak{w}({\bf k},{\bf l})\mathfrak{w}({\bf g},{\bf h})\mathfrak{w}({\bf gh},{\bf kl})}}
	\frac{F^{\mathfrak{w}({\bf k},{\bf l})\mathfrak{w}({\bf h},{\bf kl})\mathfrak{w}({\bf g},{\bf hkl})}}
	{F^{\mathfrak{w}({\bf h},{\bf k})\mathfrak{w}({\bf hk},{\bf l})\mathfrak{w}({\bf g},{\bf hkl})}}
	\frac{F^{\mathfrak{w}({\bf h},{\bf k})\mathfrak{w}({\bf g},{\bf hk})\mathfrak{w}({\bf ghk},{\bf l})}}
	{F^{\mathfrak{w}({\bf g},{\bf h})\mathfrak{w}({\bf gh},{\bf k})\mathfrak{w}({\bf ghk},{\bf l})}}
	R^{\mathfrak{w}({\bf g},{\bf h})\mathfrak{w}({\bf k},{\bf l})}.
	\end{equation} 
	\end{widetext}
	
	Assuming Eq. \eqref{F0g0h0kCondition} is solvable, we can now finally write down the general solution to the $G$- crossed consistency equations, using the gauge choices made above. It is as follows: 
	\begin{widetext}
		\begin{align}
	[F_{[d\mathfrak{w}({\bf g},{\bf h})\mathfrak{w}({\bf gh},{\bf k})]_{{\bf ghk}}}^{a_{{\bf g}}b_{{\bf h}}c_{{\bf k}}}]_{[e\mathfrak{w}({\bf g},{\bf h})]_{{\bf gh}}[f\mathfrak{w}({\bf h},{\bf k})]_{{\bf hk}}} &= [F_{[d\mathfrak{w}({\bf g},{\bf h})\mathfrak{w}({\bf gh},{\bf k})]_{\bf 0}}^{a_{\bf 0}[bw({\bf g},{\bf h})]_{\bf 0}[cw({\bf gh},{\bf k})]_{\bf 0}}]_{[e\mathfrak{w}({\bf g},{\bf h})]_{\bf 0}[f\mathfrak{w}({\bf g},{\bf h})\mathfrak{w}({\bf gh},{\bf k})]_{\bf 0}} \nonumber \\
	& \times \frac{F_{[f\mathfrak{w}({\bf g},{\bf h})\mathfrak{w}({\bf gh},{\bf k})]_{\bf 0}}^{b_{\bf 0}[w({\bf g},{\bf h})]_{\bf 0}[cw({\bf gh},{\bf k})]_{\bf 0}}}{F_{[f\mathfrak{w}({\bf g},{\bf h})\mathfrak{w}({\bf gh},{\bf k})]_{\bf 0}}^{b_{\bf 0}[cw({\bf h},{\bf k})]_{\bf 0}[w({\bf gh},{\bf k})]_{\bf 0}}}   \frac{F^{c_{\bf 0}[w({\bf g},{\bf h})]_{\bf 0}[w({\bf gh},{\bf k})]_{\bf 0}}}
	{F^{[w({\bf g},{\bf h})]_{\bf 0}c_{\bf 0}[w({\bf gh},{\bf k})]_{\bf 0}} F^{c_{\bf 0}[w({\bf h},{\bf k})]_{\bf 0}[w({\bf g},{\bf hk})]_{\bf 0}}}  \frac{F^{0_{{\bf g}}0_{{\bf h}}0_{{\bf k}}}}
	{R^{[w({\bf g},{\bf h})]_{\bf 0} c_{\bf 0}}} \label{FsymbolEqn}\\
	R_{[c\mathfrak{w}({\bf gh})]_{{\bf gh}}}^{a_{{\bf g}}b_{{\bf h}}} &= R_{c_{\bf 0}}^{a_{\bf 0}b_{\bf 0}} \frac{F_{[c\mathfrak{w}({\bf gh})]_{\bf 0}}^{a_{\bf 0}b_{\bf 0}[w({\bf g},{\bf h})]_{\bf 0}}}{F_{[c\mathfrak{w}({\bf gh})]_{\bf 0}}^{b_{\bf 0}a_{\bf 0}[w({\bf g},{\bf h})]_{\bf 0}}}\label{RSymEq}\\
	U_{{\bf k}}(a_{{\bf g}},b_{{\bf h}};[cw({\bf g},{\bf h})]_{{\bf gh}}) &= \frac{F_{[c\mathfrak{w}({\bf g},{\bf h})\mathfrak{w}({\bf gh},{\bf k})]_{{\bf ghk}}}^{a_{{\bf g}}0_{{\bf k}}
			[b\mathfrak{w}({\bf h},{\bf k})\overline{\mathfrak{w}({\bf k},{\bf \bar{k}hk})}]_{{\bf \bar{k}hk}}}}{F_{[c\mathfrak{w}({\bf g},{\bf h})\mathfrak{w}({\bf gh},{\bf k})]_{{\bf ghk}}}^{a_{{\bf g}}b_{{\bf h}}0_{{\bf k}}}F_{[c\mathfrak{w}({\bf g},{\bf h})\mathfrak{w}({\bf gh},{\bf k})]_{{\bf ghk}}}^{0_{{\bf k}}
			[a\mathfrak{w}({\bf g},{\bf k})\overline{\mathfrak{w}({\bf k},{\bf \bar{k}gk})}]_{{\bf \bar{k}gk}}
			[b\mathfrak{w}({\bf h},{\bf k})\overline{\mathfrak{w}({\bf k},{\bf \bar{k}hk})}]_{{\bf \bar{k}hk}}}} \label{UkEqn}\\
	\eta_{c_{{\bf k}}}({\bf g},{\bf h}) &= \frac{F^{0_{{\bf g}}
			[c\mathfrak{w}({\bf k},{\bf g})\overline{\mathfrak{w}({\bf g},{\bf \bar{g}kg})}]_{{\bf \bar{g}kg}}
			0_{{\bf h}}}}{F^{c_{{\bf k}}0_{{\bf g}}0_{{\bf h}}}F^{0_{{\bf g}}0_{{\bf h}}
			[c\mathfrak{w}({\bf k},{\bf gh})\overline{\mathfrak{w}({\bf gh},{\bf \bar{h}\bar{g}kgh})}]_{{\bf \bar{h}\bar{g}kgh}}}} R^{c_{{\bf k}}[w({\bf g},{\bf h})]_{{\bf gh}}} \label{etaEqn}	
	\end{align}
	\end{widetext}

	This set of data completely specifies the SET phase, and contains the information about SET invariants characterising the phase, up to gauge transformations which alter the form of $F^{0_{{\bf g}}0_{{\bf h}}0_{{\bf k}}}$. The above solution gives additional insight into the meaning of the defect obstruction and of defect classes. It can be shown that $\mathcal{O}$ is an element of $Z^4(G,U(1))$. Moreover, it can be shown that gauge transformations, as well as defect relabelings that change the form of $\mathfrak{w}$ by a 2-coboundary in $B^2(G,\mathcal{A})$, only modify the defect obstruction $\mathcal{O}$ by an element of $B^4(G,U(1))$. If $[\mathcal{O}]$ is the trivial obstruction class, we can thus find a solution $F^{0_{{\bf g}}0_{{\bf h}}0_{{\bf k}}}$ to the equation $dF = \mathcal{O}^{-1}$. Starting with this solution, we can produce a complete set of physically distinct solutions by multiplying $F$ with cocycle representatives of $\H^3(G,U(1))$. Therefore the distinct choices of $F^{0_{{\bf g}}0_{{\bf h}}0_{{\bf k}}}$ determine the defect classes associated to $[\mathfrak{w}]$; these defect classes form a torsor over $\H^3(G,U(1))$. 
	
	To summarize, our objective is to completely determine the data for $\{F, R, U, \eta\}$ in terms of the defect $F$-symbols $F^{0_{\bf g} 0_{\bf h} 0_{\bf k}}$, the symmetry fractionalization cocycle $\mathfrak{w}({\bf g},{\bf h})$, and the data of the UMTC $\mathcal{C}_{\bf 0}$. The defect $F$-symbols $F^{0_{\bf g} 0_{\bf h} 0_{\bf k}}$ are determined by solving Eq. \eqref{F0g0h0kCondition}, in terms of a defect obstruction cocycle representative $\mathcal{O}$ (Eq. \eqref{DefObsGeneralForm}), assuming that the defect obstruction class is trivial. The symmetry fractionalization cocycle is obtained from group cohomology calculations, and we assume that the data of $\mathcal{C}_{\bf 0}$ are known. Finally, $\{F, R, U, \eta\}$ are obtained using Eqs. \eqref{FsymbolEqn}, \eqref{RSymEq},\eqref{UkEqn},\eqref{etaEqn}.
	\subsection{General invariant for symmetry fractionalization}
	\label{Sec:SymFracInvt}
	Having reviewed the general aspects of $G$-crossed BTC theory in the preceding subsections, we now derive the main identities that are new to this paper. Using the mathematical relations listed above, it is possible to write down a general formula which can be used to derive invariants for all the symmetry fractionalization classes studied in this work. The intuition behind the formula can be summarized in two steps. First, the insertion of a symmetry flux through the symmetry operations ${\bf  g}_1, \dots , {\bf  g}_n$ induces a certain Abelian anyon $f$. Second, if we know the mutual braiding statistics $M_{a,f}$ for all anyons $a$, we can identify $f$ uniquely, since braiding is nondegenerate. Doing this for all possible nontrivial symmetry fluxes will give us the complete set of anyons that define the symmetry fractionalization class. 
	
	The first step can be made mathematically precise as follows. For an arbitrary group $G$, consider the symmetry operations ${\bf  g}_i, 1 \le i \le n$. Each defect $0_{{\bf  g}_i}$ is a $G$-flux. Then, repeatedly using the fusion rules $0_{{\bf  g}_i}\times 0_{{\bf  g}_j} = 0_{{\bf  g}_1{\bf  g}_j} \times [\mathfrak{w}({\bf  g}_{i},{\bf  g}_j)]_{\bf 0}$, we see that
	\begin{equation}
	0_{{\bf  g}_n}\times 0_{{\bf  g}_{n-1}} \times \dots 0_{{\bf  g}_1} = \prod\limits_{i=1}^{n-1}[\mathfrak{w}({\bf  g}_{i+1},{\bf  g}_i\dots {\bf  g}_1)]_{\bf 0} \times 0_{{\bf  g}_n \dots {\bf  g}_1}.
	\end{equation}  
	Now we define 
	\begin{equation}
	f = \prod\limits_{i=1}^{n-1}[\mathfrak{w}({\bf  g}_{i+1},{\bf  g}_i\dots {\bf  g}_1)]_{\bf 0}.
	\end{equation}
	This relation identifies $f$ as the anyon associated to the insertion of $G$ flux via the symmetry operations ${\bf g}_1,\dots , {\bf g}_n$. Note that $f$ is nontrivial only for certain suitable choices of the symmetry operations ${\bf g}_i$, and when the system has nontrivial symmetry fractionalization.
	
	The second step is to observe that the quantity
	
\begin{align}
 \mathcal{F}_a({\bf g}_1, \dots , {\bf g}_n) &= \frac{\left(\prod\limits_{i=1}^{n} R^{a, 0_{{\bf g}_i}} R^{ 0_{{\bf g}_i},a}\right)}{ R^{a, 0_{{\bf  g}_n \dots {\bf  g}_1}} R^{ 0_{{\bf  g}_n \dots {\bf  g}_1,a}}} \prod\limits_{i=1}^{n-1}\eta_{a}({\bf g}_{i+1},{\bf g}_i\dots {\bf g}_1)
\end{align} 
is gauge invariant under the vertex basis and symmetry action gauge transformations (the product of $R$ symbols and the product of $\eta$ symbols are separately invariant under vertex basis transformations, while under symmetry action transformations these two factors transform in an equal and opposite manner).

Using the general consistency equations, we show below that this quantity is equal to $M_{a,f}$, with $f$ defined as above. First, we can use Eq. \eqref{AbelianHeptEqn} to write

\begin{align}
\eta_a({\bf g},{\bf h}) &= \frac{R^{a, 0_{{\bf g}}\times 0_{{\bf h}}} R^{ 0_{{\bf g}}\times 0_{{\bf h}},a}}{R^{a, 0_{{\bf g}}} R^{ 0_{{\bf g}},a}R^{a, 0_{{\bf h}}} R^{ 0_{{\bf h}},a}} \\
&= \frac{R^{a, 0_{{\bf gh}}} R^{ 0_{{\bf gh}},a}M_{a, [\mathfrak{w}({\bf g},{\bf h})]_{\bf 0}}}{R^{a, 0_{{\bf g}}} R^{ 0_{{\bf g}},a}R^{a, 0_{{\bf h}}} R^{ 0_{{\bf h}},a}}.
\end{align}
Here we used the result $R^{0_{\bf g}0_{\bf h},a}R^{a,0_{\bf g}0_{\bf h}} = R^{0_{\bf gh},a} R^{a,0_{\bf gh}} M_{a,\mathfrak{w}({\bf g},{\bf h})}$, which can in turn be derived as a special case of Eq. \eqref{AbelianHeptEqn}. Applying the last result repeatedly, we obtain
\begin{widetext}
	\begin{align}
	\prod\limits_{i=1}^{n-1}\eta_{a}({\bf g}_{i+1},{\bf g}_i\dots {\bf g}_1) &= \prod\limits_{i=1}^{n-1}\frac{R^{a, 0_{{\bf g}_{i+1}\dots {\bf g}_1}} R^{0_{{\bf g}_{i+1}\dots {\bf g}_1} ,a}  M_{a,[\mathfrak{w}({\bf g}_{i+1},{\bf g}_i\dots {\bf g}_1)]_{\bf 0}}}{R^{a, 0_{{\bf g}_{i+1}}} R^{ 0_{{\bf g}_{i+1}},a}R^{a, 0_{{\bf g}_i\dots {\bf g}_1}} R^{ 0_{{\bf g}_i\dots {\bf g}_1},a}}  R^{a, 0_{{\bf  g}_n \dots {\bf  g}_1}} R^{ 0_{{\bf  g}_n \dots {\bf  g}_1,a}} \\
	&= \frac{\prod\limits_{i=1}^{n-1} M_{a,[\mathfrak{w}({\bf g}_{i+1},{\bf g}_i\dots {\bf g}_1)]_{\bf 0}}}{\prod\limits_{i=1}^{n}R^{a, 0_{{\bf g}_i}} R^{ 0_{{\bf g}_i},a}}  R^{a, 0_{{\bf  g}_n \dots {\bf  g}_1}} R^{ 0_{{\bf  g}_n \dots {\bf  g}_1,a}} \\
\implies \mathcal{F}_a({\bf g}_1, \dots , {\bf g}_n) &:= \frac{\left(\prod\limits_{i=1}^{n} R^{a, 0_{{\bf g}_i}} R^{ 0_{{\bf g}_i},a}\right)}{ R^{a, 0_{{\bf  g}_n \dots {\bf  g}_1}} R^{ 0_{{\bf  g}_n \dots {\bf  g}_1,a}}} \prod\limits_{i=1}^{n-1}\eta_{a}({\bf g}_{i+1},{\bf g}_i\dots {\bf g}_1) = \prod\limits_{i=1}^{n-1} M_{a,[\mathfrak{w}({\bf g}_{i+1},{\bf g}_i\dots {\bf g}_1)]_{\bf 0}} \\
&= M_{a,\prod_{i=1}^{n-1}[\mathfrak{w}({\bf g}_{i+1},{\bf g}_i\dots {\bf g}_1)]_{\bf 0}} = M_{a,f} .
\end{align}
\end{widetext}

By moving the $R$-symbols on the second line to the lhs, we obtain an expression on the rhs of the third line purely in terms of anyons. In the last line, we have used the relation that for anyons $a,b_1,b_2$ where $b_1,b_2$ are Abelian, $M_{a,b_1\times b_2} = M_{a,b_1} M_{a,b_2}$. Crucially, $a$ does not have to be Abelian in any of the above steps. In general, the quantity $\mathcal{F}_a({\bf g}_1, \dots , {\bf g}_n)$ defines a fractional quantum number of $a$. By choosing different sets of group elements ${\bf  g}_1,\dots , {\bf  g}_n$, we can determine the complete set of anyons that characterize the symmetry fractionalization class. In the special case where ${\bf  g}_n\dots{\bf  g}_1 = {\bf 0}$, the factor of $R^{a, 0_{{\bf  g}_n \dots {\bf  g}_1}} R^{ 0_{{\bf  g}_n \dots {\bf  g}_1,a}}$ in the denominator becomes trivial. This special case will apply to many of our examples.

\subsection{Invariants measuring the defect class}
\label{Sec:DefectInvts}

In this section we prove general identities that will be used to derive the defect invariants discussed in this paper. All the derivations shown here are new results of this work.
\subsubsection{Derivation of $\mathcal{I}_n(0_{\bf g})$} 

The first invariant is related to the $G$-crossed modular $\mathcal{T}$-matrix \cite{Barkeshli2019}, and is defined as follows:

\begin{align}
\mathcal{I}_n (0_{{\bf g}}):= [\mathcal{T}^n]^{ ({\bf g}, {\bf 0})}_{0_{\bf g} 0_{\bf g}} = \theta_{0_{\bf g}}^n \prod_{j = 0}^{n-1} \eta_{0_{\bf g}}({\bf g}, {\bf g}^j) ,
\end{align}
where we require ${\bf g}^n = {\bf 0}$. We can check that this is invariant under both vertex basis and symmetry action gauge transformations. Its relation to defect responses can be stated as follows. Suppose $0_{{\bf  g}}^n = f$, i.e. inserting $n$ copies of ${\bf  g}$-flux induces the anyon $f$. Then the invariant satisfies

\begin{align}
(\mathcal{I}_n)^{2n}(0_{{\bf  g}}) &= (\theta_f)^2 = M_{f,f}
\end{align}

The proof is given below. Let $\theta_f = e^{2\pi i h_f}$ for some $h_f \in [0,1)$. Note that if we fix one of the $2n$th roots of $M_{f,f}$ and denote it as $e^{2\pi i h_f/n}$, the invariant associated to a general defect class will be $\mathcal{I}_n(0_{{\bf  g}}) = e^{i \pi \frac{2h_f + k}{n}}$, were the integer $k \mod 2n$ parametrizes the defect class. 

\textit{Proof: } We begin the proof by deriving the following expressions in terms of $\eta$ and $U$ variables using the heptagon equations:
\begin{align}
&\eta_{0_{{\bf g}}}({{\bf g}},j{{\bf g}}) = \frac{R^{0_{{\bf g}},0_{{\bf g}}^{j+1}}}{R^{0_{{\bf g}},0_{{\bf g}}}R^{0_{{\bf g}},0_{{\bf g}}^{j}}} \frac{1}{F^{0_{{\bf g}} 0_{{\bf g}}^j 0_{{\bf g}}}} \\
\implies \qquad &\left(R^{0_{{\bf g}} 0_{{\bf g}}}\right)^n \prod\limits_{j=0}^{n-1} \eta_{0_{{\bf g}}}({{\bf g}},j{{\bf g}})= R^{0_{{\bf g}},f} \prod\limits_{j=0}^{n-1}\frac{1}{F^{0_{{\bf g}} 0_{{\bf g}}^j 0_{{\bf g}}}}, \label{TmatrixWithEtas}
\end{align}
and
\begin{align}
&U_{{\bf g}}(0_{{\bf g}},0_{{\bf g}}^j) = \frac{R^{0_{{\bf g}} 0_{{\bf g}}}R^{0_{{\bf g}}^j 0_{{\bf g}}}}{R^{0_{{\bf g}}^{j+1} 0_{{\bf g}}}}\frac{1}{F^{0_{{\bf g}} 0_{{\bf g}}^j 0_{{\bf g}}}} \\
\implies \qquad&\frac{1}{\left(R^{0_{{\bf g}} 0_{{\bf g}}}\right)^n} \prod\limits_{j=0}^{n-1}U_{{\bf g}}(0_{{\bf g}},0_{{\bf g}}^j) = \frac{1}{R^{f, 0_{{\bf g}}}}\prod\limits_{j=0}^{n-1}\frac{1}{F^{0_{{\bf g}} 0_{{\bf g}}^j 0_{{\bf g}}}}. \label{TmatrixWithUs}
\end{align}
Now for group elements ${\bf k},{\bf l} \in G$ and Abelian topological charges $a_{\bf g},b_{\bf h}$ such that ${^{\bf k}} a_{\bf g} = a_{\bf g}$ and ${^{\bf k}}b_{\bf h} = b_{\bf h}$, we can simplify Eq. \eqref{EtaUEqn} as follows:
\begin{equation}
\frac{\eta_a({\bf k},{\bf l})\eta_b({\bf k},{\bf l})}{\eta_c({\bf k},{\bf l})} = \frac{U_{{\bf kl}}(a,b)}{U_{{\bf k}}(a,b)U_{{\bf l}}(a,b)}
\end{equation}
where we now have $c_{\bf gh} = a_{\bf g} \times b_{\bf h}$. Setting $a = 0_{{\bf g}}, b = 0_{{\bf g}}^j$ and ${\bf k}={\bf g}, {\bf l}=i{\bf g}$ for integers $i$ and $j$, so that the above equation is valid, we can then take its product over all $0 \le i,j \le n-1$. This gives
\begin{equation}\label{U1IntEq1}
\prod\limits_{i,j=0}^{n-1} \frac{\eta_{0_{{\bf g}}}({\bf g},i{\bf g})\eta_{0_{{\bf g}}^j}({\bf g},i{\bf g})}{\eta_{0_{{\bf g}}^{j+1}}({\bf g},i{\bf g})} = \prod\limits_{i,j=0}^{n-1} \frac{U_{(i+1)g}(0_{{\bf g}},0_{{\bf g}}^j)}{U_{g}(0_{{\bf g}},0_{{\bf g}}^j)U_{ig}(0_{{\bf g}},0_{{\bf g}}^j)}.
\end{equation}

The lhs can be simplified as follows:
\begin{align}
\prod\limits_{i,j=0}^{n-1} \frac{\eta_{0_{{\bf g}}}({\bf g},i{\bf g})\eta_{0_{{\bf g}}^j}({\bf g},i{\bf g})}{\eta_{0_{{\bf g}}^{j+1}}({\bf g},i{\bf g})} &= \prod\limits_{i=0}^{n-1} \frac{(\eta_{0_{{\bf g}}}({\bf g},i{\bf g}))^n}{\eta_{f}({\bf g},i{\bf g})}. 
\end{align}
Now the general symmetry fractionalization formula for a sequence of $n$ symmetry operations ${\bf  g}_1 = \dots = {\bf g}_n = {\bf g}$, which we derived above, is applied as follows:
\begin{align}
\prod\limits_{i=0}^{n-1}\eta_{f}({\bf g},i{\bf g}) &= \frac{M_{f,f}}{\left(R^{f,0_{{\bf g}}} R^{0_{{\bf g}},f}\right)^n} \\
\implies \prod\limits_{i=0}^{n-1} \frac{(\eta_{0_{{\bf g}}}({\bf g},i{\bf g}))^n}{\eta_{f}({\bf g},i{\bf g})} &= \frac{\left(R^{f,0_{{\bf g}}} R^{0_{{\bf g}},f}\right)^n}{M_{f,f}}\left(\prod\limits_{i=0}^{n-1} \eta_{0_{{\bf g}}}({\bf g},i{\bf g}) \right)^n.
\end{align}  
Furthermore, the rhs of Eq \eqref{U1IntEq1} can be simplified as follows:
\begin{align}
\prod\limits_{i,j=0}^{n-1} \frac{U_{(i+1)g}(0_{{\bf g}},0_{{\bf g}}^j)}{U_{g}(0_{{\bf g}},0_{{\bf g}}^j)U_{ig}(0_{{\bf g}},0_{{\bf g}}^j)} &= \left(\prod\limits_{j=0}^{n-1} \frac{1}{U_{g}(0_{{\bf g}},0_{{\bf g}}^j)}\right)^n.
\end{align} 
Equating the two sides and comparing with the product of Eqs. \eqref{TmatrixWithEtas} and \eqref{TmatrixWithUs} gives
\begin{align}\label{TmatrixBraidingRelation}
\left(\frac{R^{0_{{\bf g}},f}}{R^{f,0_{{\bf g}}}}\prod\limits_{j=0}^{n-1}\left(\frac{1}{F^{0_{{\bf g}} 0_{{\bf g}}^j 0_{{\bf g}}}}\right)^2 \right)^n &= \frac{M_{f,f}}{\left(R^{f,0_{{\bf g}}} R^{0_{{\bf g}},f}\right)^n}
\end{align}
which means that 
\begin{align}
M_{f,f} &=  \left(R^{0_{{\bf g}},v}\prod\limits_{j=0}^{n-1}\left(\frac{1}{F^{0_{{\bf g}} 0_{{\bf g}}^j 0_{{\bf g}}}}\right)\right)^{2n} \\
&= (I_n(0_{{\bf  g}}))^{2n}, 
\end{align}
where the last line follows from the relation derived at the start of the proof. Therefore we have proved that $M_{f,f} = \theta_f^2= (I_n(0_{{\bf  g}}))^{2n}$.    

\subsubsection{Derivation of mixed defect invariant for $G=\Z_m\times\Z_n$}

We can use this result to identify invariants corresponding to mixed SPTs (or more generally, defect invariants which depend on a combination of symmetry groups). We will derive the mixed defect invariant assuming that $G=\Z_m \times \Z_n$, and that there is no mixed symmetry fractionalization class. We apply similar ideas to extract mixed defect invariants associated to different $\Z_m\times\Z_n$ subgroups of a larger symmetry group $G$, if such nontrivial defect classes actually exist within the group cohomology classification.

Let the generators of $\Z_m\times\Z_n$ be given by ${\bf g}$ and ${\bf h}$. Let $0_{{\bf  g}}^n = p$ and $0_{{\bf  h}}^m = q$. Then, ${\bf gh}$ is an element of order $L=mn/l$, where $l = \gcd(m,n)$ and $L = \text{ lcm}(m,n)$. Moreover, we have $0_{{\bf gh}}^{L} = \prod_{k=1}^{L-1} \mathfrak{w}({\bf gh},{\bf gh}^k)$.  

Now we write the identity 
\begin{align}
&(0_{{\bf gh}}\times \mathfrak{w}({\bf g},{\bf h}))^{L} = (0_{{\bf g}}\times 0_{\bf h})^{L} \\
\implies &\prod_{k=1}^{L-1} \mathfrak{w}({\bf gh},{\bf gh}^k) \times \mathfrak{w}({\bf g},{\bf h})^{L} = \prod_{k=1}^{L-1} \mathfrak{w}({\bf g},{\bf g}^k) \nonumber \\ &\times \prod_{k=1}^{L-1} \mathfrak{w}({\bf h},{\bf h}^k) \\
&= \prod_{k=1}^{n-1} \mathfrak{w}({\bf g},{\bf g}^k)^{L/m}\times \prod_{k=1}^{m-1} \mathfrak{w}({\bf h},{\bf h}^k)^{L/n} \\
&= p^{L/m} \times q^{L/n}.
\end{align}
Since there is no mixed symmetry fractionalization class, we can set $\mathfrak{w}({\bf g},{\bf h}) = 1$, and we therefore have

\begin{equation}
\prod_{k=1}^{L-1} \mathfrak{w}({\bf gh},{\bf gh}^k) = p^{L/m} \times q^{L/n} = p^{n/l}\times q^{m/l}
\end{equation}

As a result, we have 

\begin{equation}
\mathcal{I}_L^L(0_{{\bf gh}}) = \theta_{p^{n/l}\times q^{m/l}} = \theta_p^{n^2/l^2}\theta_q^{m^2/l^2} M_{p,q}^{mn/l^2}
\end{equation} 

To expand $\theta_{p^{n/l}\times q^{m/l}}$, we used the $G$-crossed ribbon property for Abelian anyons $\theta_{ab} = \theta_a \theta_b M_{a,b}$, along with the identities $\theta_{a^m} = (\theta_a)^{m^2}$, and $M_{a^m,b} = M_{a,b^m} = M_{a,b}^m$. 

Now we can use the relations $\mathcal{I}_n^n(0_{\bf g}) = \theta_p$ and $\mathcal{I}_m^m(0_{\bf h}) = \theta_q$ to isolate the braiding term as follows:  
\begin{align}
\frac{\mathcal{I}_L^L(0_{{\bf gh}})}{\mathcal{I}_m^{mn^2/l^2}(0_{{\bf h}})\mathcal{I}_n^{m^2n/l^2}(0_{{\bf g}})} &= M_{p,q}^{mn/l^2} \\
\implies \left(\frac{\mathcal{I}_L(0_{{\bf gh}})}{\mathcal{I}_m^{n/l}(0_{{\bf h}})\mathcal{I}_n^{m/l}(0_{{\bf g}})}\right)^{mn/l}  &= M_{p,q}^{mn/l^2}
\end{align}
Defining $M_{p,q} = e^{2\pi i p \star q}$, and taking the $l/(mn)$th power of the last equation, we can finally write

\begin{equation}
\frac{\mathcal{I}_L(0_{{\bf gh}})}{\mathcal{I}_m^{n/l}(0_{{\bf h}})\mathcal{I}_n^{m/l}(0_{{\bf g}})} = e^{2\pi i \frac{p \star q + k}{l}}
\end{equation}
for some $k \in \Z_l$. This integer $k$ parametrizes the mixed defect class. Since from group cohomology we expect exactly $l$ mixed defect classes, this invariant provides a complete characterization.

\subsubsection{Derivation of the filling invariant} 

The second defect invariant is a generalized filling per magnetic unit cell, which is written as $\nu_H$, where $H$ is some subgroup of $G$. Let ${\bf g}$ be an element of $H$ such that ${\bf g}^n = {\bf 0}$. Also, let $0_{{\bf g}}^n = f$. Let ${\bf r_1},{\bf r_2}$ be elementary \textit{magnetic translations}, i.e. ${\bf r_1}$ and ${\bf r_2}$ must commute. Then we define

\begin{equation}
e^{2\pi i\frac{\nu_H}{n}} := \frac{\eta_{0_{{\bf g}}}({\bf r_1},{\bf r_2})}{\eta_{0_{{\bf g}}}({\bf r_2},{\bf r_1})}
\end{equation}	

The rhs is a gauge-invariant quantity whenever ${^{{\bf r}_i}}0_{{\bf g}} = 0_{{\bf g}}$. For this to happen, it is necessary that ${\bf g}$ commutes with the ${\bf r}_i$. If $\mathfrak{b}({\bf r_1},{\bf r_2})$ is the anyon associated to each magnetic unit cell, then we have the general identity

\begin{equation}
e^{2\pi i \nu_H} = \left(\frac{\eta_{0_{{\bf g}}}({\bf r_1},{\bf r_2})}{\eta_{0_{{\bf g}}}({\bf r_2},{\bf r_1})}\right)^n = M_{f,\mathfrak{b}({\bf r_1},{\bf r_2})}.
\end{equation}

Note that the proof below will only make use of the translation symmetry and the additional $H$ symmetry.

\textit{Proof}: The proof is as follows. First we recall the combined heptagon equation for $\eta$ and $U$, Eq. \eqref{AbelianHeptEqn}:
\begin{align}
\eta_a({\bf g},{\bf h}) &= \frac{M_{a,0_{\bf g}\times 0_{\bf h}}}{M_{a,0_{\bf g}}M_{a,0_{\bf h}}} \\
&= \frac{M_{a,0_{\bf gh}}\times M_{a,\mathfrak{w}({\bf g},{\bf h})}}{M_{a,0_{\bf g}}M_{a,0_{\bf h}}}.
\end{align} 
Taking ${\bf g} = {\bf r_1},{\bf h} = {\bf r_2}$ and $a=f$, we obtain
\begin{equation}
\eta_f({\bf r_1},{\bf r_2}) = \frac{M_{f,0_{{\bf r_1}+ {\bf r_2}}}\times M_{f,\mathfrak{w}({\bf r_1},{\bf r_2})}}{M_{f,0_{\bf r_1}}M_{f,0_{\bf r_2}}};
\end{equation}
now we can interchange ${\bf r_1}$ and ${\bf r_2}$ and divide the two expressions. Since magnetic translations commute, ${\bf r_1}+ {\bf r_2} = {\bf r_2} + {\bf r_1}$. Therefore, upon simplifying, we obtain
\begin{align}
\frac{\eta_{f}({\bf r_1},{\bf r_2})}{\eta_{f}({\bf r_2},{\bf r_1})}\ &= M_{f,\mathfrak{b}({\bf r_1},{\bf r_2})} = M_{f,f}.
\end{align}

Next, we use the relation between $\eta$ and $U$ variables, Eq. \eqref{EtaUEqn}, to write

\begin{align}
\frac{\eta_{0^{k+1}_{{\bf g}}}({\bf r_1},{\bf r_2})}{\eta_{0^{k}_{{\bf g}}}({\bf r_1},{\bf r_2})\eta_{0_{{\bf g}}}({\bf r_1},{\bf r_2})} &= \frac{U_{{\bf r_1}}(0_{{\bf g}}^k,0_{{\bf g}})U_{{\bf r_2}}(0_{{\bf g}}^k,0_{{\bf g}})}{U_{{\bf r_1}+ {\bf r_2}}(0_{{\bf g}}^k,0_{{\bf g}})} 
\end{align}

The rhs is invariant under the substitution ${\bf r_1} \leftrightarrow {\bf r_2}$. This means that

\begin{align}
\frac{\eta_{0^{k+1}_{{\bf g}}}({\bf r_1},{\bf r_2})}{\eta_{0^{k}_{{\bf g}}}({\bf r_1},{\bf r_2})\eta_{0_{{\bf g}}}({\bf r_1},{\bf r_2})} &= \frac{\eta_{0^{k+1}_{{\bf g}}}({\bf r_2},{\bf r_1})}{\eta_{0^{k}_{{\bf g}}}({\bf r_2},{\bf r_1})\eta_{0_{{\bf g}}}({\bf r_2},{\bf r_1})} 
\end{align}
Now we take the product of the above relations for $k=0,1,\dots , n-1$. This cancels out all the $\eta_{0_{{\bf g}}^k}$ factors except for $k=0,n$. Noting that $0_{{\bf g}}^n = f$, we obtain

\begin{align}
\left( \frac{\eta_{0_{{\bf g}}}({\bf r_1},{\bf r_2})}{\eta_{0_{{\bf g}}}({\bf r_2},{\bf r_1})} \right)^n  &= \frac{\eta_{f}({\bf r_1},{\bf r_2})}{\eta_{f}({\bf r_2},{\bf r_1})} \\
&= M_{f,f} 
\end{align} 
which completes the proof. When $H = U(1)$, $\nu_H$ is the usual filling per magnetic unit cell, denoted as $\nu$, and $f$ is denoted as the vison $v$. In the context of $U(1)$ symmetry, it is important to emphasize that we have not made any additional assumptions of Galilean invariance in this argument.

\section{Consistency of SET classification with piecewise continuous and measurable data}
\label{Sec:Cont_meas}

In situations where $G$ has continuous components, it is not clear \textit{a priori} what continuity properties should be imposed on the $G$-crossed BTC data in order to fully capture all possible SETs. It is clear that requiring the data to be continuous is not sufficient, as it does not capture all possible SETs. The next possible choices are to require the data to be either piece-wise continuous or measurable functions on the group manifold. In the context of SPT states, Ref. \cite{Chen2013} suggested that Borel cohomology, which requires the group cochains to be measurable functions, should be used to obtain a proper classification. Here we will briefly explain how these two choices are equivalent for the symmetry groups of interest in our paper. In practice we are therefore justified in working entirely with piecewise continuous data for the $G$-crossed theory, at least for the symmetry groups of interest in this paper (see Appendix \ref{Sec:Gxsolns} below). 

Ref. \cite{Barkeshli2019} showed that, when the symmetries do not permute anyons, the $G$-crossed BTC solutions can be fully parameterized by elements of $\mathcal{H}^2(G, \mathcal{A})$ and $\mathcal{H}^3(G, U(1))$, where the continuity properties of the cochains defining the above cohomology groups are inherited from those of the $G$-crossed data. Let $\mathcal{H}_{\text{meas}}^2(G, \mathcal{A})$ and $\mathcal{H}_{\text{meas}}^3(G, U(1))$ denote the cohomology groups defined using measurable cochains, while $\mathcal{H}_{\text{pc}}^2(G, \mathcal{A})$ and $\mathcal{H}_{\text{pc}}^3(G, U(1))$ denotes cohomology groups defined using piece-wise continuous cochains. Below we will explain that, for the symmetry groups of interest in this paper, $\mathcal{H}_{\text{meas}}^2(G, \mathcal{A}) \cong \mathcal{H}_{\text{pc}}^2(G, \mathcal{A})$ and $\mathcal{H}_{\text{meas}}^3(G, U(1)) \cong \mathcal{H}_{\text{pc}}^3(G, U(1))$. Moreover, $\mathcal{H}_{\text{meas}}^2(G, \mathcal{A})$ and $\mathcal{H}_{\text{meas}}^3(G, U(1))$ always have representative cocycles that are piecewise continuous. Therefore we find that assuming piecewise continuous (p.c.) $G$-crossed data is sufficient and equivalent to assuming measurable $G$-crossed data, for the symmetry groups of interest in this paper. 

Since p.c. functions are also measurable, there is a natural map $\iota: \mathcal{H}^3_{\text{pc}}(G,U(1)) \rightarrow \H_{\text{meas}}^3(G,U(1))$. For the $G$ studied in this paper, our calculations involving spectral sequences are associated with $\H_{\text{meas}}^3(G,U(1))$, and  we have explicitly found p.c. cocycle representatives for every element of $\H_{\text{meas}}^3(G,U(1))$. Therefore, $\iota$ is surjective. Next, let us prove that $\iota$ is injective. Suppose the contrary, that there are two distinct classes $[w]_{\text{pc}},[v]_{\text{pc}} \in \mathcal{H}^3_{pc}(G, U(1))$ which map to the same class in $\H_{\text{meas}}^3(G,U(1))$. This would imply that the representative cocycles $w$ and $v$ are related by a measurable coboundary. But since $w$ and $v$ are themselves p.c., this coboundary would itself have to be p.c.; thus $w$ and $v$ must represent the same element of  $\H^3_{\text{pc}}(G,U(1))$ as well, which is a contradiction. Therefore the map $\iota$ is both injective and surjective, and this completes the proof of our claim.  A similar argument can be made to show that $\mathcal{H}^2_{\text{meas}}(G, \mathcal{A}) \cong \mathcal{H}^2_{\text{pc}}(G, \mathcal{A})$.   

Finally, let us consider the specific formulas for the symmetry fractionalization and defect invariants presented in this paper. These are defined in terms of some defect denoted as $0_{{\bf  g}}$. In fact the symmetry fractionalization invariant does not depend on the specific choice of $0_{{\bf  g}}$. However, some of the defect invariants we study are meaningful only if we have a canonical way in which to identify the elementary defect $0_{{\bf g}}$.  

While working with p.c. data, we can canonically define $0_{{\bf  g}}$ for continuous ${\bf g}$ by demanding that it is the unique defect for which the functions $R^{0_{{\bf g}},a}, a \in \A,$ are continuous for sufficiently small ${\bf  g}$ and approach 1 as ${\bf  g} \rightarrow {\bf 0}$ (such a defect will always exist). This formalizes our intuition that $0_{{\bf g}}$ represents the unique {\bf g}- flux that can be turned on continuously from zero. 

When we work with measurable data, however, the cocycle representative of $\mathfrak{w}$ may be highly discontinuous, and the choice of $0_{{\bf  g}}$ can be unclear. In this case, we first have to perform gauge transformations which convert the data into a p.c. form. We expect that the possible gauge transformations will only allow the braiding data for a unique ${\bf  g}$-defect to be continuously connected to the identity. However, we do not fully understand how to identify the necessary gauge transformations in practice, or how to directly identify $0_{{\bf  g}}$ canonically if the data is highly discontinuous. 

Finally, if ${\bf g}$ is discrete, we cannot fix $0_{\bf g}$ canonically: there may always be relabelings of the defects that leave the fusion rules unchanged, and the defects involved will be indistinguishable. This is the origin of the redundancy in the defect classes measured by $\H^3(G,U(1))$, as discussed in Appendix \ref{Sec:relabeling}.   
\section{$G$-crossed solutions}
\label{Sec:Gxsolns}
In this section we will present solutions to the $G$-crossed consistency equations for each of the symmetry groups studied in this paper. We will rely on the general solution to the $G$-crossed BTC equations for non-permuting symmetries, which is given in Appendix \ref{Sec:GeneralBTCSoln}. Since the solutions discussed below are not completely general, we will not use them to derive invariants in subsequent Appendices. However, the solutions give useful insight into how the SET parameters discussed in the main text are related to the $G$-crossed data. One important application of these explicit solutions is that whenever they can be written down, we can use them to prove stronger LSM constraints involving the filling than have been proven previously. For continuum FQH systems we can prove the relation $\nu = \sigma_H$ . Without using explicit solutions, we can only establish that $\nu = \sigma_H \mod 1$. Analogously, for lattice FQH systems we can prove that $\nu = q(v \star m + k_6) + q\phi \bar{\sigma}_H$ (Appendix \ref{Sec:LSM}). Without using explicit solutions, we can only prove this result modulo 1. 

First let us establish notation. Suppose the Abelian anyons form a group $\mathcal{A} = \Z_{n_1}\times\dots\times\Z_{n_r}$. If an anyon is written as $a = (a_1 \mod n_1,\dots , a_r \mod n_r)$, the fusion rule for two anyons is given by $a \times b := [a+b]_{\A} = ((a_1+b_1) \mod n_1,\dots , (a_r + b_r) \mod n_r)$. The anyons form the defect category $\mathcal{C}_{\bf 0}$. Since the symmetry does not permute anyons, the ${\bf g}$-defects corresponding to each ${\bf g} \in G$ form a category $\mathcal{C}_{{\bf g}}$, which is in bijection with $\mathcal{C}_{\bf 0}$. Also recall the definitions 

\begin{equation}
h: \A \rightarrow [0,1), \quad e^{2\pi i h_a} := \theta_a; 
\end{equation}

\begin{equation}
\star: \A\times \A \rightarrow [0,1), \quad e^{2\pi i a \star b} := M_{a,b} = R^{ab} R^{ba}.
\end{equation}

When $\mathcal{C}_{\bf 0}$ consists entirely of Abelian anyons, we can write $a \star b = \vec{a}^T K^{-1} \vec{b}$, where the nongedenerate, symmetric $K$ matrix with even diagonal entries now specifies the topological order. Here $\vec{a}$ is the integer vector that is associated to the anyon $a$. Two vectors $\vec{a},\vec{a}'$ represent the same anyon if and only if $\vec{a}' = \vec{a} + K \vec{\Lambda}$ for some integer vector $\vec{\Lambda}$. Then the $F$- and $R$- symbols for operations within $\mathcal{A}$ are given by \cite{Barkeshli2020Anomaly} 
\begin{align}
F_{a+b+c}^{a b c} &= e^{i \pi a \star (b + c - [b+c]_{\A})}\label{Fdata} \\
R_{a+b}^{a b} &= e^{i \pi a \star b}\label{Rdata}
\end{align}
Here we use the symbol $[a + b]_{\A}$ for an anyon which has been translated back into its fundamental domain of definition, analogous to the expression $[a+b]_N \equiv a + b \mod N$ for $\Z_N$ anyons. 

When $\mathcal{C}_{\bf 0}$ has non-Abelian anyons, the Abelian sector of $\mathcal{C}_{\bf 0}$ need not be modular, and in that case a nondegenerate $K$ matrix associated to $\A$ might not exist. However, given any Abelian braided fusion category $\A$, we can always consider the Drinfeld center $Z(\A)$, which is an Abelian UMTC with $\A$ as a subcategory. Therefore $Z(\A)$ will have a well-defined bosonic $K$ matrix, which can be used to define the $F$ and $R$ symbols within $\A$. We can then give an equivalent definition of the $h$ symbol and the $\star$ operation using the $K$ matrix of $Z(\A)$, as follows:
\begin{align}
e^{i 2\pi h_a} &:= e^{i \pi \vec{a}^T K^{-1} \vec{a}} \\
e^{i \pi a \star b} &:= e^{i \pi \vec{a}^T K^{-1} \vec{b}}.
\end{align} 
Using these definitions, the $F$- and $R$-symbols within $\A$ can also be defined using Eqs. \eqref{Fdata},\eqref{Rdata}, when $\A$ is not modular.

Our strategy to obtain solutions is the following. A complete solution to the $G$-crossed equations consists of the following information: (i) a symmetry fractionalization cocycle representative for each element of $\H^2(G,\A)$; (ii) a solution to the obstruction equation, Eq. \eqref{DefObsGeneralForm}, which determines the contribution to the defect $F$-symbols from symmetry fractionalization, and (iii) a representative cocycle for each element of $\H^3(G,U(1))$, which determines the part of the defect $F$-symbols corresponding to the different defect classes. Using these results, the $G$-crossed data for the full defect sector can be written down using the formulas in Eqs. \eqref{FsymbolEqn},\eqref{UkEqn}, and \eqref{etaEqn}. In the following sections, steps (i) and (iii) will be addressed in full generality. However, in general, step (ii) will only be solved by making the assumption that the obstruction equation written in the gauge choices of Appendix \ref{Sec:GeneralBTCSoln} reduces to $\mathcal{O}({\bf g},{\bf h},{\bf k},{\bf l}) = R^{\mathfrak{w}({\bf g},{\bf h}) \mathfrak{w}({\bf g},{\bf h})}$. This is the case, for example, when the $F$-symbols of the anyons corresponding to certain fusion paths can all be set to 1 if those fusion paths are allowed, and zero otherwise. Solving step (ii) in full generality is an interesting direction which we leave for future work. Note, however, that the assumptions made to carry out step (ii) in fact always hold for $G=U(1)$ and $G=\Z_M$.

\subsection{$G=U(1)$}
\label{Sec:Gxsolns-U1}
Here the symmetry fractionalization classes can be described by the following set of representative cocycles in $\H^2(U(1),\A) \cong \A$: with ${\bf g_i} = e^{2\pi i z_i}$, 
\begin{equation}\label{RepresentativeCocycleH2}
\mathfrak{w}({\bf g_1},{\bf g_2}) =v^{z_1+z_2-[z_1+z_2]} 
\end{equation} 
where $[z_1+z_2] = z_1+z_2 \mod 1$. The same cocycle can also be written as follows:
\begin{equation}\label{RepresentativeFracClassesH2}
\mathfrak{w}({\bf g_1},{\bf g_2}) = \begin{cases*}
v, &$\quad z_1+z_2\ge 1$ \\
0, &$\quad \text{otherwise}$
\end{cases*}
\end{equation}
where $v \in \A$ and $z_1,z_2 \in [0,1)$ are added as real numbers. In general, the fusion of symmetry defects is given by the rule 
\begin{equation}\label{DefectFusionRule}
a_{{\bf g_1}} \times b_{{\bf g_2}} = ([a+b]_{\A})_{{\bf g_1g_2}} \times \mathfrak{w}({\bf g_1},{\bf g_2}). 
\end{equation}

The equation Eq.\eqref{F0g0h0kCondition} defining the defect obstruction has a solution only if the 4-cocycle $\mathcal{O}({\bf g},{\bf h},{\bf k},{\bf l})$ is trivial, i.e. it is a 4-coboundary in $B^4(G,U(1))$. In this case, $\H^4(G,U(1)) \cong \Z_1$, so $\mathcal{O}$ is indeed trivial. Therefore we look for explicit solutions. When $G=U(1)$, with the above representation of the $F$-symbols and of $\mathfrak{w}$, one can show that the contribution to the defect obstruction from the $F$-symbols cancels out even if the $F$-symbols are nonzero. Thus in this particular case, the expression for $\mathcal{O}$ always simplifies to give 
\begin{equation}\label{DefObsSoln}
\mathcal{O}({\bf g},{\bf h},{\bf k},{\bf l}) = R^{\mathfrak{w}({\bf g},{\bf h})\mathfrak{w}({\bf k},{\bf l})} = e^{i \pi \mathfrak{w}({\bf g},{\bf h})\star \mathfrak{w}({\bf k},{\bf l})}
\end{equation}  
Let ${\bf g}_i = e^{2\pi i z_i}, {\bf h} = e^{2\pi i z_2}, {\bf k} = e^{2\pi i z_3}$. For the form of $\mathfrak{w}({\bf g},{\bf h})$ in Eq. \eqref{RepresentativeCocycleH2}, it can be checked that the following functions form a set of solutions to Eq. \eqref{F0g0h0kCondition}: 

\begin{equation}\label{U1DefectFSym}
F^{0_{{\bf g}}0_{{\bf h}}0_{{\bf k}}} = 
e^{-2\pi i\left( \frac{v \star v}{2} + k_1 \right) z_1(z_2+z_3-[z_2+z_3])}	
\end{equation} 
Here $k_1$ is an integer. As required by the theory, changing the value of $k_1$ changes the solution by an element of $\H^3(U(1),U(1)) \cong \Z$, and physically this corresponds to stacking a $U(1)$ SPT state atop the existing SET state. 

\subsection{$G=U(1)\leftthreetimes \mathbb{E}^2$}
\label{Sec:Gxsolns-continuum}
Denote a general group element as ${\bf  g}_i = (e^{2\pi i z_i},{\bf  r}_i,e^{2\pi i h_i})$. As we derive in Appendix \ref{Sec:Coho_calc_cont}, a general symmetry fractionalization cocycle is parametrized by $v,s \in \A$, and takes the form 

\begin{align}
\mathfrak{w}({\bf  g}_1,{\bf  g}_2) &= v^{z_1+z_2+\frac{{\bf  r_1}\times {^{h_1}} {\bf  r_2}}{2l_B^2}-[z_1+z_2+\frac{{\bf  r_1}\times {^{h_1}}{\bf  r_2}}{2l_B^2}]}\nonumber \\ & \times s^{h_1+h_2-[h_1+h_2]} .
\end{align} 

Define $A_{12} := \frac{{\bf  r_1}\times {^{h_1}} {\bf  r_2}}{2l_B^2}$, $z_{12} := z_1+z_2-[z_1+z_2+\frac{{\bf  r_1}\times {^{h_1}} {\bf  r_2}}{2l_B^2}] = z_{1} + z_2 - [z_1 + z_2 + A_{12}]$, and $h_{12} := h_1+h_2-[h_1+h_2]$.  

In this case it is not true that the contribution from the $F$-symbols to Eq. \eqref{F0g0h0kCondition} identically vanishes; therefore the solution below will be written only for those topological orders where the contribution from $F$-symbols is indeed trivial. In solving the obstruction equation we encounter a subtlety which, for convenience, we will discuss in the $K$-matrix formalism introduced at the beginning of this Appendix (as discussed there, when the topological order is non-Abelian we will use the nondegenerate bosonic $K$ matrix associated to $Z(\A)$, which contains $\A$ as a subcategory, to represent the $F$- and $R$- symbols within $\A$). In the $K$ matrix formalism, the anyon $a$ is written as a vector $\vec{a}$, and the symmetry fractionalization anyon $\mathfrak{w}({\bf g}_1,{\bf g}_2)$ is written as the vector $\vec{\mathfrak{w}}({\bf g}_1,{\bf g}_2)$. 

In order to solve the obstruction equation, we would like to decompose the exponent of  $R^{\mathfrak{w}({\bf g}_1,{\bf g}_2), \mathfrak{w}({\bf g}_3,{\bf g}_4)} $ into terms proportional to $v\star v$, $s \star s $ and $v \star s$, and evaluate the contribution of each term of this type to the defect $F$-symbol. The subtlety is that we cannot directly write $R^{v^{a}\times s^b, v^c \times s^d} = e^{i\pi (a \vec{v}+b \vec{s})^T K^{-1} (c \vec{v}+d \vec{s})}$. This is because a braiding term involving $v^a\times s^b$ is defined using a particular representative vector which is \textit{equivalent} to $(a \vec{v}+b \vec{s})$ as an anyon up to some additional bosonic particle, but may not be \textit{equal} to the vector $(a \vec{v}+b \vec{s})$. In general we have
\begin{align}
R^{v^{a}\times s^b, v^c \times s^d} &= e^{i \pi [a \vec{v}+b \vec{s}]_{\A}^T K^{-1} [c \vec{v}+d \vec{s}]_{\A}}\\
&= e^{i\pi (a \vec{v}+b \vec{s} + K \vec{\Lambda}_1)^T K^{-1} (c \vec{v}+d \vec{s} + K \vec{\Lambda}_2)} 
\end{align}
for some integer vectors $\vec{\Lambda}_1,\vec{\Lambda}_2$. Therefore we write 
\begin{widetext}
	\begin{align}
	\mathcal{O}({\bf g}_1,{\bf g}_2,{\bf g}_3,{\bf g}_4) &= e^{\pi i (\vec{\mathfrak{w}}({\bf g}_1,{\bf g}_2) + K \vec{\Lambda}({\bf g}_1,{\bf g}_2))^T K^{-1} (\vec{\mathfrak{w}}({\bf g}_3,{\bf g}_4)+ K \vec{\Lambda}({\bf g}_3,{\bf g}_4))} \\
	&= e^{\pi i (\vec{\mathfrak{w}}({\bf g}_1,{\bf g}_2)^T K^{-1} \vec{\mathfrak{w}}({\bf g}_3,{\bf g}_4) + \vec{\Lambda}({\bf g}_1,{\bf g}_2)^T\vec{\mathfrak{w}}({\bf g}_3,{\bf g}_4) + \vec{\mathfrak{w}}({\bf g}_1,{\bf g}_2)^T \vec{\Lambda}({\bf g}_3,{\bf g}_4) + \vec{\Lambda}^T({\bf g}_1,{\bf g}_2) K \vec{\Lambda}({\bf g}_3,{\bf g}_4))} \\
	&= e^{ \pi i \{ \vec{v}^T K^{-1}\vec{v} (z_{12}+A_{12})(z_{34}+A_{34}) +\vec{v}^T K^{-1}\vec{s} \left((z_{12}+A_{12}) h_{34} + h_{12} (z_{34}+A_{34}) \right) + \vec{s}^T K^{-1}\vec{s}(h_{12}h_{34})\}} \nonumber \\ &\times e^{i \pi \left(\vec{\Lambda}_{12}^T (\vec{v}(z_{34}+A_{34}) + \vec{s} h_{34}) + (\vec{v}(z_{12}+A_{12}) + \vec{s} h_{12})^T \vec{\Lambda}_{34} \right)} \\
	\implies F^{0_{{\bf g_1}}0_{{\bf g_2}}0_{{\bf g_3}}} &= e^{- i \pi \left(v \star v (z_{12}z_3 + A_{12}z_3 + z_1 A_{23}) + v \star s (z_{12} h_3 + h_{12} z_3 + A_{12}h_3 + h_1 A_{23}) + s \star s h_1 h_{23}\right)}\nonumber \\& \times e^{-i \pi \left(\vec{\Lambda}^T_{12} (\vec{v}z_3 + \vec{s}h_3) + (\vec{v}z_1 + \vec{s}h_1)^T \vec{\Lambda}_{23} \right)} \times e^{-i \pi v \star v \lambda({\bf g}_1,{\bf g}_2,{\bf g}_3)}.
	\end{align}
\end{widetext}
Here, $\lambda$ is a function which satisfies $d\lambda({\bf g}_1,{\bf g}_2,{\bf g}_3,{\bf g}_4) =A_{12}A_{34}$. This function always exists, but we will not need to determine it explicitly for our subsequent applications. Note that the $K$ matrix is symmetric with even diagonal entries, and therefore the contribution $e^{\pi i \vec{\Lambda}^T({\bf g}_1,{\bf g}_2) K \vec{\Lambda}({\bf g}_3,{\bf g}_4)}$ is trivial and does not appear in the expression for the defect $F$-symbol. This is important because the quantity $\vec{\Lambda}^T({\bf g}_1,{\bf g}_2) K \vec{\Lambda}({\bf g}_3,{\bf g}_4)$ cannot be easily expressed as a coboundary of some 3-variable function.

Having obtained this particular solution for the defect $F$-symbols, the full set of $G$-crossed solutions can now be parametrized by the integers $k_1,k_2,k_3$ as follows:
\begin{widetext}
	\begin{align}
	F^{0_{{\bf g_1}}0_{{\bf g_2}}0_{{\bf g_3}}} &= e^{-2\pi i\left(\frac{v\star v}{2}+k_1\right)(z_1z_{23}+z_1 \frac{{\bf  r_2}\times {^{h_2}}{\bf  r_3}}{2l_B^2}+ \frac{{\bf  r_1}\times {^{h_1}}{\bf  r_2}}{2l_B^2}z_3)} \times e^{-2\pi i\left(\frac{v\star s}{2}+k_2\right)(h_1 (z_{23}+\frac{{\bf  r_2}\times {^{h_2}}{\bf  r_3}}{2l_B^2})+ (z_{12}+\frac{{\bf  r_1}\times {^{h_1}}{\bf  r_2}}{2l_B^2})h_3)} \nonumber \\& 
	\times e^{-2\pi i\left(\frac{s\star s}{2}+k_3\right)h_1 h_{23}} \times e^{-i \pi \left(\vec{\Lambda}^T_{12} (\vec{v}z_3 + \vec{s}h_3) + (\vec{v}z_1 + \vec{s}h_1)^T \vec{\Lambda}_{23}  \right)} \times e^{-i \pi (v \star v+k_1) \lambda({\bf g}_1,{\bf g}_2,{\bf g}_3)} \\
	\eta_{a}({\bf g_1},{\bf g_2}) &= M_{a \mathfrak{w}({\bf g_1},{\bf g_2})}
	\end{align}
\end{widetext}
The contributions from $k_1,k_2,k_3$ are precisely the defect $F$-symbols associated to the SPT states with $G$ symmetry, derived in Appendix \ref{Sec:Coho_calc_cont}.
\subsection{$G=\Z_M$}
\label{Sec:Gxsolns-ZM}
From standard results on the cohomology of cyclic groups (see Appendix \ref{Sec:GrpCohIntro}, Eq.\eqref{CohomologyCyclicGrp_trivial}), the group of symmetry fractionalization classes is $\mathcal{H}^2(\Z_M,\mathcal{A}) \cong \mathcal{A}/M\mathcal{A}$. $M\mathcal{A}$ is the group of $M$th powers of all anyons in $\mathcal{A}$. We write a general symmetry fractionalization cocycle as follows. Let ${\bf g_i} = e^{2\pi i h_i/M}$ where $0 \le h_i < M$. Then we can write
\begin{align}
\mathfrak{w}({\bf g_1},{\bf g_2}) &= \begin{cases*}
s, &$\quad h_1+h_2\ge M$ \\
0, &$\quad \text{otherwise}$
\end{cases*} \\
&= s^{\frac{h_1+h_2-[h_1+h_2]_M}{M}}
\end{align}   
With this choice, we understand symmetry fractionalization as inducing an anyon $s$ when $M$ elementary units of $\Z_M$ flux are inserted. Now, suppose $s$ is of the form $s'^M$, for some anyon $s'$. Then the above cocycle is
\begin{align}
\mathfrak{w}({\bf g_1},{\bf g_2}) &= s'^{h_1+h_2-[h_1+h_2]_M} \\
&= d\chi({\bf g_1},{\bf g_2})
\end{align} 
where $\chi({\bf g_i}) = s'^{{h_i}}$. Therefore, symmetry fractionalization cocycles associated to anyons of the form $s'^M$ are in fact 2-coboundaries. 

The set of symmetry fractionalization classes is therefore the full set of anyons modded out by the subgroup of all anyons of the form $s'^M$. If we specialize to $\mathcal{A} = \Z_N$ with the generator $\psi$, the group $\mathcal{A}^M$ is just the group generated by $\psi^{d}$, where $d = (M,N)$. Therefore there are $d$ different fractionalization classes, and for a given system, the class labelled by $q \in \Z_d$ is associated to an anyon in the set ${\psi^{q+kd}}$. The above discussion shows that unlike the $U(1)$ case, there is no unique anyon that can be associated to the insertion of a $\Z_M$ flux quantum. 

The $G$-crossed solution in our usual choice of gauge is very similar to the solution for $U(1)$ symmetry. If the $F$- and $R$- symbols of the Abelian anyons are written in the usual notation, the obstruction is given by 	
\begin{align}
\mathcal{O}({\bf g_1},{\bf g_2},{\bf g_3},{\bf g_4}) &= e^{\frac{2\pi i s \star s}{2 M^2}(h_1+h_2-[h_1+h_2]_M)(h_3+h_4-[h_3+h_4]_M)}
\end{align}
and so the defect $F$-symbols take the form
\begin{align}
F^{0_{{\bf g}}0_{{\bf h}}0_{{\bf k}}} &= e^{-\frac{2\pi i}{M^2} \left(\frac{s \star s}{2} + k_3 \right)h_1(h_2+h_3-[h_2+h_3]_M)}.
\end{align}
The obstruction equation for $G=\Z_M$ does not rely on any assumptions on the anyon $F$-symbols, which always cancel out from the definition of $\mathfrak{w}$. The above exponent can now be written as $e^{-2\pi i \frac{s \star s + 2 k_3}{2M}\frac{h_1(h_2+h_3-[h_2+h_3]_M)}{M}}$. Note that the topological twist $\theta_s$ of $s$ is well-defined up to multiples of $2\pi$, and consequently $s \star s = \frac{1}{\pi} \arg(\theta_s)$ is defined up to an even integer. In the equation above, we define $\frac{s \star s}{2M} := \frac{1}{2\pi M} \arg (\theta_s)$ assuming $0 \le \theta_s < 2\pi$. Therefore, if we change $s \star s$ by the integer $2k$, we must correspondingly shift $k_3 \rightarrow k_3 - k$ in order to describe the same defect class. 
\subsection{$G=U(1)\leftthreetimes_{\phi} \Z^2$}
\label{Sec:Gxsolns-magtrans}
The cohomology calculations for $G=U(1)\leftthreetimes_{\phi} \Z^2$ are outlined in Appendix \ref{Sec:Coho_calc_magtrans}. Using the results from that section, we see that symmetry fractionalization is classified by 

\begin{equation}
\H^2(G,\A) \cong \A \times \A
\end{equation}
with a general symmetry fractionalization cocycle taking the form

\begin{equation}
\mathfrak{w}({\bf g}_1,{\bf g}_2) = v^{z_1 + z_2 + \phi w({\bf  r_1},{\bf r_2})  - [z_1 + z_2 + \phi w({\bf  r_1},{\bf r_2})]} \times m^{w({\bf  r_1},{\bf r_2})}. 
\end{equation}
Here we have defined a magnetic translation group element as ${\bf g}_i = (e^{2\pi i z_i},{\bf r_i})$, and in symmetric gauge we have $w({\bf  r_1},{\bf r_2}) = \frac{{\bf r_1} \times {\bf r_2}}{2}$. The anyon $v$ is associated to $U(1)$ charge fractionalization, while $m$ refers to the anyon per unit cell and is associated to the fractionalization of $\Z^2$ translational symmetry. The symmetry fractionalization cocycle gives fractional values of the anyon $\mathfrak{w}({\bf g}_1,{\bf g}_2)$ when we work in symmetric gauge. Therefore only gauge-invariant combinations of $\mathfrak{w}$ are guaranteed to be well-defined anyons. Alternatively, in this example we can choose to work throughout in an integer-valued gauge such as the Landau gauge; in that case $\mathfrak{w}({\bf g}_1,{\bf g}_2)$ will always be well-defined.

Next we consider the defect data. Defining $z_{12} = z_1+z_2-[z_1+z_2+\phi w({\bf  r_1},{\bf  r_2})]$, we can write the above symmetry fractionalization cocycle as $\mathfrak{w}({\bf g}_1,{\bf g}_2) = v^{z_{12} +\phi w({\bf  r_1},{\bf  r_2}) }\times m^{w({\bf  r_1},{\bf  r_2})}$. We assume that the defect obstruction simplifies to the form 
\begin{align}
\mathcal{O}({\bf g}_1,{\bf g}_2,{\bf g}_3,{\bf g}_4) &= R^{\mathfrak{w}({\bf g}_1,{\bf g}_2), \mathfrak{w}({\bf g}_3,{\bf g}_4)}.
\end{align}
Note that if $h({\bf g}) = z$ is a projection map, then $dh({\bf g}_1,{\bf g}_2) = h({\bf g}_1)+h({\bf g}_2) - h({\bf g}_1{\bf g}_2) = z_{12}$. As we showed in Appendix \ref{Sec:Gxsolns-continuum}, when we use a $K$ matrix to represent braiding data and when the symmetry fractionalization anyon $\mathfrak{w}({\bf g}_1,{\bf g}_2)$ is a product of the form $v^a \times m^b$, we need to be careful about the choice of the vector corresponding to $v^a \times m^b$. In general the anyon $\vec{\mathfrak{w}}({\bf g}_1,{\bf g}_2)$ (written as a vector in the $K$ matrix formalism) does not lie in the fundamental domain of anyons, which is used to define the $R$ symbols; therefore the correct choice of vector is actually $\vec{\mathfrak{w}}({\bf g}_1,{\bf g}_2) + K \vec{\Lambda}({\bf g}_1,{\bf g}_2)$, for some integer vector $\vec{\Lambda}({\bf g}_1,{\bf g}_2)$. With this, the obstruction takes the form 
\begin{align}
& \mathcal{O}({\bf g}_1,{\bf g}_2,{\bf g}_3,{\bf g}_4) \nonumber \\
&= e^{i \pi (\vec{\mathfrak{w}}({\bf g}_1,{\bf g}_2) + K \vec{\Lambda}({\bf g}_1,{\bf g}_2))^T K^{-1} (\vec{\mathfrak{w}}({\bf g}_3,{\bf g}_4) + K \vec{\Lambda}({\bf g}_3,{\bf g}_4))} \\
\end{align}
In this notation the obstruction equation becomes
\begin{widetext}
	\begin{align}
	\mathcal{O}({\bf g}_1,{\bf g}_2,{\bf g}_3,{\bf g}_4)
	&= \exp \pi i\left( v\star v(z_{12} z_{34})  + v\star (\phi v+m)(z_{12} w({\bf r_3},{\bf r_4})+w({\bf  r_1},{\bf  r_2})z_{34}) + m\star m (d\lambda)({\bf r_1},{\bf r_2},{\bf r_3},{\bf r_4})\right) \nonumber \\
	&\times  \mathcal{O}_{\vec{\Lambda}}({\bf g}_1,{\bf g}_2,{\bf g}_3,{\bf g}_4)
	\end{align}
\end{widetext}
Here we have absorbed the extra contributions from the bosonic particles corresponding to $K\vec{\Lambda}({\bf g}_1,{\bf g}_2)$ and $K\vec{\Lambda}({\bf g}_3,{\bf g}_4)$ into the definition of $\mathcal{O}_{\vec{\Lambda}}$. Also, $\lambda$ is some 3-variable function such that $d\lambda({\bf r_1},{\bf r_2},{\bf r_3},{\bf r_4}) = w({\bf r_1},{\bf r_2})w({\bf r_3},{\bf r_4})$. A possible choice for $\lambda$, assuming $w({\bf r_1},{\bf r_2}) = \frac{1}{2}(x_1y_2-x_2y_1)$, is
\begin{align}
&4\lambda({\bf r}_1,{\bf r}_2,{\bf r}_3) \nonumber \\ &= x_1y_1(x_2y_3-x_3y_2) - x_1y_2^2x_3-x_1^2y_2y_3 - 2 x_1x_2y_2y_3.
\end{align}
However, we will not need to know this explicit form in subsequent calculations. 

The symbols $F^{0_{{\bf g}_1}0_{{\bf g}_2}0_{{\bf g}_3}}$ are the solutions to $dF = (\mathcal{O})^{-1}$. Specifically, we have 
\begin{widetext}
	\begin{align}
	F^{0_{{\bf g}_1}0_{{\bf g}_2}0_{{\bf g}_3}} &= \exp -\pi i\left( v\star v(z_{1} z_{23})  + v\star (\phi v+m)(z_{1} w({\bf r_2},{\bf r_3})+w({\bf  r_1},{\bf  r_2})z_{3}) + m\star m (\lambda({\bf r_1},{\bf r_2},{\bf r_3}))\right) \nonumber \\
	& \times F^{0_{{\bf g}_1}0_{{\bf g}_2}0_{{\bf g}_3}}_{\vec{\Lambda}}.
	\end{align}
\end{widetext}
Here, the function $F^{0_{{\bf g}_1}0_{{\bf g}_2}0_{{\bf g}_3}}_{\vec{\Lambda}}$ is obtained as a solution to the equation $dF_{\vec{\Lambda}}^{0_{{\bf g}_1}0_{{\bf g}_2}0_{{\bf g}_3}} = \mathcal{O}^{-1}_{\vec{\Lambda}}({\bf g}_1,{\bf g}_2,{\bf g}_3)$: the steps are similar to those explained in Appendix \ref{Sec:Gxsolns-continuum}. 

As a last step, it is necessary to look for solutions to $dF = 1$ which yield representatives of $\H^3(G,U(1))$. From the spectral sequence (Appendix \ref{Sec:Coho_calc_magtrans}), we can see that $\H^3(G,U(1))\cong \H^4(G,\Z) \cong \H^4(U(1),\Z) \times \H^2(\Z^2,\Z) \cong \Z \times \Z$. The first term corresponds to SPTs of pure $U(1)$ symmetry, i.e. bosonic IQH states (with Hall conductivity $2k_1$). The second term corresponds to a mixed SPT of $U(1)$ and $\Z^2$ symmetry, where an integer number of $U(1)$ charges (given by $k_6 \in \Z$) is placed in each unit cell. Using the results of Appendix \ref{Sec:Coho_calc_magtrans}, a general SPT cocycle can be written in terms of these parameters, leading to the desired expression for the full defect $F$-symbol, which we distinguish as $F_{\text{mag}}$ (this terminology will be of use in the next section):
\begin{widetext}
	\begin{align}
	f_{\text{SPT}}({\bf g}_1,{\bf g}_2,{\bf g}_3) &= e^{2\pi i k_1 (z_{12} + \phi w({\bf  r_1},{\bf  r_2})) z_{3} + (\phi k_1 + k_6) (z_{1} w({\bf  r_2},{\bf  r_3}) + \phi (\phi k_1 + k_6) \lambda({\bf  r_1},{\bf  r_2},{\bf  r_3}))} \label{MagTransSPTCocycle}\\
	\implies F^{0_{{\bf g}_1}0_{{\bf g}_2}0_{{\bf g}_3}}_{\text{mag}} &= \exp\left(-2\pi i \left(\frac{v\star v}{2}+k_1\right)
	z_{1}z_{23}  +\left(\frac{v\star (\phi v + m)}{2}+(\phi k_1 + k_6)\right)z_{1} w({\bf  r_2},{\bf  r_3}) \right. \nonumber \\
	& \left. +\left(\frac{(\phi v + m)\star v}{2}+(\phi k_1)\right) w({\bf  r_1},{\bf  r_2})z_3 +\left(\frac{m\star m}{2} + \phi (\phi k_1+k_6) \right)\lambda({\bf r_1},{\bf r_2},{\bf r_3})
	\right) \nonumber \\
	& \times F^{0_{{\bf g}_1}0_{{\bf g}_2}0_{{\bf g}_3}}_{\vec{\Lambda}}
	\end{align}
\end{widetext}

\subsection{$G=U(1)\leftthreetimes_{\phi} [\Z^2\rtimes \Z_M]$}
\label{Sec:Gxsolns-gspace}

Let $h$ be the generator of point group rotations. We will abuse the notation for point group elements slightly, as follows: the quantities $h_i$ written in line should be understood as integers mod $M$, corresponding to rotations by the angle $2\pi h_i/M$, while in an expression such as ${^h} {\bf r}$ or ${^{1-h^k}}{\bf r}$, $h$ is understood as the $2\times 2$ matrix generator of point group rotations, and thus the symbol 1 in ${^{1-h^k}}{\bf r}$ denotes the identity $2\times 2$ matrix. In this section we will also denote ${\bf g_i} = (z_i,{\bf r_i},h_i)$ for better readability.

In Appendix \ref{Sec:Coho_calc_gspace} we derive 
\begin{align}
\H^2(G,\mathcal{A}) &\cong \mathcal{A}^2 \times (\A/M\A)\times (K_M \otimes \mathcal{A}).
\end{align}
This classification is independent of the flux $\phi$. There we also show that a general cocycle representative of $\H^2(G,\A)$ is given by
\begin{align}\label{Eq:GspaceSFcocycle}
	&\mathfrak{w}((z_1,{\bf r_1},h_1),(z_2,{\bf r_2},h_2)) \nonumber \\&=
	v^{z_1+z_2+\phi w({\bf  r_1},^{h_1}{\bf  r_2})-[z_1+z_2+\phi w({\bf  r_1},^{h_1}{\bf  r_2})]}  \nonumber\\ 
	&\times m^{w({\bf  r_1},^{h_1}{\bf  r_2})} \nonumber\\
	&\times t_x^{\left((1-h)^{-1}(1-h_1) {\bf r_2}\right)_x}\times t_y^{\left((1-h)^{-1}(1-h_1) {\bf r_2}\right)_y}\nonumber \\
	&\times s^{\frac{h_1+h_2-[h_1+h_2]_M}{M}} 
	\end{align} 
The four distinct contributions are each associated to anyons. We have already seen three of them in previous sections: the vison $v$, the anyon per unit cell $m$, and the anyon $s$ associated to the insertion of $M$ elementary disclinations. Note that in the presence of rotation symmetry, the term $w({\bf  r_1},{\bf  r_2})$ is replaced by $w({\bf  r_1},^{h_1}{\bf  r_2})$ in order for the function $\mathfrak{w}$ to satisfy the 2-cocycle condition for $U(1)\leftthreetimes_{\phi} [\Z^2\rtimes \Z_M]$. The fourth contribution, which is given by a pair of anyons $t_x,t_y$, describes a form of mixed fractionalization of translational and rotational symmetry.

We can also write down the full solution to the defect obstruction and hence the defect $F$-symbols. In doing so, we must remember the subtlety that arose while handling the obstruction equation for magnetic translation symmetry. This requires us to add an additional term in order to split the obstruction term into products such as $v \star v$, $v \star s$ and so on. Define $z_{ij} = z_i + z_j - [z_i + z_j + w({\bf  r_i},{^{h_i}}{\bf  r_j})]$,  $h_{ij} = h_i + h_j - [h_i + h_j]$ and $\vec{\Theta}_{ij} = {^{(1-h)^{-1}(1-h_i)}} {\bf r_j}$. Then, the final result is
\begin{widetext}
	\begin{align}\label{FullSpaceGrpDefectFSyms}
	& F^{0_{{\bf g}_1}0_{{\bf g}_2}0_{{\bf g}_3}} = \tilde{F}^{0_{{\bf g}_1}0_{{\bf g}_2}0_{{\bf g}_3}}_{\text{mag}} \times e^{-2\pi i \alpha({\bf g}_1,{\bf g}_2,{\bf g}_3)}\times F^{0_{{\bf g}_1}0_{{\bf g}_2}0_{{\bf g}_3}}_{\vec{\Lambda}},\nonumber \\
	&\alpha({\bf g}_1,{\bf g}_2,{\bf g}_3) = \left(\frac{q \star t_i}{2}+k_{4,i}\right)\cdot (z_1 \vec{\Theta}_{23} + \phi w({\bf  r_1},^{h_1}{\bf  r_2}) {\bf  r_3}) + \left(\frac{q\star t_i}{2}\right)\cdot (\vec{\Theta}_{12} z_3 + \phi  {\bf  r_1} w({\bf  r_2},^{h_2}{\bf  r_3}) ) \nonumber \\
	&+ \left(\frac{q\star s}{2M}+\frac{k_2}{M}\right) (z_1h_{23} + \phi w({\bf  r_1},^{h_1}{\bf  r_2}) h_3) + \left(\frac{q\star s}{2M}\right) (h_{12}z_3 + \phi h_1 w({\bf  r_2},^{h_2}{\bf  r_3}))  \nonumber \\
	&+ \left(\frac{m \star s}{2M}+\frac{k_7}{M}\right) h_1 w({\bf  r_2},^{h_2}{\bf  r_3}) + \left(\frac{m\star s}{2M}\right) w({\bf  r_1},^{h_1}{\bf  r_2}) h_3 \nonumber \\
	&+ \left(\frac{s\star t_i}{2M}+\frac{k_{5,i}}{M}\right) \cdot h_1 \vec{\Theta}_{23} + \left(\frac{s\star t_i}{2M}\right)\cdot \vec{\Theta}_{12} h_3 \nonumber \\
	&+ \left(\frac{s\star s}{2M^2}+\frac{k_3}{M^2}\right) h_1 h_{23} \nonumber \\
	&+ {^{(1-h)^{-1}}}\left(\vec{t} \cdot \vec{\Theta}_{12}\right) \star \left(\vec{t} \cdot {\bf r}_3\right) + (m + \phi v)\star \left(\vec{t} \cdot {\bf r}_1\right)\left( w({\bf  r_2},^{h_2}{\bf r_3})\right)  + (m+\phi v)\star (m+\phi v) \lambda({\bf  g}_1,{\bf  g}_2,{\bf  g}_3).
	\end{align}
\end{widetext}
(See Appendix \ref{Sec:Gxsolns-magtrans} for the definition of $F^{0_{{\bf g}_1}0_{{\bf g}_2}0_{{\bf g}_3}}_{\vec{\Lambda}}$.) The symbol $\tilde{F}_{\text{mag}}$ is the same as $F_{\text{mag}}$, with one modification: all occurrences of the function $w({\bf r_i},{\bf r_j})$ are replaced by $w({\bf r_i},{^{h_i}}{\bf r_j})$. The integers $k_1$ through $k_7$ parametrize $G$-SPT phases, and hence shifting the values of $k_i$ while keeping all the other data fixed will change the defect class. Notice that the last line contains three expressions that are not associated to any SPT parameters. These expressions are needed in order to solve the obstruction equation. They describe quantized responses that are completely fixed by the symmetry fractionalization class, and are therefore not associated to SPT states. The first term is related to the quantized torsional response $\Pi_{ij}$, the second is related to the momentum per unit cell $\vec{\nu}_p$, and the third is related to the response coefficient $\alpha$ for which we do not have any meaningful physical interpretation. These coefficients can be read off from an effective action written using crystalline gauge fields (see Appendix \ref{Sec:Eff_Ac-gspace}). However, we have not been able to find gauge invariant quantities in the $G$-crossed BTC that return these coefficients exactly. We can only find an expression which gives $M \Pi_{ij} \mod 1$ as opposed to $\Pi_{ij} \mod 1$, while we have not found a gauge-invariant expression which returns $\vec{\nu}_{p}$.
\section{Specific invariants for fractional symmetry quantum numbers and fractionally quantized responses}
\label{Sec:Invts}

In this section, we will examine each of the symmetry groups studied in the paper and derive invariants for the corresponding symmetry fractionalization and defect classes. The $G$-crossed identities used in deriving these invariants are proved in Appendix \ref{Sec:SymFracInvt} and \ref{Sec:DefectInvts}. To evaluate these formulas for a particular choice of SET parameters, we will use the $G$-crossed solutions stated in Appendix \ref{Sec:Gxsolns}, in which particular gauge choices are made for the $G$-crossed data. However, since the solutions are not completely general, we will independently verify that the invariants completely characterize the SET, using $G$-crossed identities.
\subsection{$G=U(1)$}
\label{Sec:Invts-U1}
The invariant for the $U(1)$ symmetry fractionalization class gives the fractional charge $Q_a$ of an arbitrary anyon $a$. First we prove that it is a gauge-invariant quantity. Fix an integer $n$ and consider a group element ${\bf g} = e^{2\pi i /n}$. Pick an Abelian defect $0_{{\bf g}}$. We can always choose $n$ so that $(0_{{\bf g}})^n = v$, where $v$ is the fixed anyon which completely determines the symmetry fractionalization class. Informally, this is because the insertion of $n$ copies of $2\pi/n$ flux is equivalent to the insertion of a $2\pi$ flux, which induces the unique anyon $v$. Formally we argue as follows. Using the fusion rule for defects, we have the identity
\begin{equation}
v = 0_{{\bf g}}^n = \prod\limits_{j=0}^{n-1} \mathfrak{w}({\bf g},{\bf g}^j).
\end{equation}
If we choose a different defect as $0_{{\bf g}}$, we must modify the 2-cocycle $\mathfrak{w}$ by a 2-coboundary $d\chi$ for some $\chi: U(1) \rightarrow \mathcal{A}$, and the above equation will send $v \rightarrow v \times \chi({\bf g})^n$. Therefore, $n$ must be chosen such that $\chi({\bf g})^n$ is always trivial. Since $\A = \prod_{i=1}^r \Z_{n_i}$, this happens when $n$ is a common multiple of each $n_i$. Therefore it suffices to choose $n = p := \text{lcm}(n_1,\dots , n_r)$. Alternatively, we can demand that the braiding data of $0_{\bf g}$ are continuously connected to that of the identity particle as ${\bf g}\rightarrow {\bf 0}$. This requirement canonically determines $0_{{\bf g}}$.

Now consider the gauge-invariant quantity 
\begin{equation}\label{ChargeFracInvt}
\F_a({\bf g}_1 = {\bf g}, \dots , {\bf g}_p = {\bf g}) = \left(R^{0_{{\bf g}},a}R^{a,0_{{\bf g}}}\right)^p\prod\limits_{k=0}^{p-1} \eta_{a}({\bf g},{\bf g}^j)
\end{equation} 
where $a$ is an anyon, possibly non-Abelian. In Appendix \ref{Sec:SymFracInvt} it is shown that
\begin{align}
\left(R^{0_{{\bf g}},a}R^{a,0_{{\bf g}}}\right)^n\prod\limits_{j=0}^{p-1} \eta_a({\bf g},{\bf g}^j) &= M_{v,a}.
\end{align}
The quantity $M_{v,a}$ obtained using this formula is manifestly gauge-invariant; for our choice of $p$, the anyon $v$ is also invariant under arbitrary relabelings of the defects. We define $M_{v,a} = e^{2\pi i Q_a}$, where $Q_a$ is the $U(1)$ charge of $a$. 

We can obtain the same result by plugging in the solution discussed in Appendix \ref{Sec:Gxsolns-U1}. With this solution, $R^{0_{{\bf g}},a} = R^{a,0_{{\bf g}}} = 1$, and $\eta_a({\bf g},{\bf g}^j) = 1$ for $0 \le j < p-1$, while $\eta_a({\bf g},{\bf g}^{p-1}) = M_{v,a}$. So the direct evaluation is consistent with the general result. 

In order to measure the defect class, we have the following gauge-invariant quantity:

\begin{align}
\mathcal{I}_n (0_{{\bf g}}) := \theta_{0_{\bf g}}^n \prod_{j = 0}^{n-1} \eta_{0_{\bf g}}({\bf g}, {\bf g}^j) ,
\end{align}
where ${\bf g}=e^{2\pi i/n}$, and $n$ is arbitrary. The particular solution gives that $\theta_{0_{\bf g}} = 1$ and $\eta_{0_{\bf g}}({\bf g}, {\bf g}^j) = \frac{1}{F^{0_{{\bf g}} 0^j_{{\bf  g}} 0_{{\bf g}}}}$. Using this we compute
\begin{align}
\mathcal{I}_n (0_{{\bf  g}})&= \prod\limits_{j=0}^{n-1} \frac{1}{F^{0_{{\bf g}} 0^j_{{\bf  g}} 0_{{\bf g}}}} \\
&= e^{2\pi i \left(\frac{v \star v}{2}+k_1\right)\frac{1}{n} \times \sum_j(\frac{1}{n}+\frac{j}{n}-\frac{[j+1]_n}{n})} \\
&= e^{i\pi \left(v \star v+2k_1\right)\frac{1}{n}}
\end{align} 
This gives the value of $k_1$ modulo $n$; we can choose different values of $n$ to fix $k_1$. Note that $v \star v + 2k_1 = \bar{\sigma}_H$. Therefore when $k_1$ is changed by +1, $\bar{\sigma}_H$ is changed by $+2$. This is consistent with the general formula for the defect response, Eq. \eqref{PureSPTInvtFormula}, which showed that $\mathcal{I}_n(0_{{\bf  g}})^{2n} = M_{v,v}$. Eq. \eqref{PureSPTInvtFormula} is derived in Appendix \ref{Sec:DefectInvts}.

The quantity $\mathcal{I}_n(0_{\bf g})$ is manifestly gauge-invariant. We can also make this quantity invariant under relabelings by fixing $0_{\bf g}$ canonically: we demand that $R^{0_{\bf g},a},R^{a,0_{\bf g}} \rightarrow 1$ as ${\bf g} \rightarrow {\bf 0}$. For $n$ sufficiently large, the defect $0_{\bf g}$ is thus defined canonically, so $\mathcal{I}_n(0_{\bf g})$ is an absolute SET invariant.

We note that in our choice of gauge the defect $F$-symbols $F^{0_{{\bf  g}}0_{{\bf  h}}0_{{\bf k}}}$ are directly related to the effective response theory obtained by integrating out the internal gauge fields describing the topological order. In fact, evaluating the response theory $\frac{\bar{\sigma}_H}{2\pi} A \cup dA$ on a 3-simplex of a triangulation $[0123]$ with a flat configuration of the $U(1)$ gauge field $A$, such that $A_{01} = {\bf  g}, A_{12} = {\bf h}$, and $A_{23} = {\bf k}$, gives precisely the value of $F^{0_{{\bf  g}}0_{{\bf  h}}0_{{\bf k}}}$ stated in Eq. \eqref{U1DefectFSym}. A similar relationship can be observed between the response theory and the defect $F$-symbols in later sections.

\subsection{$G=U(1)\leftthreetimes \mathbb{E}^2$}
\label{Sec:Invts-cont}
For SET phases with continuous magnetic translation and spatial rotation symmetry, all the SET invariants can be extracted by using the subset of the data in which the translation group elements are set to zero. The two anyons $v$ and $s$ giving the symmetry fractionalization class are determined by using the formula for a single $U(1)$ symmetry, in the following way. With the usual notation, we first set $ {\bf g}= (e^{2\pi i/p},{\bf 0},1)$ and evaluate $\mathcal{F}_a({\bf g}_1={\bf g},\dots, {\bf g}_p={\bf g}) = M_{v,a}$. This allows us to determine the vison $v$, as we showed above. Then we set ${\bf h}= (1,{\bf 0},e^{2\pi i/p})$ and evaluate $\mathcal{F}_a({\bf g}_1={\bf h},\dots, {\bf g}_p={\bf h}) = M_{s,a}$, and thus obtain $s$. It is emphasized that this result defines the fractional charge and topological spin of $a$ even when $a$ is non-Abelian.  

The defect $F$-symbols, which are useful for evaluating the defect invariants, were calculated explicitly in Appendix \ref{Sec:Gxsolns-continuum}. The SPT coefficient $k_1$ corresponding to IQH states is found by setting their $SO(2)$ components to zero and evaluating $\mathcal{I}_n(0_{{\bf  g}})$ for ${\bf g} = (e^{2\pi i /n},{\bf  0},1)$. This calculation gives 
\begin{equation}
\mathcal{I}_n(0_{{\bf  g}})=e^{2\pi i \left( \frac{v\star v}{2} + k_1\right)\frac{1}{n}},
\end{equation} 
in agreement with the general identities from Appendix \ref{Sec:DefectInvts}. Similarly, the parameter $k_3$, which corresponds to SPT states associated to the $SO(2)$ plane rotation symmetry, is determined by setting ${\bf h} = (1,{\bf  0},e^{2\pi i /n})$, and computing 
\begin{equation}
\mathcal{I}_n(0_{\bf h}) = e^{2\pi i \left( \frac{s\star s}{2} + k_3\right)\frac{1}{n}}.
\end{equation}
Therefore we can obtain $k_1$ and $k_3$ once we know $v$ and $s$. It remains to find the mixed defect invariant (i.e. the shift) parametrized by $k_2$. To that end, we consider ${\bf k} = (e^{2\pi i /n},{\bf  0},e^{2\pi i /n}), {\bf g} = (e^{2\pi i /n},{\bf  0},1), {\bf h} = (1,{\bf  0},e^{2\pi i /n})$ and evaluate
\begin{align}
\frac{\mathcal{I}_n(0_{{\bf  k}})}{\mathcal{I}_n(0_{{\bf  g}})\mathcal{I}_n(0_{{\bf h}})} &=  \prod\limits_{j=0}^{n-1} \frac{F^{0_{{\bf h}} 0^j_{{\bf  h}} 0_{{\bf h}}}F^{0_{{\bf g}} 0^j_{{\bf  g}} 0_{{\bf g}}}}{F^{0_{{\bf k}} 0^j_{{\bf  k}} 0_{{\bf k}}}} \\ 
&= e^{2\pi i \left(v\star s+k_2 \right)\frac{1}{n} \times \sum_j(\frac{1}{n}+\frac{j}{n}-\frac{[j+1]_n}{n})} 
\\ &= e^{2\pi i\left(v\star s+k_2 \right)\frac{1}{n}}	
\end{align} 
This result is consistent with the general derivation of mixed defect invariants discussed in Appendix \ref{Sec:DefectInvts}. The lhs is a product of three gauge-invariant quantities, and gives $k_2$ once $v$ and $s$ are known, upon choosing different values of $n$. 

In the above analysis we have omitted the invariant for the filling per magnetic unit cell $\nu$, which is not an independent invariant (its fractional part is constrained to equal that of $\bar{\sigma}_H$). The derivation of the invariant for the filling, and the proof of this constraint, are discussed in Appendix \ref{Sec:LSM_cont}.

\subsection{$G=\Z_M$}
\label{Sec:Invts-ZM}
The calculations for $G=\Z_M$ closely resemble those performed above for $G=U(1)$. The invariant for the symmetry fractionalization class is obtained with the same identity used for $U(1)$ symmetry, with the $\Z_M$ generator here denoted as ${\bf h} = e^{2\pi i/M}$:
\begin{align}
e^{2\pi i L_a} &:= \F_a({\bf g}_1 = {\bf h},\dots , {\bf g}_M = {\bf h})\nonumber \\ 
&= \left(R^{0_{{\bf h}},a}R^{a,0_{{\bf h}}}\right)^M\prod_{j=1}^{M-1} \eta_a( {\bf h},{\bf h}^j )  = M_{s,a}.
\end{align}
Here the anyon $s$ is defined as $s = 0_{\bf h}^M$. Knowing the value of $M_{s,a}$ as a function of $a$ allows us to determine $s$, and hence the symmetry fractionalization class. However, we must keep in mind the fact that the same symmetry fractionalization class can be described by diferent values of $s$. If we choose a different ${\bf  h}$-defect, namely $s'_{{\bf  h}}$ instead of $0_{{\bf  h}}$, the above identity will instead measure the anyon given by $(s'_{{\bf  h}})^M = s \times s'^M$. However, as we saw from group cohomology calculations as well as physical arguments, this anyon is associated to the same symmetry fractionalization class. 

To distinguish the $M$ defect classes, we use the formula
\begin{align}
\mathcal{I}_M(0_{{\bf h}})  &= \theta_{0_{\bf h}}^M \prod_{j = 0}^{M-1} \eta_{0_{\bf h}}({\bf h}, {\bf h}^j)
\end{align}
When evaluated against the $G$-crossed solution in Appendix \ref{Sec:Gxsolns-ZM}, which is parametrized by the anyon $s$ and the integer $k_3 \in \Z_M$, the formula returns the value
\begin{align}
\mathcal{I}_M(0_{{\bf h}})  &= e^{2\pi i \frac{s \star s + 2 k_3}{2M}} := e^{i \pi \ell_s/M }.
\end{align}
However, different values of $k_3 \mod M$ may corespond to the same SET phase. This is because the above formula depends on a choice of $0_{{\bf h}}$, and returns different values upon relabeling the defects; under such relabelings, solutions corresponding to different values of $k_3$ become interchangeable. Therefore, properly counting the number of SET phases with $\Z_M$ symmetry becomes a nontrivial exercise. The origin of the redundancy, as well as a general counting strategy, are discussed in detail in Appendix \ref{Sec:relabeling_ZM}. 

\subsection{$G=U(1)\leftthreetimes_{\phi} \Z^2$}
\label{Sec:Invts-magtrans}

For a system with discrete magnetic translation symmetry, the invariant giving the vison $v$ is obtained by setting all translations to zero and using the standard formula for $U(1)$ symmetry fractionalization. Let us now discuss the invariant which measures the anyon $m$ per unit cell. The desired formula can be obtained by considering the following sequence of symmetry operations applied to the anyon $a$: ${\bf g}_1 = (1,{\bf x}); {\bf g}_2 = (1,{\bf y}); {\bf g}_3 = (1,{\bf -x}); {\bf g}_4 = (1,{\bf -y})$. The effect of the symmetry operations is to take $a$ around a unit cell, and in doing so also around an overall $\phi$ flux; this latter contribution  must be cancelled away. Using the results of Appendix \ref{Sec:SymFracInvt}, we see that this is done using the gauge-invariant quantity
\begin{widetext}
	\begin{align}
	\F_a({\bf g}_1,{\bf g}_2,{\bf g}_3,{\bf g}_4) &= \frac{\left(\prod\limits_{i=1}^{4} R^{a, 0_{{\bf g}_i}} R^{0_{{\bf g}_i},a}\right)}{ R^{a, 0_{{\bf g}_4{\bf g}_3{\bf g}_2{\bf g}_1}} R^{  0_{{\bf g}_4{\bf g}_3{\bf g}_2{\bf g}_1},a}} \eta_{a}({\bf g}_4,{\bf g}_1{\bf g}_2{\bf g}_3)\eta_{a}({\bf g}_3,{\bf g}_1{\bf g}_2)\eta_{a}({\bf g}_2,{\bf g}_1).
	\end{align}
\end{widetext}
Using the notation of Appendix \ref{Sec:Gxsolns-magtrans}, we can explicitly evaluate this in the gauge where $R^{a, 0_{{\bf g}_i}} R^{ 0_{{\bf g}_i},a} = 1$, and $\mathfrak{w}({\bf g}_1,{\bf  g}_2) = v^{z_{12} + \frac{\phi}{2} {\bf r_1}\times {\bf r_2}}\times m^{\frac{1}{2} {\bf r_1}\times {\bf r_2}}$. A direct calculation then gives the anyon induced by the above symmetry operations:

	\begin{align}
	& \left(\mathfrak{w}({\bf g}_2,{\bf g}_1)\mathfrak{w}({\bf g}_3,{\bf g}_1{\bf  g}_2)\mathfrak{w}({\bf g}_4,{\bf g}_1{\bf g}_2{\bf  g}_3)\right) \\
	&= (v^{\frac{\phi}{2}-\frac{\phi}{2}}\times m^{\frac{1}{2}})\times (v^{\frac{\phi}{2}-\frac{\phi}{2}}\times m^{\frac{1}{2}})\times (v^0\times m^0)\\
	&  = m
	\end{align}
As a result, we obtain the relation $\F_a({\bf g}_1,{\bf g}_2,{\bf g}_3,{\bf g}_4)= M_{a,m}$. Notice that the individual terms in parentheses on the middle line of the above computation are not well-defined anyons. This is because we are working in symmetric gauge. We can do the same calculation in Landau gauge and obtain the same answer, with each $\mathfrak{w}$ factor well-defined. However, the above invariant gives the same answer for any gauge choice. Moreover, from the computation of $\H^2(G,\A)$ (Appendix \ref{Sec:Coho_calc_magtrans}), we see that different choices of $m$ are all physically distinct, and therefore $m$ will remain invariant under defect relabelings.

Next we study the invariants characterizing the defect class. The $U(1)$ SPT index $k_1$ is determined by setting all the ${\bf r}_i$ to zero and choosing an element ${\bf g} = (e^{2\pi i /n},{\bf 0})$. The invariant $\mathcal{I}_n(0_{{\bf g}})$ then gives the value of $k_1 \mod n$, as we have seen previously.  

We also have an invariant which measures the filling $\nu$ per magnetic unit cell. From the cohomology calculations in Appendix \ref{Sec:Coho_calc_magtrans}, we find that each $q \times 1$ magnetic unit cell encloses the anyon
\begin{align}
\mathfrak{b}({\bf r_1},{\bf r_2}) &= \mathfrak{w}({\bf r_1},{\bf r_2})\overline{\mathfrak{w}({\bf r_2},{\bf r_1})}= v^p \times m^q.
\end{align}
Consider the quantity 
\begin{align}
\frac{\eta_{0_{{\bf g}}} ({\bf  r_2},{\bf  r_1})}{\eta_{0_{{\bf g}}}({\bf  r_1},{\bf  r_2})}
\end{align}
This quantity is manifestly invariant under both vertex basis and symmetry action gauge transformations when ${\bf r_1}$ and ${\bf r_2}$ commute, i.e. if we consider a magnetic unit cell instead of the standard one. In the general discusion of Appendix \ref{Sec:DefectInvts}, we prove that when ${\bf g} = (e^{i 2\pi/n},{\bf 0})$,
\begin{align}
\left(\frac{\eta_{0_{{\bf g}}} ({\bf  r_2},{\bf  r_1})}{\eta_{0_{{\bf g}}}({\bf  r_1},{\bf  r_2})}\right)^n &= \frac{\eta_{v} ({\bf  r_2},{\bf  r_1})}{\eta_{v}({\bf  r_1},{\bf  r_2})} = M_{v,\mathfrak{b}({\bf  r_2},{\bf  r_1})}.
\end{align}
Note that when ${\bf r_1}$ and ${\bf r_2}$ span a $q \times 1$ magnetic unit cell, $M_{v,\mathfrak{b}({\bf  r_2},{\bf  r_1})} = e^{2\pi i v \star (v^p \times m^q)} = e^{2\pi i \nu}$, where $\nu$ is the filling per magnetic unit cell. Therefore the above invariant measures $\nu/n \mod 1$. To verify this explicitly, we use the solution in Appendix \ref{Sec:Gxsolns-magtrans} and write, for those specific gauge choices, 
\begin{align}
\frac{\eta_{0_{{\bf g}}} ({\bf  r_2},{\bf  r_1})}{\eta_{0_{{\bf g}}}({\bf  r_1},{\bf  r_2})} &= \frac{F^{0_{\bf r_1}0_{\bf g}0_{\bf r_2}}}{F^{0_{\bf r_2}0_{\bf g}0_{\bf r_1}}} \frac{F^{0_{\bf g}0_{\bf r_2}0_{\bf r_1}}}{F^{0_{\bf g}0_{\bf r_1}0_{\bf r_2}}}\frac{F^{0_{\bf r_2}0_{\bf r_1}0_{\bf g}}}{F^{0_{\bf r_1}0_{\bf r_2}0_{\bf g}}}
\end{align}
Evaluating the $F$- symbols, we finally obtain
\begin{align}
\frac{\eta_{0_{{\bf g}}} ({\bf  r_2},{\bf  r_1})}{\eta_{0_{{\bf g}}}({\bf  r_1},{\bf  r_2})} &= e^{2\pi i \left((v \star v + 2 k_1) \frac{1}{n} (\phi q - [\phi q]) + (v\star m +k_6) \frac{1}{n} q\right)} \nonumber\\
&= e^{2\pi i (p \bar{\sigma}_H + q (v \star m + k_6)) \frac{1}{n}} = e^{ 2\pi i \nu \frac{1}{n}},
\end{align}
where we have identified $\frac{\nu}{n} = \frac{p \bar{\sigma}_H + q (v \star m  + k_6)}{n} \mod 1$. With this result, we can verify a stronger form of the filling LSM relation than the one stated in Eq. \eqref{DiscreteMagTransFillingRelation}, in terms of the $G$-crossed data. The usual LSM relation and its stronger counterpart are discussed using $G$-crossed identities separately in Appendix \ref{Sec:LSM_lattice}.

\subsection{$G=U(1)\leftthreetimes_{\phi} [\Z^2\rtimes \Z_M]$}
\label{Sec:Invts-gspace}

First we make the following remark regarding notation. Let $h$ be the generator of point group rotations. We will abuse the notation for rotation point group elements slightly, as follows: the quantities $h_i$ written in line should be understood as integers mod $M$, corresponding to rotations by the angle $2\pi h_i/M$, while in an expression such as ${^h} {\bf r}$ or ${^{1-h}}{\bf r}$, $h$ is understood as the $2\times 2$ matrix generator of point group rotations, and thus the symbol 1 in ${^{1-h}}{\bf r}$ denotes the identity $2\times 2$ matrix.

\subsubsection{Symmetry fractionalization invariants}

We have previously discussed invariant formulas to measure the symmetry fractionalization anyons $v$, $s$ and $m$. The same formulas hold in the present case, once we set the translation components of the group elements involved to zero.  
Below we will discuss how to measure the anyons $t_x,t_y$ separately for $M=2,3,4$. \\

$\underline{M=2}$: Using the formulas for group cohomology given in Appendix \ref{Sec:Coho_calc_gspace}, we see that this symmetry fractionalization is classified by $(\Z_2\times\Z_2)\otimes \A$, corresponding to a pair of anyons $(t_x,t_y)$ modulo the equivalence relation $(t_x,t_y) \sim (t_x\times a^2,t_y\times b^2)$, where $a,b \in \A$. We define ${\bf g} = (1,{\bf r},e^{i \pi})$. Thus ${\bf g}^2 = (e^{2\pi i \frac{1}{2}{\bf r}\times -{\bf r}},{\bf 0},1) = (1,{\bf 0},1)$. Now using the explicit symmetry fractionalization cocycle given in Appendix \ref{Sec:Coho_calc_gspace}, the anyon induced is given by

\begin{equation}
\mathfrak{w}({\bf g},{\bf g}) = v^0 \times m^0 \times s^1 \times (\vec{t}\cdot {^{\frac{1-h}{1-h}}}{\bf r}) = s \times (\vec{t}\cdot {\bf r}).
\end{equation} 

Hence the invariant $\mathcal{F}_a({\bf g},{\bf g})$ will measure $M_{a,s \times (\vec{t}\cdot {\bf r})}$. We can remove the contribution from $s$ by defining ${\bf k} = (1,{\bf 0},e^{i \pi})$ and using the identity
\begin{equation}
\frac{\mathcal{F}_a({\bf g},{\bf g})}{\mathcal{F}_a({\bf k},{\bf k})} = \frac{M_{a,s\times \vec{t}\cdot {\bf r}}}{M_{a,s}} = M_{a,\vec{t}\cdot {\bf r}}.
\end{equation}
Finally, we examine how a relabeling of defects changes the anyon measured by the above symmetry operation. A relabeling $0_{{\bf g}}\rightarrow 0_{{\bf g}} \chi({\bf g})$ takes $\mathfrak{w}({\bf g},{\bf g}) \rightarrow \mathfrak{w}({\bf g},{\bf g}) (\chi({\bf g}))^2$. Similarly, a relabeling $0_{{\bf k}}\rightarrow 0_{{\bf k}} \chi({\bf k})$ takes $\mathfrak{w}({\bf k},{\bf k}) \rightarrow \mathfrak{w}({\bf k},{\bf k}) (\chi({\bf k}))^2$. Therefore the anyon measured by $\frac{\mathcal{F}_a({\bf g},{\bf g})}{\mathcal{F}_a({\bf k},{\bf k})}$ transforms under a relabeling as $\vec{t}\cdot {\bf r}\rightarrow \vec{t}\cdot {\bf r} (\chi({\bf g}))^2 \overline{\chi({\bf k})}^2$. Since the function $\chi$ is arbitrary, the redundancies in the above invariant reproduce the equivalence relation on $(t_x,t_y)$ obtained using group cohomology. \\

$\underline{M=3}$: The symmetry fractionalization is classified by $\Z_3\otimes \A$, corresponding to a pair of anyons $(t_x,t_y)$ modulo the equivalence relation $(t_x,t_y) \sim (t_x\times a \times b,t_y\times b^2 \times \overline{a})$, where $a,b \in \A$. (This is equivalent to saying that $t_x\overline{t_y^2}$ is defined modulo anyons of the form $a^3$.) Define ${\bf g} =(1,{\bf x},e^{i 2\pi/3})$ and ${\bf k} =(1,{\bf y},e^{i 2\pi/3})$. Three successive operations of ${\bf g}$ are associated to a combination of different anyons. The translations cover an area

\begin{equation}
A = \frac{1}{2}({\bf r}\times {^h}{\bf r}+({\bf r}+ {^h}{\bf r})\times {^{h^2}}{\bf r}) = \frac{1}{2}{\bf r}\times {^h}{\bf r} + 0
\end{equation}
where we have used the fact that for any ${\bf r}$, ${^{1+h+h^2}}{\bf r}={\bf 0}$. Therefore an evaluation of $\mathcal{F}_a({\bf g},{\bf g},{\bf g})$ will have a contribution from the anyon per unit cell, which is given by $m^A$. There will also be a contribution $v^{\alpha}$, where $\alpha$ is the integer part of $\phi A$ and is associated to the total number of flux quanta contained in the area $A$. 
The desired anyon can be extracted by plugging in the form of the general symmetry fractionalization cocycle from Eq. \eqref{Eq:GspaceSFcocycle} into the expression below:
\begin{align}
& \mathfrak{w}({\bf g},{\bf g}) \times \mathfrak{w}({\bf g},{\bf g}^2)\times \overline{\mathfrak{w}({\bf k},{\bf k})} \times \overline{\mathfrak{w}({\bf k},{\bf k}^2)}   \nonumber \\
&= (v^{\alpha}\times m^{A} \times \vec{t}\cdot {\bf x} \times s^0) \times (v^{0}\times m^{0} \times \vec{t}\cdot {^{1+h}}{\bf x} \times s^1) \nonumber \\
& \times (v^{\alpha}\times m^{A} \times \vec{t}\cdot {\bf y} \times s^0)^{-1} \times (v^{0}\times m^{0} \times \vec{t}\cdot {^{1+h}}{\bf y} \times s^1)^{-1} \\
&= \vec{t}\cdot {^{2+h}}({\bf x}-{\bf y}) = t_x \overline{t_y^2}
\end{align} 
Note, in particular, that the contributions proportional to the area vanish from the final expression. This implies that
\begin{equation}
\frac{\mathcal{F}_a({\bf g},{\bf g},{\bf g})}{\mathcal{F}_a({\bf k},{\bf k},{\bf k})}= M_{a, t_x\overline{t^2_y}}.
\end{equation}
Here we used $h = \begin{pmatrix}
0 & 1 \\ -1 & -1
\end{pmatrix}$.  A relabeling will shift the observed anyon by the amount $\chi({\bf g}))^3 \overline{\chi({\bf k})^3}$, and therefore, since $\chi$ is arbitrary, the above identity measures $t_x\overline{t^2_y}$ modulo anyons of the form $a^3$. This exactly reproduces the equivalence relation from group cohomology.

$\underline{M=4}$: Here the computations closely resemble those performed for $M=2$. The symmetry fractionalization is classified by $\Z_2\otimes \A$, corresponding to a pair of anyons $(t_x,t_y)$ modulo the equivalence relation $(t_x,t_y) \sim (t_x\times a \times b,t_y\times a \times \overline{b})$, where $a,b \in \A$. Define ${\bf g} = (1,{\bf r},e^{i \pi})$. Thus ${\bf g}^2 = (e^{2\pi i \frac{1}{2}{\bf r}\times -{\bf r}},{\bf 0},1) = (1,{\bf 0},1)$. Then we compute 

\begin{align}
\mathfrak{w}({\bf g},{\bf g}) = v^0 \times m^0 \times s^1 \times (\vec{t}\cdot {^{\frac{1-h^2}{1-h}}}{\bf r}) = s \times (\vec{t}\cdot {^{1+h}} {\bf r}).
\end{align} 
With the definition $h = \begin{pmatrix}
0 & 1 \\ -1 &0 
\end{pmatrix}$, we have ${^{1+h}} {\bf r} = (r_1 + r_2,r_2-r_1)^T$. Therefore the anyon measured is $s \times t_x^{r_1+r_2}\times t_y^{r_2-r_1}$. Repeating the arguments for $M=2$, we can eliminate the contribution from $s$ and obtain 

\begin{equation}
\frac{\mathcal{F}_a({\bf g},{\bf g})}{\mathcal{F}_a({\bf k},{\bf k})} = M_{a, t_x^{r_1+r_2}\times t_y^{r_2-r_1}} = M_{a, (t_xt_y)^{r_2}\times (t_x\overline{t_y})^{r_1}}
\end{equation}
where ${\bf k} = (1,{\bf 0},e^{i \pi})$. 

Under a relabeling of defects, the anyon $(t_xt_y)^{r_2}\times (t_x\overline{t_y})^{r_1}$ gets shifted by $(\chi({\bf g}))^2 \overline{\chi({\bf k})}^2$, which is an anyon of the form $a^2$. Therefore the symmetry fractionalization is specified by $t_xt_y$ (or $t_x\overline{t_y}$) up to anyons $a^2, a \in \A$. This reproduces the group cohomology equivalence relation stated above.

$\underline{M=6}$: In this case there is no nontrivial symmetry fractionalization class. It is important to note that this does not mean that $t_x$ and $t_y$ will always be trivial when measured using $G$-crossed identities. However, every value of $t_x$ and $t_y$ can be shifted to the trivial value $(0,0)$ through an appropriate relabeling of defects. Therefore there are no invariants associated to symmetry fractionalization in this case.

\subsubsection{Defect invariants}

Next we study the invariants giving the physical response properties, and whose integer parts parametrize the defect class. In our discussion of magnetic translation symmetry we have already seen how to isolate the integer coefficients $k_1$ and $k_6$. The remaining invariants are discussed below. Note that the fractional charge/angular momentum/crystal momentum of a defect as determined by these invariants is well-defined only up to the charge/angular momentum/crystal momentum of an anyon. This is because we can always attach an anyon to an elementary discrete symmetry defect such as a dislocation or a disclination by adjusting the local energetics of the underlying Hamiltonian, without changing the topological phase to which the system belongs. The remaining invariants can be described as follows.

\textbf{Discrete shift} : The coefficient $\mathscr{S}= q\star s +k_2$ is a discrete analog of the shift. The term with $\mathscr{S}$ assigns the fractional $U(1)$ charge $\frac{\mathscr{S}}{M}$ to elementary $2\pi/M$ disclinations and to corners of angle $2\pi/M$. To find the corresponding $G$-crossed invariant, we use the prescription for mixed defect invariants discussed in Appendix \ref{Sec:DefectInvts}. Namely, we set all translations to zero and take ${\bf g} = (e^{2\pi i/M},{\bf 0},1),{\bf h} = (1,{\bf 0},e^{2\pi i/M})$, and then evaluate
\begin{align}
\frac{\mathcal{I}_M(0_{\bf gh})}{\mathcal{I}_M(0_{\bf g})\mathcal{I}_M(0_{\bf h})} &= e^{\frac{2\pi i}{M} \mathscr{S}}.
\end{align}    
The three factors in the lhs are manifestly invariant under vertex basis and symmetry action gauge transformations. The denominator terms cancel out the pure $U(1)$ and $\Z_M$-SPT terms arising in the numerator, leaving only the desired mixed term. We have not yet accounted for how the rhs transforms under relabelings of the defects: we will do this in Appendix \ref{Sec:relabeling_full}, where we discuss how to obtain the full SET classification for $G=U(1)\leftthreetimes_{\phi} [\Z^2\rtimes \Z_M]$ after accounting for such relabelings.

\textbf{Angular momentum of an elementary disclination}: The quantity $\frac{\ell_s}{2M}=\frac{s \star s + 2k_3}{2M} \mod 1$ measures the angular momentum of an elementary disclination (up to a contribution from the chiral central charge), where $k_3$ is the 'strong' rotation SPT index. It is calculated via the usual formula $e^{i \pi \ell_s/M} := \mathcal{I}_M(0_{{\bf h}})$, where ${\bf h}=(1,{\bf 0},e^{2\pi i/M})$.

\textbf{Quantized electric polarization}: This quantity defines an electric polarization given by $\vec{\mathscr{P}}_c \times \hat{z}$, which is quantized by the rotational symmetry of the lattice. From the response theory (Appendix \ref{Sec:Eff_Ac-gspace}), we obtain $\mathscr{P}_{c,i} = {^{(1-h)^{-1}}} \left(q \star t_i+k_{4,i}\right)$. We will describe the formulas for each $M$ separately below.

$\underline{M=2}$: In this case, define ${\bf g} = (e^{i \pi},{\bf r},e^{2\pi i/2})$ and ${\bf k} = (1,{\bf r},e^{2\pi i/2})$; they satisfy ${\bf g}^2 = {\bf k}^2 = {\bf 0}$. We use the general symmetry fractionalization cocycle in Appendix \ref{Sec:Gxsolns-gspace} to write

\begin{align}
\mathfrak{w}({\bf g},{\bf g}) &= v\times s\times \vec{t} \cdot {\bf r} \nonumber \\
\mathfrak{w}({\bf k},{\bf k}) &= s\times \vec{t} \cdot {\bf r} \label{checkP_2_a}
\end{align}
Using this and the ribbon property of anyons, we can deduce that
\begin{align}
\left(\frac{\mathcal{I}_2(0_{{\bf g}})}{\mathcal{I}_2(0_{{\bf k}})}\right)^2 = \frac{\theta_{v\times s\times \vec{t} \cdot {\bf r}}}{\theta_{s\times \vec{t} \cdot {\bf r}}}=  \theta_v \times M_{v,s \times  \vec{t} \cdot {\bf r}}.
\end{align}
Crucially, the rhs has a factor of $M_{v,\vec{t}\cdot {\bf r}}$, which is associated to the anyonic contribution to the polarization. Therefore we expect that the invariant itself measures the anyonic contribution as well as the desired SPT contribution to the polarization, along with other responses that need to be subtracted away. Indeed, a direct evaluation using the $G$-crossed solution in Appendix \ref{Sec:Gxsolns-gspace} gives

\begin{equation}\label{checkP_2}
\frac{\mathcal{I}_2(0_{{\bf g}})}{\mathcal{I}_2(0_{{\bf k}})} = e^{i\pi \frac{\bar{\sigma}_H}{2}}\times e^{2\pi i \frac{\mathscr{S}}{2}}\times e^{2\pi i \vec{\mathscr{P}}_c\cdot {\bf r}}
\end{equation} 

The explicit calculation therefore agrees with the above general identity. Knowing $\bar{\sigma}_{H}$ and $\mathscr{S}$, we can extract $\vec{\mathscr{P}}_c$. The four defect classes form the group $\Z_2 \times \Z_2$ and can be distinguished by taking ${\bf r} = {\bf x},{\bf y}$. 

$\underline{M=3}$: Here we have three defect classes. We define ${\bf g},{\bf k}$ so that ${\bf g}^3 = {\bf k}^3 = {\bf 0}$. Now the first component $e^{2\pi i z}$ depends on $\phi$, and must satisfy $3z + \frac{\phi}{2} {\bf r} \times {^h}{\bf r} = 0 \mod 1$. Assuming ${\bf r} = {\bf x}$ or ${\bf y}$, we obtain $3z + \phi = 0$. Hence we choose ${\bf g} = (e^{2\pi i \frac{2-\phi}{3}},{\bf r},e^{2\pi i/3})$ and ${\bf k} = (e^{2\pi i \frac{1-\phi}{3}},{\bf r},e^{2\pi i/3})$.
We now use the explicit forms of the symmetry fractionalization cocycles to calculate

\begin{align}
& \mathfrak{w}({\bf g},{\bf g}) \times \mathfrak{w}({\bf g},{\bf g}^2) \nonumber \\ &= (v^1\times s^0\times m^{1/2}\times\vec{t} \cdot {\bf r})  \times (v^1\times s^1\times m^{1/2}\times\vec{t} \cdot {^{1+h}}{\bf r}) \nonumber \\ &= v^2 \times s \times m \times \vec{t} \cdot {^{2+h}} {\bf r}, \label{checkP_3_a1}\\
&\mathfrak{w}({\bf k},{\bf k}) \times \mathfrak{w}({\bf k},{\bf k}^2) \nonumber \\ &= (v^0\times s^0\times m^{1/2}\times\vec{t} \cdot {\bf r})  \times (v^1\times s^1\times m^{1/2}\times\vec{t} \cdot {^{1+h}}{\bf r}) \nonumber \\ &= v \times s \times m \times \vec{t} \cdot {^{2+h}} {\bf r}. \label{checkP_3_a2}
\end{align}

From this we see that

\begin{align}
\left(\frac{\mathcal{I}_3(0_{{\bf g}})}{\mathcal{I}_3(0_{{\bf k}})}\right)^3 = \theta_v \times M_{v,v \times s \times m \times \vec{t} \cdot {^{2+h}} {\bf r}}.
\end{align}

Now a direct evaluation using the $G$-crossed solution gives

\begin{equation}\label{checkP_3}
\frac{\mathcal{I}_3(0_{{\bf g}})}{\mathcal{I}_3(0_{{\bf k}})} = e^{i \pi \bar{\sigma}_H}\times e^{2\pi i \frac{\mathscr{S}}{3}}\times e^{2\pi i \frac{\nu}{3}}\times e^{2\pi i \frac{1}{3}\vec{\mathscr{P}}_c\cdot {^{2+h}}{\bf r}},
\end{equation}
in agreement with the expected result. After subtracting the contributions from $\bar{\sigma}_H$, $\mathscr{S}$ and $\nu$, this invariant gives us the components of $\frac{1}{3}\vec{\mathscr{P}}_c\cdot {^{2+h}} {\bf r} \mod 1$; i.e. the values of $\frac{1}{3}(2\mathscr{P}_{c,x} + \mathscr{P}_{c,y})$ and $\frac{1}{3}(-\mathscr{P}_{c,x} + \mathscr{P}_{c,y})$ modulo 1. The defect classes differ from each other by shifting the above quantities by $\pm 1/3$. Therefore the above identity can indeed distinguish all the defect classes. 

We observe that the filling $\nu$ appears in Eq. \eqref{checkP_3}, but not in Eq. \eqref{checkP_2}. This is related to the fact that the anyons measured in Eqs. \eqref{checkP_3_a1},\eqref{checkP_3_a2} receive a contribution from $m$, while those measured in Eq. \eqref{checkP_2_a} do not. The simplification observed in Eq. \eqref{checkP_2} occurs for order 2 rotations, and will also be seen in Eq. \eqref{checkP_4} below.

$\underline{M=4}$: Now there are two defect classes. Define ${\bf g} = (e^{i \pi},{\bf r},e^{2\pi i/2})$ and ${\bf k} = (1,{\bf r},e^{2\pi i/2})$; they satisfy ${\bf g}^2 = {\bf k}^2 = {\bf 0}$. Then we use similar reasoning as above to evaluate

\begin{equation}\label{checkP_4}
\frac{\mathcal{I}_2(0_{{\bf g}})}{\mathcal{I}_2(0_{{\bf k}})} = e^{i \pi \frac{\bar{\sigma}_H}{2}}\times e^{2\pi i \frac{\mathscr{S} }{2}}\times e^{2\pi i \frac{1}{2}\vec{\mathscr{P}}_c\cdot {^{1+h}} {\bf r}}
\end{equation} 
After subtracting the contributions from $\bar{\sigma}_H$ and $\mathscr{S}$, this invariant gives us the components of $\frac{1}{2}\vec{\mathscr{P}}_c\cdot {^{1+h}} {\bf r}$; i.e. the values of $\frac{1}{2}(\mathscr{P}_{c,x} \pm \mathscr{P}_{c,y})$, instead of $\mathscr{P}_{c,i}$ directly. However, the two defect classes differ by a shift of $\frac{1}{2}(\mathscr{P}_{c,x} \pm \mathscr{P}_{c,y})$ by $1/2$. Therefore, in the above expression, the two defect classes will give values that differ by a sign, so we can indeed distinguish them.

\textbf{Angular momentum polarization}: The quantity $\vec{\mathscr{P}}_s$ defines a fractional angular momentum polarization, analogous to the charge polarization studied above. From the response theory we obtained $\mathscr{P}_{s,i} = {^{(1-h)^{-1}}}\left(s\star t_i + k_{5,i}\right)$. The corresponding invariant is extracted in a similar manner to $\vec{\mathscr{P}}_c$. The diferent cases are worked out separately below:

$\underline{M=2}$: Take ${\bf g} = (1,{\bf r},e^{i\pi})$; thus ${\bf g}^2 = {\bf 0}$. Also define ${\bf k} = (1,{\bf 0},e^{i\pi})$. Now we can evaluate

\begin{align}
\mathfrak{w}({\bf g},{\bf g}) &= s\times \vec{t} \cdot {\bf r} \nonumber \\
\mathfrak{w}({\bf k},{\bf k}) &= s
\end{align}
which implies that
\begin{equation}
\left(\frac{\mathcal{I}_2(0_{\bf g})}{\mathcal{I}_2(0_{\bf k})}\right)^2 = \theta_{\vec{t} \cdot {\bf r} } \times M_{s,\vec{t} \cdot {\bf r} }.
\end{equation}
Crucially, there is a contribution from $M_{s,\vec{t} \cdot {\bf r} }$, which is associated to the anyonic contribution to $\vec{\mathscr{P}}_s$. Using the explicit solution in Appendix \ref{Sec:Gxsolns-gspace}, we verify that
\begin{equation}
\frac{\mathcal{I}_2(0_{\bf g})}{\mathcal{I}_2(0_{\bf k})} = e^{2\pi i \frac{\vec{\mathscr{P}}_s\cdot {\bf r}}{2}} \times e^{\frac{i \pi}{2} (\vec{t}\cdot {\bf r})\star ((\vec{t}\cdot {\bf r}))}.
\end{equation}

Note that we now have to subtract the contribution $e^{\frac{i \pi}{2} (\vec{t}\cdot {\bf r})\star ((\vec{t}\cdot {\bf r}))}$, related to the quantized torsional response, in order to isolate the desired invariant. We will have to subtract similar contributions for $M=3,4$ as well.

$\underline{M=3}$: Take ${\bf g} = (e^{2\pi i z},{\bf x},e^{i2\pi/3})$, where $z$ is chosen so that ${\bf g}^3 = {\bf 0}$. Also define ${\bf k} = (e^{2\pi i z},{\bf y},e^{i2\pi/3})$. Here ${\bf x},{\bf y}$ are elementary translations. We find that

\begin{align}
\mathfrak{w}({\bf g},{\bf g})\times \mathfrak{w}({\bf g},{\bf g}^2) &= v \times s\times m \times \vec{t} \cdot {^{2+h}}{\bf x} \nonumber \\
\mathfrak{w}({\bf k},{\bf k})\times \mathfrak{w}({\bf k},{\bf k}^2) &= v \times s\times m \times \vec{t} \cdot {^{2+h}}{\bf y},
\end{align}
which implies that 

\begin{equation}
\left(\frac{\mathcal{I}_3(0_{\bf g})}{\mathcal{I}_3(0_{\bf k})}\right)^3 = \theta_{\vec{t} \cdot {^{2+h}}({\bf x-y}) } \times M_{s,v \times s\times m \times \vec{t} \cdot {^{2+h}}({\bf x-y})}.
\end{equation}
In agreement with this, we can perform an explicit computation and see that

\begin{align}
\frac{\mathcal{I}_3(0_{\bf g})}{\mathcal{I}_3(0_{\bf k})} &= e^{2\pi i \frac{1}{3}\vec{\mathscr{P}}_s\cdot {^{2+h}}({\bf x}-{\bf y})} \times e^{2\pi i \frac{\mathscr{S}}{3}} \times e^{\frac{\pi i}{3} (t_x\star t_y + t_y \star t_x + t_y \star t_y)} \nonumber \\
& \times e^{2\pi i z \vec{\mathscr{P}}_c\cdot {^{2+h}}({\bf x}-{\bf y})}
\end{align}
Having determined the charge polarization and the shift, we can extract the angular momentum polarization as well.  

$\underline{M=4}$: Take ${\bf g} = (1,{\bf r},e^{i\pi})$; thus ${\bf g}^2 = {\bf 0}$. Also define ${\bf k} = (1,{\bf 0},e^{i\pi})$. Now we can evaluate
\begin{equation}
\frac{\mathcal{I}_2(0_{\bf g})}{\mathcal{I}_2(0_{\bf k})} = e^{2\pi i \frac{\vec{\mathscr{P}}_s\cdot {^{1+h}}{\bf r}}{2}} \times e^{\frac{\pi i}{2} (\vec{t}\cdot {\bf r})\star (\vec{t}\cdot {^{1+h}}{\bf r})}.
\end{equation}
The invariants for the two defect classes will now differ from each other by a sign.

\textbf{Quantized linear and angular momentum per magnetic unit cell}: We expect from group cohomology, and from the defect $F$-symbols calculated in Eq. \eqref{FullSpaceGrpDefectFSyms}, that there are quantized responses associated to a fractional linear and angular momentum per unit cell. However, we have not been able to find $G$-crossed invariants for these responses. This problem will be left for future work.

In the expression for the defect $F$-symbols, Eq. \eqref{FullSpaceGrpDefectFSyms}, we observe that there are terms which are not associated to any parameters of the defect class, but instead are completely determined by the symmetry fractionalization data. These terms correspond to additional quantized responses, and have invariants of their own; however, their values are fixed by the symmetry fractionalization class. They also appear in the effective response theory associated to the SET phase once the internal gauge fields have been integrated out. We discuss one such response below:  

\textbf{Quantized torsional response $\Pi_{ij} $}: This invariant is associated to the components of the crystal momentum associated to a dislocation, through the relation $p_i = \sum_j\Pi_{ij} b_j$, where ${\bf b}$ is the dislocation Burgers vector.

We will state the formula in the simpler case where $v=m=0$. In the most general case, this formula will have several additional contributions from other responses, and so we will need to carefully isolate $\Pi_{ij}$ from the additional contributions. We will not perform the general calculation here. 

Consider the four defects
\begin{align}
d_1 &= 0_{(1,{\bf 0},e^{2\pi i/M})}; d_1^M = s \nonumber\\ 
d_2 &= 0_{(1,{\bf r'},e^{2\pi i/M})}; d_2^M = s \times \vec{t} \cdot {^{M(1-h)^{-1}}} {\bf r'} \nonumber\\ 
d_3 &= 0_{(1,{\bf r},e^{2\pi i/M})}; d_3^M = s  \times \vec{t} \cdot {^{M(1-h)^{-1}}} {\bf r} \nonumber\\ 
d_4 &= 0_{(1,{\bf r+r'},e^{2\pi i/M})}; d_4^M = s  \times \vec{t} \cdot {^{M(1-h)^{-1}}}({\bf r} + {\bf r'})
\end{align}

Using the ribbon property for anyons, we can show that

\begin{align}\label{THVderiv}
\left(\frac{\mathcal{I}_M(d_4)\mathcal{I}_M(d_1)}{\mathcal{I}_M(d_2)\mathcal{I}_M(d_3)} \right)^M &= \frac{\theta_{s  \times \vec{t} \cdot {^{M(1-h)^{-1}}}({\bf r} + {\bf r'})}\theta_s}{\theta_{s  \times \vec{t} \cdot {^{M(1-h)^{-1}}} {\bf r}}\theta_{ s \times \vec{t} \cdot {^{M(1-h)^{-1}}} {\bf r'}}}\nonumber \\ & = M_{\vec{t} \cdot {^{M(1-h)^{-1}}} {\bf r}, \vec{t} \cdot {^{M(1-h)^{-1}}} {\bf r'}}
\end{align}
Using the field theory formalism we had obtained $\Pi_{ij} = {({^{(1-h)^{-1}}} \vec{t})_i \star ({^{(1-h)^{-1}}} \vec{t})_j}$ as the coefficient of the response term $\frac{\Pi_{ij}}{2\pi} R_i \cup dR_j$. Consistent with this, we define

\begin{equation}
e^{2\pi i M {\bf r}^T \Pi {\bf r'}} := \frac{\mathcal{I}_M(d_4)\mathcal{I}_M(d_1)}{\mathcal{I}_M(d_2)\mathcal{I}_M(d_3)}.
\end{equation}

Note that this formula only gives the value of $M \Pi_{ij} \mod 1$. It is not fully clear whether this is the finest possible characterization of the torsional response, or if a more precise formula exists. 

\section{Generalized LSM constraints from filling}
\label{Sec:LSM}
\subsection{Continuum FQH systems}
\label{Sec:LSM_cont}
It is possible to extract all the SET invariants for $G=U(1)\leftthreetimes\mathbb{E}^2$ using only the $G$-crossed defect data which are restricted to the $U(1)\times SO(2)$ subgroup of $G$. This simplified picture, in which the translations are ignored, misses one important feature, namely the filling per magnetic unit cell $\nu$, which is well-defined only in the presence of translation symmetry. It is well known that the Hall conductivity and filling are in fact equal in continuum quantum Hall systems. However, past proofs of this relationship have usually relied on Galilean invariance. In the following, we show that the relationship $\nu = \bar{\sigma}_H \mod 1$ can be proved using $G$-crossed invariants associated to the magnetic translation symmetry alone. We will first give a heuristic argument to motivate the relation, and then give a more rigorous argument in terms of $G$-crossed invariants.

The heuristic argument is as follows (these arguments have been made previously in Ref \cite{Lu2017fillingenforced}, and a simplified proof of the same relation which also uses rotational symmetry has been discussed in Ref. \cite{Matsugatani2018}). Suppose we divide the continuum into magnetic unit cells defined by the vectors ${\bf x}$ and ${\bf y}$. In units of the flux quantum, the group multiplication law for magnetic translations is
\begin{equation}
(1,T_{{\bf x}}) (1,T_{{\bf x}})(1,T_{{\bf x}}^{-1})(1,T_{{\bf x}}^{-1}) = (e^{2\pi i},{\bf 0})
\end{equation}
This implies that performing a sequence of translations around a magnetic unit cell is equivalent to performing a $U(1)$ symmetry operation which inserts a $2\pi$ flux. Now the consequence of $U(1)$ symmetry fractionalization is that a $2\pi$-flux is always associated to a vison $v$. Therefore, taking an anyon around a magnetic unit cell results in braiding with the vison associated to the enclosed $2\pi$-flux. Next, we note that the fractional $U(1)$ charge per unit cell (i.e. the fractional filling) is obtained by adiabatically transporting a $2\pi$-flux around a unit cell, giving a braiding phase $e^{2\pi i \nu}$. Therefore, the same phase $e^{2\pi i \nu}$ can be interpreted as the braiding phase of the vison associated to the original $2\pi$-flux with the vison associated to the $2\pi$ flux within the unit cell. This leads to the following relation: 
\begin{equation} \label{ContFillingHcondRelation}
e^{2\pi i \nu} = M_{v,v}
\end{equation}
Finally we can use standard arguments to see how the rhs relates to the Hall conductivity. First, given an anyon $a$, its charge $Q_a$ is simply the phase obtained by braiding a $2\pi$-flux around $a$: $e^{2\pi i Q_a} = M_{v,a}$. Therefore $e^{2\pi i \nu} = e^{2\pi i Q_v}$, i.e. the fractional part of the filling equals the charge of the vison. But the Hall conductivity is defined as the charge introduced into a region by the addition of $2\pi$-flux into that region. Therefore we can conclude that
\begin{equation}
\nu \equiv Q_v \equiv \bar{\sigma}_H \mod 1.
\end{equation}

Now we rigorously derive Eq. \eqref{ContFillingHcondRelation} using $G$-crossed invariants. We will set all the rotation group elements to zero and write a magnetic translation as $(e^{2\pi i z},{\bf r})$. First we recall that the Hall conductivity is related to $M_{v,v}$ due to the properties of the corresponding $G$-crossed invariant, as summarized by the relation $M_{v,v} = e^{2\pi i \bar{\sigma}_H} = \mathcal{I}^{2n}_n(0_{{\bf g}})$, with ${\bf g} = (e^{2\pi i/n}, {\bf 0})$. Here $0_{{\bf g}}$ is an elementary $U(1)$ flux satisfying $0_{{\bf g}}^n = v$.

Now we develop another invariant which is also equal to $M_{v,v}$, and which we identify with the filling. Define ${\bf g} = (e^{2\pi i/n},0)$. Assume that ${\bf x},{\bf y}$ are magnetic translations. We can use the symmetry fractionalization cocycle and the fusion rules to see that
\begin{align}
0_{{\bf x}}\times 0_{{\bf y}} &= 0_{{\bf y}}\times 0_{{\bf x}}\times \mathfrak{w}({\bf x},{\bf y}) \times \overline{\mathfrak{w}({\bf y},{\bf x})} \\
&= 0_{{\bf y}}\times 0_{{\bf x}}\times v 
\end{align}
This means that translations which circumscribe a magnetic unit cell also enclose a $U(1)$ flux quantum, which is associated to the anyon $\mathfrak{b}({\bf x},{\bf y}) = \mathfrak{w}({\bf x},{\bf y}) \times \overline{\mathfrak{w}({\bf y},{\bf x})} = v$. Now consider the quantity

\begin{equation}
\frac{\eta_{0_{{\bf g}}}({\bf x},{\bf y})}{\eta_{0_{{\bf g}}}({\bf y},{\bf x})}.
\end{equation}	
Since ${\bf x}$ and ${\bf y}$ commute, the numerator and the denominator transform in the same way under symmetry action gauge transformations (see Appendix \ref{Sec:GCReview}), while $\eta$ symbols are automatically invariant under vertex basis gauge transformations. Therefore this expression is gauge-invariant. It can be thought of as the total projective phase associated to transporting a ${\bf g}$-defect around a magnetic unit cell. The $n$th power of this invariant therefore corresponds to the total projective phase associated to transporting a $U(1)$ flux quantum around a magnetic unit cell, i.e. it defines the filling per magnetic unit cell. Indeed, our main result is that

\begin{equation}
e^{2\pi i \nu} := \left(\frac{\eta_{0_{{\bf g}}}({\bf x},{\bf y})}{\eta_{0_{{\bf g}}}({\bf y},{\bf x})}\right)^n = M_{v,\mathfrak{b}({\bf x},{\bf y})} = M_{v,v},
\end{equation}
which proves that the fractional part of the filling equals $\bar{\sigma}_H$. 

This is a special case of the general result derived in Appendix \ref{Sec:DefectInvts}. Using the known identity $M_{v,v} = e^{2\pi i \bar{\sigma}_H}$, we conclude that $\nu=\bar{\sigma}_H \mod 1$. In the cases where we can write down explicit solutions to the $G$-crossed equations, we can actually verify the stronger result that $\nu/n = \bar{\sigma}_H/n \mod 1$ for an integer $n$ which can be made arbitrarily large. This implies that $\nu=\bar{\sigma}_H$. We use the particular solution given in Appendix \ref{Sec:Gxsolns-continuum} to evaluate the $G$-crossed invariants associated to $\nu/n$ and $\bar{\sigma}_H/n$, and verify that they equal each other. Note that this particular solution was written assuming the anyon $F$ symbols can all be set to 1. The steps are given below:

\begin{align}
e^{i \pi \bar{\sigma}_H/n} &:= \theta_{0_{{\bf  g}}}^n \prod_{j=1}^{n-1} \eta_{0_{{\bf  g}}}({\bf  g},{\bf  g}^j) \quad, {\bf  g} = (e^{2\pi i/n},{\bf  0},1) \\
&= 1^n \prod_{j=1}^{n-1} \frac{1}{F^{0_{{\bf  g}}0_{{\bf  g}^j}0_{{\bf  g}}}} \\
&= \prod_{j=1}^{n-1} e^{i 2\pi \frac{v \star v + 2 k_1}{2n} (1/n + j/n - [j+1]/n)} \\
&= e^{i 2\pi \frac{v \star v + 2 k_1}{2n}}
\end{align}

and, with $|{\bf x}\times{\bf y}|=l_B^2$,

\begin{align}
e^{2\pi i \nu/n} &:= \frac{\eta_{0_{{\bf g}}}({\bf x},{\bf y})}{\eta_{0_{{\bf g}}}({\bf y},{\bf x})} \quad, {\bf  g} = (e^{2\pi i/n},{\bf  0},1)\\
&= \frac{F^{0_{{\bf  x}}0_{{\bf  g}}0_{{\bf  y}}} F^{0_{{\bf  g}}0_{{\bf y}}0_{{\bf  x}}} F^{0_{{\bf  y}}0_{{\bf  x}}0_{{\bf  g}}}}{F^{0_{{\bf  x}}0_{{\bf y}}0_{{\bf  g}}} F^{0_{{\bf  y}}0_{{\bf  g}}0_{{\bf  x}}} F^{0_{{\bf  g}}0_{{\bf  x}}0_{{\bf  y}}}} \\
&= \frac{1\times e^{\pi i \frac{v \star v + 2 k_1}{2n}}\times e^{\pi i \frac{v \star v + 2 k_1}{2n}}}{ e^{-\pi i \frac{v \star v + 2 k_1}{2n}}\times 1\times e^{-\pi i \frac{v \star v + 2 k_1}{2n}}} \\
&= e^{2\pi i \frac{v \star v + 2 k_1}{n}}.
\end{align}
This strongly suggests that there is a proof of $\nu= \bar{\sigma}_H$ which is completely algebraic and independent of any particular solution; however, we have not been able to find such a proof. We leave this issue for future study.

\subsection{Lattice FQH systems}
\label{Sec:LSM_lattice}

For lattice systems (with or without point group rotation symmetry), there is an additional possibility of an anyon $m$ in each unit cell due to symmetry fractionalization. Consider a flux $\phi = p/q$ per unit cell. Assume that a \textit{magnetic} unit cell is spanned by the vectors ${\bf r_1},{\bf r_2}$. Then, we have the relation

\begin{align}
0_{{\bf r_1}}\times 0_{{\bf r_2}} &= 0_{{\bf r_2}}\times 0_{{\bf r_1}}\times \mathfrak{w}({\bf r_1},{\bf r_2}) \times\overline{ \mathfrak{w}({\bf r_2},{\bf r_1})} \\
&= 0_{{\bf r_1}}\times 0_{{\bf r_2}}\times \mathfrak{b}({\bf r_1},{\bf r_2}). 
\end{align}

Noting that the most general symmetry fractionalization cocycle for magnetic translation symmetry (see Appendix \ref{Sec:Gxsolns-magtrans}) can be written as
\begin{equation}
\mathfrak{w}({\bf r_1},{\bf r_2}) = v^{\frac{\phi}{2}{\bf r_1}\times {\bf r_2}} \times m^{\frac{1}{2}{\bf r_1}\times {\bf r_2}},
\end{equation}
we can use ${\bf r_1}\times {\bf r_2} = q$ and obtain that

\begin{align}
\mathfrak{b}({\bf r_1},{\bf r_2}) &= (v^{p/2}\times m^{q/2}) (v^{-p/2}\times m^{-q/2})^{-1} \\
&= v^p \times m^q.
\end{align}

This result for $\mathfrak{b}$ also holds in the presence of rotation symmetry. (Note that when we work in symmetric gauge, it is necessary to fully simplify the expression for $\mathfrak{b}({\bf r_1},{\bf r_2})$ in order to have a meaningful result. In an integer-valued gauge such as the Landau gauge, the expressions for $\mathfrak{w}({\bf r_1},{\bf r_2})$ will also be meaningful. However, both gauge choices result in the same value of $\mathfrak{b}({\bf r_1},{\bf r_2})$.)  

Now, setting ${\bf g} = (e^{2\pi i/n},{\bf 0})$, we define the filling per magnetic unit cell $\nu$ through the following invariant:

\begin{equation}
e^{2\pi i \nu/n}:=\frac{\eta_{0_{{\bf g}}}({\bf r_1},{\bf r_2})}{\eta_{0_{{\bf g}}}({\bf r_2},{\bf r_1})}.
\end{equation}

With this definition, $\nu/q$ is the usual filling per unit cell. The general result proved in Appendix \ref{Sec:DefectInvts} now shows that

\begin{align}
e^{2\pi i \nu} &= \left(\frac{\eta_{0_{{\bf g}}}({\bf r_1},{\bf r_2})}{\eta_{0_{{\bf g}}}({\bf r_2},{\bf r_1})}\right)^n = M_{v,m^q \times v^p} \\
&= M_{v,m^q} \times e^{2\pi i p \bar{\sigma}_H} 
\end{align}

From this we can conclude that
\begin{align}
\nu = q v \star m + p \bar{\sigma}_H \mod 1 .
\end{align}
This result is consistent with that of Ref. \cite{Lu2017fillingenforced}, and is in fact a Lieb-Schultz-Mattis (LSM) constraint. For systems where a particular solution is available we can verify the stronger claim that
\begin{align}
\nu/q = (v \star m + k_6) + \phi \bar{\sigma}_H.
\end{align}
Here $k_6$ refers to the number of integer $U(1)$ charges placed in each unit cell (i.e. it is the index for the filling SPT associated to magnetic translation symmetry). To verify this explicitly, we use the solution in Appendix \ref{Sec:Gxsolns-magtrans} and write, assuming ${\bf g} = (e^{i 2\pi/n},{\bf 0})$, 
\begin{align}
\frac{\eta_{0_{{\bf g}}} ({\bf  r_2},{\bf  r_1})}{\eta_{0_{{\bf g}}}({\bf  r_1},{\bf  r_2})} &= \frac{F^{0_{\chi_1}0_z0_{\chi_2}}}{F^{0_{\chi_2}0_z0_{\chi_1}}} \frac{F^{0_z0_{\chi_2}0_{\chi_1}}}{F^{0_z0_{\chi_1}0_{\chi_2}}}\frac{F^{0_{\chi_2}0_{\chi_1}0_z}}{F^{0_{\chi_1}0_{\chi_2}0_z}}
\end{align}
From the $F$- symbols, we obtain
\begin{align}
\frac{\eta_{0_{{\bf g}}} ({\bf  r_2},{\bf  r_1})}{\eta_{0_{{\bf g}}}({\bf  r_1},{\bf  r_2})} &= e^{2\pi i \left((v \star v + 2 k_1) z (\phi q - [\phi q]) + (v\star m +k_6) z q\right)} \nonumber\\
&= e^{2\pi i (p \bar{\sigma}_H + q (v \star m + k_6))  z} =: e^{ 2\pi i \nu \frac{1}{n}},
\end{align}
Thus we have $\nu/n = q/n((v \star m + k_6) + \phi \bar{\sigma}_H)$ for an arbitrary integer $n$. This is possible only if we have $\nu/q = (v \star m + k_6) + \phi \bar{\sigma}_H$, as claimed.
\section{Incorporating relabelings in the SET classification}
\label{Sec:relabeling}
\subsection{$G=\Z_M$}
\label{Sec:relabeling_ZM}

Let ${\bf h}$ be the generator of the group $\Z_M$; let ${\bf g_i} = e^{2\pi i h_i/M}$ be generic group elements. In the analysis of SET phases with $\Z_M$ symmetry the following complication arises: there is generally no unique way to specify which defect is referred to as $0_{\bf h}$. For anyons $s'$ such that $s'^M$ is the trivial particle, we can replace $0_{\bf h} \rightarrow s'_{\bf h}$; this implies that $0_{\bf g_i} \rightarrow s'^{h_i}_{\bf g_i}$. Upon such a relabeling, the symmetry fractionalization cocycle defined by $0_{\bf g_1} 0_{\bf g_2} \overline{0_{{\bf g_1g_2}}}= \mathfrak{w}({\bf g_1},{\bf g_2})$ will be changed only by the amount $s'^{h_1+h_2-[h_1+h_2]_M}$, which is trivial. For this reason, the anyon $s = 0_{\bf h}^M$ associated to flux insertion will also remain unchanged under this relabeling. This becomes important because in general, $\mathcal{I}_M(0_{{\bf  h}}) \ne \mathcal{I}_M(s'_{{\bf  h}})$, as we show below.
\subsubsection{Calculation of change in defect invariant due to relabelings}
 Before we describe our strategy to handle relabeling equivalences, we prove the following result: given two defects $0_{\bf h}$ and $a_{\bf h}$ such that $0_{\bf h}^M = a_{\bf h}^M = s$, we have
\begin{equation}
\frac{\mathcal{I}_M(a_{{\bf  h}})}{\mathcal{I}_M(0_{{\bf  h}})} = M_{s,a}\times (R^{a,a})^M.
\end{equation}
\textit{Proof:} Let ${\bf h}$ be the generator of the group $\Z_M$. First note that for Abelian anyons, we can combine the heptagon equations in the following way (see Appendix \ref{Sec:GCReview}, Eq. \eqref{AbelianHeptEqn}):
\begin{equation}
\frac{U_{\bf k}(a_{\bf g_1},b_{\bf g_2})}{\eta_{c_{\bf k}}({\bf g_1},{\bf g_2})} = \frac{R^{ac}R^{ca}R^{bc}R^{cb}}{R^{ab,c}R^{c,ab}}.
\end{equation}
Here the group element subscripts associated to $a_{\bf g_1},b_{\bf g_2},c_{\bf k}$ are not written explicitly, for clarity of presentation. Using this we can write
\begin{align}
\frac{\eta_{a_{\bf h}}({\bf h},{\bf h}^j)}{U_{\bf h}(0_{\bf h},0_{\bf h}^j)} &= \frac{R^{0_{\bf h}^{j+1},a_{\bf h}}R^{a_{\bf h},0_{\bf h}^{j+1}}}{R^{0_{\bf h},a_{\bf h}}R^{a_{\bf h},0_{\bf h}}R^{0_{\bf h}^{j},a_{\bf h}}R^{a_{\bf h},0_{\bf h}^{j}}} \\
\text{and} \quad\frac{\eta_{0_{\bf h}}({\bf h},{\bf h}^j)}{U_{\bf h}(0_{\bf h},0_{\bf h}^j)} &= \frac{R^{0_{\bf h}^{j+1},0_{\bf h}}R^{0_{\bf h},0_{\bf h}^{j+1}}}{R^{0_{\bf h},0_{\bf h}}R^{0_{\bf h},0_{\bf h}}R^{0_{\bf h}^{j},0_{\bf h}}R^{0_{\bf h},0_{\bf h}^{j}}} \\
\implies \quad \prod\limits_{j=0}^{M-1}\frac{\eta_{a_{\bf h}}({\bf h},{\bf h}^j)}{\eta_{0_{\bf h}}({\bf h},{\bf h}^j)} &= \frac{R^{s,a_{\bf h}}R^{a_{\bf h},s}}{R^{s,0_{\bf h}}R^{0_{\bf h},s}} \left(\frac{R^{0_{\bf h},0_{\bf h}}R^{0_{\bf h},0_{\bf h}}}{R^{0_{\bf h},a_{\bf h}}R^{a_{\bf h},0_{\bf h}}}\right)^M.
\end{align}
We wish to replace the mixed braiding terms $R^{a_{\bf h},0_{\bf h}}R^{0_{\bf h},a_{\bf h}}$ by terms that depend only on $a_{\bf h}$ and $0_{\bf h}$ separately. To that end, we use Eq.\eqref{AbelianHeptEqn} again to write
\begin{align}
U_{\bf h}(0_{\bf h},a_{\bf 0}) &= \frac{R^{0_{\bf h},0_{\bf h}} R^{0_{\bf h},0_{\bf h}}R^{a_{\bf 0},0_{\bf h}} R^{0_{\bf h},a_{\bf 0}}}{R^{a_{\bf h},0_{\bf h}} R^{0_{\bf h},a_{\bf h}}}
\end{align}
and the $G$-crossed ribbon property to write 
\begin{equation}
R^{a_{\bf 0},0_{\bf h}} R^{0_{\bf h},a_{\bf 0}} = \frac{R^{a_{\bf h},a_{\bf h}}}{R^{a_{\bf 0},a_{\bf 0}}R^{0_{\bf h},0_{\bf h}}} U_{\bf h}(0_{\bf h},a_{\bf 0})
\end{equation}
These two equations are combined to give
\begin{align}\label{Eq:G8}
R^{a_{\bf h},0_{\bf h}} R^{0_{\bf h},a_{\bf h}} &= \frac{R^{0_{\bf h},0_{\bf h}}R^{a_{\bf h},a_{\bf h}}}{R^{a_{\bf 0},a_{\bf 0}}}
\end{align} 
which implies that
\begin{align}
&\prod\limits_{j=0}^{M-1}\frac{\eta_{a_{\bf h}}({\bf h},{\bf h}^j)}{\eta_{0_{\bf h}}({\bf h},{\bf h}^j)} = \frac{R^{s,a_{\bf h}}R^{a_{\bf h},s}}{R^{s,0_{\bf h}}R^{0_{\bf h},s}} \left(\frac{R^{0_{\bf h},0_{\bf h}}R^{a_{\bf 0},a_{\bf 0}}}{R^{a_{\bf h},a_{\bf h}}}\right)^M \\
\implies \qquad &\frac{\left(R^{a_{\bf h},a_{\bf h}}\right)^M \prod\limits_{j=0}^{n-1}\eta_{a_{\bf h}}({\bf h},{\bf h}^j)}{\left(R^{0_{\bf h},0_{\bf h}}\right)^M \prod\limits_{j=0}^{M-1}\eta_{0_{\bf h}}({\bf h},{\bf h}^j)} = \frac{R^{s,a_{\bf h}}R^{a_{\bf h},s}}{R^{s,0_{\bf h}}R^{0_{\bf h},s}} \left(R^{a_{\bf 0},a_{\bf 0}}\right)^M \\
&= M_{a,s} \left(R^{a,a}\right)^M,
\end{align} 
as desired.

\subsubsection{Classification strategy}
Our main assertion is that two $G$-crossed solutions in different defect classes are physically equivalent if and only if the gauge-invariant quantities describing those classes are interchanged by some relabeling that preserves the function $\mathfrak{w}({\bf g_1},{\bf g_2})$. In particular, we do not need to know the explicit set of gauge transformations linking the $G$-crossed data in these two solutions. Moreover, since the defect invariant is written purely in terms of Abelian defects, we only consider relabelings of Abelian defects: allowed relabelings involving the non-Abelian defect sector will only introduce equivalences between defect classes that can anyway be described purely within the Abelian sector.

This assertion is justified by the following argument. In the specific gauge choice of Appendix \ref{Sec:GeneralBTCSoln}, the $G$-crossed BTC data can be determined entirely from the UMTC data and the symmetry fractionalization cocycle $\mathfrak{w}({\bf g_1},{\bf g_2})$. Therefore in this gauge, any relabeling that relates different defect classes but also fixes the other $G$-crossed data must leave $\mathfrak{w}({\bf g_1},{\bf g_2})$ invariant. Conversely, if two defect classes in $\H^3(\Z_M,U(1))$ can be related by such a relabeling, those defect classes must be physically equivalent since all the data are invariant except for the change in the defect parameter.

First, we show that a relabeling preserves the function $\mathfrak{w}({\bf g_1},{\bf g_2})$ if and only if it is of the form $0_{\bf h}\rightarrow s'_{\bf h}$, with $s'^M=0$. If a general relabeling is given by $\mathfrak{w}({\bf g_1},{\bf g_2}) \rightarrow \mathfrak{w}({\bf g_1},{\bf g_2}) d\chi({\bf g_1},{\bf g_2})$, then we must have $d\chi=0$, i.e. $\chi$ is an element of $\H^1(\Z_M,\A) \cong \{s' \in \A | s'^M = 0\}$. Allowed functions $\chi({\bf g_i})$ are of the form $\chi({\bf g_i}) = s'^{h_i}$, where ${\bf g_i}=e^{2\pi i h_i/M}$. The cohomology group $\H^1(\Z_M,\A)$ can be computed using the general result Eq. \eqref{CohomologyCyclicGrp_trivial} stated in Appendix \ref{Sec:GrpCohIntro}.

Motivated by these results we state again our strategy: we consider a solution corresponding to some fixed symmetry fractionalization class $[\mathfrak{w}] \in \H^2(\Z_M,\A)$ and a defect class with parameter $k \in \Z_M$. We will only consider the subset of relabelings that takes $0_{\bf h}\rightarrow s'_{\bf h}$ where $s'^M = 0$. As argued above, this set accounts for all possible redundancies in $k$. Then we evaluate $\mathcal{I}_M(s'_{\bf h})$ for all defects $s'_{\bf h} = 0_{\bf h}\times s'$ such that $s'^M = 0$. From the computation carried out previously, the set of quantities thus obtained is $\{\mathcal{I}_M(0_{\bf h})\times M_{s',s} \theta_{s'}^M | s'^M = 0\}$. It is this complete set which characterizes the SET. 

It is however cumbersome to evaluate the defect invariant for a potentially large number of defects $s'_{{\bf h}}$. To make the analysis simpler, we would like to find a single gauge-invariant quantity which is independent of the specific choice of $s'_{\bf h}$, and is therefore an absolute invariant. This can be obtained as follows. Consider the smallest power $P$ such that the phases $(\mathcal{I}_M(s'_{\bf h}))^P$ are all equal, thereby collapsing the set of invariants onto a single phase modulo $2\pi$. By construction, this quantity is gauge-invariant and also invariant under relabelings.

Moreover, we now claim that this invariant also gives different values for physically inequivalent defect classes, i.e. the process of taking $P$th powers does not lose any information. We show below that the set of phase differences $\{M_{s,s'} (R^{s',s'})^M | s'^M = 0\}$ forms a cyclic group of order $P$, where the integer $P$ depends sensitively on the structure of $\A$ and also on the anyon $s$. Due to this cyclic group structure, two defect classes can have the same value of $(\mathcal{I}_M(0_{\bf h}))^P$ if and only if there is a relabeling which relates them, thus completing our argument, which applies to both Abelian and non-Abelian topological orders. 

The invariant $(\mathcal{I}_M(0_{\bf h}))^P$ determines a set of defect classes which form a torsor over $\Z_{M/(M,P)}$. To see this, note that for some fixed $0_{{\bf h}}$, the distinct defect invariants will take values of the form $\mathcal{I}^P_M(0_{\bf h})\times e^{2\pi i k P /M}$ for $0 \le k \le M-1$. The factors $e^{2\pi i k P /M}$ generate a group $P \Z_M \cong \Z_{M/(M,P)}$. Note that because the physically equivalent defect classes are associated to a cyclic subgroup of $\Z_M$, the invariants corresponding to two physically distinct defect classes will never collapse onto each other when we take their $P$th powers.

Finally, we prove the assertion regarding the cyclic group structure of the defect invariants related by relabelings. Suppose $\A$ is the group of Abelian anyons, with $R^{a b}_{[a+b]} = e^{i \pi \vec{a}^T K^{-1} \vec{b}}$. Here we assume a nondegenerate, symmetric $r\times r$ $K$ matrix with even integers along the diagonal. (As discussed in Appendix \ref{Sec:Gxsolns}, the Abelian anyons in a general non-Abelian bosonic topological order, where $\A$ may not be modular, can still be described by the $K$ matrix associated to the UMTC $Z(\A)$, which contains $\A$.) We will show that the set of phase differences $\{M_{\vec{s},\vec{s}'} (\theta_{\vec{s}'})^M, \vec{s}'^M = 0\}$ forms a cyclic group of order $P$, and determine $P$ explicitly. The calculation below is fully consistent with analogous calculations for Abelian topological orders using crystalline gauge fields, as outlined in Ref \cite{Manjunath2020}.

Define the vectors 
\begin{equation}
K_i = (K_{i1},\dots , K_{ir})^T = K \cdot (0,0,\dots, 1, \dots , 0)^T
\end{equation}
(the 1 is in the $i$th position). $K_i$ can be written as $K \cdot e_i$ where $[e_i]_j = \delta_{ij}$. The anyon represented by $\vec{\Lambda}$, where $\Lambda_i = \sum_j K_{ij} n_j$, is considered trivial for every $n_j \in \Z$. We define a fundamental domain consisting of all integer vectors lying within the convex region formed by the vectors $K_i$. The $R$ symbols are only defined for anyons whose representative vectors lie in this region. 

The first step in the computation is to find the anyons $s'$ for which $s'^M$ is trivial. Let the generators of $\A = \Z_{n_1}\times\dots\times\Z_{n_r} $ be given by the vectors $\vec{a}_i$, which we assume lie in the fundamental domain. The $\vec{a}_i$ each satisfy the condition $n_i \vec{a}_i = K \vec{w}_i$ for some integer vectors $\vec{w}_i$. The order $M$ elements of the $\Z_{n_i}$ factor are given by the vectors $\frac{k_i n_i}{d_i} \vec{a}_i$, where  $d_i = (M,n_i)$ and $0 \le k_i < d_i$. A general form of the relabeling anyon $\vec{s}'$ is then $\vec{s}' = \sum_{i}\frac{n_i k_i}{d_i} \vec{a}_i - K \vec{\Lambda}_i= K \left(\sum_i \frac{k_i \vec{w}_i}{d_i} - \vec{\Lambda}_i\right)$. Here the integer vector $\vec{\Lambda}$ is chosen so that $\vec{s}'$ lies within the fundamental domain. 

Upon relabeling with this $\vec{s}'$, the value of the invariant $\mathcal{I}_M(0_{\bf h})$ changes by the amount
\begin{widetext}
\begin{align}
M_{\vec{s},\vec{s}'}\theta_{\vec{s}'}^M &= \exp\left(2\pi i \vec{s}^T \left(\sum\limits_i\frac{k_i}{d_i}  \vec{w}_i - \vec{\Lambda}_i\right) +2\pi i \frac{M}{2} \left(\sum\limits_i\frac{k_i}{d_i}  \vec{w}_i - \vec{\Lambda}_i\right)^T K \left(\sum\limits_j\frac{k_j}{d_j}  \vec{w}_j - \vec{\Lambda}_j\right)\right) \\
&= \exp\left(2\pi i \sum\limits_i\frac{k_i}{d_i} \vec{s}^T \vec{w}_i +2\pi i\sum\limits_{i,j} \frac{M k_ik_j}{2d_id_j}\vec{w}_i^T K \vec{w}_j\right) \\
&= \exp\left(2\pi i \sum\limits_i\frac{k_i}{d_i} \vec{s}^T \vec{w}_i +2\pi i\sum\limits_{i} \frac{M k_i^2}{2d_i^2}\vec{w}_i^T K \vec{w}_i + 2\pi i\sum\limits_{i<j} \frac{M k_ik_j}{d_i}\vec{w}_i^T \frac{n_j \vec{a_j}}{d_j}\right) 
\end{align}
\end{widetext}
In the first line, all terms containing $\vec{\Lambda}_i$ drop out because they contribute an overall integer to the exponent. Going from the second line to the third, we used the symmetry property $K_{ij} = K_{ji}$ as well as the fact that $K_{ii}$ is even. The third term on the last line now vanishes because $d_i$ divides $M$ and $d_j$ divides $n_j$, so the overall contribution from this term is a sum of integers. As a result, only the first two terms can give nontrivial contributions.

From the definition of $\vec{w}_i$, we see that $d_i$ is a factor of both $M$ and $K \cdot \vec{w}_i$; therefore each summand of the second term is either an integer or a half integer. Therefore the above expression becomes
\begin{align}
\exp\left(2\pi i \sum\limits_i k_i\left(\frac{\vec{s}}{d_i} + \frac{\vec{c}_i}{2}\right)\right)^T \vec{w}_i \qquad , \vec{c}_i = \frac{M}{d_i^2} K \cdot \vec{w}_i 
\end{align}
Note that $\vec{c}_i$ is an integer vector, so the term with $\vec{c}_i$ contributes an integer or a half-integer to the phase difference. This allows us to replace the coefficient $k_i^2$ of $\vec{c}_i/2$ with $k_i$, since the two expressions give the same value modulo 1. Importantly, the phase differences form a subgroup of $\Z_M$ upon varying the $k_i$ from $0$ to $d_i-1$: this can be seen by setting $k_i = d_i$ and observing that the phase difference is 1. The order of this subgroup is $\frac{2d_i}{(2d_i,(2\vec{s}_i + \vec{c}_i d_i)\cdot \vec{w}_i)}$. This implies that the subgroup formed by all possible linear combinations of $n_i$ is of order 
\begin{widetext}
	\begin{equation}\label{Eq:ZMrelabeling_KMat}
	P = \text{lcm}\left(\frac{2d_1}{(2d_1,(2\vec{s}_1 + \vec{c}_1 d_1)\cdot \vec{w}_1)},\dots,\frac{2d_r}{(2d_r,(2\vec{s}_r + \vec{c}_r d_r)\cdot \vec{w}_r)}\right).    
	\end{equation}
\end{widetext}

Finally, as we argued previously, the physically distinct defect classes form a torsor over $\Z_{M/(M,P)}$. This classification does not depend on the particular basis in which the $K$ matrix and the vectors $\vec{w}_i$ are defined. 

\subsubsection{Example: $\Z_N$ Laughlin anyons with $\Z_2$ symmetry}

We now discuss some specific examples illustrating the ideas developed above. Consider a $1/N$ Laughlin system with $\Z_N$ anyons denoted by $\psi^k, k = 0,1,\dots , N-1$, where $N$ is even. Let $G=\Z_2$, with ${\bf h}$ being the nontrivial $\Z_2$ element. Since $N$ is even, there are precisely two fractionalization classes, correpsonding to the even/odd parity of the spin vector $s = \psi^q$. The defect classes are parametrized by $k \in \Z_2$. Suppose the system realizes the nontrivial symmetry fractionalization class, so that $q$ is odd. Consider a defect $0_{\bf h}$ such that $0_{\bf h}^2 = \psi^q$. There is preciswely one other defect with this property, namely $s'_{\bf h} = 0_{\bf h} \times \psi^{N/2}$. If we demand that $0_{\bf h}$ is the defect that has trivial braiding with all anyons, we can directly use the defect $F$-symbols from the particular solution in Appendix \ref{Sec:Gxsolns-ZM} to calculate
\begin{equation}
\mathcal{I}_M(0_{\bf h}) = e^{2\pi i\left(\frac{q^2}{4N}+\frac{k}{2}\right)}
\end{equation}
where $k \in \{0,1\}$ denotes the defect class. Similarly, we can use the solution with $s'_{\bf h}$ to obtain
\begin{align}
\mathcal{I}_M(s'_{\bf h}) &= \mathcal{I}_M(0_{\bf h})\times M_{\psi^q,\psi^{N/2}}\times (R^{\psi^q,\psi^q})^2 \\
&= e^{2\pi i \left(\frac{q^2}{4N}+\frac{N/2 + q+k}{2}\right)}
\end{align}	
Therefore the set of phase differences between the two invariants is $\{1,e^{2\pi i\frac{N+2q}{4}}\}$. Note that these two elements always form a group, because the second element is always $\pm 1$. Now fix $q$ odd and suppose $N$ is a multiple of 4. Then $N/2 + 1$ is odd, so the second term in the expression for $\mathcal{I}_M(s'_{\bf h})$ is equal to $k+1$ modulo 2. Therefore a solution with a defect class $k$ is equivalent, after the above relabeling, to a solution with defect class $k+1$. This means that the two defect classes are physically the same. The conclusion is different when $N$ is even, but not a multiple of 4. In that case, $N/2 + 1$ is even, so the second term in the exponent of $\mathcal{I}_M(s'_{\bf h})$ remains $k$ modulo 2, implying that the defect classes $k$ and $k+1$ are distinct. These results are in agreement with edge CS theory calculations in Ref. \cite{Lu2016} and also with calulations using crystalline gauge fields outlined in Ref. \cite{Manjunath2020}. 

An absolute SET invariant for the general case is given by $\mathcal{I}_M(0_{\bf g})^{P}$ where $P$ is the smallest integer such that $P\times (N+2q)/4$ is an integer. This implies that $P = 4/(4,N+2q)$. The classification is given by $\Z_{M/(M,P)} = \Z_{(4,N+2q)/2}$. For $q$ odd, this gives $\Z_1$ when $N = 4k$, and $\Z_2$ when $N = 4k+2$. 

This can also be seen from the general formula we derived above, Eq. \eqref{Eq:ZMrelabeling_KMat}, if we plug in $K = N, w=1, s = q, M=2$ and $d=2$. In that case we also have $c = MN/d^2 = N/2$. We thus obtain
\begin{equation}
P = \frac{2d}{(2d,2q+N)} = \frac{4}{(4,2q+N)}.
\end{equation} 

Therefore we have arrived at the same result in two ways: first by a direct calculation using the defect $F$-symbols, and then by a special case of the general formula derived above. The two derivations agree with each other. 

\subsubsection{Example: $\Z_2$ gauge theory with $\Z_2$ symmetry}

Next we discuss another example where the general $K$ matrix classification result can be immediately applied. Consider a $\Z_2$ gauge theory with toric code topological order (i.e. $\A = \Z_2\times\Z_2$) with $G=\Z_2$. We assume $K = \begin{pmatrix}
0 & 2 \\
2 & 0
\end{pmatrix}$,
where the anyons are denoted as $I = (0,0),e = (1,0), m = (0,1),\psi = (1,1)$. The symmetry fractionalization class is an element of the group $\Z_2\times\Z_2$ and is given by the spin vector $\vec{s} = (s_1,s_2)$. There are two defect classes for each symmetry fractionalization class. Here we have $d_1 = d_2 = 2$; we also have $\vec{w}_1 = e, \vec{w}_2 = m$, and $\vec{c}_i = \frac{M}{d_i^2} K \vec{w}_i = (0,0)$ for $i=1,2$. From the above formula we then obtain 
\begin{equation}
P = \text{lcm}\left(\frac{4}{(4,2s_1)},\frac{4}{(4,2s_2)}\right) 
\end{equation}
If either $s_1$ or $s_2$ is zero, we get $P = 1$, therefore the defect classes are distinct. This happens when $v = I,e,m$. When $s_1=s_2=1$, i.e. when $s = \psi$, we have $P=2$, so the two defect classes are equivalent in this case. It is instructive to compare this treatment with the one given in Ref. \cite{Barkeshli2019} (Sec. X, Example I). There, the $G$-crossed solutions are written in a different choice of gauge, and an explicit gauge transformation is written down to relate the two equivalent defect classes when $s = \psi$.   

As mentioned previously, all these results can be generalized to non-Abelian topological orders, using the fact that the Abelian anyons in $\A$ can still be described by a bosonic $K$-matrix associated to $Z(\A)$. The results for some well-known topological orders describing FQH states are summarized in Table \ref{Table:ZMrelabelings} in the main text.

\subsection{$G=U(1)\leftthreetimes_{\phi} [\Z^2\rtimes \Z_M]$}
\label{Sec:relabeling_full}
In this section we wish to perform the same analysis as above, after accounting for the full symmetry of the magnetic space group. We saw that in order to preserve the form of the $G$-crossed data without carrying out additional gauge transformations, we must choose relabelings $\chi: G \rightarrow \A$ that leave the symmetry fractionalization cocycle $\mathfrak{w}({\bf g_1},{\bf g_2})$ invariant, i.e. $d\chi = 0$. Such relabelings must therefore be described by elements of $\H^1(G,\A)$. 

The group $\H^1(U(1)\leftthreetimes_{\phi} [\Z^2\rtimes \Z_M],\A)$ has three contributions. The first comes from the group $\H^1(U(1),\A)$, and describes relabelings of $U(1)$ symmetry defects. Let ${\bf g}=(e^{2\pi i z},{\bf 0},1)$. However, if we demand that the functions $\chi({\bf g})$ classified by this group are continuous in some neighbourhood around ${\bf g} = {\bf 0}$, then we can show that $\chi$ must be the zero homomorphism. This assumption amounts to canonically defining the defect $0_{\bf g}$, as we assumed in our analysis of $U(1)$ symmetry. The canonical choice is made by defining $0_{\bf g}$ as the unique defect such that $R^{a,0_{\bf g}}$ and $R^{0_{\bf g},a}$ are continuously connected to the identity as a function of ${\bf g}$. We conclude that there are no nontrivial relabelings associated to the $U(1)$ component of the defects. 

Another contribution to $\H^1(G,\A)$ comes from the group $\H^1(\Z_M,\A)$, corresponding to the relabeling function 
\begin{equation}
\chi({\bf g}) = s'^h
\end{equation}
where ${\bf g} = (e^{2\pi i z},{\bf r},e^{2\pi i h/M})$ and $s'^M = 0$. Such functions formed the basis for our analysis of relabelings for $G=\Z_M$, and must also be considered here.

There is now a third contribution, coming from the group $\H^0(\Z_M,\H^1(\Z^2,\A))$, i.e. elements of $\H^1(\Z^2,\A)$ that are invariant under rotations. A general element of $\H^1(\Z^2,\A)$ is given by the function $\chi({\bf r}) = \vec{t}' \cdot {\bf r}$. Such a function is rotationally invariant if $\chi({\bf r}) = \chi({^h}{\bf r})$ for each ${\bf r}$, i.e. if $\vec{t}' \cdot {^{1-h}}{\bf r}$ is the trivial anyon for each ${\bf r}$. These three contributions completely determine $\H^1(G,\A)$. Therefore the most general relabeling function that leaves the symmetry fractionalization cocycle invariant is
\begin{equation}
\chi({\bf g}) = s'^h \times \vec{t}' \cdot {\bf r}
\end{equation} 
where $s'^M = 0=\vec{t}' \cdot {^{1-h}}{\bf r}$ for each ${\bf r}$. The same result was obtained using crystalline gauge theory arguments in Ref. \cite{Manjunath2020}.

Our strategy is now the following. First we write down the full set of invariants characterizing a defect class. Then, we calculate how these coefficients get transformed under all possible relabelings $\chi({\bf g})$. We will find that the physically distinct defect classes are related by a \textit{subgroup} of $\H^3(G,U(1))$, for each symmetry fractionalization class.  

We use the formulas for the invariants given in Table \ref{Table:LatticeInvts} and derived in Appendix \ref{Sec:Invts}. Under the relabeling $\chi({\bf g}) = s'^h \times \vec{t}' \cdot {\bf r}$, the different defect invariants transform as follows:

\begin{enumerate}
	\item The invariant for $\mathscr{S}$ is calculated using 
	\begin{equation}
	e^{i 2\pi \frac{\mathscr{S}}{M} } =\frac{\mathcal{I}_M(0_{\bf gh})}{\mathcal{I}_M(0_{\bf g})\mathcal{I}_M(0_{\bf h})},
	\end{equation} 
	with $0_{\bf g}^M = v,0_{\bf h}^M = s,0_{\bf gh}^M = v \times s$. Thus after a relabeling, this invariant transforms as
	\begin{align}
	e^{i 2\pi \frac{\mathscr{S}}{M} } & \rightarrow \frac{M_{v\times s,s'}}{M_{s,s'}} e^{i 2\pi \frac{\mathscr{S}}{M} } \nonumber \\ &\implies k_2 \rightarrow k_2 + M v \star s'. 
	\end{align}
	\item The invariant for $\ell_s$ is calculated using $ e^{i \pi \ell_s/M} =\mathcal{I}_M(0_{\bf h})$, with $0_{\bf h}^M = s$. Thus after a relabeling, it transforms as
	\begin{align}
	e^{\frac{i 2\pi\ell_s}{2M}} &\rightarrow M_{s,s'}\theta_{s'}^M e^{\frac{i 2\pi\ell_s}{2M}} \nonumber \\
	\implies k_3 &\rightarrow k_3+ M(s \star s' + \frac{M}{2} s' \star s').
	\end{align}
	\item The invariant for $\vec{\mathscr{P}}_c$ is calculated for $M=2$ using \begin{equation}
	e^{2\pi i \vec{\mathscr{P}}_c \cdot {\bf r}} =\frac{\mathcal{I}_2(0_{\bf g})\mathcal{I}_2(0_{\bf h})}{\mathcal{I}_2(0_{\bf k})\mathcal{I}_2(0_{\bf l})},
	\end{equation} 
	with $0_{\bf g}^2 = v\times s\times \vec{t}\cdot {\bf r},0_{\bf h}^2 = s,0_{\bf k}^2 =v \times s, 0_{\bf l}^2 = s\times \vec{t}\cdot {\bf r}$. After a relabeling, this invariant transforms as
	\begin{align}
	e^{2\pi i \vec{\mathscr{P}}_c \cdot {\bf r}} &\rightarrow \frac{M_{ v\times s\times \vec{t}\cdot {\bf r},s'\times\vec{t}'\cdot {\bf r}}M_{s,s'}}{M_{v\times s,s'}M_{s\times \vec{t}\cdot {\bf r},s'\times\vec{t}'\cdot {\bf r}}} e^{2\pi i \vec{\mathscr{P}}_c \cdot {\bf r}} \nonumber \\ &= M_{v,\vec{t}'\cdot {\bf r}}e^{2\pi i \vec{\mathscr{P}}_c \cdot {\bf r}} \nonumber \\
	\implies \vec{k}_4\cdot {\bf r} &\rightarrow \vec{k}_4\cdot {\bf r} +  2 v \star (\vec{t}'\cdot {\bf r}). 
	\end{align}
	The $\theta$ factors from the relabeling transformation cancel out and do not contribute to the change in $\vec{k}_4$. The above relation holds for each ${\bf r}$. We can repeat the calculation for $M=3,4$ and find the following general result:
	\begin{align}
	\vec{k}_4\cdot {\bf r} &\rightarrow \vec{k}_4\cdot {\bf r} +  v \star (\vec{t}'\cdot {^{1-h}}{\bf r}).
	\end{align}
	\item The invariant for $\vec{\mathscr{P}}_s$ is calculated for $M=2$ using \begin{equation}
	e^{2\pi i \vec{\mathscr{P}}_s \cdot {\bf r}} =\frac{\mathcal{I}_2(0_{\bf g})}{\theta_{\vec{t}\cdot{\bf r}}^{1/2}\mathcal{I}_2(0_{\bf h})},
	\end{equation} 
	with $0_{\bf g}^2 = s\times \vec{t}\cdot {\bf r},0_{\bf h}^2 = s$. We will only analyze the transformation behaviour of the $\mathcal{I}_2$ factors in this expression, because the transformation of the additional $\theta_{\vec{t}\cdot{\bf r}}$ factor is accounted for by a change in the quantized torsional response. We find that
	\begin{align}
	e^{2\pi i \vec{\mathscr{P}}_s \cdot {\bf r}} &\rightarrow \frac{M_{ s\times \vec{t}\cdot {\bf r},s'\times\vec{t}'\cdot {\bf r}}}{M_{s,s'}} \frac{\theta^2_{s'\times \vec{t}t\cdot {\bf r}}}{\theta^2_{s'}} e^{2\pi i \vec{\mathscr{P}}_s \cdot {\bf r}} \nonumber \\ &= M_{s,\vec{t}'\cdot {\bf r} }M_{s',\vec{t}\cdot {\bf r}} M^2_{s',\vec{t}'\cdot {\bf r}}e^{2\pi i \vec{\mathscr{P}}_s \cdot {\bf r}} \nonumber \\
	\implies \vec{k}_{5}\cdot {\bf r} &\rightarrow \vec{k}_{5}\cdot {\bf r} +  2 s \star (\vec{t}'\cdot {\bf r}) + 2 s' \star \vec{t}\cdot {\bf r} + 4 s' \star \vec{t}' \cdot {\bf r}. 
	\end{align}
	The last line is obtained by observing that the SPT contribution to $\vec{\mathscr{P}}_s$ is $\vec{k}_5/2$. We can perform a similar computation for $M=3,4$, and obtain the following general transformation rule:
	\begin{align}
	&\vec{k}_{5}\cdot {\bf r} \rightarrow \vec{k}_5\cdot {\bf r} +   s \star (\vec{t}'\cdot {^{1-h}}{\bf r}) \nonumber \\&+ M s' \star \vec{t}\cdot {\bf r} + M s' \star \vec{t}' \cdot {^{1-h}}{\bf r}.
	\end{align}
\end{enumerate}

The above equivalences for $k_2,k_3,\vec{k}_4,\vec{k}_5$ can all be deduced directly from the corresponding $G$-crossed invariants. Although we do not have an invariant for the angular momentum per unit cell, from an analogous relabeling exercise in the response theory for Abelian topological orders we expect that the SPT index $k_7$ transforms as follows:

\begin{equation}
k_7 \rightarrow k_7+ M m \star a  
\end{equation}

However, we will not include $k_7$ in our relabeling calculations henceforth, since we do not have a $G$-crossed invariant with which to verify this. Instead of working with the full torsion subgroup of $\H^3(G,U(1))$ while performing the calculation, we will assume that $k_7$ is fixed and work with the torsion subgroup of $\H^3(G,U(1))/\Z_M$. Therefore this calculation will be incomplete: a full analysis within the $G$-crossed theory which includes the transformation of $k_7$ will be possible only after we identify the invariant for $k_7$.

To summarize, under the above relabeling, several of the coefficients $k_i$ get transformed as $k_i \rightarrow k_i + M a \star b$ for different anyons $a,b$. We can use the constraints on $s',\vec{t}'$ to show that the quantities $M a \star b$ are indeed integers.

When we have several parameters that transform under relabelings, we cannot obtain absolute SET invariants by taking powers of the invariants for the defect class, because we will lose information in doing so. However, we can still obtain a count of SETs: the results for some common topological orders are summarized in Table \ref{Table:relabelingFullExamples} in the main text, assuming that $k_1,k_6$ and $k_7$ are fixed. We note that when we specialize to Abelian topological orders, the above relabeling analysis can be reproduced exactly using crystalline gauge theory \cite{Manjunath2020}, by relabeling the internal gauge fields associated to the topological order.

\section{Calculation of cohomology groups and symmetry cocycles}
\label{Sec:Coho_calc}

A detailed calculation of symmetry cocycles for the groups $G=U(1)$ and $G=\Z_M$ can be found in Ref. \cite{Manjunath2020}.
Here we will discuss how to compute the cohomology groups $\mathcal{H}^2(G, \mathcal{A})$ and $\mathcal{H}^3(G, U(1))$ and to obtain representative cocycles for $G = U(1) \leftthreetimes G_{\text{space}}$, with
$G_{\text{space}} = \R^2$, $\R^2 \rtimes SO(2)$, $\Z^2$, and $\Z^2 \rtimes \Z_M$. 

We will use several technical results from group cohomology theory: these are reviewed in Appendix \ref{Sec:GrpCohIntro}. Throughout this section, $\mathcal{H}^n$ refers to group cohomology with Borel (measurable) cochains. The measurability assumption is required to pass from the group cohomology of $G$ to the cohomology of its classifying space $BG$ when $G$ is continuous, as we will explain.  

\subsection{Generalities}
\label{Sec:Coho_calc_gen}

Before discussing specific examples, we describe the general strategy that we will use. In each of the examples below, we will first compute the groups $\H^2(G,\Z)$ and
$\H^4(G,\Z)$, and obtain a set of representative cocycles for these groups. 

The computation of $\H^2(G,\A)$ follows from $\H^2(G,\Z)$ from an application of the universal coefficient theorem. Moreover, it is straightforward to use the results for $\H^2(G, \Z)$ to obtain 
representative cocycles for $\H^2(G,\A)$, as will be clear from the examples.

To understand the relation between $\H^3(G, U(1))$ and $\H^4(G, \Z)$, we will show that for the $G$ we consider in this paper, we can define an injective map
\begin{align}
  \label{sigmaMap}
\sigma: \H^4(G, \Z) \rightarrow \H^3(G, U(1)).
\end{align}
This map will allow us to obtain all relevant representative cocycles of $\H^3(G, U(1))$ in terms of those of $\H^4(G, \Z)$. 

Furthermore, we will show that the injectivity of $\sigma$ implies that
\begin{align}
\H^4(G,\Z) \cong \H^3(G, U(1)) / \text{continuous part}.
\end{align}
This explains why $\H^4(G,\Z)$ is more directly relevant for classifying SETs than $\H^3(G, U(1))$, at least for the $G$ we consider in this paper.

Finally, we will explain why, with a mild assumption on the applicability of spectral sequences, we can show that $\H^n(G, \R) = 0$ for $n \geq 3$ for the $G$ we consider in this paper. This allows us to establish an equivalence 
\begin{align}
\H^4(G,\Z) \cong \H^3(G, U(1))
\end{align}
for the $G$ considered in this paper. 

\subsubsection{Obtaining cocycles of $\H^3(G,U(1))$ from those of $\H^4(G,\Z)$}

Here we will explain the map $\sigma$ referred to in Eq. \ref{sigmaMap}. 

From the long exact sequence associated to the group extension $1\rightarrow U(1) \rightarrow \R \rightarrow \Z \rightarrow 1$ (Theorem \ref{thm: long-exact sequence}), we have the piece
\begin{equation}\label{Eq:les}
\rightarrow \H^3(G,\R) \xrightarrow[]{\text{exp}} \H^3(G,U(1)) \xrightarrow[]{d} \H^4(G,\Z) \rightarrow \H^4(G,\R) \rightarrow
\end{equation}

The map denoted as ``exp" takes a cochain $f_3 \in C^3(G,\R)$ to a cochain $f^{\star}_3 = f_3 \mod 1$ (or, in multiplicative notation, $f^{\star}_3 = e^{2\pi i f_3}$); the usual coboundary map denoted as ``$d$" takes $f^{\star}_3 = f_3 \mod 1$ to the cochain $w_4 = df_3 \in C^4(G, \Z)$. Note that this requires a choice of lift from $f_3^\star$ to $f_3$; changing the lift $f_3 \rightarrow f_3 + n$ for $n \in C^3(G, \Z)$ takes $w_4 \rightarrow w_4 + dn$, which keeps the cohomology class of $w_4$ invariant, so that $d$ is well-defined for cohomology.

Now the map $\sigma$ is defined as follows. Given a representative 4-cocycle $w_4 \in Z^4(G, \Z)$, we search for a real-valued $3$-cochain $f_3 \in C^3(G, \R)$ that satisfies $d f_3 = w_4$. The fact that $w_4$ is a 4-cocyle then follows because $d w_4 = d^2f_3 = 0$. Then we set
\begin{align}
\sigma(w_4) = f_3^\star = f_3 \mod 1.
\end{align}
The fact that $\sigma$ is well-defined for every element in $\H^4(G,\Z)$, which may not be true for general $G$, will be verified on a case-by-case basis for the $G$ we consider in this paper. 

We can see that $\sigma$ maps non-trivial $4$-cocycles to non-trivial $3$-cocycles. Consider $w_4$ to be a non-trivial $4$-cocyle. We now argue that $f_3^\star := \sigma(w_4)$ will be non-trivial in $\H^3(G, U(1))$. 
Suppose the contrary, i.e. $f^{\star}_3 = df^{\star}_2$ for some $f^{\star}_2 \in C^2(G,U(1))$. We also have $f_3 = f^{\star}_3 + n$ where $n \in C^3(G,\Z)$. Thus $w_4 = df_3 = df^{\star}_3 + dn = d(df^{\star}_2) + dn = dn$. Then $w_4$ itself would be trivial, which is a contradiction. Since $\sigma$ maps non-trivial $4$-cocycles to non-trivial $3$-cocycles, it follows that it is an injective map from $\H^4(G, \Z)$ to $\H^3(G, U(1))$.

Finally, we note that $\sigma$ is a right inverse of the map $d: \H^3(G, U(1)) \rightarrow \H^4(G, \Z)$. It is easily verified that, given any element $w_4 \in Z^4(G, \Z)$,
\begin{align}
d \sigma(w_4) = w_4 + d n , 
\end{align}
therefore $d \circ \sigma$ is the identity map on $\H^4(G, \Z)$. On the other hand,
$\sigma \circ d$ is not necessarily the identity map on $\H^3(G, U(1))$, because $\text{ker } d$ may be non-trivial in principle.

Since we have found that $\sigma$ is an injective map in the examples that we have discussed, and that $d \circ \sigma$ is the identity, it follows that $d: \H^3(G, U(1)) \rightarrow \H^4(G, \Z)$ must be a surjective map. 
  
\subsubsection{$\H^4(G,\Z) = \H^3(G, U(1)) / \text{continuous part}$ when $d$ is surjective}
\label{H4H3cont}

For finite or compact Lie groups $G$, one can show that $\H^4(G, \Z) \cong \H^3(G, U(1))$. In general, however, this is not the case and $\H^4(G, \Z)$, $\H^3(G, U(1))$ are simply part of a long exact sequence associated with the short exact sequence $1 \rightarrow \Z \rightarrow \R \rightarrow U(1) \rightarrow 1$. This raises the important question of which of these groups actually gives the right SET classification when they do not coincide. Below we will show that, at least when $d: \mathcal{H}^4(G, \Z) \rightarrow \H^3(G, U(1))$ is a surjective map, we find that $\H^4(G,\Z) \cong \H^3(G,U(1))/G_c$, where $G_c$ classifies a continuous family of topological terms. Therefore $\H^4(G,\Z)$ naturally removes the continuous part associated with $\H^3(G, U(1))$, as must be done to classify topological phases of matter. 

To show this, consider the long exact sequence in Eq. \eqref{Eq:les}. We assume $d$ is surjective; as we discussed above, this is the case for the $G$ that we considered in this paper, because the map $\sigma$ defined above can be shown to be well-defined and injective for the $G$ that we consider. 

Now for any homomorphism $f: A \rightarrow B$ we have $A/\ker f \cong \im f$. In our case, with $f=d$, we have $\H^3(G,U(1))/\ker d \cong \im d \cong \H^4(G,\Z)$. From exactness, we also have $\ker d \cong \im (\text{exp})$. The subgroup $G_c := \im(\text{exp})$ is in fact continuous, as we now argue.

The argument is in two parts: (i) the group $\H^3(G,\R)$ is a continuous group, such that every element of $\H^3(G, \R)$ is part of a continuous family, and (ii) the exponential map is continuous. To see (i), we observe that if $f \in C^3(G,\R)$ is a cocycle, i.e. $df = 0$, so is $\alpha f$, for any $\alpha \in \R$: this is because the coboundary operation $d$ is linear. Moreover, $f$ and $\alpha f$ are in the same cohomology class if and only if $(1-\alpha) f = dg$ for some $g \in C^2(G,\R)$. But by linearity, this is possible for $\alpha \ne 1$ only if $f$ is itself a coboundary. Hence every cohomology class in $\H^3(G,\R)$ is associated to a continuous family.

Since the exponential map is continuous, the group $\im(\text{exp})$ is also continuous, as claimed. Note that these arguments depend on the assumption that the map $d$ is surjective so that $\text{im } d \cong \H^4(G,\Z)$; this assumption is verified in our examples but may not hold in general. If this is not the case, $\H^3(G,U(1))$ need not contain $\H^4(G,\Z)$ as a subgroup. It would be interesting to find examples, or to rule out, such a situation for physically relevant groups $G$. 

\subsubsection{$\H^3(G,U(1)) \cong \H^4(G,\Z)$ in our examples}
\label{LHSmeasurable}

In our examples, it is possible to directly show using the Lyndon-Hochschild-Serre spectral sequence that for all the groups $G$ considered in this paper,
\begin{align}
  \label{HnR}
\H^n(G,\R) \cong 0 \;\; \text{for } n \ge 3.
\end{align}
This implies that $d: \H^3(G, U(1)) \rightarrow \H^4(G,\Z)$ is both injective and surjective, and hence $\H^3(G,U(1)) \cong \H^4(G,\Z)$ for all the examples considered in this work.

The calculation of Eq. \ref{HnR} has some subtleties, as the LHS spectral sequence needs to be applied to continuous groups with continuous coefficients. However many techniques in algebraic topology, including the LHS spectral sequence, implicitly assume the discrete topology on the coefficients. While versions of the LHS sequence for continuous cohomology (cohomology with continuous cochains) have been derived \cite{Stasheff1978}, we are not aware of a rigorous formulation involving measurable (Borel) or piece-wise continuous cohomology. 
 
\subsection{$G=U(1)\leftthreetimes\mathbb{E}^{2}$}
\label{Sec:Coho_calc_cont}

In this section we will compute the cohomology of the group $G=U(1)\leftthreetimes\mathbb{E}^{2}$, which is the symmetry group of continuum FQH systems, as derived in Appendix \ref{Sec:FQHSymm}. 

The group $U(1)\leftthreetimes\mathbb{E}^{2}$ can be written as a central extension of the Euclidean group $\mathbb{E}^2 = \R^2 \rtimes SO(2)$ by $U(1)$ according to the short exact sequence
\[
1\rightarrow U(1)\rightarrow G=U(1)\leftthreetimes\mathbb{E}^{2}\rightarrow \mathbb{E}^2\rightarrow1
\]
Then from Theorem \ref{thm:lhss} there is a first quadrant cohomological
spectral sequence which gives the cohomology: 
\begin{equation}
E_{2}^{p,q}=\H^{p}(\mathbb{E}^2;\H^{q}(U(1),M))\implies \H^{p+q}(U(1)\leftthreetimes\mathbb{E}^{2},M),
\end{equation}
where $M$ is a $U(1)\leftthreetimes\mathbb{E}^{2}$-module (see Appendix \ref{app_sseq} for a review of spectral sequences). Here we will compute the cohomology groups $\H^{2}(U(1)\leftthreetimes\mathbb{E}^{2},\A)$ and $\H^{4}(U(1)\leftthreetimes\mathbb{E}^{2},\Z)$, where $\H$ denotes cohomology with measurable cochains. The main result of this section is that these cohomology groups give the same classification that we would obtain upon replacing $G$ by the group $U(1)\times SO(2)$.

\subsubsection{Calculation of $\H^n(U(1)\leftthreetimes\mathbb{E}^{2},\Z)$}

To compute $\H^{n}(U(1)\leftthreetimes\mathbb{E}^{2},\mathbb{Z})$, we first note that $\H^{n}(U(1)\leftthreetimes\mathbb{E}^{2},\mathbb{Z})\cong H_{\mathrm{top}}^{n}(B(U(1)\leftthreetimes\mathbb{E}^{2}),\mathbb{Z})$, where $H_{\text{top}}$ denotes the singular cohomology of the classifying space $BG$ (Theorem \ref{thm:Hborel-HBG}). Note that the assumption of measurable cochains arises here, in applying Theorem \ref{thm:Hborel-HBG}. 

Now, using Theorem \ref{thm:coho_noncompact}, we see that the cohomology of $G\cong U(1)\leftthreetimes\mathbb{E}^{2}$ is the same as that of its
maximal compact subgroup $K\cong U(1)\times SO(2)$ (to prevent confusion, we distinguish the continuous $U(1)$ rotation group from the $U(1)$ charge conservation group by denoting the former as $SO(2)$). In addition, we have the standard result that $BU(1)=\mathbb{CP}^{\infty}$. Therefore,
\begin{align}
BG_{\mathrm{}}=B\left(U(1)\times SO(2)\right)=\mathbb{C}\mathrm{P}^{\infty}\times\mathbb{C}\mathrm{P}^{\infty}.
\end{align}
To summarize, we have 
\begin{align}
&\H^{n}(U(1)\leftthreetimes\mathbb{E}^{2},\mathbb{Z})\cong H_{\mathrm{top}}^{n}(B(U(1)\leftthreetimes\mathbb{E}^{2}),\mathbb{Z})\nonumber \\
&\cong H_{\mathrm{top}}^{n}(\mathbb{C}\mathrm{P}^{\infty}\times\mathbb{C}\mathrm{P}^{\infty},\mathbb{Z}).
\end{align}

Now using the result 
\[
H_{\mathrm{top}}^{n}(\mathbb{C}\mathrm{P}^{\infty},\mathbb{Z})=\begin{cases}
0 & n\text{ odd}\\
\mathbb{\mathbb{Z}} & n\text{ even}
\end{cases}
\]
we can apply the decomposition derived from the K\"unneth formula and the Universal Coefficient Theorem, as stated in Theorem \ref{thm:kunneth}, to obtain
\begin{align}
\H^{n}(U(1)\leftthreetimes\mathbb{E}^{2},\mathbb{Z})&\cong H_{\mathrm{top}}^{n}(\mathbb{C}\mathrm{P}^{\infty}\times\mathbb{C}\mathrm{P}^{\infty},\mathbb{Z})\nonumber \\
&=\begin{cases}
0, & n \text{ odd}\\
\Z^{n/2+1}, & n \text{ even}
\end{cases}
\end{align}

\subsubsection{Calculation of $\H^2(U(1)\leftthreetimes\mathbb{E}^{2},\A)$ and $\H^3(U(1)\leftthreetimes\mathbb{E}^{2},U(1))$}

From the Universal Coefficient Theorem (Theorem \ref{thm: universalcoefficienttheorem}),
we have
\begin{align}
&\H^{2}(U(1)\leftthreetimes\mathbb{E}^{2},\A) \nonumber \\ &= \H^{2}(U(1)\leftthreetimes\mathbb{E}^{2},\Z) \otimes \A \\
&= \Z^2 \otimes \A = \A\times\A.
\end{align}

We also have

\begin{equation}
\H^{4}(U(1)\leftthreetimes\mathbb{E}^{2},\mathbb{Z}) \cong \Z^3.
\end{equation}

As argued in Appendix \ref{Sec:Coho_calc_gen}, in order to obtain all possible quantized topological terms associated with $\H^3(G,U(1))$, we will only need to determine the cocycle representatives of
a subgroup of $\H^3(G,U(1))$ which is isomorphic to $\H^4(G,\Z) \cong \Z^3$, and this will be done below. 

We can also go further and demonstrate that
\begin{align}
  \H^n( U(1)\leftthreetimes\mathbb{E}^{2} ,U(1)) \cong \H^{n+1}( U(1)\leftthreetimes\mathbb{E}^{2},\Z)
\end{align}
by showing that $\H^n(U(1)\leftthreetimes\mathbb{E}^{2} ,\R) \cong \H^{n+1}(U(1)\leftthreetimes\mathbb{E}^{2} ,\R) \cong 0$, although we will not present the calculation here. As discussed in Appendix \ref{LHSmeasurable} this calculation requires applying the LHS spectral sequence to measurable cochains with continuous groups and continuous coefficients, for which we are not aware of a mathematical theorem.

\subsubsection{Explicit cocycle representatives for $\H^2(U(1)\leftthreetimes\mathbb{E}^{2},\A)$}

Let us first consider the group of magnetic translations alone. To obtain a representative 2-cocycle of the group $U(1)\leftthreetimes\mathbb{R}^{2}$, we note that the inclusion map $U(1)\xhookrightarrow{}U(1)\leftthreetimes\mathbb{R}^{2}$ induces a map between group cohomology classes $\H^2(U(1)\leftthreetimes\mathbb{R}^{2},\A) \overset{\mathrm{res}}{\rightarrow} \H^2(U(1),\A)$. This map is called the restriction map, which means that a representative cocycle $[\omega']\in \H^2(U(1)\leftthreetimes\mathbb{R}^{2},\A)$ should be reduced to a cocycle $[\omega]\in \H^2(U(1),\A)$ when we evaluate the function only within the domain of the subgroup $U(1)$. When considering symmetry fractionalization in the group $U(1)$, a representative 2-cocycle is given by $\mathfrak{w}(e^{2\pi i z_1},e^{2\pi i z_2})=v^{z_1+z_2-[z_1+z_2]}$, where $v$ is the vison \cite{Manjunath2020}. 

From the group multiplication law, the cocycle associated to the central extension is $w({\bf  r_1},{\bf  r_2}) = \frac{{\bf  r_1}\times {\bf  r_2}}{2l_B^2}$. It can indeed be verified that the following symmetry fractionalization cocycle, which depends on $w({\bf  r_1},{\bf  r_2})$, satisfies the group multiplication law for $U(1)\leftthreetimes\mathbb{R}^{2}$ and also takes the correct form when restricted to the $U(1)$ subgroup:

\begin{equation}
\mathfrak{w}((e^{2\pi i z_1},{\bf  r_1}),(e^{2\pi i z_2},{\bf  r_2})) = v^{z_1+z_2+\frac{{\bf  r_1}\times {\bf  r_2}}{2l_B^2}-[z_1+z_2+\frac{{\bf  r_1}\times {\bf  r_2}}{2l_B^2}]}
\end{equation} 

If we now include $SO(2)$ rotations, the most general 2-cocycle representatives for the group $\H^2(U(1)\leftthreetimes[\mathbb{R}^{2}\rtimes SO(2)],\A)$ can now be written as follows:
 
\begin{align}
\mathfrak{w}({\bf  g}_1,{\bf  g}_2) &= v^{z_1+z_2+\frac{{\bf  r_1}\times {^{h_1}} {\bf  r_2}}{2l_B^2}-[z_1+z_2+\frac{{\bf  r_1}\times {^{h_1}}{\bf  r_2}}{2l_B^2}]} \nonumber \\
&\times s^{h_1+h_2-[h_1+h_2]}.
\end{align}
Here we have set ${\bf g}_i = (e^{2\pi i z_i}, {\bf r}_i, e^{2\pi i h_i})$. The contribution from $SO(2)$ symmetry appears as an independent factor; the only change in the exponent of $v$ appears in the term $\frac{{\bf  r_1}\times {^{h_1}} {\bf  r_2}}{2l_B^2}$, where the group element $h_1$ explicitly appears. This is necessary in order to satisfy the cocycle condition for the full symmetry group with rotations. For Abelian topological phases, the anyon $s$ corresponds to the spin vector. The anyons $v,s$ uniquely determine the symmetry fractionalization class, which is therefore an element of $\A\times\A$, in agreement with the calculation of $\H^2(U(1)\leftthreetimes\mathbb{E}^{2},\A)$.

\subsubsection{Explicit cocycle representatives for $\H^3(U(1)\leftthreetimes\mathbb{E}^{2},U(1))$}

Next we determine a set of SPT cocycle representatives associated to the group $\H^3(U(1)\leftthreetimes\mathbb{E}^{2},U(1))$. Here we will write the  SPT cocycles as functions of the form $f({\bf g_1},{\bf g_2},{\bf g_3}) \mod 1$. This cocycle will then contribute a phase $e^{2\pi i f({\bf g_1},{\bf g_2},{\bf g_3})}$ to the defect $F$-symbol $F^{0_{\bf g_1} 0_{\bf g_2} 0_{\bf g_3}}$. Moreover, the function $df({\bf g_1},{\bf g_2},{\bf g_3},{\bf g_4})$ can be seen to be a representative cocycle of $\H^4(U(1)\leftthreetimes\mathbb{E}^{2},\Z)$. Indeed, we can find representatives $f$ such that the corresponding functions $df$ generate the full $\H^4(U(1)\leftthreetimes\mathbb{E}^{2},\Z)$ classification.

These cocycles should reduce to the following known forms when restricted to the subgroup $U(1)\times SO(2)$:
\begin{align}
&f_{\text{res}}({\bf  g}_1,{\bf  g}_2,{\bf  g}_3) = k_1 z_1(z_2 + z_3-[z_2+z_3]) \nonumber \\
&+ k_2 h_1(z_2 + z_3-[z_2+z_3])+ k_3 h_1(h_2 + h_3-[h_2+h_3])
\end{align}
where $k_i \in \Z$. Define the integer $F_{ij} = z_i + z_j + \frac{{\bf  r}_i\times {^{h_i}}{\bf  r}_j}{2l_B^2}-[z_i + z_j + \frac{{\bf  r}_i\times {^{h_i}}{\bf  r}_j}{2l_B^2}]$. Also define the integer $h_{ij} = h_i + h_j - [h_i + h_j]$. Consider the following functions:

\begin{align}
&f({\bf  g}_1,{\bf  g}_2,{\bf  g}_3) = k_1 (z_1 F_{23} + \frac{{\bf  r}_1\times {^{h_1}}{\bf  r}_2}{2l_B^2} z_3 + \lambda({\bf  g}_1,{\bf  g}_2,{\bf  g}_3))\nonumber \\  &+ k_2 h_1 F_{23} + k_3 h_1 h_{23}.
\end{align}
Here we have chosen the function $\lambda$ so that $d\lambda({\bf  g}_1,{\bf  g}_2,{\bf  g}_3,{\bf  g}_4) = \frac{{\bf  r}_1\times {^{h_1}}{\bf  r}_2}{2l_B^2}\frac{{\bf  r}_3\times {^{h_3}}{\bf  r}_4}{2l_B^2} $. Observe that when we restrict to the subgroup $U(1)\times SO(2)$, $f \equiv f_{\text{res}}$. Moreover, we now show that $df = 0$, i.e. $f$ satisfies the cocycle condition for $U(1)\leftthreetimes\mathbb{E}^{2}$, when $k_i \in \Z$. Using the properties satisfied by $\lambda$, a direct computation gives

\begin{align}
df({\bf  g}_1,{\bf  g}_2,{\bf  g}_3,{\bf  g}_4) &= k_1 F_{12} F_{34} + k_2 h_{12} F_{34} + k_3 h_{12} h_{34} \nonumber \\&= 0 \mod 1,
\end{align}
implying that these functions are indeed nontrivial cocycles in $\H^3(U(1)\leftthreetimes\mathbb{E}^{2},U(1))$. In fact, the function $df$ written above, with parameters $k_1,k_2,k_3$, is the most general cocycle representative that can be written for $\H^4(U(1)\leftthreetimes\mathbb{E}^{2},\Z)$. Therefore, by varying the parameters $k_i$, we obtain a $\Z^3$ classification of SPTs.

\subsection{$G=U(1)\leftthreetimes_{\phi} \Z^2$}
\label{Sec:Coho_calc_magtrans}

The group of magnetic translations is a central extension of the discrete translation group $\Z^2$ by $U(1)$. For a derivation of the group multiplication law, and a discussion of certain ambiguities in it which are related to different gauge choices for the vector potential, see Appendix \ref{Sec:FQHContSymm}.

\subsubsection{Calculation of $\H^2(U(1)\leftthreetimes_{\phi} \Z^2,\A)$ and $\H^3(U(1)\leftthreetimes_{\phi} \Z^2,U(1))$}
Since $G$ is not a direct product, but a central extension of the translation group $\Z^2$ by $U(1)$, we cannot immediately apply the K\"unneth decomposition to evaluate the group $\H^2(U(1)\leftthreetimes_{\phi} \Z^2,\mathcal{A})$ of symmetry fractionalization classes. To evaluate this group cohomology it is appropriate to use the Lyndon-Hochschild-Serre spectral sequence (LHSS). An introduction to the LHSS is given in Appendix \ref{app_sseq}. It will be convenient to first compute the cohomology with $\Z$ coefficients. We can then use Theorem \ref{thm: universalcoefficienttheorem} to obtain $\H^2(U(1)\leftthreetimes_{\phi} \Z^2,\A)$. Furthermore, the arguments from Appendix \ref{Sec:Coho_calc_gen} show that we only need a subgroup of $\H^3(U(1)\leftthreetimes_{\phi} \Z^2,U(1))$ isomorphic to $\H^4(U(1)\leftthreetimes_{\phi} \Z^2,\Z)$, along with the corresponding cocycle representatives.

The spectral sequence for the cohomology $\H^n(G,\Z)$ has the following nonzero terms in its $E_2$-page:

\begin{center}
	\begin{tikzpicture}
	\matrix (m) [matrix of math nodes,
	nodes in empty cells,nodes={minimum width=5ex,
		minimum height=5ex,outer sep=-5pt},
	column sep=1ex,row sep=1ex]{
		
		4&\mathbb Z   &\mathbb Z^2   &\mathbb Z  &0&0 \\
		3&0   &0   &0  &0 &0 \\
		2&\mathbb Z   &\mathbb Z^2   &\mathbb Z  &0 &0\\
		1&0   &0   &0  &0&0 \\
		0&\Z  &\Z^2 &\Z  &0 &0\\
		\quad\strut &   0  &  1  &  2  &3 &4  \strut \\};
	\draw[thick] (m-1-1.north east) -- (m-6-1.east) ;
	\draw[thick] (m-6-1.north) -- (m-6-6.north east) ;
	\node[] at (3,-2.5) {p};
	\node[] at (-2.3,3) {q};
	\end{tikzpicture}
\end{center}

It is clear that the spectral sequence stabilizes at the $E_2$-page; moreover, there is no extension problem for $\H^2(G,\Z)$, because the only group extension of $\Z$ by $\Z$ is $\Z^2$. As discussed in Appendix \ref{app_sseq}, the physical implication of this statement is that each cohomology class $\H^p(\Z^2,\H^{q}(U(1),\Z))$ on the diagonal $p+q=2$ of this page contributes a separate factor to the group $\H^2(G,\A)$, and furthermore, there are no symmetry localization anomalies. We obtain the desired classification by an application of the Universal Coefficient Theorem, Theorem \ref{thm: universalcoefficienttheorem}, and the properties of Tor, discussed in Appendix \ref{Sec:TenTorExt}:
\begin{align}
\H^2(G,\mathcal{A}) &= \H^2(G,\Z) \otimes \mathcal{A} \oplus \text{Tor}(\H^3(G,\Z),\mathcal{A}) \nonumber \\
&= (\Z^2 \otimes \mathcal{A}) \oplus \text{Tor}(\Z^2,\mathcal{A}) \nonumber \\
&= (\mathcal{A}\times \mathcal{A}) \oplus \Z_1  
\end{align}
Therefore we have a group $\mathcal{A}\times \mathcal{A}$ of fractionalization classes, irrespective of the value of flux per unit cell. To further interpret this answer, we use the spectral sequence decomposition to write
\begin{align}
&\H^2(U(1)\leftthreetimes_{\phi} \Z^2,\mathcal{A}) \nonumber \\ &\cong \H^0(\Z^2,\H^2(U(1),\mathcal{A})) \times \H^2(\Z^2,\H^0(U(1),\mathcal{A}))
\end{align}
Now $\H^0(G,A) \cong A$ whenever the symmetry does not permute anyons. Therefore we can interpret the two $\mathcal{A}$ factors as coming from fractionalization classes in $\H^2(U(1),\mathcal{A})$ and $\H^2(\Z^2,\mathcal{A})$ repectively. 

Next we compute $\H^3(U(1)\leftthreetimes_{\phi} \Z^2,U(1))$. By inspecting the $E_2$-page above, we see that
\begin{equation}
\H^4(G,\Z) \cong \Z\times \Z.
\end{equation} 

We can similarly decompose the $\Z$ factors as follows:
\begin{align}
&\H^4(U(1)\leftthreetimes_{\phi} \Z^2,\Z) \nonumber \\ &\cong \H^4(U(1),\Z) \times \H^2(\Z^2,\H^2(U(1),\Z)) \nonumber\\
&\cong \H^3(U(1),U(1)) \times \H^2(\Z^2,\H^1(U(1),U(1))) \nonumber \\ &\cong \H^3(U(1)\leftthreetimes_{\phi} \Z^2,U(1)).
\end{align}

\subsubsection{Explicit cocycle representatives for $\H^2(U(1)\leftthreetimes_{\phi} \Z^2,\A)$}
The first piece in the decomposition of $\H^2(U(1)\leftthreetimes_{\phi} \Z^2,\A)$ is familiar: it describes $U(1)$ symmetry fractionalization, in which there is a unique vison $v$ that is associated to $2\pi$ flux insertion. The corresponding symmetry property is the fractional charge of an anyon $a$, which is given by $e^{2\pi i Q_a} = M_{v,a}$. However there is a technical subtlety: the pure $U(1)$ symmetry fractionalization cocycle $\mathfrak{w}(e^{2\pi i z_1},e^{2\pi i z_2})=v^{z_1+z_2-[z_1+z_2]}$ does not satisfy the 2-cocycle condition for the group $U(1)\leftthreetimes_{\phi} \Z^2$. In fact, the group multiplication law, which mixes translations with the $U(1)$ group variables, ensures that choosing $\mathfrak{w}((e^{2\pi i z_1},{\bf r}_1)(e^{2\pi i z_2},{\bf r}_2))$ as a function of the $z$ variables alone is impossible. Therefore this symmetry fractionalization cocycle has some dependence on the translation variables as well. Indeed, it can be checked that a slight modification of the cocycle in Eq. \eqref{RepresentativeCocycleH2} gives us a valid cocycle of $U(1)\leftthreetimes_{\phi} \Z^2$ valued in $\mathcal{A}$:
\begin{align}
\mathfrak{w}({\bf g}_1,{\bf g}_2)
&= v^{z_1+z_2+\phi w({\bf r_1},{\bf r_2}) - [z_1+z_2+\phi w({\bf r_1},{\bf r_2})]}
\end{align}
Here the function $w$ satisfies $w({\bf r_1},{\bf r_2})-w({\bf r_2},{\bf r_1}) = {\bf r_1}\times {\bf r_2}$. These cocycles are all nontrivial. To see that they are indeed cocycles, it suffices to show that the exponent of $v$ satisfies the 2-cocycle condition for $\H^2(U(1)\leftthreetimes_{\phi} \Z^2,\Z)$. First note that the exponent is indeed an integer-valued function of ${\bf g}_1,{\bf g}_2$, since it is of the form $x -[x]$ where $x =z_1+z_2+\phi w({\bf r_1},{\bf r_2})$. Next, observe that the coboundary operator applied to the $z$-independent function $\phi w({\bf r_1},{\bf r_2})$ gives zero. Moreover, the remaining terms can be collectively written as $P({\bf g}_1)+P({\bf g}_2)-P({\bf g}_1 {\bf g}_2)$, where $P$ is the projection map: $P((e^{2\pi i z},{\bf r})) = z$. The coboundary operator applied to any function of this form also gives zero. Hence the sum of the two parts is an integer-valued 2-cocycle. Since this expression reduces to the familiar $U(1)$ cocycle when we restrict to the $U(1)$ subgroup, it cannot be expressed as a coboundary, i.e. it is a representative of a nontrivial cohomology class. 

The above symmetry fractionalization cocycles depend on the value of $\phi$. This corresponds to the physical fact that translations around a single unit cell enclose a $\phi$ flux; and the insertion of $2\pi$ flux via any combination of $U(1)$ or $\Z^2$ operations induces a vison $v$. However, the \textit{classification} of symmetry fractionalization only depends on the number of distinct choices for $v$, and is therefore flux-independent. 

The second symmetry fractionalization factor corresponding to $\H^2(\Z^2,\mathcal{A})$ arises as a result of the discrete translation symmetry. Since the group multiplication does not introduce $z$ terms into the ${\bf r}$ component, the corresponding cocycles are independent of the $z$ variables:
\begin{align}
\mathfrak{w}({\bf g}_1,{\bf g}_2)) = m^{w({\bf  r_1},{\bf r_2})}
\end{align}
for some $m \in \mathcal{A}$. This notation assumes that $w$ is an \textit{integer} valued cocycle, i.e. it is expressed in the Landau gauge or some other integer-valued gauge. However, we can use any other gauge to calculate symmetry fractionalization invariants, as long as we ultimately measure gauge-invariant combinations of $w$. This symmetry fractionalization class is associated to the presence of an anyon $m$ in each unit cell of the lattice. When $\phi=0$, this anyon can be measured by the gauge-invariant quantity associated to elementary translations, $\mathfrak{b}({\bf x},{\bf y}) = \mathfrak{w}({\bf x},{\bf y})\overline{\mathfrak{w}({\bf y},{\bf x})} = m$. 

The most general cocycle describing symmetry fractionalization is therefore
\begin{equation}
\mathfrak{w}({\bf g}_1,{\bf g}_2) = v^{z_1 + z_2 + \phi w({\bf  r_1},{\bf r_2})  - [z_1 + z_2 + \phi w({\bf  r_1},{\bf r_2})]} \times m^{w({\bf  r_1},{\bf r_2})}   
\end{equation}
There is also a gauge-invariant expression for $m$ when the flux is a nonzero rational number $\phi=p/q$, with $p$ and $q$ coprime. Define magnetic translations ${\bf r_1},{\bf r_2}$ spanning a magnetic unit cell, for example a cell of size $1 \times q$. Then, we can directly compute
\begin{align}
\mathfrak{b}({\bf r_1},{\bf r_2}) &= \mathfrak{w}({\bf r_1},{\bf r_2})\overline{\mathfrak{w}({\bf r_2},{\bf r_1})} = v^{\phi q - [\phi q]} \times m^{q} \nonumber \\
&= v^p \times m^q
\end{align}
This is intuitively understood as follows: the magnetic translations cover $q$ unit cells, each of which is pierced by a flux $p/q \mod 1$. Therefore they surround $q$ copies of $m$, as well as inducing the anyon $v^p$ associated to $p$ $U(1)$ flux quanta which are contained within each magnetic unit cell. This is the quantity that will be measured by the relevant $G$-crossed invariants. Note that the gauge-invariant combination is not $m$ alone, but $v^p \times m^q$. This quantity is directly associated to the filling per magnetic unit cell $\nu$, as we have calculated in Appendix \ref{Sec:LSM_lattice}.

\subsubsection{Explicit cocycle representatives for $\H^3(U(1)\leftthreetimes_{\phi} \Z^2,U(1))$}

Finally, we compute a set of representative cocycles for $\H^3(U(1)\leftthreetimes_{\phi} \Z^2,U(1)) \cong \Z\times\Z$. (Here we will write the  SPT cocycles as functions of the form $f({\bf g_1},{\bf g_2},{\bf g_3}) \mod 1$. This cocycle will then contribute a phase $e^{2\pi i f({\bf g_1},{\bf g_2},{\bf g_3})}$ to the defect $F$-symbol $F^{0_{\bf g_1} 0_{\bf g_2} 0_{\bf g_3}}$.)

The spectral sequence decomposition of $\H^3(U(1)\leftthreetimes_{\phi} \Z^2,U(1))$ shows that one factor of $\Z$ is associated to the $U(1)$ symmetry alone, i.e. to bosonic IQH states, while the other $\Z$ factor is associated to a mixed SPT of $U(1)$ and $\Z^2$ symmetry. Cocycles corresponding to the first factor must reduce to the functions $f(e^{2\pi i z_1},e^{2\pi i z_2},e^{2\pi i z_3}) = k_1 z_1 (z_2 + z_3 - [z_2 + z_3]) \mod 1$ when all translations are set to zero. Define $F({\bf g}_1,{\bf g}_2) = z_1 + z_2 + \phi w({\bf  r_1},{\bf r_2})  - [z_1 + z_2 + \phi w({\bf  r_1},{\bf r_2})])$. Consider the function $f_{\text{IQH}}$ such that 

\begin{equation}
df_{\text{IQH}}({\bf g}_1,{\bf g}_2,{\bf g}_3,{\bf g}_4) = k_1 F({\bf g}_1,{\bf g}_2) F({\bf g}_3,{\bf g}_4) = 0 \mod 1.
\end{equation} 

When $k_1 \in \Z$, this function satisfies the cocycle condition $df = 0$, since $F$ is integer-valued. An explicit form for $f_{\text{IQH}}$ is
\begin{align}
&f_{\text{IQH}}({\bf g}_1,{\bf g}_2,{\bf g}_3) \nonumber \\ &= k_1 (z_1 F({\bf g}_2,{\bf g}_3) + \phi w({\bf r}_1,{\bf r}_2) z_3 + \phi^2\lambda({\bf r}_1,{\bf r}_2,{\bf r}_3))
\end{align}
where $d\lambda({\bf r}_1,{\bf r}_2,{\bf r}_3,{\bf r}_4) = w({\bf r}_1,{\bf r}_2)w({\bf r}_3,{\bf r}_4)$. A possible choice for $\lambda$ is
\begin{align}
&\lambda({\bf r}_1,{\bf r}_2,{\bf r}_3) \nonumber \\ &= x_1y_1(x_2y_3-x_3y_2) - x_1y_2^2x_3-x_1^2y_2y_3 - 2 x_1x_2y_2y_3.
\end{align}
However, we will not need to know this explicit form in subsequent calculations. Since $f_{\text{IQH}}$ restricts to the known expression for nontrivial cocycle representatives of $\H^3(U(1),U(1))$, it is indeed a nontrivial cocycle.

In a similar way, the mixed SPT cocycles $f_{\text{mixed}}$, representing 'atomic insulators' with integer filling $k_6$ per unit cell, can be obtained by imposing the condition
\begin{equation}
df_{\text{mixed}}({\bf g}_1,{\bf g}_2,{\bf g}_3,{\bf g}_4) = k_6 F({\bf g}_1,{\bf g}_2) w({\bf r}_3,{\bf r}_4) = 0 \mod 1.
\end{equation} 

An explicit form of $f_{\text{mixed}}$ is then given by

\begin{equation}
f_{\text{mixed}}({\bf g}_1,{\bf g}_2,{\bf g}_3) = k_6(z_1 w({\bf r}_2,{\bf r}_3) + \phi \lambda({\bf r}_1,{\bf r}_2,{\bf r}_3)).
\end{equation}

As observed previously, the functions $df$ represent elements of $\H^4(G,\Z)$.

\subsection{$G=U(1)\leftthreetimes_{\phi} [\Z^2\rtimes \Z_M]$}
\label{Sec:Coho_calc_gspace}

Here the space group $G_{\text{space}}$ is a semidirect product of the point group $ \mathbb{Z}_M$ by $\mathbb{Z}^2$. The full symmetry group is then a central extension of $G_{\text{space}}$ by $U(1)$. The group multiplication law is
\begin{align}
&(e^{2\pi i z_1},{\bf r_1},e^{2\pi i h_1/M}) (e^{2\pi i z_2},{\bf r_2},e^{2\pi i h_2/M}) \nonumber \\ &= (e^{2\pi i (z_1+z_2+w({\bf  r_1},^{h_1}{\bf  r_2}))}, {\bf r_1}+^{h_1}{\bf r_2},e^{2\pi i [h_1+h_2]_M/M})    
\end{align}

In the group multiplication law we will now specifically assume that $w({\bf  r_1},{\bf  r_2}) = \frac{1}{2}{\bf  r_1}\times {\bf  r_2}$, so that the function $w$ is invariant under equal rotations of ${\bf  r_1},{\bf  r_2}$.

Let $h$ be the generator of point group rotations. We will abuse the notation for rotation point group elements slightly, as follows: the quantities $h_i$ written in line should be understood as integers mod $M$, corresponding to rotations by the angle $2\pi h_i/M$, while in an expression such as ${^h} {\bf r}$ or ${^{1-h}}{\bf r}$, $h$ is understood as the $2\times 2$ matrix generator of point group rotations, and thus the symbol 1 in ${^{1-h}}{\bf r}$ denotes the identity $2\times 2$ matrix. 

We can form the $E_2$-page of the LHSS for $\H^n(G,\Z)$ corresponding to the short exact sequence 
\begin{equation}
1 \rightarrow U(1) \rightarrow G=U(1)\leftthreetimes_{\phi} [\Z^2 \rtimes \Z_M] \rightarrow G_{\text{space}} \rightarrow 1
\end{equation}
where $G_{\text{space}} \cong \Z^2 \rtimes \Z_M$ is the space group, whose cohomology groups have been computed numerically as well as analytically \cite{Thorngren2018,Manjunath2020}. See Appendix \ref{app_sseq} for a review of spectral sequences. The $E_2$-page of the spectral sequence is

\begin{center}
	\begin{adjustbox}{width=0.5\textwidth}
		\begin{tikzpicture}
		\matrix (m) [matrix of math nodes,
		nodes in empty cells,nodes={minimum width=5ex,
			minimum height=5ex,outer sep=-5pt},
		column sep=1ex,row sep=1ex]{
			4&\H^0(G_{\text{space}},\Z)  &0 &\H^2(G_{\text{space}},\Z)  &0 &\H^4(G_{\text{space}},\Z) &0 \\
			3&0 &0  &0 &0 &0 &0 \\
			2&\H^0(G_{\text{space}},\Z)  &0 &\H^2(G_{\text{space}},\Z)  &0 &\H^4(G_{\text{space}},\Z) &0 \\
			1&0 &0  &0 &0 &0 &0 \\
			0&\H^0(G_{\text{space}},\Z)  &0 &\H^2(G_{\text{space}},\Z)  &0 &\H^4(G_{\text{space}},\Z) &0 \\
			\quad\strut &   0  &  1  &  2  &3 &4 & \strut \\};
		\draw[thick] (m-1-1.north east) -- (m-6-1.east) ;
		\draw[thick] (m-6-1.north) -- (m-6-7.north east) ;
			\node[] at (4.5,-2.5) {p};
		\node[] at (-5,3) {q};
		\end{tikzpicture}
	\end{adjustbox}
	
\end{center}
We can make several immediate observations: first, the spectral sequence stabilizes at the $E_2$-page, because all differentials are zero. Moreover, the cohomology of $G$ vanishes in odd degree. The fact that $\H^4(G,U(1)) \cong \H^5(G,\Z)$ is trivial means that there are no anomalies. Therefore we are assured of a consistent set of solutions to the $G$-crossed equations.
\subsubsection{Calculation of $\H^2(U(1)\leftthreetimes_{\phi} [\Z^2\rtimes \Z_M],\A)$}

Let us first determine $\H^2(G,\A)$. When $G_{\text{space}}$ acts trivially on the coefficients we have $\H^0(G_{\text{space}},\Z) \cong \Z$, so $\H^2(G,\Z)$ is an extension of $\Z$ by $\H^2(G_{\text{space}},\Z)$. There is only one possible extension, namely $\Z \times \H^2(G_{\text{space}},\Z)$: this is because any group extension of $\Z$ by a group $N$ must be the direct product extension $\Z\times N$. To determine the group of fractionalization classes we use the fact that $\H^3(G,\Z)$ is trivial, along with Theorem \ref{thm: universalcoefficienttheorem}:
\begin{align}
\H^2(G,\mathcal{A}) &\cong (\H^2(G,\Z)\otimes \mathcal{A}) \oplus \text{Tor}(\H^3(G,\Z),\mathcal{A}) \\
&\cong (\Z \times \H^2(G_{\text{space}},\Z)) \otimes \mathcal{A} \\
&\cong \mathcal{A} \times (\H^2(G_{\text{space}},\Z) \otimes \mathcal{A})
\end{align}

In turn, we can use the spectral sequence for the semidirect product extension defining $G_{\text{space}}$ along with numerical computations to obtain (see Ref. \cite{Manjunath2020})
\begin{equation}
\H^2(G_{\text{space}},\Z) \cong \Z\times \Z_M \times K_M,
\end{equation}
and consequently, 

\begin{align}
\H^2(G,\mathcal{A}) &\cong \mathcal{A} \times (\H^2(G_{\text{space}},\Z) \otimes \mathcal{A}) \\
&\cong \A\times \A\times (\A/M\A) \times (K_M \otimes \A).
\end{align}

Thus $\H^2(G,\A)$ splits into four separate factors. When $\A = \prod_i \Z_{n_i}$, we have, from the definition of $\otimes$ and $K_M$ (Appendix \ref{Sec:GrpCohIntro}),
\begin{align}
\A/M\A &\cong \prod_i \Z_{(n_i,M)} \\
K_2 \otimes \A & \cong \prod_i \Z^2_{(n_i,2)} \\
K_3 \otimes \A & \cong \prod_i \Z_{(n_i,3)} \\
K_4 \otimes \A & \cong \prod_i \Z_{(n_i,2)} \\
K_6 \otimes \A & \cong \Z_1.
\end{align}

\subsubsection{Explicit cocycle representatives of $\H^2(U(1)\leftthreetimes_{\phi} [\Z^2\rtimes \Z_M],\A)$}

The cocycle representatives for symmetry fractionalization can be found as follows. Denote a general group element as a triple ${\bf g} = (z,{\bf r},h)$ where $z \in \mathbb{R}/\Z, {\bf r} \in \Z^2$ and $h \in \Z/M\Z$. (For notational convenience, in this section we represent the groups $U(1)$  and $\Z_M$ as $\R/\Z$ and $\Z/M\Z$ respectively.) First we will write down cocycle representatives $f$ for $\H^2(U(1)\leftthreetimes_{\phi} [\Z^2\rtimes \Z_M],\Z)$ satisfying the condition
\begin{widetext}
	\begin{align}\label{H2CocycleCond}
	&f((z_1,{\bf r_1},h_1),(z_2,{\bf r_2},h_2)) + 
	f((z_1+z_2+\phi w({\bf r_1},{^{h_1}} {\bf r_2}),{\bf r_1}+{^{h_1}}{\bf r_2},h_1h_2),(z_3,{\bf r_3},h_3)) \nonumber \\ &= f((z_1,{\bf r_1},h_1),(z_2+z_3+\phi w({\bf r_2},{^{h_2}}{\bf r_3}),{\bf r_2}+{^{h_2}}{\bf r_3},h_2h_3)) + f((z_2,{\bf r_2},h_2),(z_3,{\bf r_3},h_3)).
	\end{align}
\end{widetext}

Let us consider the first term on the $E_2$-page. Now a set of representative cocycles of $\H^2(U(1),\Z)$ is given by 
\begin{equation}
g_n(z_1,z_2) =n(z_1+z_2 - [z_1+z_2]).
\end{equation} 
Here we have assumed that the $U(1)$ group element is given by $[z] = z \mod 1$ where $z$ is a lift of the group element to $\mathbb{R}$. However, these functions do not satisfy Eq. \eqref{H2CocycleCond}, since they do not properly account for the mixing of $z$ and ${\bf r}$ variables as a result of the central extension. Instead, it can be verified that the following function satisfies Eq. \eqref{H2CocycleCond}:

\begin{align}
&f_2({\bf g}_1,{\bf g}_2) \nonumber \\ &= n_v(z_1+z_2+\phi w({\bf  r_1},^{h_1}{\bf  r_2})-[z_1+z_2+\phi w({\bf  r_1},^{h_1}{\bf  r_2})])
\end{align}   

This function is an integer-valued cocycle when $n_v \in \Z$, being of the form $n_v(a - [a])$, where $a = z_1+z_2+\phi w({\bf  r_1},^{h_1}{\bf  r_2})$. Moreover, it cannot be written as an integer coboundary as it reduces to the nontrivial cocycle $g_n(z_1,z_2)$ when restricted to the $U(1)$ subgroup of $G$. We will see that this function is associated to the cohomology class of $\H^2(U(1)\leftthreetimes_{\phi} [\Z^2\rtimes \Z_M],\A)$ corresponding to the anyon $v$.  

Next, consider the term $\H^2(G_{\text{space}},\H^0(U(1),Z)) \cong \H^2(G_{\text{space}},\Z)$. In Ref \cite{Manjunath2020} it was shown that a general cocycle corresponding to this term takes the form 
\begin{widetext}
	\begin{equation}
	f_2({\bf g}_1,{\bf g}_2) = \frac{n_s}{M} (h_1 + h_2 - [h_1+h_2]_M) + \vec{n_t} \cdot {^{\frac{1-h_1}{1-h}}} {\bf r_2} + n_m w({\bf  r_1},^{h_1}{\bf  r_2}) 
	\end{equation}
\end{widetext}
Note that there are multiple gauge choices for $w$ corresponding to the same group multiplication law. For the above representation to be well-defined, $w$ must be integral: this constrains the allowed gauge choices. However, if we always compute invariants in terms of the gauge-invariant quantity $w({\bf  r_1},{\bf  r_2}) - w({\bf  r_2},{\bf  r_1}) = {\bf r_1}\times{\bf  r_2}$, after setting $h_1=h_2=0$, we can formally make non-integral gauge choices while doing these computations. This function is also a nontrivial 2-cocycle representative of $\H^2(U(1)\leftthreetimes_{\phi} [\Z^2\rtimes \Z_M],\Z)$. The coefficient $n_s \in \Z$ is associated to the anyon $s$; it is trivial if $n_s \in M \Z$. It contributes a factor of $\Z_M$ to $\H^2(G,\Z)$. The coefficient $\vec{n}_t \in \Z^2$ is associated to the anyons $\vec{t}$; it is trivial if $\vec{n}_t \in {^{1-h}} \Z^2$. It contributes a factor of $K_M$, which is defined as the group $\Z^2/{^{(1-h)}}\Z^2$. The coefficient $n_m \in \Z$ is associated to the anyon $m$ and is trivial if $n_m=0$. It contributes a factor of $\Z$. 

From the above discussion, a general symmetry fractionalization cocycle is given by
\begin{align}
&\mathfrak{w}((z_1,{\bf r_1},h_1),(z_2,{\bf r_2},h_2)) \nonumber \\&=
v^{z_1+z_2+\phi w({\bf  r_1},^{h_1}{\bf  r_2})-[z_1+z_2+\phi w({\bf  r_1},^{h_1}{\bf  r_2})]}  \nonumber\\ 
&\times m^{w({\bf  r_1},^{h_1}{\bf  r_2})} \nonumber\\
&\times t_x^{\left({^{\frac{1-h_1}{1-h}}} {\bf r}_2\right)_x}\times t_y^{\left({^{\frac{1-h_1}{1-h}}} {\bf r}_2\right)_y}\nonumber \\
&\times s^{\frac{h_1+h_2-[h_1+h_2]_M}{M}}.
\end{align} 
Here, a cocycle representative of $\H^2(G,\Z)$ which takes the form $n_a f({\bf g_1},{\bf g_2})$ yields a cocycle representative of $\H^2(G,\A)$ by defining $\mathfrak{w}({\bf g_1},{\bf g_2}) = a^{f({\bf g_1},{\bf g_2})}$. Thus there are four independent anyons associated to symmetry fractionalization. 

\subsubsection{Calculation of $\H^3(U(1)\leftthreetimes_{\phi} [\Z^2\rtimes \Z_M],U(1))$} 

For $G=U(1)\leftthreetimes_{\phi}[\Z^2\rtimes\Z_M]$, we can explicitly check that $\H^3(G,U(1)) \cong \H^4(G,\Z)$. However, in this case, we find that even the computation with $\Z$ coefficients is not direct. 

In the spectral sequence with $\Z$ coefficients that we introduced above, there are three terms contributing to $\H^4(G,\Z)$. The first is a $\Z$ factor given by $\H^0(G_{\text{space}},\Z) \cong \Z$, corresponding to bosonic IQH states. The second term, given by $\H^2(G_{\text{space}},\H^2(U(1),\Z))$ corresponds to mixed SPT phases of $U(1)$ and $G_{\text{space}}$. The third term $\H^4(G_{\text{space}},\Z)$ corresponds to pure space group SPT phases. It is not clear, however, that the group $\H^4(G,\Z)$ is a direct product of these three terms when there is nonzero flux. Specifically, we have to complete the following sequence of group extensions:

\begin{align}
1 &\rightarrow \H^4(G_{\text{space}},\Z)\rightarrow K \rightarrow \H^2(G_{\text{space}},\H^2(U(1),\Z)) \rightarrow 1 \nonumber \\
1 &\rightarrow K\rightarrow \H^4(U(1)\leftthreetimes_{\phi} [\Z^2\rtimes \Z_M],\Z) \rightarrow \Z \rightarrow 1.
\end{align}
Now the only possible extension of $\Z$ by $K$ on the second line is the group $\Z\times K$. However, $K$ can in principle be a nontrivial extension of the middle term $\H^2(G_{\text{space}},\H^2(U(1),\Z))$ by $\H^4(G_{\text{space}},\Z)$. In other words, the middle term may not enter the classification as a subgroup. The physical meaning of this statement is the following. Suppose there is an SPT associated to the middle term which is classified on the $E_2$-page by the factor $\Z_q$. Then, stacking $q$ identical SPTs of this type is not guaranteed to give the trivial SPT. Instead, we might obtain a nontrivial SPT classified by $\H^4(G_{\text{space}},\Z)$. This possibility is referred to as a nontrivial spectral sequence extension problem. 

The correct group structure can in principle be determined abstractly by computing the differentials in the spectral sequence. Below we will use the more direct approach of writing down a set of nontrivial cocycles of $G=U(1)\leftthreetimes_{\phi} [\Z^2\rtimes \Z_M]$ corresponding to the three groups described above, and showing that the group structure of the cocycles is indeed a direct product of the three terms, implying that the extension problem is trivial in this case.

After the calculations done below, the classification of SPTs is found to be

\begin{align}
\H^4(G,\Z) &\cong \Z\times \H^2(G_{\text{space}},\Z)\times \H^4(G_{\text{space}},\Z) \nonumber \\
&\cong \Z\times (\Z\times K_M \times \Z_M) \times (\Z_M^2\times K_M)  \nonumber \\
&\cong \Z^2 \times \Z_M^3 \times K_M^2.\label{Eq:fullCohResult}
\end{align}

\subsubsection{Explicit cocycle representatives of $\H^3(U(1)\leftthreetimes_{\phi} [\Z^2\rtimes \Z_M],U(1))$}	 

In this section we represent a generic group element as ${\bf g_i} = (z_i,{\bf r},h_i)$ where $z \in \R/\Z, {\bf r} \in \Z^2, h_i \in \Z/M\Z$. We will write down cocycle reperesentatives of $\H^3(G,U(1))$ as $f({\bf g_1},{\bf g_2},{\bf g_3}) \mod 1$ where $f$ is real-valued modulo integers (and hence, in order to satisfy the cocycle condition, $df$ is required to be integer-valued). This cocycle will then contribute a phase $e^{2\pi i f({\bf g_1},{\bf g_2},{\bf g_3})}$ to the defect $F$-symbol $F^{0_{\bf g_1} 0_{\bf g_2} 0_{\bf g_3}}$. Moreover, in each case the function $df({\bf g_1},{\bf g_2},{\bf g_3},{\bf g_4})$ represents an element of $\H^4(U(1)\leftthreetimes_{\phi} [\Z^2\rtimes \Z_M],\Z)$.

Let us consider the first term of the spectral sequence decomposition, which classifies the IQH states that are invariant under space group symmetry transformations. From the spectral sequence, we see that these states are classified by the group $\Z$, and are not affected by the group extension problem. The usual cocycle representatives of $\H^3(U(1),U(1))$ are given by $g_{\text{IQH}} \mod 1$, where $g_{\text{IQH}}(z_1,z_2,z_3) = k_1 z_1(z_2+z_3-[z_2+z_3])$. They satisfy 
\begin{align}
&d g_{\text{IQH}}(z_1,z_2,z_3,z_4) \nonumber \\ &= k_1 (z_1+z_2-[z_1+z_2])(z_3+z_4-[z_3+z_4]) \nonumber \\ &\equiv 0 \mod 1.
\end{align}
The last expression is a cup product of two cocycle representatives of $\H^2(U(1),\Z)$. However the functions $g_{\text{IQH}}$ do not directly satisfy the 3-cocycle condition for the group $G=U(1)\leftthreetimes_{\phi} [\Z^2\rtimes \Z_M]$ when $\phi \ne 0$. 

To identify the correct expressions, we follow the reasoning used in previous sections, and look for functions $f_{\text{IQH}}$ such that $df_{\text{IQH}}$ is a cup product of two cocycle representatives of $\H^2(G,\Z)$. Such functions exist, and can be written as
\begin{widetext}
	\begin{equation}
	f_{\text{IQH}}({\bf g}_1,{\bf g}_2,{\bf g}_3) = k_1 \left(z_1(z_2+z_3 + \phi w({\bf  r_2},^{h_2}{\bf  r_3}) -[z_2+z_3+\phi w({\bf  r_2},^{h_2}{\bf  r_3})]) + \phi w({\bf r_1},^{h_1}{\bf r_2}) z_3 + \phi^2 \lambda({\bf g}_1,{\bf g}_2,{\bf g}_3) \right).
	\end{equation}
\end{widetext}

Here the function $\lambda$ satisfies $d\lambda({\bf g}_1,{\bf g}_2,{\bf g}_3,{\bf g}_4) = w({\bf  r_1},^{h_1}{\bf r_2}) w({\bf  r_3},^{h_3}{\bf  r_4})$. The precise form of $\lambda$ depends on the gauge choice for the function $w$, but such $\lambda$ always exists. Now it can be verified directly that 
\begin{align}
&df_{\text{IQH}} ({\bf g}_1,{\bf g}_2,{\bf g}_3,{\bf g}_4) \nonumber \\ &= k_1 (z_1+z_2 + \phi w({\bf r_1},^{h_1}{\bf  r_2}) -[z_1+z_2+\phi w({\bf  r_1},^{h_1}{\bf r_2})])\nonumber \\ &\times(z_3+z_4 + \phi w({\bf r_3},^{h_3}{\bf  r_4}) -[z_3+z_4+\phi w({\bf  r_3},^{h_3}{\bf r_4})]) \nonumber \\
&= 0 \mod 1,
\end{align}
thus it is indeed a product of two cocycle representatives of $\H^2(G,\Z)$. Hence for $k_1 \in \Z$, $f_{\text{IQH}}$ is a $U(1)$-valued 3-cocycle of $G$, which reduces to the correct form $g_{\text{IQH}}$ when restricted to the $U(1)$ subgroup of $G$. In fact, the functions $df_{\text{IQH}}$ represent the $\Z$ factor in $\H^4(G,\Z)$ which is associated to bosonic IQH states. 

Next, we consider the term $ \H^2(G_{\text{space}},\H^1(U(1),U(1))) \cong \H^2(G_{\text{space}},\Z)$. This term classifies mixed SPTs of $U(1)$ and space group symmetry. In the case $\phi = 0$, cocycle representatives of this group are given by a cup product of a cocycle representative of $\H^2(G_{\text{space}},\Z)$, which was described above, and a representative of the class that generates $\H^1(U(1),U(1))$. They can thus be written as $g_{\text{mixed}} \mod 1$, where $g_{\text{mixed}}({\bf g}_1,{\bf g}_2,{\bf g}_3) = z_1 f_2({\bf g}_2,{\bf g}_3)$ for some $f_2({\bf g}_2,{\bf g}_3) \in Z^2(G_{\text{space}},\Z)$. Moreover, they satisfy 
\begin{align}
dg_{\text{mixed}}({\bf g}_1,{\bf g}_2,{\bf g}_3,{\bf g}_4) &= (z_1+z_2-[z_1+z_2])f_2({\bf g}_3,{\bf g}_4) \nonumber \\ &= 0 \mod 1.
\end{align}

In the case $\phi \ne 0$, we use similar reasoning as above to identify functions $f_\text{mixed}$ which satisfy 

\begin{widetext}
\begin{align}
df_\text{mixed}({\bf g}_1,{\bf g}_2,{\bf g}_3,{\bf g}_4) &= ((z_1+z_2 + \phi w({\bf r_1},^{h_1}{\bf  r_2}) -[z_1+z_2+\phi w({\bf  r_1},^{h_1}{\bf r_2})])) f_2({\bf g}_3,{\bf g}_4).
\end{align}
\end{widetext} 

In going from $dg_{\text{mixed}}$ to $df_{\text{mixed}}$, we have replaced the representative cocycle of $\H^2(U(1),\Z)$ by its lift to the group $\H^2(G,\Z)$ after accounting for the extra $\phi$-dependent term, but retained the function $f_2$. The three pieces contributing to $\H^2(G_{\text{space}},\H^1(U(1),U(1)))$ can each be analyzed in this way, and the functions $df_{\text{mixed}}$ are nontrivial elements of $\H^4(G,\Z)$. Using the general expression for $f_2$ we obtained in the previous section, the modified expressions can finally be written together as follows: 
\begin{widetext}
	\begin{align}
	f_\text{mixed}({\bf g}_1,{\bf g}_2,{\bf g}_3) &=  k_6 ( z_1 w({\bf  r_2},^{h_2}{\bf  r_3}) + \phi \lambda({\bf g}_1,{\bf g}_2,{\bf g}_3))\nonumber \\
	&+ \vec{k}_{4} \cdot (z_1+z_2 + \phi w({\bf r_1},^{h_1}{\bf  r_2}) -[z_1+z_2+\phi w({\bf  r_1},^{h_1}{\bf r_2})]) {^{(1-h)^{-1}}} {{\bf r_3}} \nonumber \\
	&+ k_2 (z_1+z_2 + \phi w({\bf r_1},^{h_1}{\bf  r_2}) -[z_1+z_2+\phi w({\bf  r_1},^{h_1}{\bf r_2})]) \frac{h_3}{M} \mod 1
	\end{align}
\end{widetext}

We now discuss each line separately. Define $(z_1+z_2 + \phi w({\bf r_1},^{h_1}{\bf  r_2}) -[z_1+z_2+\phi w({\bf  r_1},^{h_1}{\bf r_2})]) = F_{12}$. The coefficient $k_6$ in the first line can be any integer. If the function on the first line is written as $k_6 f_{3,1}({\bf g}_1,{\bf g}_2,{\bf g}_3)$, then we observe that 
\begin{align}
df_{3,1}({\bf g}_1,{\bf g}_2,{\bf g}_3,{\bf g}_4) &= F_{12} w({\bf r_3},^{h_3}{\bf  r_4}).
\end{align}
This function is integer-valued; in fact, it is a generating element of one of the $\Z$ factors in $\H^4(G,\Z)$. Hence if $k_6$ is also integer-valued, the function $k_6 f_{3,1}({\bf g}_1,{\bf g}_2,{\bf g}_3) \mod 1$ is a 3-cocycle. Moreover, a nonzero value of $k_6$ is always nontrivial, and hence the classification of distinct cocycle representatives is given by choosing $k_6 \in \Z$.

If the function on the second line is written as $\vec{k}_4 \cdot \vec{f}_{3,2}({\bf g}_1,{\bf g}_2,{\bf g}_3)$, then we observe that 

\begin{align}
d\vec{f}_{3,2}({\bf g}_1,{\bf g}_2,{\bf g}_3,{\bf g}_4) &= F_{12}{^{\frac{1-h_3}{1-h}}} {\bf r}_4 .
\end{align}

This function is integer-valued; therefore $\vec{k}_4 \cdot \vec{f}_{3,2}({\bf g}_1,{\bf g}_2,{\bf g}_3) \mod 1$ is a 3-cocycle whenever $\vec{k}_4$ is integer-valued. On the other hand, if $\vec{k}_4$ is an integer multiple of the matrix $(1-h)$, we can see that $\vec{k}_4 \cdot \vec{f}_{3,2}({\bf g}_1,{\bf g}_2,{\bf g}_3) $ is an integer-valued function; hence these values of $k_4$ are trivial. The distinct choices of $\vec{k}_4$ are hence classified by the group $K_M$.

Finally, if the function on the third line is written as $k_2 f_{3,3}({\bf g}_1,{\bf g}_2,{\bf g}_3)$, then we observe that 

\begin{align}
df_{3,3}({\bf g}_1,{\bf g}_2,{\bf g}_3,{\bf g}_4)  &= k_2F_{12}\frac{h_3+h_4-[h_3+h_4]_M}{M} .
\end{align}

For $k_2 \in \Z$, the 3-cocycle condition is satisfied. Moreover, if $k_2$ is a multiple of $M$, the cocycle is in fact equal to the integer-valued function
\begin{equation}
M f_{3,3}({\bf g}_1,{\bf g}_2,{\bf g}_3) =F_{12}  h_3 \equiv 0 \mod 1.
\end{equation}

Thus the cocycles on the third line fall into a $\Z_M$ classification. Since the three terms are independent, they together give a $\Z\times K_M \times \Z_M$ classification. Importantly, we can verify that these cocycles are well-defined and form a group for each term separately up to the addition of coboundaries: this implies that the term $ \H^2(G_{\text{space}},\H^1(U(1),U(1)))$ indeed enters the classification as a separate piece, and not as part of a group extension. 

Finally, we look at the pure space group cocycles which represent elements of the group $H^3(G_{\text{space}},U(1))$. We can copy the results from Ref. \cite{Manjunath2020} to obtain
\begin{widetext}
	\begin{align}
	&f_{\text{space}}({\bf g}_1,{\bf g}_2,{\bf g}_3) = 
	\frac{k_7}{M}  w({\bf  r_1},^{h_1}{\bf r_2}) h_3 \nonumber \\
	&+ \vec{k}_5 \cdot {^{\frac{1-h_1}{1-h}}}{\bf r_2} \frac{h_3}{M} \nonumber \\
	&+ \frac{k_3}{M^2} h_1(h_2+h_3-[h_2+h_3]_M) \mod 1
	\end{align} 
\end{widetext}

These functions can be verified as cocycles in $Z^3(G,U(1))$ without requiring any modification. We can see that choosing $k_7 \in M\Z$, $\vec{k}_5 \in {^{1-h}}\Z^2$, or $k_3 \in M\Z$ implies that the associated cocycle is integer-valued and hence trivial. Thus we obtain $\H^3(G_{\text{space}},U(1)) \cong \Z_M^2\times K_M$. Combining the results for the three terms on the $E_2$-page, we obtain Eq. \eqref{Eq:fullCohResult}.
\section{Derivation of effective actions}
\label{Sec:Eff_Ac}

The symmetry cocycles derived in Appendix \ref{Sec:Coho_calc} are useful not only to construct explicit solutions to the $G$-crossed consistency equations, but also to construct topological field theories involving flat background gauge fields. Since the allowed topological terms are intimately related to the above symmetry fractionalization and SPT cocycles, we will see that their derivation below essentially follows that of the group cocycles explained previously. Since the continuum FQH theory can be studied entirely in terms of the usual vector potential and the $SO(2)$ spin connection \cite{Cho2014,Gromov2014,Gromov2015}, and the continuum translation symmetry does not introduce any qualitatively new phenomena, we will only discuss the derivation of effective actions for lattice FQH systems.

\subsection{$G=U(1)\leftthreetimes_{\phi} \Z^2$}
\label{Sec:Eff_Ac-magtrans}

We will work in the usual simplicial formulation described in Ref. \cite{Manjunath2020}. First assume $\phi=0$, in which case $G = U(1)\times \Z^2$ and hence the $U(1)$ and $\Z^2$ gauge fields are defined independently. In this case we have the following effective action for Abelian topological orders \cite{Manjunath2020}:
\begin{align}
\mathcal{L}_{frac} &= \frac{q_I}{2\pi} a^I \cup dA +  \frac{m_I}{2\pi} a^I \cup A_{XY} \\
\mathcal{L}_{SPT} &= \frac{k_1}{2\pi} A \cup dA +  \frac{k_6}{2\pi} A \cup A_{XY}
\end{align}
Here we have defined the area element $A_{XY}$ in terms of the translation gauge field $\vec{R} = (X,Y)$ as follows: $A_{XY} = \frac{1}{4\pi}(X \cup Y - Y \cup X)$.

We will now treat the general case in which $\phi \ne 0$. A gauge field for magnetic translation symmetry is a pair $(A,\vec{R})$ of gauge fields (corresponding to the $U(1)$ and $\Z^2$ components respectively) obeying the multiplication law
\begin{equation}
(A_{ij},\vec{R}_{ij}) (A_{kl},\vec{R}_{kl}) = (A_{ij}+A_{kl}+\phi w(\vec{R}_{ij},\vec{R}_{kl}),\vec{R}_{ij} + \vec{R}_{kl})
\end{equation}
(see Appendix \ref{Sec:FQHSymm} for the definition of $w$). For a flat gauge field $(A,\vec{R})$, the group multiplication implies that on a 2-simplex $[012]$, $\vec{R}$ remains flat while we have $A_{02} = A_{01} + A_{12} + \phi w(\vec{R}_{01},\vec{R}_{12}) \mod 2\pi$. Therefore it is not $dA$, but the modified flux $F[012] = dA[012] + \phi w(\vec{R}_{01},\vec{R}_{12})$, which is trivial when evaluated on 2-simplices. In symmetric gauge, we in fact have $w(\vec{R}_{01},\vec{R}_{12}) = \frac{\phi}{2} \vec{R}_{01}\times \vec{R}_{12} = \phi A_{XY}[012]$. Therefore we can write $F = dA + \phi A_{XY}$. We will work in the symmetric gauge for the rest of this discussion; however, the same physical results will be obtained in any gauge.

In order to have an action which is retriangulation-invariant, it is necessary to couple the internal gauge fields to the flux $F$ instead of $dA$. For the same reason, we need to modify the SPT action as well. Specifically, we replace $\frac{k_1}{2\pi} A \cup dA$ by a term $\delta\mathcal{L}$ such that $d\delta\mathcal{L} = \frac{k_1}{2\pi} F \cup F$, which is always trivial when $F \in 2\pi\Z$. It can be verified that $\delta\mathcal{L} = \frac{k_1}{2\pi} (A \cup dA + 2\phi A \cup A_{XY} + \phi^2 d^{-1}(A_{XY} \cup A_{XY}))$. The precise form of the term formally written as $ \phi^2 d^{-1}(A_{XY} \cup A_{XY})$ is not important for our discussion, however such a term always exists. 

In terms of the original fields $A$ and $\vec{R}$, the correct effective action for $\phi \ne 0$ is finally
\begin{widetext}
	\begin{align}
	\mathcal{L}_{frac} &= \frac{q_I}{2\pi} a^I \cup F +  \frac{m_I}{2\pi} a^I \cup A_{XY} = \frac{q_I}{2\pi} a^I \cup dA + \frac{m_I + \phi q_I}{2\pi} a^I \cup A_{XY}\\
	\mathcal{L}_{SPT} &= \frac{k_1}{2\pi} A \cup dA + \frac{k_6 + 2\phi k_1}{2\pi} A \cup A_{XY} + \frac{k_1 \phi^2}{2\pi} d^{-1}(A_{XY} \cup A_{XY})
	\end{align}
\end{widetext}

The Hall conductivity $\sigma_H$ appears as the coefficient $\frac{\sigma_H}{2} A \cup dA$ in the response theory obtained by integrating out the gauge fields $a^I$: it equals $\vec{q}^T K^{-1} \vec{q} + 2k_1$, as expected. The filling per unit cell is the coefficient of $A \cup A_{XY}$ in the response theory; it equals $\vec{q}^T K^{-1} (\phi\vec{q} + \vec{m}) + (k_6 + 2\phi k_1)$. In fact, the response theory coefficients obtained in this way satisfy the LSM constraint Eq. \eqref{DiscreteMagTransFillingRelation}, as we prove using $G$-crossed identities in Appendix \ref{Sec:LSM_lattice}. In fact, the above results suggest a stronger result than the one proved in Eq.\eqref{DiscreteMagTransFillingRelation}; however, this stronger result has not as yet been verified in full generality within the $G$-crossed theory (see Appendix \ref{Sec:LSM_lattice}).

\subsection{$G=U(1)\leftthreetimes_{\phi} [\Z^2\rtimes \Z_M]$}
\label{Sec:Eff_Ac-gspace}

In this case the gauge field is given by a triple $B = (A,\vec{R},C)$ where $A \in \R/2\pi\Z, \frac{1}{2\pi}\vec{R} \in \Z^2$, and $C \in \frac{2\pi}{M}\Z$. The three components transform according to the group multiplication law for$U(1)\leftthreetimes_{\phi} [\Z^2\rtimes \Z_M]$. For the special case $\phi=0$ we quote from Ref \cite{Manjunath2020} the following effective action:
\begin{widetext}
	\begin{align} 
	\label{Eff_Action}
	\mathcal{L}_{frac} &= \frac{1}{2\pi} a^I \cup (q_I dA + s_I dC + \vec{t}_I \cdot d \vec{\cancel{R}} + m_I A_{XY} ) \nonumber \\
	\mathcal{L}_{SPT} &= \frac{k_1}{2\pi} A \cup dA + \frac{k_2}{2\pi} A \cup dC + \frac{k_3}{2\pi} C \cup dC 
	+ \frac{1}{2\pi} A \cup (\vec{k}_4 \cdot  d \vec{\cancel{R}})+\frac{1}{2\pi} C \cup (\vec{k}_5 \cdot d \vec{\cancel{R}}) 
	+ \left(\frac{k_6}{2\pi} A + \frac{k_7}{2\pi} C\right) \cup A_{XY}.
	\end{align}
\end{widetext}

When $\phi \ne 0$, we need to add certain terms to $\mathcal{L}$ such that the condition $d\mathcal{L} \in 2\pi\Z$ holds on any 4-simplex. If we consider $\mathcal{L}_{frac}$, this means that the coefficient $m_I$ of the area flux term  $a \cup A_{XY}$ is modified as $m_I \rightarrow m_I + \phi q_I$. The terms in $\mathcal{L}_{SPT}$ also have to be modified. The overall principle is that the appropriate flux which is valued in $2\pi\Z$ for a flat gauge field configuration is no longer $dA$, but $F = dA + \phi A_{XY}$ (assuming symmetric gauge while writing down the group multiplication law of magnetic translations). Therefore a term such as $\frac{k_1}{2\pi} A \cup dA$ is replaced by a term $\mathcal{L}_1$ such that $d\mathcal{L}_1 = \frac{k_1}{2\pi} F \cup F$, which is indeed valued in $2\pi\Z$, implying that $\mathcal{L}_1$ is retriangulation-invariant and hence a valid topological action. Similarly, a term such as $\frac{k_2}{2\pi} dA \cup C$ is replaced by a term $\mathcal{L}_2$ such that $d\mathcal{L}_2 = \frac{k_2}{2\pi} F \cup C$. One can check that we must have $\mathcal{L}_2 = \frac{k_2}{2\pi} (dA \cup C + \phi A_{XY} \cup C)$. (Formally, we are merely replacing 3-cocycle representatives of the group $\H^3(U(1)\times G_{space},U(1))$, corresponding to zero flux, with those from $\H^3(G,U(1))$, corresponding to nonzero flux.) After suitably modifying the necessary terms, we obtain the following action:
\begin{widetext}
	\begin{align}
	\mathcal{L}_{frac} &= \frac{1}{2\pi} a^I \cup (q_I dA + s_I dC + \vec{t}_I \cdot d \vec{\cancel{R}} + (m_I + \phi q_I) A_{XY} ) \nonumber \\
	\mathcal{L}_{SPT} &= \frac{k_1}{2\pi} A \cup dA + \frac{k_2}{2\pi} A \cup dC + \frac{k_3}{2\pi} C \cup dC 
	+ \frac{1}{2\pi} A \cup (\vec{k}_4 \cdot  d \vec{\cancel{R}})+\frac{1}{2\pi} C \cup (\vec{k}_5 \cdot d \vec{\cancel{R}}) 
	+ \left(\frac{k_6 + 2\phi k_1}{2\pi} A + \frac{k_7 + \phi k_2}{2\pi} C\right) \cup A_{XY} \nonumber \\
	&+ \frac{\phi^2 k_1}{2\pi} d^{-1}(A_{XY} \cup A_{XY}) + \frac{1}{2\pi} A_{XY} \cup (\phi\vec{k}_4 \cdot  \vec{\cancel{R}})
	\end{align}
\end{widetext}

It can be verified that these actions give precisely the group cocycles written in Appendix \ref{Sec:Coho_calc_gspace} when integrated on 3-simplices.
\section{Group cohomology theorems and formulas}
\label{Sec:GrpCohIntro}

\subsection{Definition of Tensor, Tor and Ext}
\label{Sec:TenTorExt}
In this section, we will define some mathematical operations that are necessary for the cohomology calculations in this paper. The first two operations $-\otimes-$ and $\mathrm{Tor}(-,-)$ will be used in the context of the Universal Coefficient Theorem discussed below. The last operation $\mathrm{Ext}(-,-)$ is used to classify extensions of Abelian groups. These definitions are taken from Ch.7, Ref. \cite{Sato1999}.
\begin{enumerate}
\item Tensor Product $-\otimes-$\\
The tensor product $\otimes$ of two Abelian groups $G$ and $H$
(specifically, the tensor product over the ring $\mathbb{Z}$) is a binary operation
that gives an Abelian group as output, written as $G\otimes H$. It
satisfies the following properties:
\begin{align*}
G_{1}\otimes G_{2} & \cong G_{2}\otimes G_{1}\\
(\prod_{i}G_{i})\otimes(\prod_{j}H_{j}) & \cong\prod_{i,j}G_{i}\otimes H_{j}\\
G\otimes\mathbb{Z} & \cong\mathbb{Z}\otimes G\cong G\\
\mathbb{Z}_{m}\otimes\mathbb{Z}_{n} & \cong\mathbb{Z}_{d},d=\gcd(m,n)
\end{align*}
Here $\prod$ denotes a direct product of groups, and $\cong$ an
isomorphism of groups. The above properties define the tensor product
for any finitely generated Abelian group. 
\item Torsion Product $\mathrm{Tor}(-,-)$\\
The torsion product of two Abelian groups $G$ and $H$ (specifically,
the degree 1 torsion product over the ring $\mathbb{Z}$), is a binary operation
that gives an Abelian group as output, written as $\mathrm{Tor}(G,H)$.
It satisfies the following properties:
\begin{align*}
\text{Tor}(G,H) & \cong\text{Tor}(H,G)\\
\text{Tor}(\prod_{i}G_{i},\prod_{j}H_{j}) & \cong\prod_{i,j}\text{Tor}(G_{i},H_{j})\\
\text{Tor}(G,\mathbb{Z}) & \cong\text{Tor}(\mathbb{Z},G)\cong\mathbb{Z}_{1}=0\\
\text{Tor}(\mathbb{Z}_{m},\mathbb{Z}_{n}) & \cong\mathbb{Z}_{d},d=\gcd(m,n)
\end{align*}
Once again, these properties define $\mathrm{Tor}(-,-)$ for all finitely
generated Abelian groups. 
\item Ext functor $\mathrm{Ext}(-,-)$\\
The Ext functor $\mathrm{Ext}(-,-)$ takes two Abelian groups $G$ and $H$ as arguments and outputs an Abelian group denoted as $\mathrm{Ext}(G,H)$. It
satisfies the following properties:
\begin{align*}
\text{Ext}(\prod_{i}G_{i},\prod_{j}H_{j}) & \cong\prod_{i,j}\text{Ext}(G_{i},H_{j})\\
\text{Ext}(\mathbb{Z},G) & \cong\mathbb{Z}_{1} \cong \text{Ext}(\mathbb{Z}_m,\mathbb{R})=0\\
\text{Ext}(\mathbb{Z}_{m},\mathbb{Z}) & \cong  \mathbb{Z}_{m}\\
\text{Ext}(\mathbb{Z}_{m},\mathbb{Z}_{n}) & \cong  \mathbb{Z}_{d},d=\gcd(m,n)
\end{align*}
Here $\prod$ denotes a direct product of groups, and $\cong$ an
isomorphism of groups. The above properties completely determine the Abelian group $\text{Ext}(G,H)$
for any two finitely generated Abelian groups, $G$ and $H$. However, we will not call it a product since it is not symmetric if we exchange $G$ and $H$. As the notation suggests, the functor has the useful interpretation of classifying inequivalent group extensions of two Abelian groups $G$ and $H$ described by the following short exact sequence:
\begin{align*}
1\rightarrow H \rightarrow K \rightarrow G \rightarrow 1.
\end{align*}
If $\text{Ext}(G,H)=0$, then the only possible extension is the trivial extension with $K\cong G\times H$.
\end{enumerate}

\subsection{Useful results in group cohomology}

In this section we state several theorems that are useful for the explicit cohomology calculations done in this work. A comprehensive development can be found, for example, in Refs. \cite{Hatcher,Rotman2009}, while a relatively detailed development directed at physicists is available in Ref. \cite{Chen2013}. 
 
In order to perform computations, it is crucial to first specify the particular cohomology theory being used. For finite groups, different cohomology theories generally agree. However, for continuous and/or noncompact groups, such as those being studied in this paper, the choice of cohomology theory strongly influences the obtained classification of topological phases. We will mostly be concerned with two different cohomology theories. The first, denoted as $\H^n(G,M)$, where $M$ is some $G$-module, is cohomology with measurable (or Borel) cochains. It has been conjectured \cite{Chen2013} that this cohomology theory is the right choice for classifying the quantized topological terms that characterize SPT phases. A second cohomology theory, widely used in computations, is the singular cohomology of the classifying space $BG$ of $G$, denoted as $H_{\text{top}}^{n}(BG,M)$.

To perform calculations, it is useful to clarify the relationship between $\H(G)$ and $H_{\text{top}}(BG)$, and also to understand the relationship between cohomology groups with different coefficients, particularly $U(1)$ and $\Z$. To that end, we state the following theorems:   

\begin{theorem}\label{thm:Hborel-HBG}
	\textit{For $M$ discrete and $G$ finite-dimensional, locally compact, $\sigma$-compact, (see page 522 of \cite{Stasheff1978})} 
	\[
	\H^{n}(G,M)\cong H_{{\rm top}}^{n}(BG,M).
	\]
\end{theorem}

\paragraph*{Remark.} 
In particular, we have $\H^{n}(G,\Z)\cong H_{{\rm top}}^{n}(BG,\Z)$ for $n \ge 0$. In various simple cases, we can look up the cohomology of the desired classifying space in the standard mathematical literature and thus obtain $\H^n(G,M)$: see Appendix \ref{Sec:coho_computations} below for examples.

We also have the following important result which relates the measurable cohomology of \textit{compact} Lie groups with $U(1)$ and $\Z$ coefficients:

\begin{theorem}\label{thm:U1vsZcoeffs}
	Let $G$ be a compact Lie group. For $n>0$, we have (see Remark IV.16, part 3 in \cite{wagemann2013cocycle})
	\[
	\H^{n}(G,U(1))\cong \H^{n+1}(G,\Z).
	\]
\end{theorem} 

The next result shows how to compute the cohomology of a cyclic group, under very general assumptions. Let $C_{m}$ be generated by the element $\mathbf{h}$, and assume that its group cohomology is calculated with coefficients in the arbitrary Abelian group $M$. Assume that $G$ has an action
on $M$ given by $\rho:G\rightarrow\text{Aut}(M)$; in our notation,
the result of $\mathbf{g}$ acting on $a\in M$ is denoted as $\mathbf{g} a=\rho_{\mathbf{g}}(a)$.
We define 
\[
M^{G}=\{a\in M|\rho_{\mathbf{g}}(a)=a,\,\forall\mathbf{g}\in G\},
\]
which is the subgroup of $M$ invariant under $G$ action. We
also define the operators $D=\mathbf{h}-1$, and $N=1+\mathbf{h}+\dots+\mathbf{h}^{m-1}$.
With this we define 
\begin{align}
DM & =\{D a|\,a\in M\}\\
NM & =\{N a|\,a\in M\}\\
_{N}M & =\{a\in M|\,N a=0\}.
\end{align}
$DM$ has the interpretation of the image of $D$; $NM$ has the interpretation
of the image of $N$; and $_{N}M$ has the interpretation of the elements
in $M$ annihilated by the operator $N$.

With these definitions, we have the following theorem (see Theorem 9.27 of \cite{Rotman2009}):

\begin{theorem}[Cyclic groups]\label{thm:coho_cyclic}
	
	The cohomology groups of $G=C_{m}$ are, for $n\ge1$, 
	\begin{align}
	\H^{0}(G,M) & \cong M^{G}\nonumber \\
	\H^{2n-1}(G,M) & \cong \,_{N}M/DM\nonumber \\
	\H^{2n}(G,M) & \cong M^{G}/NM\label{eq:CohomologyCyclicGroup}
	\end{align}
	
\end{theorem}

Several examples involving this theorem are worked out in Appendix \ref{Sec:coho_computations}.

Along with results for computing the cohomology of simple classifying spaces, Theorem \ref{thm:coho_cyclic} is the only tool we will use to perform direct computations. The rest of the theorems in this section will give us indirect ways to extend our results to more complicated groups and to more general coefficient modules.

\begin{theorem}[Long-Exact Sequence]
	
	\label{thm: long-exact sequence}
	
	Consider a short exact sequence of three $G$-modules 
	\[
	1\rightarrow M_{1}\rightarrow M_{2}\rightarrow M_{3}\rightarrow 1
	\]
	Then there is a long-exact sequence
	\begin{align}
	1\rightarrow & \H^{0}(G,M_{1})\rightarrow \H^{0}(G,M_{2})\rightarrow \H^{0}(G,M_{3})\rightarrow\nonumber \\
	& \H^{1}(G,M_{1})\rightarrow \H^{1}(G,M_{2})\rightarrow \H^{1}(G,M_{3})\rightarrow\nonumber \\
	& \H^{2}(G,M_{1})\rightarrow \H^{2}(G,M_{2})\rightarrow \H^{2}(G,M_{3})\rightarrow\nonumber \\
	& \cdots\nonumber \\
	& \H^{n}(G,M_{1})\rightarrow \H^{n}(G,M_{2})\rightarrow \H^{n}(G,M_{3})\rightarrow\cdots
	\end{align}
	
\end{theorem}

Although we have stated it for cohomology with measurable cochains, the above long exact sequence is very general and applies to any cohomology theory. A common application of this result is to compare the groups $H^n(G,U(1))$ and $H^{n+1}(G,\Z)$, where $H$ is some cohomology theory, by taking $M_1 = U(1),M_2=\R,M_3=\Z$. 

Once we know the group cohomology with one coefficient group, the group cohomology in another coefficient group is
not arbitrary. In particular, the cohomology with $\Z$ coefficients allows us to obtain the cohomology with arbitrary coefficients. 
\begin{theorem}[Universal Coefficient Theorem]
	
	\label{thm: universalcoefficienttheorem}
	
	For a group $G$, the group cohomology with $\Z$ coefficents
	determines the group cohomology with \textit{discrete} $M$ coefficients as follows, $M$ being a discrete Abelian group with trivial $G$ action (see page 246, Theorem 5.5.10 in Ref. \cite{Spanier} or Theorem 7.6 in Ref. \cite{Sato1999}):
	\[
	\H^{n}(G,M) \cong (\H^{n}(G,\mathbb{Z})\otimes M)\oplus\mathrm{Tor}(\H^{n+1}(G,\mathbb{Z}),M).
	\]
	
\end{theorem}
Our primary use of this theorem will be in computing cohomology with $\A$ coefficients, where $\A$ is a group of Abelian anyons. Note that this theorem holds only when $M$ has the discrete topology, and therefore there may be subtleties in applying the UCT when $M$ is continuous. Also note that there is no general UCT when $G$ acts nontrivially on $M$.

\begin{theorem}[K\"unneth decomposition]\label{thm:kunneth}
	
	When $M$ is a $G_1\times G_2$ module (with a possibly non-trivial action of $G_1 \times G_2$), the cohomology groups of the direct product group $G_{1}\times G_{2}$ can be decomposed as follows:
	
	\[
	\H^{n}(G_{1}\times G_{2},M)\cong\bigoplus_{p=0}^{n}\H^{p}(G_{1};\H^{n-p}(G_{2},M))
	\]
	
\end{theorem}

\paragraph*{Remark.}
This decomposition is often referred to as the K\"unneth formula in the condensed matter physics literature, and also uses the Universal Coefficient Theorem (see eg. Appendix E, Ref. \cite{wen2015SPT} for a derivation). A note on terminology: the K\"unneth formula which is standard in the mathematical literature can be found in Ref. \cite{Chen2013}. Since the version stated above is not referred to as the 'K\"unneth formula' in the mathematical literature, we use the terminology 'K\"unneth decomposition' for this result. Note that in the statement of this theorem,
there is a seemingly unequal treatment of $G_{1}$ and $G_{2}$. However, one can use the Universal Coefficient Theorem a second time to show that if we reverse the roles of $G_{1}$ and $G_{2}$ in the rhs, we will arrive at the same result. For more complicated group extensions, we have to resort to spectral sequences, which we will introduce in Appendix \ref{app_sseq}.

The next result relates to the cohomology of noncompact topological groups, and will be useful in deriving the cohomology (with discrete coefficients) of the group $G = U(1)\leftthreetimes \mathbb{E}^2$ (Appendix \ref{Sec:Coho_calc_cont}). Specifically, on Page 522 of Ref. \cite{Stasheff1978} it is stated that when the group $G$ is a connected noncompact Lie group, and $K$ is a maximal compact subgroup of $G$, then $BG=BK$. Therefore, by using Theorem \ref{thm:Hborel-HBG}, we can state the following result:

\begin{theorem}[Cohomology of noncompact Lie groups]
	
	\label{thm:coho_noncompact}
	
	When the group $G$ is a connected, locally compact, $\sigma$-compact (but not necessarily compact) Lie group, $M$ is discrete, and $K$ is a maximal compact subgroup of $G$, then we have the equalities
	\[
	\H^{n}(G,M)\cong H_{\text{top}}^{n}(BG,M) \cong H_{\text{top}}^n(BK,M) \cong \H^n(K,M).
	\]
	
\end{theorem}

Thus under some mild assumptions on $G$ (which are satisfied for the Euclidean group), the measurable cohomology of $G$ with discrete coefficients is equal to that of its maximal compact subgroup $K$.

\subsection{Explicit computations}
\label{Sec:coho_computations}
\paragraph*{Computations with classifying spaces} The computations below, which explicitly use the classifying space $BG$, can be regarded as an application of Theorem \ref{thm:Hborel-HBG}.
\begin{enumerate}
	\item $\underline{\H^{n}(\mathbb{Z}_{2},\mathbb{Z})}$:\\
	The classifying space $B\mathbb{Z}_{2}=\mathbb{RP}^{\infty}$ is the
	infinite-dimensional projective space, which is a infinite-dimensional
	sphere $S^{\infty}$ with antipodal points being identified. Therefore, using Theorem \ref{thm:Hborel-HBG}, we know that 
	\[
	\H^{n}(\mathbb{Z}_{2},\mathbb{Z})\cong \H^{n}(\mathbb{RP}^{\infty},\mathbb{Z})=\begin{cases}
	\mathbb{Z} & n=0\\
	0 & n\in\mathrm{odd}\\
	\mathbb{Z}_{2} & n\in\mathrm{even}
	\end{cases}
	\]
	\item $\underline{\H^{n}(\mathbb{Z},\mathbb{Z})}$:\\
	The classifying space $B\mathbb{Z}=S^{1}$ is a circle. Therefore,
	\[
	\H^{n}(\mathbb{Z},\mathbb{Z})\cong \H^{n}(S^{1},\mathbb{Z})=\begin{cases}
	\mathbb{Z} & n=0,1\\
	0 & n\geq2.
	\end{cases}
	\]
	\item $\underline{\H^{n}(\mathbb{Z}^{m},\mathbb{Z})}$:\\
	The classifying space satisfies
	$B(G_{1}\times G_{2})=BG_{1}\times BG_{2}$, where the first product
	is the direct product of groups, while the second product is a product
	of topological spaces. Therefore, $B\mathbb{Z}^{m}=\underbrace{S^{1}\times\cdots\times S^{1}}_{m\,\mathrm{times}}=\mathbb{T}^{m}$
	is an $n$-dimensional torus. We thus have, by using the K\"unneth decomposition,
	\begin{equation}
	\H^{n}(\mathbb{Z}^{m},\mathbb{Z})\cong \H^{n}(\mathbb{T}^{m},\mathbb{Z})=\begin{cases}
	\mathbb{Z}^{\binom{m}{n}} & 0\leq n\leq m\\
	0 & n> m
	\end{cases}\label{eq:cohomologyofZ^m_Z}
	\end{equation}
	where $\binom{m}{n}=\frac{m!}{n!(m-n)!}$ .
\end{enumerate}

Now from the Universal Coefficient Theorem (Theorem \eqref{thm: universalcoefficienttheorem})
and Eq. \eqref{eq:cohomologyofZ^m_Z}, we have
\begin{align}
\H^{n}(\mathbb{Z}^{2},M) & \cong(\H^{n}(\mathbb{Z}^{2},\mathbb{Z})\otimes M)\oplus\mathrm{Tor}(\H^{n+1}(\mathbb{Z}^{2},\mathbb{Z}),M) \\
& \cong\begin{cases}
M & n=0\\
M^{2} & n=1\\
M & n=2\\
0 & n>2
\end{cases}. \label{eq:cohomologyofZ^m_M}
\end{align}
In fact, $M$ can be chosen arbitrarily (discrete or continuous) in the above result.
\paragraph*{Computations for cyclic groups} These computations are a direct application of Theorem \ref{thm:coho_cyclic}.
\begin{enumerate}
	\item \underline{Trivial symmetry action}: $G=\Z_m$, and $M=\Z_k$. \\
	Here we assume that $m$ is arbitrary. In this case we have $M^{G}=M$; moreover $D=0$ and $N$ is just
	multiplication by $m$. Therefore $DM=\Z_{1}$. Now $NM$ is the group
	of all multiples of $m$ modulo $k$. This group is generated by $d=(m,k)$;
	there are $k/d$ such numbers forming the group $NM=\Z_{k/d}$. Conversely,
	$Na=0$ if $d a\equiv0\mod k$; there are $k/d$ such numbers
	forming the group $_{N}M=\Z_{d}$. The theorem then gives 
	\begin{align}\label{CohomologyCyclicGrp_trivial}
	\H^{0}(\Z_{m},\Z_{k}) & \cong\Z_{k}\\
	\H^{2n-1}(\Z_{m},\Z_{k}) & =\Z_{d}/\Z_{1}\cong\Z_{d}\\
	\H^{2n}(\Z_{m},\Z_{k}) & =\Z_{k}/\Z_{k/d}\cong\Z_{d}
	\end{align}
	\item \underline{Point group rotation symmetry:} $G=\Z_m$, $M = \Z \times \Z$ and $\rho$ is a point group rotation action.
	\\
	Here we assume that the group $\Z_m$ corresponds to a point group rotation symmetry, implying $m \in \{2,3,4,6\}$. The $2\times 2$ matrix generator of point group rotations is denoted as ${\bf h}$. The final result, derived below for $n \ge 0$, is
	\begin{align}\label{CohomologyCyclicGrp_lattice}
	\H^{2n+1}_{\rho}(\Z_{m},\Z^2) & = \frac{\Z^2}{(1-{\bf h})\Z^2} := K_M\\
	\H^{2n}_{\rho}(\Z_{m},\Z^2) & =0.
	\end{align}
	We obtain $K_2 \cong\Z_2^2, K_3 \cong \Z_3, K_4 \cong \Z_2, K_6 \cong \Z_1$. (We also define $K_1\cong \Z_1$.) 
	
	In this case, each element of $M$ is given by a two-dimensional vector
	of integers, and each element of $G$ is a $2\times2$ matrix. The
	only fixed point of rotations is the identity, $M^{G}=\Z_{1}$. Furthermore,
	we have $0={\bf h}^{m}-1=({\bf h}-1)(1+{\bf h}+\dots+{\bf h}^{m-1})=({\bf h}-1)N$;
	since ${\bf h}\ne1$ we have $N=0$. This immediately means that
	$NM=\Z_{1}$ and $_{N}M=M=\Z^{2}$. The form of $DM=\left\{ (1-{\bf h}) a,a\in\Z^{2}\right\} $
	depends on the value of $m$, which determines the matrix for ${\bf h}$.
	Notice that Eq. \eqref{eq:CohomologyCyclicGroup} already guarantees
	that the cohomology groups of $G$ vanish in even degree, while in
	odd degree they are equal to $\H_{\rho}^{1}(G,M)\cong\Z^{2}/D\Z^{2}$,
	which we now evaluate for different point groups. Note that in the main text we used this result to define $K_M := \H^1_{\rho}(G,M) = \frac{\Z^2}{{^{1-h}}\Z^2}$.
	\begin{enumerate}
		\item \underline{$m = 2$}: ${\bf h}=-I_{2\times2}$ and $D$ is multiplication
		by 2, so $D\Z^{2}=2\Z\times2\Z$ and $\H_{\rho}^{1}(G,M)\cong\Z_{2}\times\Z_{2}$.
		Representatives of this cohomology group are given by elements $(s,t)\in\Z^{2}$
		and are distinguished by the parity of $s$ and $t$. 
		\item \underline{$m = 3$}: A possible representation for ${\bf h}$
		is ${\bf h}=\begin{pmatrix}0 & 1\\
		-1 & -1
		\end{pmatrix}$. Therefore if $a=(s,t)$, $Da=(s-t,s+2t)$.
		Notice that the two components of $Da$ necessarily differ by a multiple
		of 3. Therefore $D\Z^{2}$ consists of all possible pairs $(s,t)$
		which differ by a multiple of 3; from this we get that $\Z^{2}/D\Z^{2}\cong\Z_{3}$,
		and representatives of this group are given by pairs $(s,t)$ distinguished
		by the value of $s-t\mod3$. 
		\item \underline{$m = 4$}: We set ${\bf h}=\begin{pmatrix}0 & 1\\
		-1 & 0
		\end{pmatrix}$, so that if $a=(s,t)$ then $Da=(s-t,s+t)$. Following the argument
		for $m=3$, we see that in this case, $\Z^{2}/D\Z^{2}\cong\Z_{2}$,
		with representatives given by $(s,t)$ and distinguished by $s-t\mod2$. 
		\item \underline{$m = 6$}: In this case, ${\bf h}=\begin{pmatrix}0 & 1\\
		-1 & 1
		\end{pmatrix}$, and for $a=(s,t)$ we have $Da=(s-t,s)$. But this means that $DM\cong\Z^{2}$,
		because there is no parity constraint. Hence $\Z^{2}/D\Z^{2}\cong\Z_{1}$,
		and there are no nontrivial elements in $\H_{\rho}^{1}(G,M)$. \\

	\end{enumerate}
\end{enumerate}

\section{Review of spectral sequences}
\label{app_sseq}
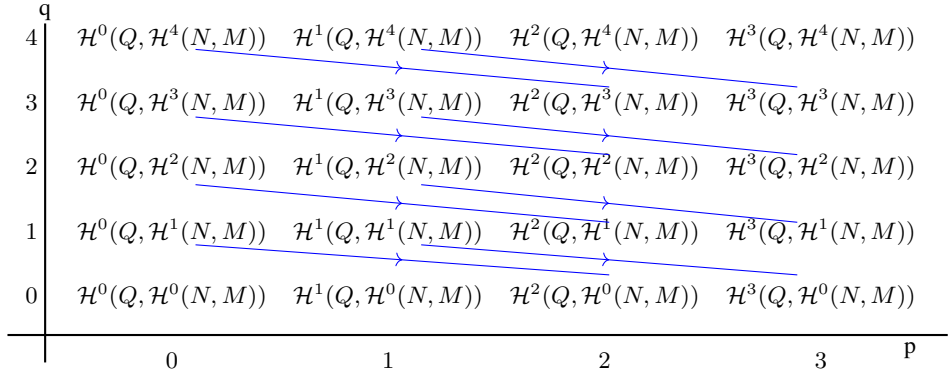
\begin{figure*}[t]
	\centering
	\begin{tikzpicture}
	\matrix (m) [matrix of math nodes,
	nodes in empty cells,nodes={minimum width=5ex,
		minimum height=5ex,outer sep=-5pt},
	column sep=1ex,row sep=1ex]{
		4&\H^0(Q,\H^4(N,M))  &\H^1(Q,\H^4(N,M)) &\H^2(Q,\H^4(N,M))  &\H^3(Q,\H^4(N,M)) \\
		3&\H^0(Q,\H^3(N,M))  &\H^1(Q,\H^3(N,M)) &\H^2(Q,\H^3(N,M))  &\H^3(Q,\H^3(N,M)) \\
		2&\H^0(Q,\H^2(N,M))  &\H^1(Q,\H^2(N,M)) &\H^2(Q,\H^2(N,M))  &\H^3(Q,\H^2(N,M)) \\
		1&\H^0(Q,\H^1(N,M))  &\H^1(Q,\H^1(N,M)) &\H^2(Q,\H^1(N,M))  &\H^3(Q,\H^1(N,M)) \\
		0&\H^0(Q,\H^0(N,M))  &\H^1(Q,\H^0(N,M)) &\H^2(Q,\H^0(N,M))  &\H^3(Q,\H^0(N,M)) \\
		\quad\strut &   0  &  1  &  2  &3  &\strut \\};
	\draw[thick] (m-1-1.north east) -- (m-6-1.east) ;
	\draw[thick] (-6.5,-1.8) -- (6,-1.8) ;
	\node[] at (5.5,-2) {p};
	\node[] at (-6,2.5) {q};
	\draw[blue,->-=.5] (-4,-0.6) -- (1.5,-1);
	\draw[blue,->-=.5] (-1,-0.6) -- (4,-1);
	\draw[blue,->-=.5] (-4,0.2) -- (1.5,-0.3);
	\draw[blue,->-=.5] (-1,0.2) -- (4,-0.3);
	\draw[blue,->-=.5] (-4,1.1) -- (1.5,0.6);
	\draw[blue,->-=.5] (-1,1.1) -- (4,0.6);
	\draw[blue,->-=.5] (-4,2) -- (1.5,1.5);
	\draw[blue,->-=.5] (-1,2) -- (4,1.5);
	\end{tikzpicture}
	\caption{The $E_{2}$ page of the LHSS. Here we only draw the first quadrant: the other three quadrants are trivial. The domain and range of the differentials $d_2$ are indicated through directed arrows. }\label{fig:E2page}
\end{figure*}

Consider a group $G$ defined by the extension 
\begin{align}
1 \rightarrow N \rightarrow G \rightarrow Q \rightarrow 1.
\end{align}
Here, $N$ is a normal subgroup of $G$ and $Q=G/N$. In the simplest example where $G = Q \times N$, we can use the K\"unneth decomposition (Theorem \ref{thm:kunneth}) to determine the cohomology of $G$ in terms of the cohomology of $N$ and $Q$. However, for more general group extensions, we have to resort to another
method to compute its group cohomology. This method is known as the \textit{Lyndon-Hochschild-Serre} spectral sequence (LHSS), and can
be thought of as a generalization of the K\"unneth decomposition. Here we will present an informal treatment directed towards physicists. For a more detailed discussion, the reader is referred to Refs \cite{Mccleary2000,Fuks,ramos2017spectral}.

We wish to compute the group cohomology $\H^{n}(G,M)$ with coefficients in the module $M$. The spectral sequence technique starts with
listing all the cohomology groups $\H^{p}(Q;\H^{q}(N,M))$ for $p,q\geq 0$ as a table. The table thus formed is given in Fig. \ref{fig:E2page}. Formally,
it is called the $E_{2}$-page, and we use the notation $E_{2}^{p,q}=\H^{p}(Q;\H^{q}(N,M))$ to label each entry in the table. We allow the groups $Q$ and $N$ to act nontrivially on $M$; we also allow $Q$ to act nontrivially on $N$. These two actions induce some action of $Q$ on the groups $\H^q(N,M)$; in examples we will explicitly specify this action if it is nontrivial. We also assign the trivial group $0$ to all entries with $p<0$ or $q<0$ and thus extend the page to all possible integral values of $p$ and $q$. Such a construction defines a 'first quadrant cohomological spectral sequence' since the nontrivial entries only appear in the first quadrant. 

Notice that the $E_{2}$ page has some resemblance to the K\"unneth decomposition, since every element of the diagonal $p+q=n$ is a
direct summand of the K\"unneth decomposition for the cohomology in degree $n$.  However, this page only gives a first approximation to the cohomology of $G$. Successively better approximations are obtained by considering maps between the entries called differentials.

The differential $d_{2}^{p,q}$ is a group homomorphism $d_{2}^{p,q}:E_{2}^{p,q}\rightarrow E_{2}^{p+2,q-1}$,
i.e., the domain is the $E_{2}$ element sharing the same label indices, and the target is two steps to the right and one below, as shown in Fig. \ref{fig:E2page}. Note that for the differentials $d_{2}^{p,q}$ whose domain is the trivial group $0$, the differential has to be the zero map since this is the only possible homomorphism. Analogous to differentials (coboundary operators) in a cochain complex, we have $d_{2}^{p+2,q-1}\circ d_{2}^{p,q}=0$
for any two consecutive differentials, which we abbreviate as $d_{2}\circ d_{2}=0$. With this differential property and the $E_{2}$ page in mind, we
can take the cohomology $\ker d_{2}^{p,q}/\text{im }d_{2}^{p-2,q+1}$ and list a table for the result, which we call the $E_{3}$ page, defined as follows:
\begin{equation}
E_{3}^{p,q}=\frac{\ker d_{2}^{p,q}}{\text{im }d_{2}^{p-2,q+1}}.
\end{equation}
Notice that we automatically have $E_{3}^{p,q}=0$ when $p<0$ or $q<0$ by virtue of the zero maps incoming or outgoing from the first quadrant of $E_2^{p,q}$. 

Now the $E_3$-page is equipped with differentials $d_{3}^{p,q}:E_{3}^{p,q}\rightarrow E_{2}^{p+3,q-2}$. These maps in turn
satisfy $d_{3}^{p+3,q-2}\circ d_{3}^{p,q}=0$, or succinctly, $d_{3}\circ d_{3}=0$. With the $d_{3}$ differentials,
we can obtain the $E_{4}$ page by defining $E_{4}^{p,q}=\ker d_{3}^{p,q}/\text{im }d_{3}^{p-3,q+2}$.
We can iterate this procedure to obtain the $E_{r+1}$ page from the $E_{r}$
page, for $r \ge 2$, by taking 
\begin{equation}
E_{r+1}^{p,q}=\frac{\ker d_{r}^{p,q}}{\text{im }d_{r}^{p-r,q+r-1}},
\end{equation}
from the differentials $d_{r}^{p,q}$ equipped with the property $d_{r}^{p+r,q-r+1}\circ d_{r}^{p,q}=0$.

For any given $(p,q)$, we have $E_{r}^{p,q}=E_{r+1}^{p,q}=E_{r+2}^{p,q}=\cdots$
when $r>\max\{p,q+1\}$ since the incoming and outgoing differentials
are zero maps. We denote this stable value as $E_{\infty}^{p,q}$
and collectively refer to these values for different $(p,q)$ as the $E_{\infty}$
page. If it happens that starting at $r= s$, all the
differentials are zeros for any page $r \ge s$, then we say the spectral
sequence collapses or stabilizes on the the $E_{s}$ page, and $E_s \equiv E_{\infty}$.

The $E_{\infty}$ page forms an approximation of $\H^{n}(G,M)$ in the following sense. There is a finite decreasing filtration of $\H^{n}(G,M)$ by Abelian groups: 
\begin{equation}
1=A^{n}\subseteq A^{n-1}\subseteq A^{n-2}\subseteq\dots A^{1}\subseteq A^{0}=\H^{n}(G,M)
\end{equation}
along with a set of short exact sequences for the $A^{i}$: 
\begin{align}
 & 1\rightarrow A^{1}\rightarrow \H^{n}(G,M)\rightarrow E_{\infty}^{0,n}\rightarrow 1\nonumber \\
 & 1\rightarrow A^{2}\rightarrow A^{1}\rightarrow E_{\infty}^{1,n-1}\rightarrow 1\nonumber \\
 & \dots\nonumber \\
 & 1\rightarrow A^{n-1}\rightarrow A^{n-2}\rightarrow E_{\infty}^{n-1,1}\rightarrow 1\nonumber \\
 & 1=A^{n}\rightarrow A^{n-1}\rightarrow E_{\infty}^{n,0}\rightarrow 1.\label{eq:extension}
\end{align}

\begin{figure}
\begin{centering}
\includegraphics[scale=0.3]{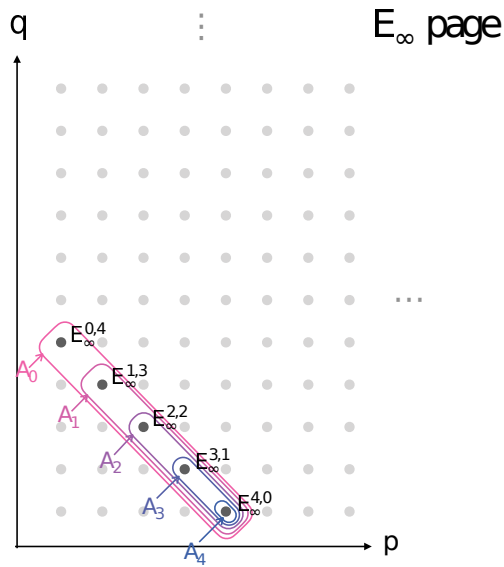}
\par\end{centering}
\centering{}\caption{We can use the $E_{\infty}$ page to retrieve the information of $\H^{4}(G,M)$.
The group $A_{4}$ is $E_{\infty}^{4,0}$. The group $A_{3}$ is $E_{\infty}^{3,1}$
extended by $A_{4}$. The group $A_{2}$ is $E_{\infty}^{2,2}$ extended
by $A_{3}$. The group $A_{1}$ is $E_{\infty}^{1,3}$ extended by
$A_{2}$. Finally, the group $A_{0}=\H^{4}(G,M)$ is $E_{\infty}^{0,4}$
extended by $A_{1}$. \label{fig:Einftyextension}}
\end{figure}

In other words, for each $p$, the group $E_{\infty}^{p,n-p} \cong A^{p}/A^{p-1}$ is a quotient group obtained from
the normal series $A_1,A_2,\dots , A_n$, and $\H^n(G,M)$ is the result of carrying out the indicated sequence of group extensions. In principle, if we are able to reconstruct $A^{n-2}$
from $A^{n-1}=E_{\infty}^{n,0}$ and $E_{\infty}^{n-1,1}$, and then
reconstruct $A^{n-3}$, $A^{n-4}$$\cdots$ , all the way to $A^{0}=\H^{n}(G,M)$,
then we have solved our problem. This procedure is illustrated in Fig.
\ref{fig:Einftyextension}.

We use the notation $E_{2}^{p,q} \implies \H^{p+q}(G,M)$ to denote the fact that $\H^{p+q}(G, M)$ is obtained from $E_2^{p,q}$ as described above. We can now summarize the above series of steps as a single theorem:

\begin{theorem}[Lyndon-Hochschild-Serre spectral sequence]\label{thm:lhss}
	 Consider a group extension 
\[
1\rightarrow N\rightarrow G\rightarrow Q\rightarrow1
\]
and let $M$ be a $G$-module. Then there is a first quadrant cohomological
spectral sequence 
\begin{equation}
E_{2}^{p,q}=\H^{p}(Q;\H^{q}(N,M))\implies \H^{p+q}(G,M).
\end{equation}
\end{theorem}

The coefficients $M$ can be arbitrary, and $Q$ may have a nontrivial action on $N$. Both $Q$ and $N$ may also have a nontrivial action on $M$. 

Let us consider a simple example, $G=N\times Q$, for which we already know that the K\"unneth decomposition should apply. First, the $E_{2}$ page has precisely the elements occurring as direct summands in the K\"unneth decomposition. If there were any nonzero differential on any page $r\geq2$, the size of the Abelian group $\H^{n}(G,M)$ will be smaller than that of the group obtained by using the K\"unneth decomposition. Therefore, all the differentials are trivial. That is, the spectral sequence collapses on the $E_{2}$ page. In addition, all the short exact sequences in Eq. \eqref{eq:extension}
should split, so that $A_{0}=\H^{n}(G,M)$ will be a simple direct sum of the terms $E_{\infty}^{p,n-p}=E_{2}^{p,n-p}=\H^{p}(Q,\H^{n-p}(N,M))$, for $0 \le p \le n$.

Note that for a group extension which is not a direct product, the differentials $d_{2}$ can be nontrivial. Taking the cohomology of $d_{2}$ and passing to the $E_3$-page will in general reduce the size of the group corresponding to each entry of the $E_{2}$ page. In this sense, the K\"unneth decomposition gives an upper bound on the size of $\H^{n}(G,M)$. The effect of higher differentials is similar to that of $d_2$: they allow us to eliminate more terms which do not ultimately contribute to $\H^n (G, M )$. Thus subsequent pages yield successively better approximations of $\H^n(G, M )$.

Using the LHSS in practice requires determining the precise form of the differentials, which is a non-trivial problem. In addition, the extensions in Eq. \ref{eq:extension} do not have to split, and generally we do not know the explicit maps in the short exact sequences: this is referred to as the extension problem in the spectral sequence literature. 

In certain cases there may be some helpful simplications. For example, if $E_{\infty}^{k,n-k}=\Z$, then $A_{k}=\Z\oplus A_{k+1}$, because $\Z$ is a free module: it can only be extended by $A_{k+1}$ as a direct sum. Also, a number of the $E_{\infty}$ terms may vanish. 

\bibliography{library,TI}

\end{document}